# Geometric Flavours of Quantum Field Theory on a Cauchy Hypersurface:

## Gaussian Analysis for the Hamiltonian Formalism and Applications to Cosmology

**David Martínez Crespo**

2024

Geometric Flavours of Quantum Field Theory

David Martínez Crespo

Director:
**Jesús Clemente Gallardo**

September 2024

**Universidad** Zaragoza

# Geometric Flavours of Quantum Field Theory on a Cauchy Hypersurface:

## Gaussian Analysis for the Hamiltonian Formalism and Applications to Cosmology

Ph.D. Thesis

Theoretical Physics

David Martínez Crespo

A mi madre, a mi padre, a mi hermana
y a mi familia.

# Contents

# List of Symbols and conventions

# Acronyms and Abbreviations

# Abstract

This thesis explores Quantum Field Theory (QFT) on curved spacetimes using a geometric Hamiltonian approach. In particular it studies the theory of the scalar field described through its configurations over a Cauchy hypersurface. It is focused on the description of a mathematically consistent framework based on analytic and geometric tools. As a result, it introduces corrections to the Schrödinger equation and explores the physical implications of these corrections, on the hybrid quantum-classical description of QFT and classical General Relativity (GR), and applications to cosmology.

The mathematical aspects of Gaussian integration theory in infinite-dimensional Topological Vector Spaces (TVS) are thoroughly reviewed. The main focus is on developing an efficient presentation and notation serving the needs of its application to the QFT on curved backgrounds. For that matter, it describes these tools over sets of functions of a Cauchy hypersurface $\Sigma$. It also reviews the complex and holomorphic versions of important results and concepts of Gaussian integration. For example, the Wiener-Itô decomposition theorem or the definition of Hida test functions.

The physical framework builds upon three interconnected levels: classical General Relativity (GR), Classical Statistical Field Theory (CSFT), and QFT. The work begins by extending the Koopman-van Hove (KvH) formalism of classical statistical mechanics to CSFT. This description is based upon prequantization theory. It reveals features inherent to both CSFT and QFT, that help delineate the genuine quantum features of a theory. In particular, the introduction of a Kähler structure in the manifold of fields, the need for a point-splitting regularization procedure, and the technique of Feynman diagrams are identified as ingredients already present in CSFT.

Upon the prequantum program, the QFT of the scalar field is built. At this stage, the kinematical aspects of the theory are discussed, mixing Geometric Quantization with the choice of Wick and Weyl orderings. In this quantum level, various quantum representations are introduced: the holomorphic, Schrödinger, field-momentum, and antiholomorphic. The relation among them is studied using integral transforms, including novel infinite-dimensional Fourier transforms.

The quantization is completed by studying the dynamics of the theory. In order to achieve a coherent kinematical description, representations obtained from geometric quantization are geometrized. This is done by identifying a Kähler structure in the manifold of pure states of the QFT. This structure depends on the ingredients of the lower levels of the construction. In turn, the structure acquires a parametric dependence in time. It is argued that this structure should be preserved by the evolution. For that matter, a covariant time derivative that modifies the evolution equations is considered.

In order to fully characterize this modification to the evolution, the covariant derivative is chosen by studying the coupling in a hybrid model of the QFT of the scalar field coupled with classical GR. In this case, the geometrodynamical description of GR selects a unique connection.

Finally, studying the dynamics of the field on fixed Friedman-Lemaître-Robertson-Walker (FLRW) spacetimes provides new phenomenological results. A novel feature of the formalism is the appearance of particle creation effects on a dynamical equation.

# Resumen

Esta tesis explora la Teoría Cuántica de Campos (QFT) en espaciotiempos curvados utilizando un enfoque hamiltoniano geométrico. Se centra en la teoría del campo escalar descrita a través de sus configuraciones sobre una hipersuperficie de Cauchy. Se centra en la descripción de un marco matemáticamente consistente basado en herramientas de análisis y geometría. Como resultado, introduce correcciones a la ecuación de Schrödinger y explora las implicaciones físicas de estas correcciones, en la descripción híbrida cuántico-clásica de la QFT y la Relatividad General (GR) clásica, y aplicaciones a la cosmología.

Se revisan exhaustivamente los aspectos matemáticos de la teoría de la integración gaussiana en Espacios Vectoriales Topológicos (TVS) de dimensión infinita. El enfoque principal es desarrollar una presentación y notación eficientes que se adapten a las necesidades de su aplicación a la QFT en *backgrounds* curvos. Para ello, describe estas herramientas sobre conjuntos de funciones de una hipersuperficie de Cauchy $\Sigma$. También se revisan las versiones complejas y holomorfas de resultados y conceptos importantes de la integración gaussiana. Por ejemplo, el teorema de descomposición de Wiener-Itô o la definición de funciones de *test* de Hida.

La descripción física del sistema se construye sobre tres niveles interconectados: la Relatividad General (GR) clásica, la Teoría Estadística de Campos Clásica (CSFT) y la QFT. El trabajo comienza extendiendo el formalismo de Koopman-van Hove (KvH) de la mecánica estadística clásica a la CSFT. Esta descripción se basa en la teoría de pre-cuantización. Revela características inherentes tanto a la CSFT como a la QFT, que ayudan a delinear las características genuinamente cuánticas de una teoría. En particular, la introducción de una estructura de Kähler en la variedad de campos, la necesidad de un procedimiento de regularización de *point splitting* y la técnica de diagramas de Feynman se identifican como ingredientes ya presentes en la CSFT.

Sobre el formlismo pre-cuántico, se construye la QFT del campo escalar. En esta etapa, se discuten los aspectos cinemáticos de la teoría, mezclando la Cuantización Geométrica con la elección de ordenamientos de Wick y Weyl. En este nivel cuántico, se introducen varias representaciones cuánticas: la holomorfa, de Schrödinger, de campo-momento y antiholomorfa. La relación entre ellas se estudia utilizando transformadas integrales, incluyendo nuevas transformadas de Fourier de dimensión infinita.

La cuantización se completa estudiando la dinámica de la teoría. Para lograr una descripción cinemática coherente, las representaciones obtenidas de la cuantización geométrica se geometrizan. Esto se hace identificando una estructura de Kähler en la variedad de estados puros de la QFT. Esta estructura depende de los ingredientes de los niveles inferiores de la construcción. Como consecuencia, la estructura adquiere una dependencia paramétrica en el tiempo. Se argumenta que esta estructura debe ser preservada por la evolución. Para ello, se considera una derivada temporal covariante que modifica las ecuaciones de movimiento.

Para caracterizar completamente esta modificación a la evolución, la derivada covariante se elige estudiando el acoplamiento en un modelo híbrido de la QFT del campo escalar acoplado con la GR clásica. En este caso, la descripción geometrodinámica de la GR selecciona una conexión única.

Finalmente, el estudio de la dinámica del campo en espaciotiempos fijos de Friedman-Lemaître-Robertson-Walker (FLRW) ofrece nuevos resultados fenomenológicos. Una característica novedosa del formalismo es la derivación de una ecuación que presenta efectos de creación de partículas en universos en expansión.

## Agradecimientos/Acknowledgements

Esta tesis es el fruto de un camino que comenzó hace cuatro años, pero cuyas raíces se extienden a lo largo de una década. Durante todo este tiempo, he compartido el trayecto con muchos amigos y compañeros, sin cuya ayuda esta tesis no habría sido posible.

En primer lugar, quiero agradecer a mi director, Jesús Clemente, por enseñarme a saborear la física con un sabor geométrico. Por concebir este proyecto de tesis y darme la confianza para llevarlo a cabo con total libertad. Sin su guía durante estos años esta tesis no habría llegado a buen puerto.

Extiendo este agradecimiento a Carlos Bouthelier, con quien he trabajado codo con codo durante estos años. Su apoyo me permitió centrarme en desgranar los aspectos matemáticos mientras él se ocupaba de la física. También agradezco a José Luis Alonso por identificar las preguntas clave que esta tesis debía responder y proporcionarme las referencias centrales para su desarrollo.

Special thanks to Cesare Tronci for hosting me in Guildford and further nurturing my geometric perspective, opening new avenues for expanding my knowledge. To our work together I owe the perspective written in Chapter 3 of this thesis. I also extend this gratitude to Werner Bauer and Paul Bergold.

No quiero olvidar a los compañeros que han compartido estos años conmigo. A Sergio Faci, por compartir tantas horas de trabajo con Carlos y conmigo mientras cocinábamos a fuego lento. A Laura González, por unirse a nuestro grupo durante el tiempo que le fue posible. A David Díez, por embarcarse conmigo en el proyecto de fundar una asociación. A Mathieu Kaltschmidt, por estar siempre dispuesto a compartir un pincho de tortilla. Y a muchos otros que se han ido sumando a las pausas del café durante estos años.

A mis amigos de la carrera de Física, Jorge, Pilar y Enrique, con quienes he compartido innumerables horas, tanto dentro como fuera del laboratorio. Me han dado cobijo en sus casas, hemos bajado montañas rusas y palas de nieve polvo dura, y nuevas historias que aparecen cada vez que nos volvemos a juntar. También quiero agradecer a mis amigos de la carrera de Matemáticas, Pablo, Violeta, Celia, Inés, Sara, Noelia, Lorién y Alberto. Gracias a todos ellos logré terminar la carrera, navegando entre los solapes de horarios con la tranquilidad de tener una red de apoyo bajo mis pies. Extiendo este agradecimiento a mis compañeros del máster, y en especial a David Alonso, siempre dispuesto a reunirnos de vez en cuando con un poco de lambrusco y a hablar de la vida y todo lo demás.

Una constante durante mis años en Zaragoza ha sido mi grupo de amigos, que me ha permitido disfrutar de esta ciudad al máximo. Quiero agradecer a Román, con quien hace once años me fui a vivir y aún seguimos viéndonos casi cada semana. A Enrique, que comparte conmigo el viaje de haber dejado Soria para convertirnos en doctores, y que siempre está disponible para cualquier plan de última hora. Más recientemente, a Elena, por añadirle su toque artístico a esta ciudad (y a esta tesis), y a Inés, por compartir esos pinchos de tortilla en sus pausas de la biblioteca.

A todas las personas que me han acompañado en estos años. Alicia, Violeta y Antonio por los vermús y momentos que hemos pasado a lo largo de esta tesis. A los sorianos Javier y los gatetes, Raúl, Nacho, María, Miguel, Andrés, Álvaro, Adrián... que siempre están, envasando el tiempo al vacío. A Misi, Blanquita y Musu que son responsables de algunas de las erratas de esta tesis.

Agradezco profundamente a mi familia por el apoyo que siempre me han brindado para seguir adelante, tanto en mis estudios como en cualquier otro proyecto en el que me haya embarcado. A mi madre Elisabel y mi padre Jose Antonio por toneladas de cariño, comprensión y buenos momentos junto con mi hermana, Sara, que con su doctorado me enseñó que poner algo en negro sobre blanco requiere contar una historia. A ellos les dedico esta tesis, pues son el pilar de mi vida. Extiendo este agradecimiento al resto de mi familia: a mi abuela Antonia y a mis tías, Reme y Charo, por abrirme siempre la puerta de sus casas con una sonrisa, a mis tíos Teodoro y Paul y a mis primos María, Julia, Pablo y Óscar. Y a los que ya no están que trabajaron toda su vida para que sus nietos pudieran ir a la universidad.

Finalmente, quiero detenerme a agradecer a Violeta todo el tiempo y la vida que pasamos juntos. Has sido mi apoyo, mi punto fijo y mi motor hacia adelante durante todos estos años. Eres una fuente interminable de risas y momentos bonitos. Una de las cosas más importantes que he aprendido durante esta tesis es que estar en casa significa estar contigo.

A todos, les doy mi más sincero agradecimiento por acompañarme en este viaje.

# Prologue

This manuscript has been written as a Ph.D. thesis in Physics. It was developed in the Department of Theoretical Physics at the Universidad de Zaragoza under the supervision of Jesús Clemente-Gallardo and in collaboration with Carlos Bouthelier-Madre and José Luis Alonso.

Large parts of the content of this thesis have been published in :

This thesis is part of a larger project on Quantum-Classical hybrid systems. In the context of this project, during the course of the thesis, the author collaborated on the following publications:

The author acknowledges financial support by Gobierno de Aragón through the grants ORDEN CUS/581/2020 and partial financial support Grant PID2021-123251NB-I00 funded by MCIN/AEI/10.13039/501100011033 and by the European Union, and of Grant E48-23R funded by Government of Aragon.

The elevator picture was designed by Elena Martínez Santiago.

# Chapter 1

# Introduction

## 1.1 Historic preliminaries

The mathematical description of Quantum Field Theory (QFT) on an arbitrary curved spacetime background has always been a complicated task. During the past century, most efforts focused on the algebraic side [HK64, DHR69a, DHR69b, DHR71, DHR74, BF82, SW89]. That approach proved to be very successful in encoding the properties of quantum operators and states in a mathematically consistent framework, leading to the state-of-the-art fields of constructive and axiomatic QFT based on the mathematical tools of microlocal analysis [GJ87, BSZ92, BF09, Sum12, Bru&15, BF20]. Nonetheless, this algebraic approach sometimes lacks the insight that differential geometry provides. These kinds of geometric techniques were not described in detail either in the other main mathematical approach to QFT, non-commutative geometry [Con95, CK00], which is more focused, again, on algebraic tools. In particular, the geometrical description of the space of fields justifying, for instance, the existence of a Poisson bracket or a symplectic form, relies on the definition of the suitable differentiable structure on the fields' phase space, which, in this case, being an infinite-dimensional manifold, is cumbersome to build.

Nonetheless, differential geometric arguments have always played a central role in Feynman path integral methods of QFT [Wei96, Nai05, Mas08] and quantum gravity [Rov04, Kie12]. In order to reconcile the geometric point of view with the subject of QFT, the main problem to address is the description, in a coherent mathematical framework, of the Feynman path integral. Even though this problem has been a well-known subject of study for many years [GJ87, Car&97, DCF97, Wes03, CD06], it is still unclear how to proceed for the physically relevant cases.

More recently, the description of the Schrödinger and holomorphic pictures of QFT in curved spacetimes has been studied with different algebraic or analytical tools [LS96, LS98, CCQ04, Oec12c, Oec12b, HS15, HS17, HSU19, EHS21, KM23]. One of the common features of each of these approaches is the breaking of covariance by introducing a time parameter in the spacetime to engage in a Hamiltonian description of QFT. These approaches also recur to path integral methods to some extent; nevertheless, it turns out that functional integration in this setting is cured of some of the problems that covariant path integrals possess [Oec12c]. This fact sets one of the first objectives of this project. In this thesis, we will address the subject of QFT from a geometric perspective. Therefore, our first task is to re-obtain those results of the Hamiltonian representations of QFT with a geometric flavour.

Hamiltonian mechanics of classical field theories are rooted in solid mathematical grounds. Some examples of these descriptions comprise the canonical Hamiltonian formalism, which can be described in terms of dual vector bundles [Got&04]. The multisymplectic formalism, defined over jet bundles and their duals [GIM04, Rom09, GR18]. Applications to continuum and fluid mechanics that include Hamiltonian Lie-Poisson reduction [Hol07, MR13] and Lyapunov stability [Hol&85]. And many others that can be found in the literature. The geometrical richness of these methods sparked the program of Geometric Quantization. This is a Hamiltonian quantization project based on symplectic geometry. It is devoted to understanding the bridge between the well-established Hamiltonian methods of classical mechanics and quantum theory [Woo97, Mos20]. In addition, geometric prequantization, which is an intermediate step of the theory, also offers a new perspective on classical statistical mechanics; the so-called Koopman-van Hove (KvH) theory brings the classical theory to the language of self-adjoint operators over Hilbert spaces resembling, in the closest possible way, the Schrödinger representation of quantum mechanics [GT20, TJ21, GT22a, GT22b]. Even though the scope of these programs was, in principle, limited to the finite-dimensional case, from the correct definition of the functional integrals straightforwardly follows its generalization to QFT [Oec12c, Oec12b, Oec12a] and Classical Statistical Field Theory (CSFT) as we will see in this thesis.

Regarding the mathematical foundations of functional integration, our construction is largely built upon the results of Gaussian analysis [Hu16] and Hida-Malliavin calculus [Hid80, Hid80, Hid&93, Oba94, Kuo96, Kon&96, Wes03, HS08] developed during the last two decades of the 20th century to study stochastic variational systems and applied with huge success to financial mathematics [NØP09, Hen09]. The relation between the tools of Malliavin calculus and operators in QFT has already been noticed within the community of stochastic calculus [Hen09]. Also, the close relation of Gaussian analysis and Hida-Malliavin calculus with Feynman integrals has been studied in detail [Hid&93, Wes03]. From the physics point of view, the necessity of Gaussian analysis has also been hinted at as a necessary tool to study the Schrödinger picture of QFT [CCQ04], the holomorphic picture [Oec12b], and their relation through integral transforms [Oec12c]. Moreover, Gaussian analysis is needed to describe star products in deformation quantization [Dit90], and in algebraic QFT, quasifree states, also called Gaussian states, are considered the class of states that reproduce the desirable physical properties of a theory [Bru&15]. Furthermore, the relation among QFT and Gaussian analysis is even more evident when we describe Wick order, which appears as an ingredient in both fields [Nai05, Hid&93], or the Wiener-Ito decomposition theorem that relates the Gaussian Hilbert space with the Bosonic Fock space and will be a central ingredient of this thesis.

Despite the plethora of examples presented above, to the best of our knowledge, the powerful and efficient formalism of Gaussian analysis developed for stochastic calculus has not been fully utilized to describe the Hamiltonian pictures of QFT.

Moreover, this mathematical foundation allows us to take this analysis one step further and study, in a systematic way, the dynamics of the theory. This is because the Hamiltonian dynamics of QFT possesses a well-known description in terms of Poisson geometry. This was introduced by Kibble (and, later, its generalization by Ashtekar *et al.*) and is called geometrical quantum mechanics [Kib79, AS99]. In this program, the Schrödinger equation is seen as a Hamiltonian system with respect to a canonical Poisson structure in the space of pure states described as a manifold.

As a final remark, we point out that the original proposal to embark on this journey was the aim to provide a mathematical framework for a hybrid quantum-classical system. Based on Poisson geometry, the project on hybrid quantum-classical molecular systems at the University of Zaragoza [Alo&11, Alo&12a, Alo&12b, Alo&15, Alo&18, Alo&20, Alo&23, Bou&23] and others [BGT19, TJ21, GT22a, GT22b, GT23, MRT23] describe finite-dimensional systems for both the classical and quantum parts. In our case, we want to deal with the infinite-dimensional systems that describe General Relativity (GR) in the classical sector and QFT in the quantum part, coupled in a dynamical way with *backreaction* from the quantum degrees of freedom to the classical ones. Using the tools developed in this thesis, and the description of Hamiltonian GR known as geometrodynamics [HKT76, Kuc74, Rom93], we have been able to successfully address this problem in [Alo&24a]. Nonetheless, this problem will only be briefly covered here. This is because the case in which gravity is treated as a background is already an interesting test bench for the mathematical analysis.

## 1.2 Geometric flavours of Hamiltonian field theories

This thesis is devoted to the study of Quantum Field Theory (QFT) on curved space-times in geometric terms. We opt for a non-covariant approach and describe our theory with the Hamiltonian approach flavoured with symplectic geometry in infinite-dimensional manifolds. Continuing along the lines written above, the aim of this work is to blend together both the mathematical tools of Gaussian analysis and their physical interpretation in QFT. As a result, we will provide new insightful perspectives about tools already known in the literature of the mathematical side, but also present new mathematical tools rooted in this relation. This relation will not be merely interpretative. The mathematical considerations will also impact the physics and the phenomenology of the systems we are considering. In this thesis, we will propose observable contributions to the dynamics of QFT we are describing that, to our knowledge, were not considered in other works.

A word of caution is to be given; even though our description will be rooted in geometric grounds, the subject of infinite-dimensional geometry, also named global analysis, is vast and technical [KM97, DGV16]. For this reason, we do not pretend to present our description in the most rigorous way possible. Instead, we say that we provide a description with a geometric flavour.

The dynamical system we are considering comprises many ingredients intertwined in non-trivial ways. In order to ease the exploration of QFT on curved spacetimes, we must set some guidelines. First, we will study only Hamiltonian descriptions of the mechanical systems in terms of Poisson or symplectic manifolds. This procedure breaks the implicit covariance of Lagrangian formalisms but eases some interpretative aspects with major conceptual implications. For instance, Hamiltonian breaking of space and time introduces a clear separation of the aspects of the theory that we understand as *kinematics* and those that we classify as *dynamics*. These are

- **(Kinematics of a Hamiltonian system)** The kinematical aspects of a theory refer to the geometric description of the manifolds and extra structures needed to describe all possible motions of a physical system. For Hamiltonian systems, the kinematics of the theory is comprised of a Poisson manifold $(\mathcal{M}, \{\cdot, \cdot\}_{\mathcal{M}})$, named the phase space of the theory. This manifold is endowed with a Poisson bracket $\{\cdot, \cdot\}_{\mathcal{M}}$. Often, we will consider a symplectic manifold $(\mathcal{M}, \omega_{\mathcal{M}})$ endowed with a symplectic form $\omega_{\mathcal{M}}$, which is a non-degenerate closed two-form and induces a well-defined Poisson bracket.

  The kinematical description also requires another ingredient, the algebra of observables. This algebra is a subset of $\mathcal{O} \subseteq C^{\infty}(\mathcal{M})$ closed under the Poisson bracket $\{\cdot, \cdot\}_{\mathcal{M}}$. In turn, this description provides the space of states of the theory as the dual to the set of observables $\rho \in \mathcal{O}'$ such that $\rho : \mathcal{O} \to \mathbb{R}$.

- **(Dynamics of a Hamiltonian system)** The dynamical aspects of a theory comprise the particular motions that a particular system undergoes along a time parameter $t$. For Hamiltonian systems, the dynamics is dictated by a single ingredient, the Hamiltonian $H \in \mathcal{O}$. The motion of the system flows along the integral curves of a vector field $X_H$ generated from $H$ with the Poisson bracket $X_H = \{\cdot, H\}_{\mathcal{M}}$. Using this, we can express the dynamics over the space of observables through Hamilton's equation; for $f \in \mathcal{O}$, we get $\partial_t f = \{f, H\}_{\mathcal{M}}$.

The covariant approaches consider either full solutions of the equations of motion or Lagrangians defined over the whole spacetime. As such, these approaches are blind to this distinction. Moreover, in most cases, such as General Relativity (GR) and Yang-Mills non-Abelian gauge theories, the Hamiltonian approach should include constraints to reduce the phase space to submanifolds of the phase space initially considered. In these cases, the covariant approach is more natural to describe the theory. In fact, the most difficult task we will engage in will be to carefully and thoroughly describe the kinematical ingredients of QFT on a curved spacetime separated from the dynamical ones.

### 1.2.1 Kinematical aspects: Geometry in three floors

The conceptual structure of the kinematic description of this thesis resembles a three-storey building in Figure 1.1. Each floor represents a theory:

**[0]** The ground floor represents General Relativity (GR). This is the theory that will govern the geometric content of the spacetime itself. This geometry will enter either as a fixed background, as a metric solution of the field equations of GR of a predefined model, or as a dynamical entity governed by the full coupling with matter fields.

**[1]** The first floor is comprised of Classical Statistical Field Theory (CSFT). This is the statistical theory of a matter field built on top of the spacetime background. We will explore its description in the Koopman-van Hove (KvH) formalism.

**[2]** The second floor is Quantum Field Theory (QFT). In this work, a quantum theory is defined from a classical one through a quantization procedure. Thus, we consider that QFT is built on top of CSFT by means of the Geometric quantization program and the treatment of ordering problems.

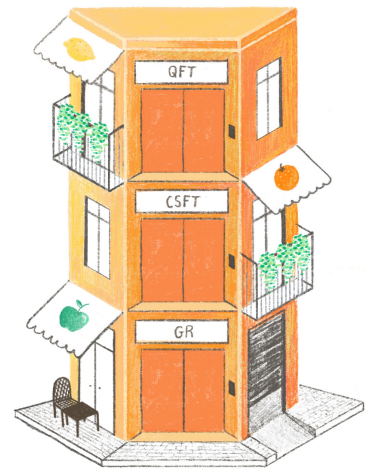

Figure 1.1: A three-storey building representing this thesis.

Each of the theories that form the building are built on top of each other. Every one of the levels of this building is comprised of a Hamiltonian theory, whereby the common structure of the building is the one described at the beginning of the section. In addition, each of the floors needs the definition of an infinite-dimensional manifold, and the first and second floors require integration theory in infinite-dimensional spaces. For this reason, these mathematical aspects are introduced in chapter 2.

In the remainder of this section, we will briefly discuss the kinematical aspects of each floor. Since the structure is common, we should identify a phase space, Poisson bracket, and algebra of observables $(\mathcal{M}, \{\cdot, \cdot\}_{\mathcal{M}}, \mathcal{O})$ at each level.

**Ground floor: General Relativity**

The ground floor in Figure 1.2 represents the description of the curved spacetime itself. In this thesis, we will assume that the curved spacetime is described by General Relativity (GR), even though other theories of gravity may be considered. As we pointed out above, this is the underlying structure supporting every other in the thesis. In spite of this fact, we will honor the Spanish saying "*empezar la casa por el tejado*", which literally translates to "build the house from the rooftop"[1], and postpone its study to chapter 6, where we will also consider its dynamics.

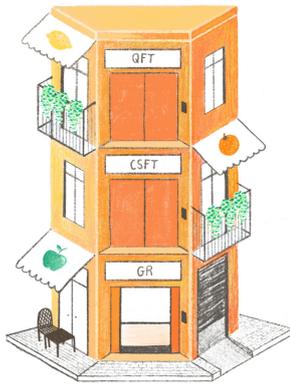

Figure 1.2: Ground Floor. General Relativity

In this thesis, we will consider only globally hyperbolic spacetimes diffeomorphic to $\Sigma \times \mathbb{R}$, where $\Sigma$ is a $d + 1$-dimensional Cauchy hypersurface. The concrete splitting into space $\Sigma$ and time $\mathbb{R}$ will be explored in section 3.5 and chapter 6. This splitting allows us to describe the evolution directly over $\Sigma$ and introduce time $t$ as an external parameter. In this manner, the phase space of General Relativity is a submanifold

$$\mathcal{M}_G \subset T^* \operatorname{Riem}(\Sigma) \tag{1.1}$$

The particular form of the submanifold will be, again, discussed in chapter 6 by means of constraints in the dynamics. This implies that the Poisson bracket of $T^* \operatorname{Riem}(\Sigma)$ will be the one that generates the evolution constrained to $\mathcal{M}_G$, and observables will be provided by $C^\infty(\mathcal{M}_G)$. We express in coordinates $(\mathbf{h}, \pi_h) \in T^* \operatorname{Riem}(\Sigma)$, where $\mathbf{h}$ is a Riemannian metric and $\pi_h$ is a symmetric bivector density. Using their matrix components $(h_{ij}, \pi_h^{ij})$ and summing over repeated indices for $f, g \in C^\infty(\mathcal{M}_G)$, we get the Poisson bracket

$$\{f, g\}_G = \int_\Sigma d^3 x \left( \frac{\partial f}{\partial h_{ij}(x)} \frac{\partial g}{\partial \pi_h^{ij}(x)} - \frac{\partial g}{\partial h_{ij}(x)} \frac{\partial f}{\partial \pi_h^{ij}(x)} \right)$$

---

[1]The correct translation would be "put the cart before the horse"

Then the kinematics of the theory is provided by

$$(\mathcal{M}_G \subset T^* \operatorname{Riem}(\Sigma),\ \{\cdot,\cdot\}_G,\ C^\infty(\mathcal{M}_G))$$

The fact that we can identify this Hamiltonian structure of gravity will be the basis in chapter 6 to minimally couple any matter field to gravity. Following [HKT76], as long as we can identify a Poisson bracket $\{\cdot,\cdot\}_m$ for matter, we postulate that $\{\cdot,\cdot\}_G + \{\cdot,\cdot\}_m$ is the Poisson bracket of the interacting theory. As we will see, this structure can be identified for the classical (first floor) and quantum (second floor) field theories. Taking advantage of this fact, in chapter 6, we will study the hybrid quantum-classical coupling of GR and QFT.

Nonetheless, our main focus in this thesis is the case in which the geometric content that describes gravity is given by a fixed background. This will be the case in chapters 3 to 5 and 7. In order to treat gravity as a background, due to the constrained nature of $\mathcal{M}_G$, it is not enough to provide $(\mathbf{h}, \pi_h)$; we also need to provide the lapse function $N$ and the shift vector field $\vec{N}$. In this case, we will specify the gravitational content through a curve

$$\gamma(t) = (\mathbf{h}(t), \pi_h(t), N(t), N^i(t)). \tag{1.2}$$

We will say that we are dealing with a parametric theory on a curved spacetime. As we can see, this concept will be central in our discussion, and we expand the details in *Remark* 3.1.

**First floor: Classical Statistical Field Theory**

The first floor of Figure 1.3 is comprised of the classical theory of fields. This is the simplest case that we consider in this thesis and is the subject of study of chapter 3. In general terms, see [GIM04, Got&04], to describe a field over the Cauchy hypersurface $\Sigma$, we need a bundle $\pi_{\Sigma L} : L \to \Sigma$, and the configuration space of classical fields is provided by the sections of this bundle $Q = \Gamma(\Sigma, L)$. In this thesis, we will study the scalar field; thus, $L$ will be a real line bundle and $Q = C^\infty(\Sigma)$. In order to get well-behaved expressions, we will consider $\Sigma$ compact, even though we can waive this restriction using Nuclear-Fréchet (NF) spaces that we will discuss in section 2.2. As a consequence, the phase space of the theory is the cotangent bundle $T^*Q$ with coordinates $(\varphi_{\mathbf{x}}, \pi^{\mathbf{x}})$, where $\varphi_{\mathbf{x}} \in Q$ and $\pi^{\mathbf{x}}$ is a vector density. The observables of the theory are comprised of $f, g \in C^\infty(T^*Q) = \mathcal{O}$, and the Poisson bracket is provided by

$$\{f, g\}_{T^*Q} = \int_\Sigma dx^d \left( \frac{\partial f}{\partial \varphi(x)} \frac{\partial g}{\partial \pi(x)} - \frac{\partial f}{\partial \pi(x)} \frac{\partial g}{\partial \varphi(x)} \right),$$

However, this description, even though simple, is not suited for the purposes of this thesis. The reason why we are introducing the classical field theory in the first place is to build a QFT on the subsequent level. We achieve this via a quantization prescription dictated (partially) by the Geometric Quantization program mentioned in the previous section. In order to have a classical field theory adapted to this quantization process, we must be able to describe prequantization theory. Therefore, we must describe the classical system in the Koopman-van Hove (KvH) formalism also mentioned above. This is, we should study a statistical theory. For these reasons, the first floor of this building is actually described by Classical Statistical Field Theory (CSFT).

The reason why the phase space presented above is not valid for KvH lies in the nuisances and technicalities of functional integration. The statistical state $\rho : \mathcal{O} \to \mathbb{R}$ should, ideally, be understood as an integral taking $f$ to

$$\langle f \rangle = \iint D\varphi_{\mathbf{x}} D\pi^{\mathbf{x}} \rho(\varphi_{\mathbf{x}}, \pi^{\mathbf{x}}) f(\varphi_{\mathbf{x}}, \pi^{\mathbf{x}}) \qquad (1.3)$$

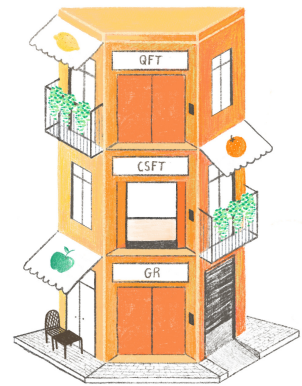

Figure 1.3: First floor. Classical Statistical Field Theory

However, the Lebesgue-like measures $D\varphi_{\mathbf{x}} D\pi^{\mathbf{x}}$ cannot be defined in proper mathematical terms. Moreover, if we restrict our attention to Gaussian states, in Appendix C, we show that there is no measure that can provide such state $\rho$ over the phase space $T^*C^\infty(\Sigma)$. Instead, in section 2.4, we develop a theory of rigorous Gaussian integration over the strong Dual of Nuclear-Fréchet (DNF) spaces. These are spaces of distributions such as $D'(\Sigma)$. As a consequence, the space of classical statistical fields should be modeled with both $\varphi^{\mathbf{x}}, \pi^{\mathbf{x}} \in D'(\Sigma)$ representing distributions.

In order to meet the conditions discussed above, in section 3.3, we opt to lift the kinematic structure of the phase space $T^*Q$ to $\mathcal{M}_F = D'(\Sigma) \times D'(\Sigma)$. For the field-momentum coordinate $\pi^{\mathbf{x}}$, this is immediate because densities are already defined as distributions. For the field coordinate $\varphi_{\mathbf{x}}$, instead, we will *raise the index* to $\varphi^{\mathbf{x}}$ by means of the inclusions $C^\infty(\Sigma) \subset L^2(d\mathrm{Vol}_h, \Sigma) \subset D'(\Sigma)$. These kinds of inclusions, denoted rigged Hilbert spaces or Gel'fand triples, are discussed in section 2.3. Notice also that $d\mathrm{Vol}_x = dx^d \sqrt{h}$ is the volume form induced by the Riemannian structure of the ground floor of this building. As a consequence, the whole kinematic structure of $\mathcal{M}_F$ will depend on the geometric content of the spacetime downstairs.

Notice that, because we want to properly define our Gaussian states, already at this classical stage we will need to introduce some extra structure. Using the results of section 2.4, we know that the only ingredient needed to describe (mean zero) Gaussian states is a symmetric and positive definite covariance. To describe the covariance, we will use the natural symplectic structure of $\mathcal{M}_F$ in the following manner. Using the fact that $T^*Q$ is a natural symplectic manifold, we obtain a symplectic structure $\omega_{\mathcal{M}_F}$. Taking advantage of this fact, this covariance can be understood as the Riemannian counterpart $\mu$ of a compatible Kähler structure on the space of fields. Thus, Gaussian states in CSFT will be completely determined by

$$(\omega_{\mathcal{M}_F}, \mu_{\mathcal{M}_F}, J_{\mathcal{M}_F}) \tag{1.4}$$

where $\omega_{\mathcal{M}_F}$ is the symplectic form, $\mu_{\mathcal{M}_F}$ the Riemannian form, and $J_{\mathcal{M}_F}$ the compatible complex structure such that $\mu_{\mathcal{M}_F}(\cdot, \cdot) = \omega_{\mathcal{M}_F}(\cdot, -J_{\mathcal{M}_F}\cdot)$. In turn, the complex structure $J_{\mathcal{M}_F}$ is the tool that we will use to define Gaussian states. This complex structure is usually regarded as a feature of the quantization procedure [AMA75, Wal94, AS99, CCQ04, MO21]. Conversely, in this thesis, we consider it as a feature of CSFT. It is important to emphasize that, in this work, the particular form of $J_{\mathcal{M}_F}$ will be dictated by dynamical considerations and will depend on the geometric content of gravity. Later, in this section, we will delve further into this topic.

Slightly departing from the standard kinematical description, in chapter 3 we will present the KvH representation of CSFT. The common feature of Koopman representations [Koo31, GT22b] is that they express the state of the statistical system $\rho$ as a function of a state $\Psi$ belonging to a Hilbert space of square-integrable functions. For instance, the original Koopman-von Neumann (KvN) prescription was simply $\Psi = \sqrt{\rho}$. As we will see in (3.2), the prescription for KvH is slightly more complicated.

In any case, the classical state of the KvH representation will be provided by $\Psi \in L^2(D'(\Sigma) \times D'(\Sigma), D\mu_c)$ where $D\mu_c$ is the Gaussian measure with covariance $\mu_{\mathcal{M}_F}$. In this picture, (1.3) is interpreted by the well-defined expression

$$\langle f \rangle = \int_{D'(\Sigma) \times D'(\Sigma)} D\mu_c(\varphi^{\mathbf{x}}, \pi^{\mathbf{x}}) \tilde{\rho}(\varphi^{\mathbf{x}}, \pi^{\mathbf{x}}) f(\varphi^{\mathbf{x}}, \pi^{\mathbf{x}}). \tag{1.5}$$

In this case, we interpret $\tilde{\rho}$ as the function formally corresponding to $\rho$ in (1.3) when the Gaussian part is factored out.

**Second floor: Quantum Field Theory**

The second floor of Figure 1.4 represents Quantum Field Theory (QFT). In this thesis, and in chapter 4, a quantum theory is obtained through a quantization procedure from a classical one. In our case, the Geometric Quantization procedure, see [Woo97], reduces the Hilbert space of prequantization (KvH) theory to half of the degrees of freedom of the domain. This reduction is performed by choosing a Lagrangian subspace, also called polarization, of $\mathcal{M}_F$. Of course, there are some ambiguities in that choice. In fact, we will find that there is no preferred choice of polarization in QFT. Instead, we will obtain different, but equivalent, Hamiltonian representations of it. In chapter 4, we find that the most relevant ones are the holomorphic, antiholomorphic, Schrödinger, and field-momentum representations. Moreover, in section 4.3, we will introduce the Segal-Bargmann and Fourier integral transforms that relate them.

For the sake of simplicity, in this introduction, let us stick to the Schrödinger representation[2]. In this case, the polarization restriction produces $\mathcal{M}_F \simeq D'(\Sigma) \times D'(\Sigma) \xrightarrow{\text{Quantization}} D'(\Sigma)$, and the states are represented by functions $\Psi(\varphi^{\mathbf{x}})$. In this case, the Hilbert space of the quantum theory is

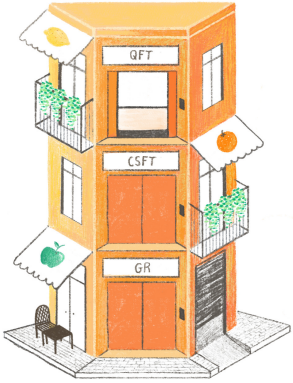

$$L^2_{Sch}(D'(\Sigma), D\mu_S).$$

The Gaussian measure $\mu_S$ is related to the measure of the prequantum Hilbert space. As a consequence, it depends on the classical Kähler structure that we selected downstairs. The observables of the theory come from a quantization mapping $Q$ that lifts the classical observables to self-adjoint operators acting on $L^2_{Sch}(D'(\Sigma), D\mu_S)$. However, geometric quantization is not enough to fully determine $Q$. In order to provide a successful prescription, we need to discuss ordering problems. For this reason, we will discuss Weyl and Wick orders in section 4.4.

Figure 1.4: Second floor. Quantum Field Theory

It is at this stage where the connections of QFT with Gaussian analysis flourish. To showcase them, we mention that Weyl relations, the starting point of most approaches to Algebraic Quantum Field Theory (AQFT) [Bru&15], naturally follow from the decomposition in trigonometric exponentials of section 2.5. The Wiener-Ito decomposition theorem of section 2.6 provides the connection of the Schrödinger representation (and the others considered here) with the Bosonic Fock space. Finally, Skorokhod and Malliavin derivatives, which will be introduced in that section, are interpreted as the creation and annihilation operators of the QFT. All these relations, and more, will be explored throughout chapter 4.

Nonetheless, this Schrödinger picture does not quite fit with the Hamiltonian kinematical structure that we are looking for, as was the case for KvH theory. In order to meet that description, in chapter 5, we will follow [Kib79] and introduce the Geometrization of QFT. In this procedure, the Hilbert space that represents pure quantum states, $L^2_{Sch}(D'(\Sigma), D\mu_S)$, is understood as the phase space manifold $\mathscr{P}$. For technical reasons, we must restrict it to a subset of smooth functions; thus, $\mathscr{P} \subset L^2_{Sch}(D'(\Sigma), D\mu_S)$ is chosen as the space of Hida test functions that we define in section 2.7.

A set of coordinates of $\mathscr{P}$ is provided by $(\Psi, \bar{\Psi})$ where $\Psi(\varphi^{\mathbf{x}})$ is a function representing a pure state. Over this set, observables are provided by the set of quadratic functions $\mathcal{F}_2(\mathscr{P})$ comprised of $f_{\hat{G}} \in \mathcal{F}_2(\mathscr{P})$ if

$$f_{\hat{G}}(\bar{\Psi}, \Psi) = \langle \Psi, \hat{G}\Psi \rangle_{\mathscr{P}},$$

---

[2]In most parts of this thesis, we will prefer instead the Holomorphic representation. As we will see, most geometric tools and analytical properties are easily handled in that case.

where $\hat{G}$ is a self-adjoint operator acting on the Hilbert space and $\langle \cdot, \cdot \rangle_{\mathscr{P}}$ is its scalar product. The symplectic form of the phase space $\Omega$ is simply the imaginary part of this scalar product. Moreover, the real part of this scalar product $\mathcal{G}$ and the trivial complex structure $\mathcal{J}$ also induce a Kähler structure on the space of pure states

$$(\mathcal{G}, \Omega, \mathcal{J})_{\mathscr{P}} \in \text{Kähl}(\mathscr{P})$$

The Poisson bracket of this phase space also exists $\{\cdot, \cdot\}_{\mathscr{P}}$. Nonetheless, we are not yet ready to write down its explicit expression. In chapter 5, we will develop a formalism based on the Gaussian analysis of the white noise measure to deal with these expressions efficiently. Here we just mention that the algebra of observables is, indeed, an algebra closed through the Poisson bracket because

$$\{f_{\hat{F}}, f_{\hat{G}}\} = f_{-i[\hat{F}, \hat{G}]}.$$

As a result, the Poisson algebra is closed because it is a representation of the Lie algebra of self-adjoint operators with Lie bracket $-i[\cdot, \cdot]$.

### 1.2.2 Dynamical aspects: Corrections to the time derivative

The interdependence of each of the kinematical levels described above produces non-trivial consequences for the dynamics. In this section, and in most of the thesis, we focus on the parametric case, that is, when the gravitational content is provided by (1.2).

Already at the classical level, we must explain how to choose a physically meaningful complex structure in (1.4). We will argue, in section 3.4.4, that the source complex structure must be dynamical in nature. The Gaussian measure $\mu_c$ works as a reference to express the evolution of the states. For this reason, even though it can have a parametric dependence on time $\mu_c(t)$, it should be unaffected by the generator of the evolution.

In more concise terms, let $H$ be the Hamiltonian that generates the evolution of the CSFT. This is the dynamical flow generated by the Hamiltonian vector field $X_H$. To write down the complex structure, we understand $X_H$ as an operator acting on $C^{\infty}(\mathcal{M}_F)$. In this thesis, we propose that $J_{\mathcal{M}_F} = |X_H|^{-1} X_H$. In this fashion, the complex structure, being a function of the generator of the evolution, is not affected by the flow of $X_H$. This proposal coincides with others in the literature [CCQ04], but our approach is novel in that it is already a requirement in the classical theory.

In any case, the dynamical choice of $J_{\mathcal{M}_C}$ introduces some issues that must be taken into account. In the first place, this choice of connection is dependent on the parameter $t$. In turn, the whole Kähler structure of $\mathcal{M}_F$ is dependent on the time parameter. Since the quantum Kähler structure (second floor) directly depends on the classical one (first floor), the measure $\mu_S$, which defines the Schrödinger Hilbert space, also depends on $t$. In a nutshell, the structure that we presented above transfers the parametric time dependence introduced in the gravitational theory (ground floor) all the way up to the QFT (second floor). This will have implications in both the classical (3.4.4) and quantum evolutions. For the purposes of this thesis, we will focus on the dependence of the quantum structures.

The immediate consequence of the arbitrary time dependence of the quantum Kähler structure $(\mathcal{G}(t), \Omega(t), \mathcal{J}(t))_{\mathscr{P}}$ is that it threatens the probabilistic interpretation of the quantum part of the theory. In fact, if the quantum evolution is generated by a self-adjoint Hamiltonian $\hat{H}$ for the Schrödinger equation $i\partial_t \Psi = \hat{H}\Psi$, then $\partial_t \langle \Psi, \Psi \rangle_{\mathscr{P}} = \hat{\mathcal{G}}(t)(\bar{\Psi}, \Psi)$, and the norm of the state $\Psi$ is not conserved. In principle, this might not be a problem if we consider that we are dealing with an open system and we are not properly taking into account all the nuances of the interaction with gravity. This kind of interpretation of the norm loss has also been studied by some authors [HS15, HS17, Sch18, HSU19, EHS21].

In spite of this, we want to describe a theory that preserves the probabilistic interpretation. Ultimately, this is a consequence of the desire to describe a consistent hybrid quantum-classical system in chapter 6. Our strategy to achieve this goal will be to introduce, in section 5.4, a connection term for the time derivative $\partial_t \to \nabla_t = \partial_t + \Gamma_t$. To preserve the probabilistic interpretation, the first criterion to select the appropriate connection will be to preserve the quantum Kähler structure

$$\nabla_t(\mathcal{G}(t), \Omega(t), \mathcal{J}(t))_{\mathscr{P}} = (0, 0, 0). \tag{1.6}$$

As we will discuss in that section, this condition does not exhaust all the possible ambiguities in the choice of the connection. In order to select a unique one, we will borrow from chapter 6 the extra requirement

$$\nabla_t Q(f) = Q(\partial_t f) \tag{1.7}$$

that will fix $\Gamma$. This connection term adds an extra factor to the Schrödinger equation that we name the modified Schrödinger equation

$$i(\partial_t + \Gamma_t)\Psi = \hat{H}\Psi$$

Modifying the Schrödinger equation in this manner leads to interesting and non-trivial phenomenological contributions. In order to illustrate these phenomena, in chapter 7, we particularize the analysis to cosmological Friedman-Lemaître-Robertson-Walker (FLRW) models and show how particle creation in dynamic universes manifests itself in our formalism.

## 1.3 Structure of the thesis and reading guide

So far, we have provided context and portrayed the structure of the project that we will undertake in this thesis. In the previous section, we already linked the important features of the project to their relevant chapters and sections. In this section, we will go in the opposite direction and briefly summarize the content of every chapter. Besides, at the beginning of every main chapter of the thesis, there will be a detailed introduction on the scope and content of the chapter. In general, the thesis is comprised of four more or less independent pieces.

- **Mathematical foundations** comprised of chapter 2 and appendices B to D. These are the mathematical foundations of the thesis.

- **Classical and Quantum Field theories in curved spacetimes** comprised of chapters 3 to 5 and appendix A. This is the main body of the thesis and represents the theoretical development of the physical models.

- **Hybrid quantum-classical coupling of QFT with GR**. This topic is only treated in chapter 6.

- **Applications to Klein-Gordon theory in cosmological FLRW models**. To this category belong section 3.5 and chapter 7. This is the principal example of the application to physical systems presented in this thesis.

In more detail, the contents of every chapter are:

**Chapter 2** presents the mathematical foundations of Gaussian Analysis. This is a rigorous method to deal with integration theory in infinite-dimensional spaces. Notational conventions that will be used throughout the thesis are established in this chapter.

The content of this chapter will be extensively used in the rest of the chapters. Nonetheless, a thorough understanding of the concepts in this chapter is not a requirement to follow the rest of the thesis. In section 1.2, we tried to indicate which specific sections and tools of chapter 2 serve as a basis for each part of the project. Moreover, each mathematical tool required in the main body of the thesis is cross-referenced when possible.

**Chapter 3** presents the Koopman-van Hove (KvH) approach to classical statistical mechanics. The chapter starts in section 3.2 with a discussion of the theory in finite dimensions. The rest of the chapter discusses kinematics and evolution in Classical Statistical Field Theory (CSFT).

**Chapter 4** develops the quantization program of a QFT of the scalar field in two steps. First, geometric quantization in sections 4.2 and 4.3 for linear observables. This introduces the Holomorphic, Schrödinger, antiholomorphic, and field-momentum representations of QFT and their relations. Second, Wick and Weyl ordering prescriptions for higher-order observables. The chapter ends with a discussion on algebraic QFT. This discussion is built upon the prequantization theory introduced in Chapter 3, thus it is convenient to read both together.

**Chapter 5** studies the evolution of the QFT in curved spacetime. For that matter, sections 5.2 and 5.3 introduce the Geometrization program for QFT and use white noise Gaussian analysis to describe the geometrization in *second quantized* (*s.q.*) coordinates. The evolution itself is introduced in sections 5.4 and 5.5, where the Schrödinger equation is modified via a connection term, which is analyzed in detail. The content of this chapter is fairly independent of chapters 3 and 4 but is better understood after reading them.

**Chapter 6** studies the hybrid coupling to gravity. In this chapter, we use geometrodynamical arguments to derive the coupling of quantum matter with classical gravity. This chapter is a standalone piece that can be understood with minimal ingredients from other chapters.

**Chapter 7** studies the physical implications of Klein-Gordon theory in cosmological FLRW models. It discusses the phenomenon of particle creation in these examples to showcase the potential applications of the general formalism. This chapter requires the reading of section 3.5.

**Chapter 8** presents the main results of this thesis and discusses some open problems left for future work.

**Appendices** To complete the discussion in the main body of the thesis, we prepared four appendices. appendix A presents some explicit computations of integral transforms from chapter 4 and, based on those, connections from chapter 5. Three of them are devoted to expanding the mathematical discussion of chapter 2. These are:

- appendix B that adapts well-known results of Hida-Malliavin calculus to the case of interest of this thesis, holomorphic Hida test functions.

- appendix C that summarizes some problems that forbid the definition of a measure over NF and even Hilbert spaces. This chapter also provides a detailed review of Gaussian integration in Topological Vector Spaces (TVSs).

- appendix D that proves the Wiener-Ito decomposition theorem in homogeneous chaos in the notation used in this thesis.

# Chapter 2

# Mathematical aspects: Gaussian analysis

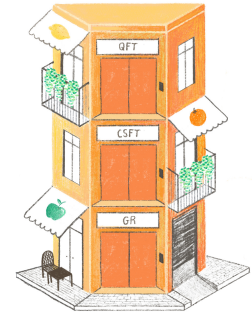

## 2.1 Introduction

The objective of Gaussian analysis is to describe the correct setting of (Gaussian) integro-differential calculus in infinite-dimensional Topological Vector Spaces (TVSs) [Hid&93, Kuo96, Hu16]. Our approach in this chapter will be strongly focused on its application to Quantum Field Theory (QFT). As Gaussian measures solve the Schrödinger equation for free field theories [LS98, HS15, HS17], this will be our preferred choice. Roughly speaking, Gaussian analysis aims to answer, among others, the following basic questions:

- What TVSs admit Gaussian Borel probability measures?

- What is the support of the measure?

- How can we define derivatives or Sobolev derivatives over spaces of functions with domain in the TVS?

The answers to these questions are often surprising, even for the less exotic case of Banach spaces; we summarize them in Appendix C. In this chapter, we present an approach to this theory strongly influenced by the field of white noise analysis. This field was originally developed in the last decades of the XXth century to provide a rigorous mathematical definition to the time derivative of Brownian motion. This is achieved by the study of the white noise Gaussian measure [Hid80, Hid&93, Oba94, Kuo96, Kon&96, Wes03, HS08].

In that approach, the natural setting to define Gaussian Borel probability measures are Nuclear-Fréchet (NF) spaces and strong Duals of Nuclear-Fréchet (DNF) spaces that we introduce in sections 2.2 and 2.3. This is also discussed thoroughly in [GV16]. In the following, we will refer to this class of measures as Gaussian measures. Our playground is then $(\mathcal{N}, \mathcal{N}')$, where $\mathcal{N}$ is a NF space and $\mathcal{N}'$ is its dual endowed with the strong topology[1].

White noise analysis differs from other approaches (see e.g. [Hu16]) in its focus on defining spaces of stochastic or *second quantized* (*s.q.*) test functions and distributions named after Hida and Kondratiev. These test functions are again NF spaces, and distributions are again an example of a DNF space. We devote a separate Appendix B to study different notions of *s.q.* test functions.

The goal of this chapter is to enlarge the study and definition of Gaussian and white noise analysis to a slightly more general setting. We are interested in cases in which $(\mathcal{N}, \mathcal{N}')$ represent spaces of classical fields. In particular, we will treat $(C_c^\infty(\Sigma), D'(\Sigma))$, the compactly supported functions and distributions over a compact $d$-dimensional manifold $\Sigma$, that will represent a Cauchy hypersurface in a $d+1$-dimensional spacetime, and $(\mathcal{S}(\mathbb{R}^d), \mathcal{S}'(\mathbb{R}^d))$, the space of Schwartz functions of fast decay and the space of tempered distributions.

We then define a Gaussian measure $\mu$ over $\mathcal{N}'$ in section 2.4 and study the space of square-integrable functions $L^2(\mathcal{N}', D\mu)$ using algebras of well-behaved functions that belong to it in section 2.5. Namely, the algebras of polynomials $\mathcal{P}$, trigonometric exponentials $\mathcal{T}$, and coherent states $\mathcal{C}$.

The study of these algebras leads to the central result of Gaussian analysis, the Wiener-Ito chaos decomposition theorem, which we analyze in section 2.6 and later in Appendix D. This theorem provides the particle interpretation in QFT. It also allows for the discussion of differential calculus, introducing the Malliavin (also named Gross-Sobolev [Hu16]) derivative. The introduction of these derivatives will also pave the way for the introduction of Hida test functions in section 2.7.

We will end this chapter in section 2.8 with a flavour of geometry. Following [KM97], both NF and DNF spaces are convenient and admit partitions of unity. Thus, we conclude that they are convenient models of geometry. We briefly discuss the precise mathematical content of this affirmation.

The concrete interdependencies of the sections of this chapter with the rest of the thesis have been presented in section 1.2.

## 2.2 Nuclear spaces

Let's briefly introduce some definitions and examples of nuclear spaces.

**Definition 2.1.  (Nuclear Space)** *A nuclear space $X$ is a topological vector space whose topology is given by a family of Hilbert seminorms $p$ such that, if $X_p$ is the Hilbert space obtained as the completion of the pre-Hilbert space $(X/p^{-1}(0), p)$, then there is a larger seminorm $q$ such that $X_q \hookrightarrow X_p$ is a Hilbert-Schmidt operator [Tre67].*

---

[1]In this case, we endow the DNF space with the strong topology, which coincides with the Mackey topology, but the same Borel $\sigma$-algebra is generated from the weak topology [Kuo96].

For infinite-dimensional topological vector spaces $X, Y$, tensor products can be endowed with a family of different topologies between the projective $X \otimes_\pi Y$ and injective $X \otimes_\epsilon Y$ tensor product induced topologies. If one or both spaces are nuclear, both topologies coincide, and therefore there is only one notion of tensor product $X \otimes Y$. We will use this fact in the following sections to define polynomials over $X$.

**Definition 2.2.** *(NF Nuclear-Fréchet Space) A nuclear space that is also Fréchet $\mathcal{N}$ can be regarded as the projective limit of a countable family of separable [2] Hilbert spaces $\{\mathcal{H}_p\}_{p=0}^\infty$, with scalar products $\langle \cdot, \cdot \rangle_p$ such that for $q > p$ the injection $\mathcal{H}_q \hookrightarrow \mathcal{H}_p$ is Hilbert-Schmidt. Thus*

$$\mathcal{N} = \bigcap_{p=0}^\infty \mathcal{H}_p$$

*with the projective limit topology.*

The Riesz representation theorem allows us to identify a Hilbert space with its dual. By convention, we set $\mathcal{H}_0' = \mathcal{H}_0^*$, where $*$ represents complex conjugation. This implies that in the inclusion $\mathcal{H}_p \hookrightarrow \mathcal{H}_0$, for $p > 0$, the space $\mathcal{H}_p$ can no longer be identified with its dual. We will denote this dual space by $\mathcal{H}_0 \hookrightarrow \mathcal{H}_{-p}$. The next definition can be regarded as a consequence of the last one.

**Definition 2.3.** *(DNF strong dual of a Nuclear Fréchet Space) The strong dual of a NF space $\mathcal{N}'$ can be regarded as the inductive limit of a countable family of separable Hilbert spaces $\{\mathcal{H}_{-p}\}_{p=0}^\infty$ such that for $q > p$ the injection $\mathcal{H}_{-q} \hookrightarrow \mathcal{H}_{-p}$ is the transpose of a Hilbert-Schmidt operator. Thus*

$$\mathcal{N}' = \bigcup_{p=0}^\infty \mathcal{H}_{-p}$$

*with the inductive limit topology. A DNF space is always nuclear. Furthermore, it is a nuclear Silva space [KM97].*

## 2.3 Rigged Hilbert spaces and main examples

With the definitions given so far, we notice that the space $\mathcal{H}_0$ plays an important role. This motivates the consideration of a triple of spaces, leading to the concept of a rigged Hilbert space.

---

[2] Every Fréchet space is complete and barreled ([Tre67] Corollary 1 Chapter 33). A complete barreled nuclear space is a Montel space ([Tre67] Corollary 3 Chapter 50). Montel spaces are reflexive ([Tre67] Corollary of Proposition 36.9), and Fréchet Montel spaces are separable ([KM97] Result 52.27). Thus, it follows that a nuclear-Fréchet space is reflexive and separable. This fact allows us to choose the chain of separable Hilbert spaces for this definition.

**Definition 2.4.  (*Rigged Hilbert Space*)**  *Assume that $\mathcal{H}_0$ is a separable Hilbert space; then we define a rigged Hilbert space or Gel'fand triple as the triple of spaces*

$$\mathcal{N} \subset \mathcal{H}_0 \subset \mathcal{N}'$$

*where $\mathcal{N}$ is a NF space and $\mathcal{N}'$ is its dual, and each inclusion is continuous.*

In the structure presented here, $\mathcal{H}_0$ is said to realize the duality, which means that the scalar product of $\mathcal{H}_0$ identifies an element of $\mathcal{N}$ with its dual in $\mathcal{N}'$. $\mathcal{N}$ is a dense subspace of the Hilbert space, and this construction allows us to, in some sense, parametrize the duality and define $\mathcal{N}$ as a subset of $\mathcal{N}'$. Notice, however, that in general, different Hilbert spaces can be used to parametrize the duality of the same pair $(\mathcal{N}, \mathcal{N}')$. In the following, unless stated otherwise, we will consider $\mathcal{N}$ as a real NF space, and we will denote by $\mathcal{N}_{\mathbb{C}}$ the complex case.

In the rigging, there are three main ingredients: the set of test functions $\mathcal{N}$, its dual space $\mathcal{N}'$, and the dual pairing of the Hilbert space $\mathcal{H}_0$. Firstly, elements $\xi_{\mathbf{x}}, \eta_{\mathbf{x}} \in \mathcal{N}$ will be typically well-defined real test functions (see examples below) and will be denoted by a bold subindex. Secondly, elements $\varphi^{\mathbf{x}}, \pi^{\mathbf{x}} \in \mathcal{N}'$ will be real distributions (with no pointwise meaning in general) and they will be denoted by a bold superindex. Sometimes we will denote the value of the function at a point $x$ with a regular subindex $\xi_x$, but notice that this cannot be done in general for distributions. With those two ingredients, the duality is denoted, mimicking Einstein summation convention,

$$\xi_{\mathbf{x}} \in \mathcal{N}, \varphi^{\mathbf{x}} \in \mathcal{N}' \text{ paired by } \xi_x \varphi^x := \langle \varphi^{\mathbf{x}}, \xi_{\mathbf{x}} \rangle \in \mathbb{R}. \tag{2.1}$$

The third and last ingredient is the dual pairing of $\mathcal{H}_0$. This element can be considered in a threefold way:

a) Dual pairing between elements in $\mathcal{N} \subset \mathcal{H}_0$ with $\langle \xi_{\mathbf{x}}^1, \xi_{\mathbf{y}}^2 \rangle = \xi_x^1 \Delta^{xy} \xi_{xy}^2$.

b) Dual pairing between elements in $\mathcal{H}_0 \subset \mathcal{N}'$ with $\langle \varphi_1^{\mathbf{x}}, \varphi_2^{\mathbf{y}} \rangle = \varphi_1^x K_{xy} \varphi_2^y$.

c) Injection of $\Delta^{\mathbf{xy}} : \mathcal{N} \hookrightarrow \mathcal{N}'$ through the pairing of $\mathcal{H}_0$ such that $\xi_x \Delta^{xy} = \xi^y$ with a densely defined inverse $\varphi^x K_{xy} = \varphi_y$ such that $\Delta^{xz} K_{zy} = \delta_y^x$ is the evaluation mapping.

We will be using the three different frameworks in the following sections.

The main examples of NF spaces, which play the roles of model spaces for the classical field theories in this work, are the following:

**Example 2.1.  (*Smooth functions of compact support on a compact $K$*)**  Let $K \subset \mathbb{R}^d$; then the space $C_c^\infty(K)$ is NF with the initial topology induced from the family of Hilbertian norms

$$\|\varphi\|_{K,n}^2 = \sum_{|\alpha| \le n} \int_K |\partial^\alpha \varphi(x)|^2 dx^d, \tag{2.2}$$

where $\alpha \in \mathbb{N}_0^d$ is a multi-index[3] and $|\alpha| = \alpha_1 + \cdots + \alpha_d$. Its dual is the DNF space $D'(K)$. This is not the family of Hilbertian seminorms of **Definition** 2.2; instead, we should use the Hamiltonian of the harmonic oscillator to introduce them as in [Hid&93].

Notice that we could also define $C_c^\infty(U)$, for $U \subset \mathbb{R}$ an open set. This is a nuclear space but not a Fréchet space; instead, it is a Nuclear LF space (see [GV16] for details). For our purposes, we will limit our attention to the compact domain in order to reduce technicalities.

**Example 2.2. (Smooth functions of fast decay)** The space $\mathcal{S}(\mathbb{R}^d)$ of smooth functions of fast decay is NF with the initial topology induced from the family of Hilbertian norms

$$\|\varphi\|_{m,n}^2 = \sum_{|\alpha| \le n, |\beta| \le m} \int_{\mathbb{R}^d} \prod_{i=1}^d (1 + x_i^2)^{2\beta_i} |\partial^\alpha \varphi(x)|^2 dx^d, \tag{2.3}$$

where $\alpha, \beta \in \mathbb{N}_0^d$ are multi-indices. This is not the family of Hilbertian seminorms of **Definition** 2.2; instead, we should use the Hamiltonian of the harmonic oscillator to introduce them as in [Hid&93]. This topology is equivalent to the one usually introduced via supremum-type seminorms [Hid80]. Its dual $\mathcal{S}'(\mathbb{R}^d)$ is the DNF space of tempered distributions. This space is the one commonly used in white noise analysis [Hid&93].

**Example 2.3. (Smooth functions of compact support on a compact manifold)** For $\Sigma$ a compact manifold, the definition of $C_c^\infty(\Sigma)$ goes on as in Example 2.1. If we endow the manifold with a Riemannian metric $h$, there is a canonical volume form

$$\int_\Sigma \xi \, d\mathrm{Vol}_h = \int_U dx^d \sqrt{|h(x)|} \xi(x), \tag{2.4}$$

for $\xi$ supported on a subset of $\Sigma$ covered by a single chart with coordinates in $U$. This, in turn, leads to a canonical Hilbert space of square integrable functions $L^2(\Sigma, d\mathrm{Vol}_h)$ and therefore to the rigged Hilbert space

$$C_c^\infty(\Sigma) \subset L^2(\Sigma, d\mathrm{Vol}_h) \subset D'(\Sigma). \tag{2.5}$$

As discussed in **Definition** 2.4, the scalar product on a rigged Hilbert space can be considered in a threefold way, as we can illustrate over a chart with coordinates in $U \subset \mathbb{R}^d$. Because of the importance of this particular example throughout this work, we detail the notation for the bilocal kernels exposed below (2.1). All three cases will be denoted by the same symbol $\delta^{\mathbf{xy}}$, $\delta_{\mathbf{xy}}$ and $\delta_{\mathbf{y}}^{\mathbf{x}}$, and the difference between them is understood by context as follows:

---

[3] $\mathbb{N}_0$ represents the natural numbers with the zero.

a) *Dual pairing between elements in $C_c^\infty(\Sigma) \subset L^2(\Sigma, \mathrm{dVol}_h)$ with*

$$\langle \xi_{\mathbf{x}}, \eta_{\mathbf{y}} \rangle = \chi_x \delta^{xy} \eta_y = \int_U dx^d \sqrt{|h(x)|} \xi(x) \eta(x). \qquad (2.6)$$

b) *Dual pairing between elements in $L^2(\Sigma, \mathrm{dVol}_h) \subset D'(\Sigma)$ with*

$$\langle \pi^{\mathbf{x}}, \varphi^{\mathbf{y}} \rangle = \pi^x \delta_{xy} \varphi^y = \int_U \frac{dx^d}{\sqrt{|h(x)|}} \pi(x) \varphi(x),$$

*where $\pi(x), \varphi(x)$ are identified with densities of weight 1. Notice that the factor $\sqrt{|h(x)|}$ appears in the denominator and $\frac{\pi(x)}{\sqrt{|h(x)|}}, \frac{\varphi(x)}{\sqrt{|h(x)|}}$ are $L^2(\Sigma, \mathrm{dVol}_h)$ representatives.*

c) *We can also see the injection $\delta^{\mathbf{xy}} : C_c^\infty(\Sigma) \hookrightarrow D'(\Sigma)$ such that*

$$\eta^y = \eta_x \delta^{xy} = \int_U dx^d \sqrt{|h(x)|} \delta^d(x-y) \eta(x) = \sqrt{|h(y)|} \eta(y)$$

*is a representative of a density of weight 1. On the other hand, $\delta_{\mathbf{xy}}$ is a densely defined inverse*

$$\varphi_y = \varphi^x \delta_{xy} = \int_U \frac{dx^d}{\sqrt{|h(x)|}} \delta^d(x-y) \varphi(x) = \frac{\varphi(y)}{\sqrt{|h(y)|}}$$

*such that $\delta^{xz} \delta_{zy} = \delta_y^x$ is the evaluation mapping*

$$\delta_y^x \eta_x = \eta_y = \eta(y)$$

*which only makes sense over $C_c^\infty(\Sigma)$.*

The differences between $\delta^{\mathbf{xy}}$, $\delta_{\mathbf{xy}}$, and $\delta_{\mathbf{y}}^{\mathbf{x}}$ are often misinterpreted in the literature since all three can be identified with the Dirac delta in $\mathbb{R}^d$ with the Lebesgue measure.

In this picture, elements of $D'(\Sigma)$ are distributions used to integrate the elements of $C_c^\infty(\Sigma)$, and this nontrivial structure is reframed in the rigged Hilbert space with a nontrivial representation of the scalar product as an integral kernel.

Notice that for different applications, the identification of the field and its dual may be different. Thus, in classical field theory, the common approach [*Got&04*, *GIM04*] is to model classical fields as elements of an $n$-th jet bundle with $n$ being the number of derivatives needed to describe the theory. We can say that, in this approach, fields are modeled over elements of $C_c^\infty(\Sigma)$. In those approaches, the dual jet bundle is built upon the concept of scalar density of weight one that is constructed using volume forms $\xi dx^1 \wedge \cdots \wedge dx^d$ with $\xi \in C_c^\infty(\Sigma)$. Thus, the topology of the model space for the dual jet bundle is that of $C_c^\infty(\Sigma)$. This approach makes the treatment significantly easier at the cost of giving up the possibility of describing the duality *with the strong topology. In those approaches, the dual pairing is described as in a).
On the other hand, in algebraic QFT, the situation is radically different. Classical
fields to be quantized are taken as a priori distributions of $D'(\Sigma)$, and equations of
motion must be implemented weakly using arbitrary test functions acted upon by the
adjoint evolution operator. For this reason, we must think of both fields and their
duals as modeled over the same space of distributions, and the dual pairing only
makes sense as in case b) [Bru&15].*

Another important family of examples of rigged Hilbert spaces relies upon the NF
space of Hida test functions, which will be the set used to model the space of pure
quantum field states, considered as Schrödinger wavefunctionals with fields in their
domain. Its construction is significantly more difficult and needs a careful study of
Gaussian integration to be developed in the following sections. For this reason, we
postpone its introduction to section 2.7.

## 2.4 Gaussian measures

As we intend to describe a theory of pure quantum field states, we must endow the
space of quantum states with a scalar product. As we mentioned above, the most
commonly used measures in QFT correspond to Gaussian measures, often referred to
as a given vacuum state. We will review below their main definition and properties,
first for the simple case of real test functions, and then in the case of complex ones,
which will be more useful for us later.

### 2.4.1 The real case

The procedure to define a Gaussian measure $\mu$ over a NF space $\mathcal{N}'$ is straightforward
once we notice that the $\sigma$-algebra generated by its cylinder subsets corresponds to
the Borel $\sigma$-algebra of the weak and strong topologies of $\mathcal{N}'$ [Kuo96]. We develop this
construction in Appendix C. The Bochner-Minlos theorem, see Theorem C.2, provides
an even straighter path to defining $\mu$. First, we consider the bilinear operation
$\xi_x \Delta^{xy} \xi_y$ to be continuous with respect to the topology of $\mathcal{N}$. Then, we consider the
characteristic functional of a Gaussian measure $\mu$

$$C(\xi_{\mathbf{x}}) = \int_{\mathcal{N}'} D\mu(\varphi^{\mathbf{x}}) e^{i\xi_x \varphi^x} := e^{-\frac{1}{2}\xi_x \Delta^{xy} \xi_y}. \tag{2.7}$$

The Bochner-Minlos theorem states that there exists a unique Borel probability
measure that fulfills (2.7). We will use this fact as the definition of $\mu$. In chapter 4,
we will interpret this measure as part of the vacuum. Considered with respect to a
Gel'fand triple,

$$\mathcal{N} \subset \mathcal{H}_\Delta \subset \mathcal{N}',$$

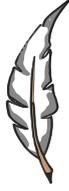

**Notational changes in Schrödinger quantization**

The real measure $\mu$ of this chapter uses a different convention than the rest of the thesis. This is because this definition is adapted to the mathematical discussion while (4.18) and (4.19) are adapted to the Schrödinger representation. The difference amounts to a scale factor of $\frac{1}{2}$ in the covariance.

the bilinear symmetric form $\Delta^{\mathbf{xy}}$ corresponds to the representation of the scalar product of the Cameron-Martin Hilbert space $\mathcal{H}_\Delta$. This space is obtained by completion of $\mathcal{N}$ with the covariance $\Delta^{\mathbf{xy}}$ and is of huge importance in Gaussian analysis because it provides the necessary tools to develop integration, as we will see below. For further mathematical details, see *Example* C.1. This triple also possesses an important physical interpretation in QFT; it represents the Hilbert space of one-particle states, as will be explained around (3.33).

### 2.4.2 The complex case

If we consider the case of complex test functions $\mathcal{N}_\mathbb{C}$ from any of the aforementioned examples of nuclear spaces, most of the properties work in the same form as for the real case. Let us consider the complexification of the space of distributions. In this work, $\phi^{\mathbf{x}}, \sigma^{\mathbf{x}} \in \mathcal{N}'_\mathbb{C}$ will often represent a complex distribution while $\alpha_{\mathbf{x}}, \chi_{\mathbf{x}}, \rho_{\mathbf{x}} \in \mathcal{N}_\mathbb{C}$ will usually represent complex test functions. We relate the complex holomorphic coordinate to the real ones as

$$\phi^{\mathbf{x}} = \frac{1}{\sqrt{2}}(\varphi^{\mathbf{x}} - i\pi^{\mathbf{x}}), \tag{2.8}$$

for $(\varphi^{\mathbf{x}}, \pi^{\mathbf{x}}) \in \mathcal{N}' \times \mathcal{N}'$, from which follows $\bar{\phi}^{\mathbf{x}} = \frac{1}{\sqrt{2}}(\varphi^{\mathbf{x}} + i\pi^{\mathbf{x}})$. We use this sign convention for the holomorphic coordinate to ease the discussion of chapter 4. Complex test functions use the dual conjugate convention; for $\rho_{\mathbf{x}} \in \mathcal{N}_\mathbb{C}$,

$$\rho_{\mathbf{x}} = \frac{1}{\sqrt{2}}(\xi_{\mathbf{x}} + i\eta_{\mathbf{x}}) \tag{2.9}$$

for $(\xi_{\mathbf{x}}, \eta_{\mathbf{x}}) \in \mathcal{N} \times \mathcal{N}$ and $\bar{\rho}_x = \frac{1}{\sqrt{2}}(\xi_{\mathbf{x}} - i\eta_{\mathbf{x}})$. With these conventions, the dual pairing is written $\langle \rho_{\mathbf{x}}, \phi^{\mathbf{y}} \rangle = \rho_x \phi^x$. If we endow each copy of the Cartesian product with the measure given by (2.7), we can define a measure $\mu_c$ on the complex domain defined by the characteristic functional.

$$C_c(\rho_{\mathbf{x}}, \bar{\rho}_{\mathbf{y}}) = \int_{\mathcal{N}'_\mathbb{C}} D\mu_c(\bar{\phi}^x, \phi^{\mathbf{x}}) e^{i(\rho_x \phi^x + \overline{\rho_x \phi^x})} = e^{-\bar{\rho}_x \Delta^{xy} \rho_y}, \tag{2.10}$$

where now the covariance $\Delta^{\mathbf{xy}}$ will be a Hermitian form on $\mathcal{N}_{\mathbb{C}}$, i.e. $\Delta^{\mathbf{xy}} = (\Delta^*)^{\mathbf{xy}}$. Notice that, by construction, the complex measure is the product measure of two real ones. In this case, the associated Cameron-Martin Gel'fand triple is

$$\mathcal{N}_{\mathbb{C}} \subset \mathcal{H}_\Delta \subset \mathcal{N}_{\mathbb{C}}'^*,$$

The conjugation $*$ follows from our conventions (2.8) and (2.9). To fix notation, we indicate that, throughout this work, the strong dual will be denoted by $\prime$ while $*$ indicates only complex conjugation.

### 2.4.3 Integral transforms on Gaussian measures

Among the different integral transforms defined over a Gaussian measure, the most relevant for QFT is the $S$ transform. Throughout this whole section, we will consider a Gaussian measure $\mu$ defined on $\mathcal{N}'$ or $\mathcal{N}_{\mathbb{C}}'$ with the associated characteristic functional (2.10). For the real case, we define:

**Definition 2.5.** *($S_\mu$ transform) For the Gaussian measure of* (2.7), *we define the $S_\mu$ transform as a generalization of the Laplace transform. For a given function $\Psi : \mathcal{N}' \to \mathbb{C}$, it takes the form*

$$S_\mu[\Psi](\xi_{\mathbf{x}}) = \int_{\mathcal{N}'} D\mu(\varphi^{\mathbf{x}})\Psi(\varphi^{\mathbf{x}})\exp\left(\xi_x\varphi^x - \frac{1}{2}\xi_x\Delta^{xy}\xi_y\right) \tag{2.11}$$

This is to be thought of as the action of a translation $\varphi^{\mathbf{x}} \to \varphi^{\mathbf{x}} + \Delta^{\mathbf{xy}}\xi_y$ on the domain of integration. For the complex case, it is readily defined in (anti-)holomorphic coordinates as

**Definition 2.6.** *($S_{\mu_c}$ complex transform) For the complex Gaussian measure of* (2.10) *and a given function $\Psi : \mathcal{N}_{\mathbb{C}}' \to \mathbb{C}$, the transform takes the form*

$$S_{\mu_c}[\Psi](\rho_{\mathbf{x}}, \bar{\rho}_{\mathbf{x}}) = \int_{\mathcal{N}_{\mathbb{C}}'} D\mu_c(\phi^{\mathbf{x}})\Psi(\phi^{\mathbf{x}}, \bar{\phi}^{\mathbf{x}})\exp\left(\overline{\rho_x\phi^x} + \rho_x\phi^x - \bar{\rho}_x\Delta^{xy}\rho_y\right) \tag{2.12}$$

Notice that this definition is nothing but the combination of two copies of **Definition** 2.5, one for each variable, and the required change to holomorphic coordinates.

Another relevant transform is the $T$ transform. It is the extension of the definition of the characteristic functional (2.10) to include a generic function besides the measure, defined as follows:

**Definition 2.7.** *($T_{\mu_c}$ complex transform) For the complex Gaussian measure of* (2.10) *and a given $\Psi : \mathcal{N}_{\mathbb{C}}' \to \mathbb{C}$, it takes the form*

$$T_{\mu_c}[\Psi](\rho_{\mathbf{x}}, \bar{\rho}_{\mathbf{y}}) = \int_{\mathcal{N}_{\mathbb{C}}'} D\mu_c(\phi^{\mathbf{x}})e^{i\left(\overline{\rho_z\bar{\phi}^z} + \rho_z\phi^z\right)}\Psi(\phi^{\mathbf{x}}, \bar{\phi}^{\mathbf{y}}) \tag{2.13}$$

Both transforms are related by $C(\rho_{\mathbf{x}}, \bar{\rho}_{\mathbf{y}})S_{\mu_c}[\Psi](\rho_{\mathbf{x}}, \bar{\rho}_{\mathbf{x}}) = T_{\mu_c}[\Psi](-i\rho_{\mathbf{x}}, -i\bar{\rho}_{\mathbf{y}})$. Each of these transformations reveals different features of $L^2(\mathcal{N}_{\mathbb{C}}', D\mu_c)$, as we will see below.

## 2.5 The Gaussian Hilbert space $L^2(\mathcal{N}'_{\mathbb{C}}, D\mu_c)$ and dense subalgebras

The Hilbert space $L^2(\mathcal{N}'_{\mathbb{C}}, D\mu_c)$ will represent, in [chapter 4], the Hilbert space of pure quantum states of a QFT of the scalar field. In this section, we introduce it through some of its subalgebras. The structure of the Hilbert space $L^2(\mathcal{N}'_{\mathbb{C}}, D\mu_c)$ may seem difficult to tackle. At first glance, we are dealing with functions with domains in an infinite-dimensional space of distributions. This is sometimes referred to as working at the *second quantized* (*s.q.*) level in the literature. In order to study the whole space, it is useful to consider families of functions that are dense in the space $L^2(\mathcal{N}'_{\mathbb{C}}, D\mu_c)$ and that simplify the definition of the objects we are interested in. Here we include the algebras of polynomials $\mathcal{P}$, trigonometric exponentials $\mathcal{T}$, and coherent states $\mathcal{C}$. We will provide a physical interpretation of each set as wavefunctions of a finite number of particles, generators of the classical algebra of observables, and generators of the quantization mapping, respectively, in [chapter 3] and [chapter 4]. We claim that they are dense subsets of the whole space; for a proof of this statement in each case, we refer to general arguments made at the beginning of [Hid&93]. Each of these families is easily and unambiguously defined thanks to the properties of nuclear spaces, and each of them shows different properties when combined with the $S$ and $T$ transforms introduced in the previous section.

### 2.5.1 The algebra of Polynomials

The most important dense subset in Gaussian analysis is polynomials. The fact that $\mathcal{N}'_{\mathbb{C}}$ is a nuclear space implies that the tensor product $\mathcal{N}'^{\otimes k}_{\mathbb{C}}$ is unambiguously defined, as well as $\mathcal{N}'^{\otimes n}_{\mathbb{C}}$ for $k, n$ positive integers. With this remark, we define the algebra of polynomials $\mathcal{P}(\mathcal{N}'_{\mathbb{C}})$, with coefficients in $\mathcal{N}_{\mathbb{C}}$, as the set of polynomials of arbitrary degrees in which a polynomial of degrees $(n, m)$ is written as

$$P_{nm}(\phi^{\mathbf{x}}, \bar{\phi}^{\mathbf{x}}) = \sum_{k=0}^{n} \sum_{l=0}^{m} C^{kl}_{\vec{x}_k, \vec{y}_l} (\phi^k)^{\vec{x}_k} (\bar{\phi}^l)^{\vec{y}_l}, \tag{2.14}$$

where $C^{kl}_{\vec{x}_k, \vec{y}_l} \in \mathcal{N}^{\otimes k+l}_{\mathbb{C}}$ and $(\phi^k)^{\vec{x}_k} \in \mathcal{N}'^{\otimes k}_{\mathbb{C}}$, $(\bar{\phi}^l)^{\vec{x}_l} \in \mathcal{N}'^{*\otimes l}_{\mathbb{C}}$ are the tensor products of copies of the same distribution. Here we introduce the notation $\alpha_{\vec{x}_n} := \alpha_{x_1, \cdots, x_n}$. To make sense of this expression, it is enough to consider a coefficient $C^{kl}_{\vec{x}_k, \vec{y}_l} = \chi^1_{x_1} \cdots \chi^k_{x_k} \bar{\rho}^1_{y_1} \cdots \bar{\rho}^l_{y_l}$ because these coefficients generate the whole space. Then, using the natural dual pairing of $\mathcal{N}_{\mathbb{C}}$ with its dual expressed as $\langle \chi, \phi \rangle = \chi_x \phi^x$, we get

$$C^{kl}_{\vec{x}_k, \vec{y}_l} (\phi^k)^{\vec{x}_k} (\bar{\phi}^l)^{\vec{y}_l} = \prod_{i=1}^{k} \langle \chi^i, \phi \rangle \prod_{j=1}^{l} \overline{\langle \rho^j, \phi \rangle}$$

Thus, polynomials defined in (2.14) have pointwise meaning because each dual pairing is finite. Nonetheless, there is still an ambiguity in this definition because the monomials are contractions of $n$ or $m$ products of the same distribution. This fact implies there are different coefficients that provide the same polynomial. To eliminate this ambiguity, we introduce the symmetrization operator

$$\alpha_{(\vec{x}_n)} = \alpha_{(x_1, \cdots, x_n)} := \frac{1}{n!} \sum_{\pi \in S_n} \alpha_{x_{\pi(1)}, \cdots, x_{\pi(n)}}. \tag{2.15}$$

Here $S_n$ is the space of permutations of $n$ elements. Then, for polynomials to be unambiguously defined through their coefficients, we ask for them to fulfill $C^{kl}_{(\vec{x}_k),(\vec{y}_l)} = C^{kl}_{\vec{x}_k, \vec{y}_l}$, and we call them Bosonic coefficients. Among polynomials, we can find an orthonormal basis of the whole space and then describe it efficiently through its Bosonic coefficients in an appropriate way using the $S$ transform. This feature is so rich and enlightening that it deserves its own separate analysis. That will be the subject of study in section 2.6.

### 2.5.2 The algebra of trigonometric exponentials: The nature of Reproducing Kernel Hilbert space

Polynomials do not exhaust the classes of dense subsets that provide valuable information on the structure of $L^2(\mathcal{N}_\mathbb{C}, D\mu_c)$. In this section, we turn our attention to the algebra of trigonometric exponentials $\mathcal{T}(\mathcal{N}'_\mathbb{C})$, which can be seen to be a dense subset of $L^2(\mathcal{N}_\mathbb{C}, D\mu_c)$. They are defined as

$$\mathcal{T}(\mathcal{N}'_\mathbb{C}) = \left\{ \text{Finite linear combinations of } \mathcal{E}_\chi = \exp\left(i\chi_x \phi^x + i\overline{\chi_x \phi^x}\right) \right\} \tag{2.16}$$

where $\chi_\mathbf{x} \in \mathcal{N}_\mathbb{C}$. Below, we perform an analysis from the point of view of a basis of trigonometric exponentials. If we consider the $T_{\mu_c}$ transform of **Definition** 2.7, its action over $\mathcal{T}(\mathcal{N}'_\mathbb{C})$ is completely determined by linearity from

$$T_{\mu_c}[\mathcal{E}_\chi](\rho_\mathbf{x}, \bar{\rho}_\mathbf{y}) = C_c(\rho_\mathbf{x} + \chi_\mathbf{x}, \overline{\rho_\mathbf{y} + \chi_\mathbf{y}}). \tag{2.17}$$

Here, $C_c$ is the characteristic functional given by (2.10). Notice that in this equation, $\rho_\mathbf{x}$ plays the role of a variable while $\chi_\mathbf{x}$ is a parameter. Hence, we interpret this expression as a function of $\rho_\mathbf{x}$ and extend the domain of $T_{\mu_c}[\mathcal{E}_\chi](\rho_\mathbf{x}, \bar{\rho}_\mathbf{y})$ to make sense in the whole Cameron-Martin Hilbert space $\rho_\mathbf{x} \in \mathcal{H}_\Delta$. The next step is to construct a reproducing kernel Hilbert space $\mathcal{R}$. In order to do so, we firstly introduce the kernel functional defined as

$$\begin{array}{rccc} RK(\cdot, \cdot): & \mathcal{H}_\Delta \times \mathcal{H}_\Delta & \longrightarrow & \mathbb{R} \\ & (\rho_\mathbf{x}, \chi_\mathbf{y}) & \longmapsto & C_c(\chi_\mathbf{x} - \rho_\mathbf{x}, \overline{\chi_\mathbf{y} - \rho_\mathbf{y}}). \end{array} \tag{2.18}$$

Using this definition, $\mathcal{R}$ is defined from the space of linear combinations of functions (of $\rho, \bar{\rho}$) spanned by the right-hand side of (2.17) as its completion with a sesquilinear scalar product $\langle , \rangle_\mathcal{R}$. It is enough to define this scalar product through its action over trigonometric exponentials under $T_{\mu_c}$ as

$$\langle T_{\mu_c}[\mathcal{E}_\rho], T_{\mu_c}[\mathcal{E}_\mathbf{x}]\rangle_\mathcal{R} = RK(\rho_\mathbf{x}, \chi_\mathbf{x}). \tag{2.19}$$

A key factor in this construction is that $RK(\rho_\mathbf{x}, \bullet) \in \mathcal{R}$ and for every $F \in \mathcal{R}$ we have $\langle RK(\rho_\mathbf{x}, \bullet), F(\bullet)\rangle_\mathcal{R} = F(\rho_\mathbf{x})$; that is, point evaluation is continuous in $\mathcal{R}$, which is the definition of a reproducing kernel Hilbert space.

The remaining part of the discussion is to show that $T_{\mu_c} : L^2(\mathcal{N}'_\mathbb{C}, D\mu_c) \to \mathcal{R}$ is an isometric isomorphism of Hilbert spaces. This easily follows from

$$\int_{\mathcal{N}'_\mathbb{C}} D\mu_c \overline{\mathcal{E}_\chi} \mathcal{E}_\rho = \langle T_{\mu_c}[\mathcal{E}_\chi], T_{\mu_c}[\mathcal{E}_\rho]\rangle_\mathcal{R}. \tag{2.20}$$

From this perspective, we may ask in which sense point evaluation is allowed for elements in $L^2(\mathcal{N}'_\mathbb{C}, D\mu_c)$. It is important to notice that the domain of the functions that form this space is much larger than that of the functions of $\mathcal{R}$, given by $\mathcal{H}_\Delta$. Therefore, there is no hope for the full notion of a reproducing kernel to be found in the former as we did in the latter, and as such, there is no notion of point evaluation to be found. The issue of point evaluation in this case is similar to the issue of defining a Dirac delta in finite-dimensional domains. Similarly to that case, there is a notion of evaluation that makes sense in a distributional sense, as we will see in *Example* 2.10 below. At the level of Hilbert space, the family of functions considered in the next section provides tools to delve further into this question.

### 2.5.3 The algebra of coherent states: Subspace of Holomorphic functions

In this case, we combine features from the two families considered above.

Extending the algebra of polynomials to the limit of infinite degree, there is a natural definition of a holomorphic function.

$$\Psi_{Hol}(\phi^\mathbf{x}) = \sum_{n=0}^\infty \psi_{\vec{x}_n}^{(n)} (\phi^n)^{\vec{x}_n}. \tag{2.21}$$

Here we can allow the Bosonic coefficients $\psi_{\vec{x}_n}$ to be elements of a larger space $\mathcal{H}_\Delta^{\bar\otimes n}$. The price to pay in that extension is that (2.21) is no longer defined pointwise. Instead, we can regard it as an element of $\Psi_{Hol} \in L^2_{Hol}(\mathcal{N}'_\mathbb{C}, D\mu_c)$ with the pertinent convergence restrictions. This is a general feature of the Wiener-Ito decomposition theorem that will be treated in detail in section 2.6. We notice that, following this reasoning, there is a decomposition in holomorphic and antiholomorphic functions as follows:

$$L^2(\mathcal{N}'_\mathbb{C}, D\mu_c) = L^2_{Hol}(\mathcal{N}'_\mathbb{C}, D\mu_c) \bar\otimes \overline{L^2_{Hol}(\mathcal{N}'_\mathbb{C}, D\mu_c)} \tag{2.22}$$

where $\bar\otimes$ represents the topological tensor product completed with the natural Hilbert topology.

Let us particularize our analysis to the subspace $L^2_{Hol}(\mathcal{N}'_\mathbb{C}, D\mu_c)$. Besides holomorphic polynomials, this space contains another dense subspace of interest. We introduce the algebra of functions generated by (holomorphic) coherent states $\mathcal{K}_\chi(\phi^\mathbf{y}) = e^{\chi_x \phi^x}$ and we denote it

$$\mathcal{C}_{Hol}(\mathcal{N}'_\mathbb{C}) = \{\text{Finite linear combinations of } \mathcal{K}_\chi = \exp(\chi_x \phi^x)\}. \qquad (2.23)$$

Using this subset, calculus can be easily implemented on the whole space by just considering a few rules. First, notice that the scalar product of two coherent states corresponds to

$$\langle \mathcal{K}_\chi, \mathcal{K}_\rho \rangle = e^{\bar{\chi}_x \Delta^{xy} \rho_y},$$

from which, by linearity, follows the following reproducing property: let $\Psi \in L^2_{Hol}(\mathcal{N}'_\mathbb{C}, D\mu_c)$, then

$$\langle \Psi, \mathcal{K}_\chi \rangle = \Psi(\chi_x \Delta^{x\mathbf{y}}).$$

Recall that we extended (2.21) to have coefficients in $\mathcal{H}_\Delta^{\otimes n}$. Thus, we conclude that the evaluation above provides a finite result and $\mathcal{K}_\chi(\phi^\mathbf{y}) \in L^2_{Hol}(\mathcal{N}'_\mathbb{C}, D\mu_c)$. Those coherent states can be extended to be parametrized by elements $\rho^\mathbf{x} \in \mathcal{H}_\Delta$ regarded as distributions with $\mathcal{K}_\rho(\phi^\mathbf{y}) = e^{\bar{\rho}^x K_{xy} \phi^x}$, the duality expressed as in option b) of the previous section in terms of the inverse of the covariance, $K_{\mathbf{xy}}$, and therefore, lacking any pointwise meaning.

This subspace possesses good analytical properties similar to those of holomorphic functions in finite dimensions. We can think about it as the Hardy space $H^2(\mathcal{H}_\Delta) \subset Hol(\mathcal{H}_\Delta)$ of square integrable holomorphic functions with domain an infinite-dimensional Hilbert space $\mathcal{H}_\Delta$ whose orthogonal directions are provided by the covariance $\Delta^{xy}$. There is a subtlety with the sense of holomorphic function that needs to be treated with care, though. In order to do that, let us first introduce a more general notion of holomorphic function than the one introduced above:

**Definition 2.8. (Holomorphic functions on locally convex spaces)** *Let $\Psi : \mathcal{E} \to \mathbb{C}$ be a function defined on an open set $U \subset \mathcal{E}$ where $\mathcal{E}$ is a locally convex space. In this work, we take $\mathcal{E} = \mathcal{H}_\Delta, \mathcal{N}_\mathbb{C}$ or $\mathcal{N}'_\mathbb{C}$. $\Psi$ is said to be Gâteaux-holomorphic on $U$ if for all $\phi^\mathbf{x}_0, \phi^\mathbf{x} \in U$ the function*

$$\begin{array}{rccc} \Psi_{\phi^\mathbf{x}_0, \phi^\mathbf{x}} : & \mathbb{C} & \longrightarrow & \mathbb{C} \\ & z & \longmapsto & \Psi(\phi^\mathbf{x}_0 + z\phi^\mathbf{x}) \end{array} \qquad (2.24)$$

*is holomorphic.*

*As it usually happens in infinite dimensions, continuity does not follow from differentiability, and it must be explicitly requested [Din81]. Thus, we will say that a function $\Psi$ is holomorphic on $U$ if it is Gâteaux-holomorphic and locally bounded. We will denote then the space of holomorphic functions as $Hol(\mathcal{E})$, assuming it is endowed with the topology of convergence on bounded subsets.*

Notice here that the definition of a holomorphic function requires the pointwise meaning of $\Psi$. This is not true if the decomposition in (2.21) is allowed to have coefficients in $\mathcal{H}_\Delta^{\otimes n}$. For $\Psi \in L^2_{Hol}(\mathcal{N}'_{\mathbb{C}}, D\mu_c)$ and arbitrary elements of $\phi^{\mathsf{x}} \in \mathcal{N}'$, we have that $\Psi(\phi^{\mathsf{x}})$ does not necessarily converge. However, consider this notion of $Hol(\mathcal{H}_\Delta)$ over the Hilbert space $\mathcal{H}_\Delta$. The convergence of (2.21) is ensured if we restrict $\phi^{\mathsf{x}} \in \mathcal{H}_\Delta \subset \mathcal{N}'$. The content of this statement can be made precise using coherent states as follows. Let us consider the map

$$\langle \cdot, \mathcal{K}_\rho \rangle : L^2_{Hol}(\mathcal{N}'_{\mathbb{C}}, D\mu_c) \to H^2(\mathcal{H}_\Delta) \subset Hol(\mathcal{H}_\Delta) \tag{2.25}$$

which is an injective continuous linear mapping [Oec12b], and $H^2(\mathcal{H}_\Delta)$ stands for its image. Moreover, the reproducing kernel

$$RK_{Hol}(\chi_{\mathbf{x}}, \rho_{\mathbf{y}}) = e^{\overline{\chi}_x \Delta^{xy} \rho_y} \tag{2.26}$$

turns $(H^2(\mathcal{H}_\Delta), RK_{Hol})$ into a reproducing kernel Hilbert space, see [Hid&93], known as the Hardy space of square integrable functions. Furthermore, (2.25) establishes an isometric isomorphism between $L^2_{Hol}(\mathcal{N}'_{\mathbb{C}}, D\mu_c)$ and $H^2(\mathcal{H}_\Delta)$.

A similar discussion can be carried out for the anti-holomorphic functions, but in that case, the isomorphism is realized by

$$\langle \cdot, \overline{\mathcal{K}_\rho} \rangle : L^2_{\overline{Hol}}(\mathcal{N}'_{\mathbb{C}}, D\mu_c) \to \overline{H^2}(\mathcal{H}_\Delta). \tag{2.27}$$

Finally, the $T_{\mu_c}$ transform of **Definition** 2.7 can be written with these objects as $T_{\mu_c} = \langle \overline{\mathcal{K}_{i\rho}} \otimes \mathcal{K}_{-i\rho}, \cdot \rangle$. Therefore, there is an isometric isomorphism of reproducing kernel Hilbert spaces such that

$$H^2(\mathcal{H}_\Delta) \otimes \overline{H^2}(\mathcal{H}_\Delta) \cong \mathcal{R}, \tag{2.28}$$

where $\mathcal{R}$ is the reproducing kernel Hilbert space described in the previous section.

### 2.5.4 Some physical implications

So far, we have seen how the dense subsets of trigonometric exponentials and coherent states unveil several properties of the space $L^2(\mathcal{N}'_{\mathbb{C}}, D\mu_c)$. In chapter 3, this kind of space will play a twofold role. On one hand, a Gaussian Hilbert space of complex domain is a prequantum Hilbert space. This space will be the Hilbert space of the Koopman-van Hove (KvH) representation of classical field theory. In chapter 4, this is also a basic ingredient in the geometric quantization procedure. On the other hand, a similar Gaussian space (with different covariance) represents the class of classical functions that can be represented as bounded operators in chapter 3 or quantized under Weyl quantization in chapter 4.

Among the properties above, the most important one for geometric quantization is the decomposition into holomorphic and antiholomorphic parts (2.22). This is because, as we will see in chapter 4, the Hilbert space of pure states in quantum field theory can be represented in several equivalent ways, from which the holomorphic representation selects the subspace $L^2_{Hol}(\mathcal{N}'_{\mathbb{C}}, D\mu_c)$. Finally, the fact that (2.25) provides an isomorphism of Hilbert spaces with $H^2(\mathcal{H}_\Delta)$, a reproducing kernel Hilbert space with a well-behaved domain $\mathcal{H}_\Delta$, ensures nice analytical properties for $L^2_{Hol}(\mathcal{N}'_{\mathbb{C}}, D\mu_c)$ that allow for efficient calculus and study of field theory while keeping technicalities under control.

For the case of Weyl quantization in section 4.4, we show that it is enough to understand the quantization of trigonometric exponentials (2.16) as plane waves of classical field theory for the machinery of reproducing kernels to quantize them according to the Weyl or the canonical commutation relations.

The features unveiled by the family of polynomials are much richer, and we devote the next section to them. We advance that the Wiener-Ito decomposition theorem will be the tool that provides the particle interpretation of the QFT.

## 2.6 Wiener-Ito Decomposition in Homogeneous Chaos

In this section, we will present the central result in Gaussian analysis, the Wiener-Ito chaos decomposition theorem. This result identifies the space $L^2(\mathcal{N}', D\mu)$ with the Bosonic Fock space built upon the Cameron-Martin Hilbert space $\mathcal{H}_\Delta$. The latter can be expressed as $\oplus_{n=0}^{\infty} \mathcal{H}_\Delta^{n\hat{\oplus}}$, where $\oplus$ represents the orthogonal direct sum and $n\hat{\otimes}$ means the symmetric tensor product of $n$ copies of the Hilbert space. In order to skip technicalities, we provided definitions and proofs of the central results of this section in Appendix D. This, in turn, provides the necessary tools to represent the concept of pure states of a QFT and their main properties and defines the concept of physical particles.

In the same vein, we will define other common tools of Gaussian analysis: Malliavin derivatives and Skorokhod integrals. These are introduced as weak Sobolev derivatives and their adjoint operators, respectively. Along the same lines, we identify these with creation and annihilation operators in QFT, defining in the process the notion of charge and particle number.

### 2.6.1 The Segal Isomorphism: Wiener-Ito Chaos Decomposition and the Fock Space

We will start by dealing with the complex case and then particularize our result to the real case. The latter is shown in detail in Appendix D. If the Gel'fand triple discussed above is considered with respect to complex spaces, we can build the triple

$$\mathcal{N}_{\mathbb{C}} \subset \mathcal{H}_\Delta \subset \mathcal{N}'^*_{\mathbb{C}}, \tag{2.29}$$

where $\mathcal{H}_\Delta$ represents the now complex Cameron-Martin Hilbert space. The conjugation $*$ follows from our conventions (2.8) and (2.9). By using this triple, we can consider $\mathcal{N}_\mathbb{C}$ as a linear subspace of $\mathcal{N}'^*_\mathbb{C}$.

Now, we move to the *second quantized* (*s.q.*) level. We therefore deal with the Hilbert space $L^2(\mathcal{N}'_\mathbb{C}, D\mu_c)$. Let us consider now an orthogonal decomposition of the algebra of polynomials $\mathcal{P}(\mathcal{N}'_\mathbb{C})$ introduced above in (2.14). Thus, we define $\mathcal{W}^{:(n,m):}_\mu$ and denote it as the space of homogeneous complex chaos of degrees $(n, m)$ as

$$\mathcal{W}^{:(n,m):}_{\mu_c} = \overline{\mathcal{P}^{(n,m)}} \cap \overline{\mathcal{P}^{(n-1,m)}}^\perp \cap \overline{\mathcal{P}^{(n,m-1)}}^\perp, \qquad (2.30)$$

where $\overline{\mathcal{P}^{(n,m)}}$ represents the completion (in $L^2(\mathcal{N}'_\mathbb{C}, D\mu_c)$) of the set of polynomials of degrees $(n, m)$, and $\overline{\mathcal{P}^{(n-1,m)}}^\perp$ is the orthogonal complement of the completion of the set of polynomials of degrees $(n - 1, m)$. The sequence of spaces $\mathcal{W}^{:(n,m):}_\mu$ contains a basis of orthogonal polynomials for the Hilbert space $L^2(\mathcal{N}'_\mathbb{C}, D\mu_c)$. In addition, these spaces are generated by complex Wick monomials of the Gaussian measure $\mu_c$ (or, equivalently, of the covariance $\Delta^{\mathbf{xy}}$), generalizing the definition of complex Hermite polynomials to this more abstract setting [Hid80] as follows:

**Definition 2.9. (Complex Wick Monomials)** *Let $\Delta^{\mathbf{xy}}$ be the covariance of the Gaussian measure $\mu$. We will use $\mathcal{W}_\Delta(\phi^n \bar\phi^m)^{\vec{\mathbf{x}}\vec{\mathbf{y}}}$, with $\vec{x} = (x_i)^n_{i=1}, \vec{y} = (y_i)^m_{i=1}$, to refer to the distribution belonging to the symmetric tensor product of $n$ copies on $\mathcal{N}'_\mathbb{C}$ and the symmetric product of $m$ copies on $\mathcal{N}'^*_\mathbb{C}$, i.e., $\mathcal{W}_\Delta(\phi^n \bar\phi^m)^{\vec{\mathbf{x}}\vec{\mathbf{y}}} \in \mathcal{N}'^{\hat\otimes n}_\mathbb{C} \otimes \mathcal{N}'^{*\hat\otimes m}_\mathbb{C}$ such that*

$$\mathcal{W}_\Delta(\phi^n \bar\phi^m)^{\vec{x}\vec{y}} := \prod_{i=1}^n \prod_{j=1}^m \frac{\delta}{\delta\rho_{x_i}} \frac{\delta}{\delta\bar\rho_{y_j}} \exp\overline{\left[ \rho_z\phi^{\bar{z}} + \rho_z\phi^z - \bar\rho_u\Delta^{uv}\rho_v \right]}\Bigg|_{\rho_{\mathbf{x}}=0}. \qquad (2.31)$$

*We will refer to this distribution as a complex Wick monomial of degrees $(n, m)$ with respect to the covariance $\Delta^{\mathbf{xy}}$.*

With this set of polynomials, we can write any state in the Hilbert space $L^2(\mathcal{N}_\mathbb{C}, D\mu_c)$ as a linear combination:

$$\Psi(\phi^{\mathbf{x}}, \bar\phi^{\mathbf{y}}) = \sum_{n,m=0}^\infty \psi^{(n,\bar m)}_{\vec{x}_n \vec{y}_m} \mathcal{W}_\Delta(\phi^n \bar\phi^m)^{\vec{x}_n \vec{y}_m} \qquad (2.32)$$

We call this decomposition the Wiener-Ito decomposition in homogeneous chaos. Notice that the set of polynomials can be decomposed into holomorphic and anti-holomorphic parts:

$$\mathcal{P}_{Hol}(\mathcal{N}'_\mathbb{C}) \otimes \mathcal{P}_{\overline{Hol}}(\mathcal{N}'^*_\mathbb{C}) \subset L^2(\mathcal{N}'_\mathbb{C}, D\mu_c) = L^2_{Hol}(\mathcal{N}'_\mathbb{C}, D\mu_c) \bar\otimes L^2_{\overline{Hol}}(\mathcal{N}'_\mathbb{C}, D\mu_c). \qquad (2.33)$$

This decomposition provides an exceptional tool for Gaussian analysis. In terms of the coefficients in the Wiener-Ito decomposition, we can write the scalar product as:

$$\int_{\mathcal{N}'_\mathbb{C}} D\mu_c(\phi^{\mathbf{x}}) \overline{\Phi}(\phi^{\mathbf{x}}) \Psi(\phi^{\mathbf{x}}) =$$

$$\sum_{n,m=0}^{\infty} n! m! \, \overline{\varphi^{(n,\bar{m})}}_{\vec{u}_n \vec{v}_m} \left[ (\Delta^*)^n \right]^{\vec{u}_n \vec{x}_n} \left[ \Delta^m \right]^{\vec{v}_m \vec{y}_m} \psi^{(n,\bar{m})}_{\vec{x}_n \vec{y}_m} =$$

$$\sum_{n,m=0}^{\infty} n! m! \langle \varphi^{(n,\bar{m})}_{\vec{x}_n \vec{y}_m}, \psi^{(n,\bar{m})}_{\vec{x}_n \vec{y}_m} \rangle_{\mathcal{H}^{(n,\bar{m})}_\Delta} \tag{2.34}$$

Here, $\varphi^{(n,\bar{m})}_{\vec{x}_n \vec{y}_m}$ represents the coefficients of the chaos decomposition of $\Phi$. Therefore, once weighted with $\sqrt{n!m!}$, the coefficients of the Wiener-Ito decomposition belong to the Bosonic Fock space $\Gamma \mathcal{H}_\Delta \otimes \Gamma \mathcal{H}^*_\Delta$, where $\Gamma \mathcal{H}^*_\Delta$ represents the complex conjugate of the Fock space. For a definition of this space, see [section D.2](#). To find the coefficients of the decomposition, it is useful to use the $S_\mu$ transform as:

$$\psi^{(n,\bar{m})}_{\vec{x}_n \vec{y}_m} = \frac{1}{n!m!} \prod_{i=1}^{n} \prod_{j=1}^{m} K_{x_i u_i} K^*_{y_j v_j} \frac{\partial}{\partial \bar{\rho}_{u_i}} \frac{\partial}{\partial \rho_{v_j}} S_\mu[\Psi](\rho_{\mathbf{x}}, \bar{\rho}_{\mathbf{x}}) \Big|_{\rho_{\mathbf{x}}=0} \tag{2.35}$$

Let us recall that $K_{\mathbf{xy}}$ is the (weak) inverse of the covariance $\Delta^{\mathbf{xy}}$. We see that the decomposition of (2.34) is such that

$$L^2(\mathcal{N}'_\mathbb{C}, D\mu_c) = L^2_{Hol}(\mathcal{N}'_\mathbb{C}, D\mu_c) \bar{\otimes} L^2_{\overline{Hol}}(\mathcal{N}'_\mathbb{C}, D\mu_c) \cong \bigoplus_{n=0}^{\infty} \left( \bigoplus_{m=0}^{n} \mathcal{H}^{(n-m,\bar{m})}_\Delta \right). \tag{2.36}$$

This leads us to build an isomorphism between the Hilbert space and the Fock space, which associates each state in $L^2(\mathcal{N}'_\mathbb{C}, D\mu_c)$ with its coordinates in the linear decomposition:

**Definition 2.10. *(Segal Isomorphism)*** *We will call the Segal isomorphism the mapping*

$$\begin{aligned} \mathcal{I}: \quad L^2(\mathcal{N}'_\mathbb{C}, D\mu_c) \quad &\longrightarrow \quad \Gamma \mathcal{H}_\Delta \otimes \Gamma \mathcal{H}^*_\Delta \\ \Psi \quad &\longmapsto \quad \mathcal{I}(\Psi) = \left( \sqrt{n!m!} \psi^{(n,\bar{m})}_{\vec{x}_n \vec{y}_m} \right)_{n,m=0}^{\infty} \end{aligned} \tag{2.37}$$

*which defines a unitary isomorphism of Hilbert spaces, considering the scalar product defined in (2.34). In this decomposition, the Fock space is further decomposed into the orthogonal direct sum $\Gamma \mathcal{H}_\Delta \otimes \Gamma \mathcal{H}^*_\Delta = \bigoplus_{n=0}^{\infty} \left( \bigoplus_{m=0}^{n} \mathcal{H}^{(n-m,\bar{m})}_\Delta \right)$.*

Two cases are particularly relevant for us:

**The Holomorphic Representation**   The holomorphic representation is described by the holomorphic subspace $L^2_{Hol}\left(\mathcal{N}'_{\mathbb{C}}, D\mu_c\right)$. Their Wick monomials are particularly simple: $\mathcal{W}_\Delta(\phi^n\bar\phi^0)^{\vec{\mathbf{x}}} = (\phi^n)^{\vec{\mathbf{x}}}$. This fact remarkably simplifies computations because holomorphic functions have a Wiener-Ito chaos decomposition given by

$$\Psi_{Hol}(\phi^{\mathbf{x}}) = \sum_{n=0}^{\infty} \psi^n_{\vec{x}_n}(\phi^n)^{\vec{x}_n} \tag{2.38}$$

independently of the Gaussian measure considered. Besides, the coefficients of each regular monomial belong to orthonormal subspaces.

**The Schrödinger Representation**   For the real case, the subspace is $L^2\left(\mathcal{N}', D\mu\right)$. In this case, we also have a Wiener-Ito decomposition. In fact, this is the case we explicitly prove in [Appendix D]. In that appendix, we work out the decomposition from the real $S$ transform. Therefore, Wick monomials are given by

$$: \varphi^n :\big|^{\vec{\mathbf{x}}}_\Delta = \prod_{i=1}^{n} \frac{\delta}{\delta\xi_{y_i}} \exp\left(\xi_x\varphi^x - \frac{1}{2}\xi_x\Delta^{xy}\xi_y\right)\bigg|_{\xi_{\mathbf{x}}=0} \tag{2.39}$$

or recursively

**Definition 2.11.  (Wick monomials)** *We denote* $: \varphi^n :\big|^{\vec{\mathbf{x}}}_\Delta = \; : \varphi^n :\big|^{\mathbf{x_1},\cdots,\mathbf{x_n}}_\Delta$, *and call the Wick monomial of degree $n$ with respect to the covariance $\Delta^{\mathbf{xy}}$, the distribution belonging to the symmetric tensor product of $n$ copies on $\mathcal{N}'$, i.e.* $: \varphi^n :\big|^{\vec{\mathbf{x}}}_\Delta \in (\mathcal{N}')^{n\hat\otimes}$ *such that*

$$: \varphi^0 :\big|_\Delta = 1$$
$$: \varphi^1 :\big|^{\mathbf{x}}_\Delta = \varphi^{\mathbf{x}}$$
$$: \varphi^n :\big|^{\vec{\mathbf{x}},\mathbf{x_{n-1}},\mathbf{x_n}}_\Delta =: \varphi^{n-1} :\big|^{\vec{\mathbf{x}},\mathbf{x_{n-1}}}_\Delta \varphi^{\mathbf{x_n}} - (n-1)\Delta^{\mathbf{x_n(x_{n-1})}} : \varphi^{n-2} :\big|^{\vec{\mathbf{x}}}_\Delta \tag{2.40}$$

*where the parenthesized indices are symmetrized according to the convention* [(2.15)].

Then the Wiener-Ito decomposition is given by

$$\Psi(\varphi^{\mathbf{x}}) = \sum_{n=0}^{\infty} \psi^{(n)}_{\vec{x}_n} : \varphi^n :\big|^{\vec{x_n}}_\Delta \tag{2.41}$$

the coefficients belong to the weighted Bosonic Fock space obtained from $\Gamma\mathcal{H}_\Delta$ with weights $\sqrt{n!}$ and the scalar product is

$$\int_{\mathcal{N}} D\mu(\varphi^{\mathbf{x}})\overline{\Phi}(\varphi^{\mathbf{x}})\Psi(\varphi^{\mathbf{x}}) = \sum_{n=0}^{\infty} n!\; \overline{\phi^{(n)}}_{\vec{u}_n}\left[\Delta^n\right]^{\vec{u}_n\vec{x}_n}\psi^{(n)}_{\vec{x}_n} = \sum_{n=0}^{\infty} n!\langle\phi^{(n)}_{\vec{\mathbf{x}}_n}, \psi^{(n)}_{\vec{\mathbf{x}}_n}\rangle_{\mathcal{H}^n_\Delta} \tag{2.42}$$

Here $\phi_{\vec{x}_n}^{(n)}$ represents the coefficients of the chaos decomposition of $\Phi$. Notice that this recursive definition of Wick powers coincides with the usual prescription for normal ordering in quantum field theory where $\Delta^{\mathbf{xy}}$ resembles the propagator of the theory considered. However, that prescription is rooted in an ordering choice for creation and annihilation operators. This is not the case for this representation as they appear as generators of orthogonal polynomials built upon one-dimensional Hermite polynomials. This is shown in [Appendix D]. Moreover, $\Delta^{\mathbf{xy}}$ only integrates over spatial coordinates and therefore is not a propagator. Nonetheless, this resemblance allows us to compute pointwise products of functions in terms of their chaos decompositions using the well-known tool of Feynman diagrams [Mos19].

### 2.6.2 Feynman diagrams

In particular problems, such as perturbative expansions, it is often useful to compute chaos decompositions of pointwise products of functions. Let us assume that we have $\Psi, \Phi \in L^2(\mathcal{N}', D\mu)$ and they are such that the pointwise product $\Theta(\varphi^{\mathbf{x}}) = \Psi(\varphi^{\mathbf{x}})\Phi(\varphi^{\mathbf{x}})$. Examples of these functions include the spaces of Hida-Kondratiev test functions introduced below and in [section B.1]. The question at hand is, given the chaos decomposition of $\Psi(\varphi^{\mathbf{x}})$ and $\Phi(\varphi^{\mathbf{x}})$, what is the chaos decomposition of $\Psi(\varphi^{\mathbf{x}})\Phi(\varphi^{\mathbf{x}})$? To answer this question, we show the following identities:

$$: \varphi^n :\big|_{\Delta}^{\vec{x}_n} = \sum_{k=0}^{\lfloor \frac{n}{2} \rfloor} (-1)^k \binom{n}{2k} (2k-1)!! (\Delta^k)^{(\vec{x}_{2k}} \big[\varphi^{n-2k}\big]^{\vec{x}_{n-2k})}$$

$$[\varphi^n]^{\vec{x}} = \sum_{k=0}^{\lfloor \frac{n}{2} \rfloor} \binom{n}{2k} (2k-1)!! (\Delta^k)^{(\vec{x}_{2k}} : \varphi^{n-2k} :\big|_{\Delta}^{\vec{x}_{n-2k})} \qquad (2.43)$$

In these expressions, we must be aware that products of distributions must be dealt with carefully. In particular, Wick monomials can be interpreted as a point-splitting regularization procedure that is needed to make sense of local products of distributions, see *Remark* 4.2 and [TW87]. Taking this into account, we show that, for $m \leq n$, we get

$$: \varphi^n :\big|_{\Delta}^{\vec{x}_n} : \varphi^m :\big|_{\Delta}^{\vec{y}_m} = \sum_{k=0}^{m} k! \binom{n}{k} \binom{m}{k} (\Delta^k)^{\vec{x}_k, \vec{y}_k} : \varphi^{n+m-2k} :\big|_{\Delta}^{\overrightarrow{(x,y)}_{n+m-2k}} \qquad (2.44)$$

To compute integrals of pointwise products of monomials, the definitions given so far are rather difficult to implement. For this reason, we will present a diagrammatic approach to deal with these equations, the well-known Feynman diagram technique. Let $\{\psi_x^i \varphi^x\}_{i=1}^4$ be functions of $H_\Delta$, the Cameron-Martin or one-particle Hilbert space,

$$
\begin{array}{ccc}
\psi_x^1 \; \xrightarrow{\quad \Delta^{xy} \quad} \; \psi_y^3 \\[2em]
\psi_u^2 \; \xrightarrow{\quad \Delta^{uv} \quad} \; \psi_v^4
\end{array}
\quad = (\psi_x^1 \Delta^{xy} \psi_y^3)(\psi_u^2 \Delta^{uv} \psi_v^4)
$$

and the set of half edges $\{\{1,2\},\{3,4\}\}$, we can construct a Feynman diagram as follows:

In the following, we skip the notation of the function by its label and skip the bilinear $\Delta^{xy}$ in each line. To be more precise, let us define rigorously what we mean by a Feynman diagram.

**Definition 2.12. (Feynman diagram)** *A tuple $(V, H)$ is called a Feynman diagram with $n$ vertices and rank $r \leq n/2$ if $V$ is a set of $n$ vertices and $H$ is the so-called set of half edges which consists of $r$ disjoint pairs of vertices. The vertices which do not belong to $H$ are called unpaired vertices and are part of the set $A$.*

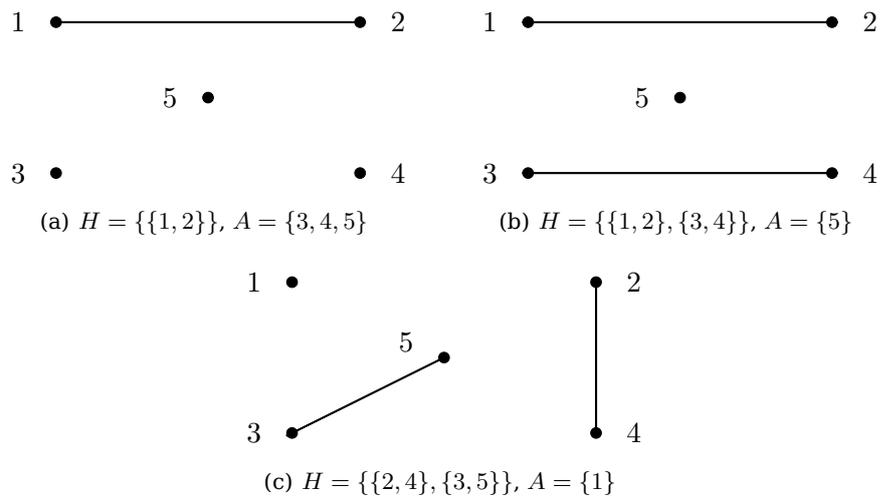

(a) $H = \{\{1,2\}\}$, $A = \{3,4,5\}$      (b) $H = \{\{1,2\},\{3,4\}\}$, $A = \{5\}$

(c) $H = \{\{2,4\},\{3,5\}\}$, $A = \{1\}$

Figure 2.1: Examples of Feynman graphs with $V = \{1,2,3,4,5\}$ and different $H$ and $A$. Figure from [Mos19].

This technique is particularly useful for computing correlations between functions in the homogeneous chaos of order one. These are identified with the Cameron-Martin Hilbert space $\mathcal{H}_\Delta$ and, provided that their product is well-defined, we get

**Theorem 2.1. (Wick theorem)** *Let $\{\psi_x^i \varphi^x\}_{i=1}^n \subset \mathcal{W}_\mu^{;1:}$ be a family of functions of the homogeneous chaos of order one. Provided that the product is well-defined, we*

*obtain*

$$\psi_x^1\varphi^x\cdots\psi_x^n\varphi^x = \sum_{k=0}^{\lfloor n/2\rfloor}\sum_{\substack{\Gamma=\\ \text{rank } k\\ n \text{ vertex diagrams}}}\Gamma_{\vec{x}_{n-2k}}:\varphi^{n-2k}:\big|_\Delta^{\vec{x}_{n-2k}} \qquad (2.45)$$

*Noticing that the integral of any Wick monomial of order greater than zero is null, we get*

$$\int D\mu(\varphi^{\mathbf{x}})\psi_x^1\varphi^x\cdots\psi_x^n\varphi^x = \sum_{\substack{\Gamma=\\ \text{complete}\\ \text{diagrams}\\ \text{of order } n}}\Gamma \qquad (2.46)$$

*where we denote diagrams with even $n$ vertices and rank $n/2$ complete Feynman diagrams.*

Using this result, we interpret the combinatorial factor $\binom{n+2k}{2k}(2k-1)!!$ of (2.43) as the number of Feynman diagrams with $n$ vertices and rank $k$ that one can draw. Notice that they become degenerate due to the symmetrization operator in the indices. With these considerations, we are in a position to answer the question posed at the beginning of the section.

**Example 2.4. (Chaos decomposition of the pointwise product)** *Choose appropriate functions such that $\Theta(\varphi^{\mathbf{x}}) = \Psi(\varphi^{\mathbf{x}})\Phi(\varphi^{\mathbf{x}})$ is well-defined. With chaos decomposition given by $\Psi(\varphi^{\mathbf{x}}) = \sum_n \psi_{\vec{x}_n}^{(n)}:\varphi^n:|_\Delta^{\vec{x}_n}$ and $\Phi(\varphi) = \sum_m \phi_{\vec{x}_m}^{(m)}:\varphi^m:|_\Delta^{\vec{x}_m}$. Then we write*

$$\theta^{(l)}\left(\begin{array}{c}\bullet\\\bullet\\\bullet\\\vdots\\\bullet\\\bullet\\\bullet\end{array}\right) = \sum_{n,m=0}^{\infty}\sum_{n+m-l=2k}k!\binom{n}{k}\binom{m}{k}\ \psi^{(n)}\left(\begin{array}{c}\bullet\\\bullet\\\vdots\\\bullet\\\bullet\\\bullet\end{array}\right.k\left\{\begin{array}{c}\bullet\\\vdots\\\bullet\end{array}\right.\phi^{(m)}\left(\left.\begin{array}{c}\bullet\\\vdots\\\bullet\end{array}\right\}k\right.\left.\begin{array}{c}\bullet\\\vdots\\\bullet\\\bullet\end{array}\right)$$

*Where the free vertices are supposed to be symmetrized. The subgraph with $k$ vertices is any that only has half edges $H$ such that $v \in H$ is of the form $v = \{x_\psi, y_\phi\}$ with $x_f$ being a vertex of $\psi^{(n)}$ and $y_\phi$ being a vertex of $\phi^{(m)}$. The combinatorial factor takes into account how many diagrams of this type can be written down and are equivalent due to the symmetry of $\psi^{(n)}$ and $\phi^{(m)}$.*

## 2.6.3 The Segal-Bargmann transform: Isomorphisms between the real and holomorphic cases.

In the case where the covariance for the complex case, $\Delta^{\mathbf{xy}}$ is real and symmetric, we can establish a unitary isomorphism between the subspace of holomorphic functions $L^2_{Hol}(\mathcal{N}'_{\mathbb{C}}, D\mu_c)$ and $L^2(\mathcal{N}', D\mu)$ just by performing an integration over the imaginary part of the coordinates:

**Definition 2.13. (Segal-Bargmann transform)** *We denote by the Segal-Bargmann transform the unitary isomorphism*

$$\mathcal{B}: \ L^2(\mathcal{N}', D\mu) \ \longrightarrow \ L^2_{Hol}(\mathcal{N}'_{\mathbb{C}}, D\mu_c)$$

*where both $\mu, \mu_c$ are Gaussian measures with a real symmetric covariance $\Delta^{\mathbf{xy}}$, adapted in each case to its domain. To write the explicit formula for this isomorphism, we take an arbitrary $\Psi_{Real} \in L^2(\mathcal{N}', D\mu)$ and denote $\mathcal{B}(\Psi_{Real}) = \Psi_{Hol}$. Then it follows that $\int_{\mathcal{N}'_{\mathbb{C}}} D\mu_c(\phi^{\mathbf{x}}) = \int_{\mathcal{N}'} D\mu(\varphi^{\mathbf{x}}) \int_{\mathcal{N}'} D\mu(\pi^{\mathbf{x}})$. Let $\Psi_{Hol}$ be a holomorphic function with chaos decomposition given by* (2.38), *integrating over the imaginary degrees of freedom*

$$\int D\mu(\pi^{\mathbf{x}})\Psi_{Hol}(\phi^{\mathbf{x}}) = \sum_{n=0}^{\infty} \int D\mu(\pi^{\mathbf{x}})\psi^{(n)}_{\vec{x}_n}[(\varphi - i\pi)^n]^{\vec{x}_n} =$$

$$\sum_{n=0}^{\infty} \psi^{(n)}_{\vec{x}_n} \frac{\partial^n}{\partial \xi^n_{\vec{x}_n}} \int D\mu(\pi^{\mathbf{x}})e^{\xi_z(\varphi - i\pi)^z}\Big|_{\xi_{\mathbf{x}}=0} =$$

$$\sum_{n=0}^{\infty} \psi^{(n)}_{\vec{x}_n} \frac{\partial^n}{\partial \xi^n_{\vec{x}_n}} \left[ e^{\xi_z \varphi^z} \int D\mu(\pi^{\mathbf{x}})e^{-i\xi_z \pi^z} \right]_{\xi_{\mathbf{x}}=0} =$$

$$\sum_{n=0}^{\infty} \psi^{(n)}_{\vec{x}_n} \frac{\partial^n}{\partial \xi^n_{\vec{x}_n}} e^{\xi_x \varphi^x - \frac{\xi_u \Delta^{uv} \xi_v}{2}}\Big|_{\xi_{\mathbf{x}}=0} = \sum_{n=0}^{\infty} \psi^{(n)}_{\vec{x}_n} : \varphi^n :|_{\Delta}^{\vec{x}_n} = \Psi_{Real}(\varphi^{\mathbf{x}})$$

*where $\xi_{\mathbf{x}} \in \mathcal{N}$ is real. With this consideration, $\mathcal{B}(\Psi_{Real})$ is computed by obtaining the chaos decomposition of $\Psi_{Real}$ and replacing Wick monomials with regular holomorphic monomials.*

*A similar isomorphism is obtained using the antiholomorphic subspace in the analogous form*

$$\overline{\mathcal{B}} : L^2(\mathcal{N}', D\mu) \longrightarrow L^2_{\overline{Hol}}(\mathcal{N}'_{\mathbb{C}}, D\mu_c).$$

Notice that this integral expression computes $\mathcal{B}^{-1}$. We will be able to introduce an integral expression for $\mathcal{B}^{-1}$ using the technique of white noise integral kernels in (5.56).

These transforms play a relevant role in the construction of physical quantum field theories, as they allow for the relation of the holomorphic and Schrödinger (or antiholomorphic and field-momentum) functional representations of quantum states with respect to their appropriate Gaussian measures. We will use this fact in chapter 4 to study the relation of different representations of the same QFT.

### 2.6.4 Malliavin calculus: Weak derivatives and Skorokhod integrals

The algebras studied in section 2.5 allow for a well-posed notion of Fréchet derivative defined as in [Hid&93]. This is because we can explicitly write down its expression and check its continuity. We can enlarge, in a weak sense, this notion of derivative to an appropriate subdomain $\mathbb{D}^{2,1}_\mu \subset L^2(\mathcal{N}', D\mu)$. This weak notion of derivative is the Malliavin derivative. We will also present its adjoint operator, the Skorokhod integral. These tools may be used to model further relevant tools in QFT such as creation and annihilation operators, or the number operator. Let us briefly introduce them in the simpler case of a real domain and generalize them later to the case of complex ones.

As in the complex case, a polynomial of a real variable $\varphi^{\mathbf{x}} \in \mathcal{N}'$, with coefficients in $\mathcal{N}_{\mathbb{C}}$ and degree $n$, can be written as

$$P_n(\varphi) = C^0 + C^1_x \varphi^x + C^2_{x_1 x_2} (\varphi^2)^{x_1 x_2} + \cdots + C^n_{\vec{x_n}} (\varphi^n)^{\vec{x_n}} \tag{2.47}$$

The algebra of polynomials of a real variable is denoted by $\mathcal{P}(\mathcal{N}')$. This set is dense in $L^2(\mathcal{N}', D\mu)$, so we start introducing the operations of Malliavin calculus on $\mathcal{P}(\mathcal{N}')$ and then extend them to the whole space by linearity. On polynomials, the Fréchet derivative has a particularly useful expression in terms of the operator $\partial_{\varphi^y} = \partial_{\mathbf{y}}$ that can be defined over monomials as

$$\partial_{\varphi^y} C_{x_1, \cdots, x_n} (\varphi^n)^{x_1, \cdots, x_n} = n C_{y, x_1, \cdots, x_{n-1}} (\varphi^{n-1})^{x_1, \cdots, x_{n-1}}. \tag{2.48}$$

By linearity, this operator is extended to $\mathcal{P}(\mathcal{N}')$ without further complications.

The Malliavin derivative is the extension of the Fréchet derivative, defined in (2.48), to an operator

$$\partial_{\mathbf{x}} : \mathbb{D}^{2,1}_\mu \longrightarrow \mathcal{H}_\Delta \otimes L^2(\mathcal{N}', D\mu) \tag{2.49}$$

where the domain of the operator is

$$\mathbb{D}^{2,1}_\mu = \left\{ \Psi(\varphi^{\mathbf{x}}) = \sum_{n=0}^{\infty} \psi^{(n)}_{\vec{x_n}} : \varphi^n : |^{\vec{x_n}}_\Delta \text{ s.t. } \sum_{n=0}^{\infty} (1+n) \cdot n! \|\psi^{(n)}_{\vec{\mathbf{x}}_n}\|^2_\Delta < \infty \right\} \tag{2.50}$$

and acts on $\Psi(\varphi^{\mathbf{x}})$ as

$$\partial_x \Psi(\varphi^{\mathbf{x}}) = \sum_{n=1}^{\infty} n \psi^{(n)}_{x, \vec{y}_{n-1}} : \varphi^{n-1} : |^{\vec{y}_{n-1}}_\Delta \tag{2.51}$$

The key feature of this derivative is that the resulting derivative makes sense as a $\mathcal{H}_\Delta$ function and does not have a pointwise meaning as in (2.48). This notion is analogous to the construction of the weak derivative in Sobolev spaces, and for this reason, it is sometimes referred to as the Gross-Sobolev derivative [Hu16]. The Malliavin derivative can thus be considered as the generator of derivatives in directions of the Cameron-Martin Hilbert space. Furthermore, through the isomorphism $\mathcal{I}$, the contraction $\Delta^{xy}\partial_y$ is mapped to the annihilation operator on the Fock space $\Gamma(\mathcal{H}_\Delta)$. We will use this in chapter 4 to construct the annihilation operator of QFT.

The adjoint operator of the Malliavin derivative is the Skorokhod integral $\partial^{*\mathbf{x}}$. It is an operator defined over $\mathrm{Dom}(\partial^{*\mathbf{x}}) \subset \mathcal{H}_\Delta \otimes L^2(\mathcal{N}', D\mu)$

$$\partial^{*\mathbf{x}}: \quad \begin{aligned} \mathrm{Dom}(\partial^{*\mathbf{x}}) &\longrightarrow L^2(\mathcal{N}', D\mu) \\ \Psi_{\mathbf{x}}(\varphi^{\mathbf{y}}) &\longmapsto \partial^{*x}\Psi_x(\varphi^{\mathbf{y}}) \end{aligned} \tag{2.52}$$

It is important to remark that the Skorokhod integral is able to integrate more elements than those obtained from (2.51). Indeed, a general element $\Psi_{\mathbf{x}} \in \mathcal{H}_\Delta \otimes L^2(\mathcal{N}', D\mu)$ admits a Wiener-Ito decomposition in which

$$\sum_{n=0}^{\infty} \psi_{x,\vec{y}_n}^{(n)} : \varphi^n :|_\Delta^{\vec{y}_n} \quad \text{in general} \quad \psi_{(x,\vec{y}_n)}^{(n)} \neq \psi_{x,\vec{y}_n}^{(n)} \tag{2.53}$$

In other words, $(\psi_{x,\vec{y}_n}^{(n)})_{n=0}^{\infty}$ is not necessarily symmetrized and hence it may not be an element of the Bosonic Fock space $\Gamma(\mathcal{H}_\Delta)$. Nonetheless, if $\tilde{\psi}_{\vec{z}_{n+1}}^{(n+1)} = \psi_{(\mathbf{x},\vec{y}_n)}^{(n)}$ is indeed the decomposition of an element of $\Gamma(\mathcal{H}_\Delta)$, then $\Psi_{\mathbf{x}} \in \mathrm{Dom}(\partial^{*\mathbf{x}})$ and

$$\partial^{*x}\Psi_x = \sum_{n=1}^{\infty} \tilde{\psi}_{\vec{z}_n}^{(n)} : \varphi^n :|_\Delta^{\vec{z}_n} \tag{2.54}$$

This, for a large class of functions in $\mathrm{Dom}(\partial^{*\mathbf{x}})$, can be written in terms of the Malliavin derivative as

$$\partial^{*x} = -\Delta^{xy}\partial_y + \varphi^x \tag{2.55}$$

Notice that the Skorokhod integral is a generalization of the Ito integral, which coincides with it when the domain has dimension one. In this case, it is identified with a time parameter, and we consider only $\Psi_x$ as a process adapted to the filtration of Brownian motion, see [NØP09]. See chapter 4 for its connection with the creation operator of QFT. It is even possible to extend these operations to the space of Hida distributions $(\mathcal{N}_\mathbb{C})'$ introduced in the next section. For further details on Hida-Malliavin calculus, we refer the interested reader to [NØP09].

Another operator that can be easily written in terms of the latter two is the particle number

$$N_\Delta = \partial^{*x}\partial_x = -\Delta^{xy}\partial_y\partial_x + \varphi^x\partial_x \tag{2.56}$$

The first application that we will find is the extension of $: \quad :|_\Delta$ of **Definition** 2.11 to an operator acting on arbitrary functions

**Example 2.5.** (*Wick order operator* $: \quad :|_\Delta$*) [CD06]*
*Let $F(\varphi^{\mathbf{x}})$ be an arbitrary but well-behaved function. Using Malliavin derivatives, we define*

$$: F(\varphi^{\mathbf{x}}) :|_\Delta = \exp\Big(-\frac{1}{2}\Delta^{xy}\partial_{\varphi^x}\partial_{\varphi^y}\Big)F(\varphi^{\mathbf{x}}) \tag{2.57}$$

*Using the Baker-Campbell-Hausdorff (BCH) formula, we can see that treating $\xi_x\varphi^x$ as a multiplicative operator*

$$\exp\Big(-\frac{1}{2}\Delta^{xy}\partial_{\varphi^x}\partial_{\varphi^y}\Big)\exp\Big(\xi_x\varphi^x\Big)=\exp\Big(\xi_x\varphi^x-\frac{1}{2}\xi_x\Delta^{xy}\xi_y\Big)\exp\Big(-\frac{1}{2}\Delta^{xy}\partial_{\varphi^x}\partial_{\varphi^y}-\xi_x\Delta^{xy}\partial_{\varphi^y}\Big)$$

*Thus, applying these operators to the constant function 1, we get*

$$: e^{\xi_x\varphi^x} :|_\Delta = e^{\xi_x\varphi^x-\frac{1}{2}\xi_x\Delta^{xy}\xi_y}$$

*which coincides with the generator of (2.39). As a consequence, this operator generalizes **Definition** 2.11 to arbitrary functions. As an operator, it is invertible, and*

$$: \; :|_\Delta^{-1} =: \; :|_{-\Delta}. \tag{2.58}$$

These notions can be easily extended to the holomorphic and antiholomorphic cases.

$$\partial_{\phi^x}\Psi(\phi^{\mathbf{x}},\bar\phi^{\mathbf{y}})=\sum_{n,m=0}^{\infty}n\mathcal{W}_\Delta(\phi^{n-1}\bar\phi^m)^{\vec z_{n-1}\vec y_m}\psi^{(n,\bar m)}_{x,\vec z_{n-1}\vec y_m} \tag{2.59}$$

$$\partial_{\bar\phi^x}\Psi(\phi^{\mathbf{x}},\bar\phi^{\mathbf{y}})=\sum_{n,m=0}^{\infty}m\mathcal{W}_\Delta(\phi^n\bar\phi^{m-1})^{\vec z_n\vec y_{m-1}}\psi^{(n,\bar m)}_{\vec z_n\vec y_{m-1},x} \tag{2.60}$$

Here, each derivative must be understood in the Wirtinger sense:

$$\partial_{\phi^x}=\frac{1}{\sqrt2}\big(\partial_{\varphi^x}+i\partial_{\pi^x}\big),\quad \partial_{\bar\phi^x}=\frac{1}{\sqrt2}\big(\partial_{\varphi^x}-i\partial_{\pi^x}\big) \tag{2.61}$$

On the other hand, the associated Skorokhod integrals act on (anti)holomorphic functionals as:

$$\partial^{*\phi^x}=-\Delta^{xy}\partial_{\bar\phi^y}+\phi^x,\qquad \partial^{*\phi^x}\mathcal{W}_\Delta(\phi^n\bar\phi^m)^{\vec z_n\vec y_m}=\mathcal{W}_\Delta(\phi^{n+1}\bar\phi^m)^{x,\vec z_n\vec y_m}$$
$$\partial^{*\bar\phi^x}=-(\Delta^*)^{xy}\partial_{\phi^y}+\bar\phi^x,\qquad \partial^{*\bar\phi^x}\mathcal{W}_\Delta(\phi^n\bar\phi^m)^{\vec z_n\vec y_m}=\mathcal{W}_\Delta(\phi^n\bar\phi^{m+1})^{\vec z_n\vec y_m,x} \tag{2.62}$$

In the same form as in the real case, we can write down the particle number operator $N$ and the $U(1)$ charge $Q$, which can be interpreted as encoders of the matter/antimatter content of the states in QFT:

$$N=\partial^{*\phi^x}\partial_{\phi^x}+\partial^{*\bar\phi^x}\partial_{\bar\phi^x}\qquad\qquad Q=\partial^{*\phi^x}\partial_{\phi^x}-\partial^{*\bar\phi^x}\partial_{\bar\phi^x} \tag{2.63}$$

With those operators, we see that the decomposition of (2.34) is such that

$$L^2\big(\mathcal{N}'_{\mathbb{C}}, D\mu_c\big) \cong \sum_{n=0}^{\infty} \oplus \left(\sum_{m=0}^{n} \oplus \mathcal{H}_{\Delta}^{(n-m,\bar{m})}\right) \tag{2.64}$$

where $\mathcal{H}_{\Delta}^{(n-m,\bar{m})}$ are simultaneous eigenspaces of $N$ and $Q$ with eigenvalues $n$ and $n-2m$, respectively.

We can also generalize *Example* 2.5 to the complex case.

**Example 2.6. (Complex Wick order operator $\mathcal{W}_{\Delta}$)**   *Let $F(\phi^{\mathbf{x}}, \bar{\phi}^{\mathbf{x}})$ be a well-behaved function. Using Malliavin derivatives, we can obtain explicit expressions for the operators $\mathcal{W}_{\Delta}$. Let*

$$\mathcal{W}_{\Delta}\Big[F(\phi^{\mathbf{x}}, \bar{\phi}^{\mathbf{x}})\Big] = \exp\Big(-\Delta^{xy}\partial_{\bar{\phi}^x}\partial_{\phi^y}\Big)F(\phi^{\mathbf{x}}, \bar{\phi}^{\mathbf{x}}). \tag{2.65}$$

*Taking the function $\exp\overline{\big[\rho_z\bar{\phi}^z} + \rho_z\phi^z\big]$ and treating it as a multiplicative operator, the BCH formula states*

$$\exp\Big(-\Delta^{xy}\partial_{\bar{\phi}^x}\partial_{\phi^y}\Big)\exp\overline{\big[\rho_z\bar{\phi}^z} + \rho_z\phi^z\big] =$$
$$\exp\overline{\big[\rho_z\bar{\phi}^z} + \rho_z\phi^z - \bar{\rho}_x\Delta^{xy}\rho_y\big]\exp\Big(-\Delta^{xy}\partial_{\bar{\phi}^x}\partial_{\phi^y} - \bar{\rho}_x\Delta^{xy}\partial_{\phi^y} - \rho_x\Delta^{xy}\partial_{\bar{\phi}^y}\Big).$$

*Thus, applying this operator to the constant function 1, we get*

$$\mathcal{W}_{\Delta}\Big[\exp\overline{\big[\rho_z\bar{\phi}^z} + \rho_z\phi^z\big]\Big] = \exp\overline{\big[\rho_z\bar{\phi}^z} + \rho_z\phi^z - \bar{\rho}_x\Delta^{xy}\rho_y\big]$$

*This is the generator appearing in **Definition** 2.9 and generalizes it to any function $F$. As an operator, it is invertible, and*

$$\mathcal{W}_{\Delta} = \mathcal{W}_{-\Delta}. \tag{2.66}$$

Let us briefly introduce, for later convenience, Gross-Sobolev spaces [Hu16] that extend the concept of (2.50) to any $L^p$-norm and $m$ number of derivatives. This represents the generalization of Gaussian analysis to the standard notion of Sobolev space, see *e.g.* [Tre67]. The $L^p$ norm $\|\cdot\|_{\mu_c,p}$ of a function $\Psi(\phi^{\mathbf{x}}, \bar{\phi}^{\mathbf{x}})$ is

$$\|\Psi\|_{\mu_c,p} = \left(\int_{\mathcal{N}'_{\mathbb{C}}} D\mu(\phi^{\mathbf{x}})|\Psi(\phi^{\mathbf{x}}, \bar{\phi}^{\mathbf{x}})|^p\right)^{\frac{1}{p}}$$

**Definition 2.14. (Gross-Sobolev spaces)**  The holomorphic Gross-Sobolev space $\mathbb{D}_{\mu_c}^{p,m}$ is the Banach space for integers $p \geq 1$ and $m \geq 0$ defined as

$$\mathbb{D}_{\mu_c,h}^{p,m} = \Big\{\Psi(\phi^{\mathbf{x}}) \text{ s.t. } \|(1+N)^{\frac{m}{2}}\Psi\|_{\mu_c,p} < \infty\Big\} \tag{2.67}$$

*For the special case of $m = 0$, we denote $\mathbb{D}^{p,0}_{\mu_c} = L^p_{Hol}(\mathcal{N}'_{\mathbb{C}}, D\mu_c)$, the holomorphic $L^p$ Banach space of the Gaussian measure $\mu_c$. It can be shown [Hu16] that $\mathbb{D}^{p,m}_{\mu_c}$ is the domain where*

$$\partial^m_{\phi^{\otimes_m}} : \mathbb{D}^{p,m}_{\mu_c} \longrightarrow \mathcal{H}^{\otimes m}_\Delta \otimes L^p_{Hol}(\mathcal{N}'_{\mathbb{C}}, D\mu_c)$$

*can be defined as a bounded operator. See Theorem B.3 and the comments below.*

Gross-Sobolev spaces fulfill $\mathbb{D}^{p,m+1}_{\mu_c} \subset \mathbb{D}^{p,m}_{\mu_c}$ and, due to Hölder's inequality for probability measures, $\mathbb{D}^{p+1,m}_{\mu_c} \subset \mathbb{D}^{p,m}_{\mu_c}$. In particular, $L^q(\mathcal{N}'_{\mathbb{C}}, D\mu_c) \subset L^p(\mathcal{N}'_{\mathbb{C}}, D\mu_c)$ for $p \leq q$.

## 2.7 Hida test functions and Hida calculus: *second quantized* test functions

The set of Hida test functions represents a *second quantized* (*s.q.*) notion of test functions that may be considered just an additional example of Nuclear-Fréchet (NF) space. What we mean by second quantized construction is to consider that these Hida test functions are defined as functions on a DNF space $\mathcal{N}'_{\mathbb{C}}$ with coefficients in its dual $\mathcal{N}_{\mathbb{C}}$. We denote $(\mathcal{N}_{\mathbb{C}})$ as the set of Hida test functions built upon $\mathcal{N}_{\mathbb{C}}$ and $(\mathcal{N}_{\mathbb{C}})'$ as its strong dual, the set of Hida distributions. The corresponding nuclear space structure can be seen in (2.72).[4]

Due to its relevance in the description of the space of pure quantum field states, we have prepared a specific section for them and their tools. This concept of *s.q.* test function is far from unique, and we refer to Appendix B for other definitions and a deeper study of their structure. In this work, we use $(\mathcal{N}_{\mathbb{C}})$ because there is a straightforward generalization of Malliavin calculus to the space of Hida distributions $(\mathcal{N}_{\mathbb{C}})'$.

We will introduce $(\mathcal{N}_{\mathbb{C}})$ using rigged Hilbert spaces. Let us first consider the Hilbert space that we will use in the construction of a Gel'fand triple at the level of $\mathcal{N}_{\mathbb{C}}$. Using **Definition** 2.2, we know that the topology of the latter is provided by a sequence of Hilbert spaces $(\mathcal{H}_p, \langle \cdot, \cdot \rangle_p)$ and embeddings $\mathcal{H}_q \hookrightarrow \mathcal{H}_p$, which are Hilbert-Schmidt. We will use these ingredients to build another sequence of spaces and Hilbertian norms at the level of the functions on $\mathcal{N}'_{\mathbb{C}}$ and define, via a projective limit, the corresponding nuclear space structure for $(\mathcal{N}_{\mathbb{C}})$, see (2.72).

Let us consider the Gel'fand triple associated with the case when $\mathcal{N}_{\mathbb{C}}$ consists of complex test functions of compact support (*i.e.* the complex counterpart of equation (2.5)) with respect to the Hilbert space $\mathcal{H}_{Vol} = L^2(\Sigma, \mathrm{dVol}_h)$. This Hilbert space is obtained as the completion of $C^\infty_c(\Sigma)$ with respect to the scalar product induced by the measure $\mathrm{dVol}_h$ associated with a Riemannian metric $h$ on the hypersurface $\Sigma$.

---

[4] Notation in this case, even though standard, could be misleading. Notice that $(\mathcal{N}_{\mathbb{C}})$ represents Hida test functions with domain $\mathcal{N}'_{\mathbb{C}}$.

Thus, we are considering the triple

$$\mathcal{N}_{\mathbb{C}} \subset L^2(\Sigma, \mathrm{dVol}_h) \subset \mathcal{N}_{\mathbb{C}}^{\prime *} \tag{2.68}$$

By using this rigged space, we can represent the space of test functions $C_c^\infty(\Sigma)_{\mathbb{C}}$ as a suitable linear subspace of the conjugate space of distributions $\mathcal{N}_{\mathbb{C}} \subset D'(\Sigma)_{\mathbb{C}}^*$. We shall use this set as the set of classical fields later in [chapter 3](). Our notion of test function is provided by a nuclear topology and the injection of that topology into a Gel'fand triple respecting continuity. This is the construction used in white noise analysis [Hid&93, Wes03, Hol&10]. The topology of the space of test functions does not really depend on the particular Gel'fand triple that we use to realize the duality. Because of this, it is common to introduce them as a subset of the white noise space of square integrable functions $L^2(\mathcal{N}_{\mathbb{C}}', D\beta)$, where $\beta$ is the white noise measure whose characteristic functional is

$$\int_{\mathcal{N}_{\mathbb{C}}'} D\beta(\phi^{\mathbf{x}}) e^{i(\bar{\rho}_x \phi^x + \rho_x \bar{\phi}^x)} = e^{-\bar{\rho}_x \delta^{xy} \rho_y} = \exp\left(-\int_{\Sigma} \mathrm{dVol}(x) \bar{\rho}(x)\rho(x)\right) \tag{2.69}$$

Moreover, from this definition, we have that $\mathcal{H}_{Vol}$ becomes the Cameron-Martin Hilbert space of $L^2(\mathcal{N}_{\mathbb{C}}', D\beta)$. Analogously, we will use the same symbol $\beta$ for the Gaussian measure over the real space $\mathcal{N}'$ and the same covariance, that is, with characteristic functional $\exp\left(\frac{\xi_x \Delta^{xy} \xi_y}{2}\right)$.

We will treat holomorphic and antiholomorphic functions separately. Thus, we start by considering holomorphic polynomials with coefficients in $\mathcal{N}_{\mathbb{C}}$ as the minimal space of test functions[5] in $L_{Hol}^2(\mathcal{N}_{\mathbb{C}}', D\beta)$. Recall that a holomorphic polynomial of degree $n$ with coefficients in $\mathcal{N}_{\mathbb{C}}$ can be written as

$$P_n(\phi^{\mathbf{x}}) = C^0 + C_x^1 \phi^x + C_{x_1 x_2}^2 (\phi^2)^{x_1 x_2} + \cdots + C_{\vec{x}_n}^n (\phi^n)^{\vec{x}_n} \tag{2.70}$$

Consequently, we must consider the algebra of holomorphic polynomials $\mathcal{P}_{Hol}(\mathcal{N}_{\mathbb{C}}')$, which we know forms a dense subset of the Hilbert space $L_{Hol}^2(\mathcal{N}_{\mathbb{C}}', D\beta)$. The next step is to enlarge the set $\mathcal{P}_{Hol}(\mathcal{N}_{\mathbb{C}}')$ to obtain a more operative notion of test function. To that end, we make use of the topology of $\mathcal{N}_{\mathbb{C}}$ to build a sequence of Hilbert spaces and injections, and define our final set as the projective limit of that sequence, as explained in **Definition** 2.2.

Remember that $\mathcal{N}_{\mathbb{C}}$ is a nuclear space that we see as a subset of $\mathcal{H}_{Vol}$, and therefore it admits the sequence of Hilbert spaces, scalar products, and injections described in **Definition** 2.2. If we denote $\mathcal{H}_{p,q}$ with $0 \leq p, q$ as the weighted Hilbert spaces with norm $\| \cdot \|_{p,q} = 2^{\frac{q}{2}} \| \cdot \|_p$ and $\| \cdot \|_0 = \| \cdot \|_{Vol}$, we can also consider a $q$-parametric

---

[5]For the case of homogeneous polynomials, we can even endow this space with a nuclear topology [Din81].

sequence of Bosonic Fock spaces $\Gamma\mathcal{H}_{p,q}$ whose projective limits define

$$
\begin{array}{ccccccc}
\Gamma\mathcal{H}_{\infty,0} & \hookrightarrow & \cdots & \hookrightarrow & \Gamma\mathcal{H}_{p,0} & \hookrightarrow & \Gamma\mathcal{H}_{Vol} & \cong L^2_{Hol}(\mathcal{N}'_\mathbb{C}, D\beta) \\
\updownarrow & & & & \updownarrow & & \updownarrow & \\
\Gamma\mathcal{H}_{\infty,q} & \hookrightarrow & \cdots & \hookrightarrow & \Gamma\mathcal{H}_{p,q} & \hookrightarrow & \Gamma\mathcal{H}_{0,q} & \\
\updownarrow & & & & \updownarrow & & \updownarrow & \\
\vdots & & & & \vdots & & \vdots & \\
\updownarrow & & & & \updownarrow & & \updownarrow & \\
(\mathcal{N}_\mathbb{C}) \cong \ \Gamma\mathcal{H}_{\infty,\infty} & \hookrightarrow & \cdots & \hookrightarrow & \Gamma\mathcal{H}_{p,\infty} & \hookrightarrow & \Gamma\mathcal{H}_{0,\infty} &
\end{array}
\tag{2.71}
$$

We refer to [Hid&93, Oba94, Kon&96, Kuo96, Wes03] for detailed studies of these projective limits. The space of Hida test functions $(\mathcal{N}_\mathbb{C})$ is then obtained as the projective limit of that sequence, and we can write

$$
(\mathcal{N}_\mathbb{C}) = \bigcap_{p,q=0}^{\infty} \mathcal{I}^{-1}\Big(\Gamma\mathcal{H}_{p,q}\Big).
\tag{2.72}
$$

Where $\mathcal{I}$ is the Seagal isomorphism (2.37) with respect to the white noise measure $\beta$. It is important to remark that the topology of $(\mathcal{N}_\mathbb{C})$ depends only on the topology of the nuclear space $\mathcal{N}_\mathbb{C}$ and not on the particularities of the white noise measure $\beta$ (see [Kon&96]). This is a consequence of the independence of the Gaussian measure on the chaos decomposition for holomorphic functions that we noted in (2.38).

With these spaces in hand, we can define the white noise Gel'fand triple by introducing Hida distributions

$$
(\mathcal{N}_\mathbb{C}) \subset L^2_{Hol}(\mathcal{N}'_\mathbb{C}, D\beta) \subset (\mathcal{N}_\mathbb{C})'^*,
\tag{2.73}
$$

We can describe the dual space $(\mathcal{N}_\mathbb{C})'$, in an analogous way to the definition of $(\mathcal{N}_\mathbb{C})$, as the injective limit

$$
(\mathcal{N}_\mathbb{C})' = \bigcup_{p=0}^{\infty} \mathcal{I}^{-1}\Big(\Gamma\mathcal{H}^*_{-p,-q}\Big).
\tag{2.74}
$$

Here, the Banach space $\mathcal{H}_{-p,-q}$ is identified with the dual of $\mathcal{H}_{p,q}$ in the Gel'fand triple $\mathcal{H}_{p,q} \subset \mathcal{H}_{Vol} \subset \mathcal{H}^*_{-p,-q}$ and $\mathcal{I}^{-1}$ is extended to provide a chaos decomposition in this distributional context using Hida Malliavin-calculus techniques that we present below. Again, we refer to [Hid&93, Oba94, Kon&96, Kuo96] for further details.

As anticipated, the role of the white noise measure $\beta$ is auxiliary in the construction. If we consider any other Gaussian measure $\mu_c$ with covariance $\Delta^{\mathbf{xy}}$ such that there exists a $q \geq 0$ and a norm of the chain $\|\cdot\|_{p,q}$ such that $|\bar{\rho}_x \Delta^{xy} \rho_y| \leq \|\rho_{\mathbf{x}}\|^2_{p,q}$, then according to the characterization theorems [Kon&96], we also have

$$
(\mathcal{N}_\mathbb{C}) \subset L^2_{Hol}(\mathcal{N}'_\mathbb{C}, D\mu_c) \subset (\mathcal{N}_\mathbb{C})'^*,
\tag{2.75}
$$

An equivalent derivation can be considered for any Gaussian measure, as for instance the one corresponding to the vacuum of the quantum theory. We call this case **the second quantization Gel'fand triple**. We must also consider the decomposition for the antiholomorphic part. Thus, we consider the triple

$$(\mathcal{N}_{\mathbb{C}}) \otimes (\mathcal{N}_{\mathbb{C}})^* \subset L^2(\mathcal{N}_{\mathbb{C}}', D\mu_c) \subset (\mathcal{N}_{\mathbb{C}})'^* \otimes (\mathcal{N}_{\mathbb{C}})'. \tag{2.76}$$

Hence, Hida test functions respect the decomposition into holomorphic and antiholomorphic parts of (2.33).

We list below some important properties that showcase the power of the Hida triple. For different presentations and proofs, we refer to [Hid&93, Oba94, Kon&96, Kuo96, Wes03].

- Let $\Psi \in (\mathcal{N}_{\mathbb{C}}) \otimes (\mathcal{N}_{\mathbb{C}})^*$, the coefficients of its chaos decomposition fulfill $\psi_{\vec{\mathbf{x}},\vec{\mathbf{y}}}^{(n,\bar{m})} \in \mathcal{N}_{\mathbb{C}}^{\hat{\otimes} n} \otimes \mathcal{N}_{\mathbb{C}}^{*\hat{\otimes} m}$. Moreover, $\Psi(\phi^{\mathbf{x}})$ is a pointwise defined continuous function.

- $(\mathcal{N}_{\mathbb{C}})$ and $(\mathcal{N}_{\mathbb{C}}) \otimes (\mathcal{N}_{\mathbb{C}})^*$ are algebras under pointwise multiplication. Moreover, the algebras $\mathcal{P}$, $\mathcal{T}$, and $\mathcal{C}$ introduced in Section section 2.5 are subalgebras of one or both spaces. See Appendix B for a proof.

- Hida test functions are infinitely differentiable, and they and their derivatives belong to $L_{Hol}^p(\mathcal{N}_{\mathbb{C}}', D\mu_c)$ for all $p \geq 1$. This is encoded in

$$(\mathcal{N}_{\mathbb{C}}) \subsetneq \bigcap_{p \geq 1, m \geq 0} \mathbb{D}_{\mu_c}^{p,m} := \mathbb{D}_{\mu_c} \tag{2.77}$$

  See Appendix B for a proof. [6]

- As a consequence of the previous property and the fact that the dual of an $L^p$-space is an $L^q$-space with $\frac{1}{p} + \frac{1}{q} = 1$, we get

$$L^p(\mathcal{N}_{\mathbb{C}}', D\mu_c) \subset (\mathcal{N}_{\mathbb{C}})'^* \otimes (\mathcal{N}_{\mathbb{C}})' \qquad \forall p \geq 1$$

Below we enumerate some examples of Hida test functions and distributions that allow us to operate in more general contexts in Gaussian analysis and will be of great importance in the next chapters.

**Example 2.7.  (Hida-Malliavin calculus)**  *The generator of the $S_{\mu_c}$ transform is a Hida test function*

$$\exp\left(\overline{\rho_x \phi^x} + \rho_x \phi^x - \bar{\rho}_x \Delta^{xy} \rho_y\right) \in (\mathcal{N}_{\mathbb{C}}) \otimes (\mathcal{N}_{\mathbb{C}})^*.$$

---
[6] $\mathbb{D}_{\mu_c}$ is often called the Meyer-Watanabe space of test functions. We refer to [Hid&93, Hu16] for further details.

As a consequence, $S_{\mu_c}[\Psi]$ can be extended to any Hida distribution $\Psi \in (\mathcal{N}_{\mathbb{C}})'^* \otimes (\mathcal{N}_{\mathbb{C}})'$. In particular, this means that we can find a chaos decomposition for any Hida distribution with a careful use of (2.35). This turns Hida distributions into easier-to-handle objects than other notions of s.q. distributions.

The generalization of the machinery of chaos decompositions to general distributions allows us to extend most of the tools of Malliavin calculus, described in section 2.6.4, to distributions in $(\mathcal{N}_{\mathbb{C}})'^* \otimes (\mathcal{N}_{\mathbb{C}})'$. This type of calculus is called Hida-Malliavin calculus, and we refer elsewhere for a thorough construction [NØP09, Hol&10].

A common problem that is solved by Hida calculus is to make sense of Radon-Nikodym derivatives. This will be particularly relevant when QFT theories with respect to evolving backgrounds are considered since these properties of Hida calculus will provide us with tools to compare the states of the quantum fields with respect to them. With these applications in mind, we briefly comment now on a few useful properties:

**Example 2.8. (Translation of a Gaussian measure)** *A general result of Gaussian analysis, known as the Cameron-Martin theorem, states that a translation in the domain by a constant element $\chi^{\mathsf{x}} \in \mathcal{N}_{\mathbb{C}}'$ provides a Gaussian measure mutually singular with the original one, unless $\chi^{\mathsf{x}} \in \mathcal{H}_K$, the Cameron-Martin Hilbert space. This implies that the would-be Radon-Nikodym derivative*

$$\frac{D\mu_c(\phi^{\mathsf{x}} - \chi^{\mathsf{x}})}{D\mu_c(\phi^{\mathsf{x}})} = \exp\left(\bar{\phi}^x K_{xy}\chi^y + \bar{\chi}^x K_{xy}\phi^y - \bar{\chi}^x K_{xy}\chi^y\right)$$

*does not exist as an $L^1(\mathcal{N}_{\mathbb{C}}', D\mu_c)$ function. The power of Hida calculus is to extend its meaning to general $\chi^{\mathsf{x}} \in \mathcal{N}_{\mathbb{C}}'$ as an element of $(\mathcal{N}_{\mathbb{C}})'^* \otimes (\mathcal{N}_{\mathbb{C}})'$ and provide a tool to perform calculus with the generalized Wiener-Ito decomposition.*

**Example 2.9. (Mean zero Gaussian measure)** *In general, Gaussian measures with different covariances, even if they both fulfill an estimate of the type $|\rho_x \Delta^{xy} \rho_y| \leq \|\rho_{\mathsf{x}}\|_{p,q}^2$, are mutually singular.*

*When working with the space of Hida test functions, if $\nu_c$ is another mean zero Gaussian measure with covariance $\Omega^{\mathsf{xy}}$, then*

$$S_{\mu_c}\left[\frac{D\nu_c}{D\mu_c}\right](\rho_{\mathsf{x}}) = e^{\bar{\rho}_x(\Omega^{xy} - \Delta^{xy})\rho_y},$$

*where $\frac{D\nu_c}{D\mu_c} \in (\mathcal{N}_{\mathbb{C}})'^* \otimes (\mathcal{N}_{\mathbb{C}})'$ makes sense as a Hida distribution even if it does not as a function.*

**Example 2.10. (Reproducing Kernel)** *Coherent states of the holomorphic subspace $\mathcal{K}_\chi(\phi^{\mathsf{y}}) \in L^2_{Hol}(\mathcal{N}_{\mathbb{C}}', D\mu_c)$ can be extended to elements $\rho^{\mathsf{x}} \in \mathcal{H}_\Delta$ as $\mathcal{K}_\rho(\phi^{\mathsf{y}}) = e^{\rho^x K_{xy}\phi^x}$ that makes sense as an element of $L^2_{Hol}(\mathcal{N}_{\mathbb{C}}', D\mu_c)$ but, moreover,*

$$S_{\mu_c}\left[e^{\bar{\sigma}^x K_{xy}\phi^x}\right](\rho_{\mathsf{x}}) = e^{\overline{\bar{\sigma}^x \rho_x}}$$

Then $e^{\bar{\sigma}^x K_{xy}\phi^x} \in (\mathcal{N}_{\mathbb{C}})'$ is a Hida distribution and works as a reproducing kernel when it is paired with $(\mathcal{N}_{\mathbb{C}}) \subset L^2_{Hol}(\mathcal{N}_{\mathbb{C}}', D\mu_c)$. Thus, for $\Psi \in (\mathcal{N}_{\mathbb{C}})$

$$\int_{\mathcal{N}_{\mathbb{C}}'} D\mu_c(\phi^{\mathbf{x}}) \overline{e^{\bar{\sigma}^x K_{xy}\phi^x}} \Psi(\phi^{\mathbf{x}}) = \Psi(\sigma^{\mathbf{x}}) \tag{2.78}$$

A question that deserves more attention is why we choose white noise instead of **the second quantization Gel'fand triple** if the latter possesses richer physical meaning when we interpret $\mu_c$ as part of the vacuum of our quantum theory. We could also present Hida test functions and distributions using this triple, but the fact that the topology of $(\mathcal{N}_{\mathbb{C}})$ only depends on that of $\mathcal{N}_{\mathbb{C}}$ turns white noise into an auxiliary but simplifying way of presenting it. Moreover, the vacuum of a theory is a particular feature of such a theory, and therefore we cannot assume prior knowledge of the measure $\mu_c$ that is part of the vacuum. In this way, in order to consider the most general setting, it is better to model pure states as the common dense subspace of Hida test functions $(\mathcal{N}_{\mathbb{C}})$. In that case, we can write down the equations of motion and the operators of the theory only referring to this set of test functions, as we will consider in chapter 5. However, these equations cannot be solved in $(\mathcal{N}_{\mathbb{C}})$, and the operators considered to describe observables will have naturally larger domains. In order to recover the full space, we will use our *a posteriori* knowledge of the Gaussian measure that endows $(\mathcal{N}_{\mathbb{C}})$ with a pre-Hilbert topology and complete it to recover the whole space. This procedure is similar to the protocol to recover the one-particle Hilbert space from a classical field theory used in [Wal94]. In that case, we can model the classical theory over a space of test functions $\mathcal{N}$ and then complete it with a Hilbertian norm that depends on the particular theory we are considering. Notice, though, that the treatment in [Wal94] is covariant, and here we aim to model a non-covariant approach. This is important to notice in regard to the treatment of the equations of motion. Our discussion on how to introduce evolution is radically different from covariant approaches and is the subject of study in chapter 5.

### 2.7.1 Integral kernel representation of operators

In chapter 3 and chapter 4, the reproducing kernel machinery of *Example* 2.10 will be used extensively to express operators acting on $(\mathcal{N}_{\mathbb{C}}) \subset L^2_{Hol}(\mathcal{N}_{\mathbb{C}}', D\mu_c)$ as integral kernels with respect to $\mu_c$ in chapter 3 and with respect to white noise $\beta$ in chapter 4. In this section, we prepare the basic tools to handle this kind of object.

Skorohod and Malliavin derivatives of a holomorphic function have simple integral kernel representations. Let $\Psi \in (\mathcal{N}_{\mathbb{C}})$ and $f_{\mathbf{x}} \in \mathcal{H}_{\Delta}$, then we can define the action of a symbol $f$ on the state $\Psi$ as:

$$f_x \Delta^{xy} \partial_{\sigma^y} \Psi(\sigma^{\mathbf{x}}) = \int_{\mathcal{N}'_{\mathbb{C}}} D\mu_c(\phi^{\mathbf{x}}) e^{\bar{\phi}^x K_{xy} \sigma^y} f_x \bar{\phi}^x \Psi(\phi^{\mathbf{x}})$$

$$f_x \sigma^x \Psi(\sigma^{\mathbf{x}}) = \int_{\mathcal{N}'_{\mathbb{C}}} D\mu_c(\phi^{\mathbf{x}}) e^{\bar{\phi}^x K_{xy} \sigma^y} f_x \sigma^x \Psi(\phi^{\mathbf{x}}). \tag{2.79}$$

This is a simple consequence of the Skorohod-Malliavin duality and *Example* 2.10. The white noise kernel will be

$$f_x \Delta^{xy} \partial_{\sigma^y} \Psi(\sigma^{\mathbf{x}}) = \int_{\mathcal{N}'_{\mathbb{C}}} D\beta(\phi^{\mathbf{x}}) e^{\bar{\phi}^x \delta_{xy} \sigma^y} f_x \Delta^x_y \bar{\phi}^y \Psi(\phi^{\mathbf{x}})$$

$$f_x \sigma^x \Psi(\sigma^{\mathbf{x}}) = \int_{\mathcal{N}'_{\mathbb{C}}} D\beta(\phi^{\mathbf{x}}) e^{\bar{\phi}^x \delta_{xy} \sigma^y} f_x \sigma^x \Psi(\phi^{\mathbf{x}}). \tag{2.80}$$

These expressions are particularly simple in the holomorphic case and, for that reason, we limit our attention to that case in this section. However, spaces with real domains will play an important role in the description of the Schrödinger and field momentum pictures in the following chapters. We will briefly discuss how to enlarge this machinery to white noise reproducing kernels in section 5.5.2.

The action of linear kernels over functions amounts to a particular order in the application of derivatives and multiplications by the monomial $\phi^{\mathbf{x}}$. This is, for arbitrary kernels, we express

$$\int D\beta(\sigma)\mathcal{K}(\phi^{\mathbf{x}}, \bar{\sigma}^{\mathbf{x}})\Psi(\sigma) = \mathcal{K}(\phi^{\mathbf{x}}, \delta^{xy}\partial_{\sigma^y})\Psi(\sigma)|_{\sigma=0}. \tag{2.81}$$

This can be easily extended to kernels for different measures. However, for white noise kernels, we can derive a composition rule that defines an algebra.

First, consider linear operators

$$A^{\mathbf{x}}_{\mathbf{y}} : \mathcal{N} \longrightarrow \mathcal{N}. \tag{2.82}$$

These linear operators can be interpreted as infinite dimensional matrices. Extending the conventions of (2.1) we introduce the conventions of multiplication $(AB)^{\mathbf{x}}_{\mathbf{y}} = A^{\mathbf{x}}_{\mathbf{z}} B^{\mathbf{z}}_{\mathbf{y}}$ and the index notation $A_{\mathbf{xy}} = \delta_{\mathbf{xz}} A^{\mathbf{z}}_{\mathbf{y}}$ and $A^{\mathbf{xy}} = A^{\mathbf{x}}_{\mathbf{z}} \delta^{\mathbf{zy}}$. Using the expressions above, we obtain

$$\int_{\mathcal{N}'_{\mathbb{C}}} D\beta(\sigma) e^{\phi^y A_{yx} \bar{\sigma}^x + \sigma^x B_{xy} \gamma^y} = e^{\phi^x (AB)_{xy} \bar{\gamma}^y}. \tag{2.83}$$

Using this computational rule, we can find an algebra of trigonometric white noise kernels

$$\int D\beta(\sigma) A(\phi, \bar\sigma) B(\sigma, \bar\gamma):$$

| B ╲ A | $\sqrt{2}\cosh(N)\cos(V)$ | $\sqrt{2}\cosh(N)\sin(V)$ | $\sqrt{2}\sinh(N)\cos(V)$ | $\sqrt{2}\sinh(N)\sin(V)$ |
|---|---|---|---|---|
| $\sqrt{2}\cosh(M)\cos(U)$ | $\cosh\boxplus\cos\ominus + \cosh\boxminus\cos\oplus$ | $0$ | $0$ | $-\sinh\boxplus\sin\ominus + \sinh\boxminus\sin\oplus$ |
| $\sqrt{2}\cosh(M)\sin(U)$ | $0$ | $\sinh\boxplus\cos\ominus - \sinh\boxminus\cos\oplus$ | $\cosh\boxplus\sin\ominus + \cosh\boxminus\sin\oplus$ | $0$ |
| $\sqrt{2}\sinh(M)\cos(U)$ | $0$ | $-\cosh\boxplus\sin\ominus + \cosh\boxminus\sin\oplus$ | $\sinh\boxplus\cos\ominus + \sinh\boxminus\cos\oplus$ | $0$ |
| $\sqrt{2}\sinh(M)\sin(U)$ | $\sinh\boxplus\sin\ominus + \sinh\boxminus\sin\oplus$ | $0$ | $0$ | $\cosh\boxplus\cos\ominus - \cosh\boxminus\cos\oplus$ |

Table 2.1: Result of the integral $\int D\beta(\sigma) A(\phi,\bar\sigma) B(\sigma,\bar\gamma)$ where the kernels $A$ and $B$ are specified in the first column and row respectively. $M = \phi^x M_{xy}\bar\sigma^y, U = \phi^x U_{xy}\bar\sigma^y, N = \sigma^x N_{xy}\bar\gamma^y$ and $V = \sigma^x V_{xy}\bar\gamma^y$. Also $\boxplus = \phi^x(MN + UV)_{xy}\bar\gamma^y$, $\boxminus = \phi^x(MN - UV)_{xy}\bar\gamma^y$, $\oplus = \phi^x(UN + MV)_{xy}\bar\gamma^y$ and $\ominus = \phi^x(UN - MV)_{xy}\bar\gamma^y$

Of course, to describe a real algebra of white noise reproducing kernels, we should ask for $M_{\mathbf{y}}^{\mathbf{x}}, N_{\mathbf{y}}^{\mathbf{x}}, U_{\mathbf{y}}^{\mathbf{x}}$ and $V_{\mathbf{y}}^{\mathbf{x}}$ to belong to an algebra of operators. We will focus on this algebra because $A(\phi,\bar\sigma)^* = A(\bar\phi,\sigma)$, so when they act on a function by $\int D\beta(\sigma)A(\phi,\bar\sigma)\Psi(\sigma)$, they do not mix the real and imaginary parts of the chaos decomposition (2.21).

## 2.8 Comments on Geometry

Nuclear-Fréchet (NF) spaces present further convenient properties for the modeling of infinite-dimensional manifolds. The topic of geometry on locally convex spaces is a vast subject; for further results, we refer to [KM97]. Roughly speaking, the problem is the following: we may want a notion of derivation compatible with the well-posed notion of the derivative of a curve on a locally convex vector space $E$. That is, let $\gamma : \mathbb{R} \to E$ be any smooth curve. Then a map $f : E \to F$, where $F$ is another locally convex space, is called smooth and denoted $f \in C^\infty(E, F)$ if

$$f \circ \gamma : \mathbb{R} \to E \to F \tag{2.84}$$

is a smooth curve. Notice that since it must work for any $\gamma$, we are asking that $f$ must preserve smooth curves.

Once this notion is established, we may ask if a smooth map is continuous with respect to the locally convex topologies of $E$ and $F$. To see this, we endow $E$ with the $c^\infty$-topology, which is the finest topology making every smooth curve $\gamma : \mathbb{R} \to E$ continuous. This is, in general, finer than the locally convex topology of $E$ and does not have to render the completion $\bar{E}^{c^\infty}$ into a locally convex space.

With this discussion at hand, we have that $f \in C^\infty(E, F)$ is continuous if $E$ is endowed with its $c^\infty$-topology. So it is convenient to perform geometry in a category in which both topologies coincide. Fréchet spaces belong to this class; so do the strong duals of Fréchet-Schwartz spaces, of which DNF spaces are a subcategory.

For NF and their dual DNF spaces, we find that they are even $C^\infty$-paracompact, which implies that there exist $C^\infty$ partitions of unity. We can even find a stronger result asserting that strict inductive limits of FN spaces are also $C^\infty$-paracompact, which is the case that allows us to extend the Bochner-Minlos theorem in [GV16].

These considerations allow us to extend the tools of integro-differential analysis described in this section from the linear spaces $\mathcal{N}, \mathcal{N}'$ to manifolds locally modeled over those spaces. This extension may be needed in more general frameworks such as Yang-Mills theories but will not be treated in this work. We limit our discussion to linear spaces and their geometrical structures, thus providing a flavour of this geometrical, or global analytic [KM97], setting.

# Chapter 3

# Classical Statistical Field Theory: Koopman-van Hove theory in infinite dimensions

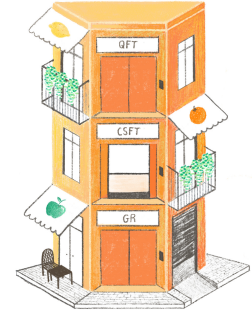

## 3.1 Introduction

Most of the problems that we will face in QFT in curved spacetimes are already present in the description of a Classical Statistical Field Theory (CSFT) in the Koopman formalism. Our objective in this chapter will be to present CSFT in the Koopman-van Hove (KvH) formalism. This formalism builds upon the Koopman-von Neumann (KvN) approach [Koo31, vNeu32] and van Hove prequantization theory [Van51] to study the representation and dynamics of classical statistical mechanics. Our focus in this chapter will be presenting KvH theory for the infinite-dimensional case. It is worth noting, though, that the main focus of this thesis is the quantum counterpart of the theory. For this reason, we will just present a brief discussion drawing some conclusions that will be useful in the quantum case and leave the most difficult parts for the quantum case.

Koopman formalisms deal with statistical descriptions in phase space. This formalism, in the finite-dimensional case, has been widely studied since its inception [Koo31, vNeu32] and has been recently revived to develop hybrid quantum-classical systems in finite-dimensional settings [BGT19, GT20, GT22a, GT22b, Bou&23, MT24]. For the sake of simplicity, consider the manifold $\mathfrak{M} = T^*\mathbb{R}^n \cong \mathbb{R}^{2n}$. This is a trivial example of the cotangent bundle of coordinates $(q^i, p_i)_{i=0}^n$ endowed with the canonical symplectic form $\omega = dp^i \wedge dq_i$, where the sum over indices is implied.

Consider the symplectic volume

$$\Lambda = \omega \wedge \overbrace{\cdots}^{n} \wedge \omega$$

; a classical statistical state of the system is described by a probability density

$$\varrho \in L^1(\mathfrak{M}, \Lambda).$$

Sometimes it is convenient to deal with the representative of this in the space of densities $\varrho(q_i, p^i) \in \mathrm{Den}(\mathfrak{M})$, this is without the volume form. The evolution of this state, for a given Hamiltonian $H(q^i, p_i)$, is provided by the Liouville equation

$$\frac{d}{dt}\varrho = -\{\varrho, H\} = \frac{\partial H}{\partial q^i}\frac{\partial \varrho}{\partial p_i} - \frac{\partial H}{\partial p_i}\frac{\partial \varrho}{\partial q^i}. \tag{3.1}$$

Here the Poisson bracket of two functions $f, g \in C^\infty(\mathfrak{M})$ is defined using their Hamiltonian vector fields $X_f, X_g$. These are the solutions to $X_f \lrcorner \omega = -df$. Therefore, we define the Poisson bracket $\{\cdot, \cdot\}$ and the Poisson tensor $\mathbb{J}$ as

$$\{f, g\} := \omega(X_f, X_g) := \mathbb{J}(df, dg). \tag{3.2}$$

In the Koopman formalisms, we look for representations of the previous dynamics as unitary transformations on the Hilbert space $L^2(\mathfrak{M}, \Lambda)$. The original Koopman-von Neumann (KvN) prescription takes the square root of $\varrho$ as the representation of the state. However, this representation does not represent the Liouvillian operator, the generator of the dynamics, as an observable [Bou&23] and leads to ill-defined variational principles to treat the dynamics [BGT19]. Instead of that prescription, we will use the Koopman-van Hove (KvH) prescription, see [GT22b], that consists of representing the algebra of observables using prequantization theory.

For that matter, we should understand the space $L^2(\mathfrak{M}, \Lambda)$ as the space of sections on a line bundle $B$ and introduce a connection $\nabla$ with curvature $\frac{\omega}{\hbar}$. Then, the classical observable $F \in C^\infty(\mathfrak{M})$ is represented as an operator acting on the Hilbert space through the prequantization mapping $\mathbb{P}$ as

$$\mathbb{P}(F) = F - i\hbar\nabla_{X_F}.$$

The Liouville equation (3.1) in this framework is provided by a Schrödinger-like equation (see [BGT19])

$$i\frac{d\psi}{dt} = \mathbb{P}(H)\psi. \tag{3.3}$$

Where $\psi \in L^2(\mathfrak{M}, \Lambda)$ is obtained from $\varrho$ using the KvH prescription that we will discuss below. The formalism can be extended to any symplectic manifold; we refer to [GT22b] for a great account of the Koopman treatment of the theory.

As we can see, this formalism presents classical statistical mechanics in a way close to the Schrödinger picture of quantum mechanics. Extending this formalism to the infinite-dimensional case allows us to determine which features of a QFT are already present at the level of CSFT. In addition, it will be the pathway used in chapter 4 to derive Hamiltonian representations of QFT, such as the Schrödinger picture, through geometric quantization.

Nonetheless, this formalism has some drawbacks that we must discuss in section 3.2. In summary, the relation between $\psi$ and $\varrho$ is not straightforward due to a gauge symmetry present in the choice of connection $\nabla$ that will permeate every aspect of the theory.

Firstly, in section 3.3, we will focus on the description of Gaussian states. In order to describe them in the infinite-dimensional case, we should deal with the nuisances of integration theory presented in chapter 2. This is, we will describe the fields and their associated field-momenta as distributions of $\mathcal{N}'$ denoted by $(\varphi^{\mathbf{x}}, \pi^{\mathbf{x}})$ and introduce the state as a Gaussian measure $\mu$ defined through a characteristic functional as in section 2.4.

Secondly, in section 3.4, we will focus our description on the KvH formalism in which we factor out Gaussian states and describe prequantization theory and the evolution equations. In this context, we will describe the evolution equation with the addition of a connection $\Gamma_t$ to the time derivative. The particular nuisances of this description are beyond the scope of this work, but we present here an informal derivation to exemplify the appearance of an object of huge importance in the quantum counterpart of the theory as we will derive in chapter 5.

Lastly, in section 3.5, we will apply some of the results over a Klein-Gordon theory on a curved spacetime with minimal coupling. In particular, we will get a prescription to choose a Kähler structure that will be the basis for quantization in chapter 7. In addition, we will see that the time-dependence of these structures reflects a dependence on the geometric content that describes gravity. This aspect will be exploited in chapter 6 to describe the coupling to gravity using geometrodynamics.

## 3.2 Koopman-van Hove theory in finite dimensions

### 3.2.1 Prequantization

As we advanced in the introduction, prequantization theory [Van51, Woo97] represents the Poisson algebra of classical observables $(C^\infty(\mathfrak{M}), \{\cdot, \cdot\})$ acting on a Hilbert space built in the following manner. First, we introduce a Hermitian line bundle $\pi : \mathfrak{B} \to \mathfrak{M}$; this is, a vector bundle whose fibers are $\mathbb{C}$. Over this fiber bundle, we consider its sections $\psi_s, \zeta_s \in \Gamma(\mathfrak{M}, \mathfrak{L})$ whose scalar product is

$$\langle \psi_s, \zeta_s \rangle = \int_{\mathfrak{M}} \psi_s^* \zeta_s(q^i, p_i) \Lambda. \tag{3.4}$$

The Hilbert space in which we will represent our classical statistical states is the Hilbert space of square-integrable sections.

$$\mathcal{H} = \overline{\{\psi_s \in \Gamma(\mathfrak{M}, \mathfrak{L}) \text{ s.t. } \langle \psi_s, \psi_s \rangle := \|\psi_s\|^2 < \infty\}}. \tag{3.5}$$

Over this bundle $B$, we introduce a connection $\nabla$ such that its curvature is $\frac{\omega}{\hbar}$. Throughout this work, we will choose natural units such that $\hbar = 1$. To study this kind of connection, we introduce its connection one-form $\theta$. We name this object the symplectic potential because it must fulfill $-d\theta = \omega$. The introduction of this connection also allows for a simplification of the expression of $\mathcal{H}$ by taking the reference section $\psi_r \in \Gamma(\mathfrak{M}, \mathfrak{L})$:

$$\nabla_X \psi_r = -i\theta(X)\psi_r, \qquad \forall\, X \in \mathfrak{X}(\mathfrak{M}). \tag{3.6}$$

where $\mathfrak{X}(\mathfrak{M})$ are vector fields over $\mathfrak{M}$. Also, we require

$$\psi_r(p_i, q^i)^* \psi_r(p_i, q^i) = 1 \qquad \forall\, (q^i, p_i) \in \mathfrak{M}. \tag{3.7}$$

Using the multiplication structure of the field $\mathbb{C}$, any other section $\psi_s \in \Gamma(\mathfrak{M}, \mathfrak{L})$ can be written as $\psi_s = \psi\psi_r$ with $\psi \in \mathrm{Den}^{\frac{1}{2}}(\mathfrak{M})$ a half density. This decomposition allows us to identify the Hilbert space of sections with a Hilbert space of densities:

$$\mathcal{H} = L^2(\mathfrak{M}, \Lambda),$$

as we were looking for. The observables are represented by the prequantization mapping

$$\mathbb{P}(F) = F - i\nabla_{X_F}, \text{ for } F \in C^\infty(\mathfrak{M}). \tag{3.8}$$

It follows that this prequantization procedure $\mathbb{P}$ meets the Dirac prequantization conditions [Hal13]:

**Q1)** $\mathbb{P}(F + G) = \mathbb{P}(F) + \mathbb{P}(G)$

**Q2)** $\mathbb{P}(F)\mathbb{P}(G) - \mathbb{P}(G)\mathbb{P}(F) = -i\mathbb{P}(\{F, G\})$

**Q3)** $\mathbb{P}(F) = F\mathbb{1}$ if $F$ is constant

for any $F, G \in C^\infty(\mathfrak{M})$. This means that the prequantization mapping $\mathbb{P}$ represents the Poisson algebra of classical observables $(C^\infty(\mathfrak{M}), \{\cdot, \cdot\})$ as a Lie subalgebra of Hermitian operators acting on $\mathcal{H}$.

### 3.2.2 Gauge freedom and observables

The prescription $-d\theta = \omega$ for the symplectic potential leaves room for a redefinition

$$\theta \to \theta + d\alpha \tag{3.9}$$

for any $\alpha \in C^\infty(\mathfrak{M})$. This redefinition does not affect $Q1) - Q3)$; thus, we obtain a family of representations of the same theory related by a gauge transformation. This gauge freedom acts on the elements of $\mathcal{H}$ in the following way. Let $\psi_s = \psi\psi_r \in \mathcal{H}$ and $\varrho$ be the probability density of (3.1) representing the same state. The outcome of a measurement must be equal in any representation. Thus, let $F \in C^\infty(\mathfrak{M})$ represent an observable; we must impose

$$\int_{\mathfrak{M}} \varrho F \Lambda = \int_{\mathfrak{M}} \psi_s^* \mathbb{P}(F) \psi_s \Lambda. \tag{3.10}$$

In terms of the function $\psi$, using (3.6), and integrating by parts, we get

$$\varrho = |\psi|^2 + \operatorname{div}\left(\theta \lrcorner \mathbb{J} |\psi|^2\right) + i\{\psi, \psi^*\}. \tag{3.11}$$

The right-hand side of this equality is sign indefinite. To deal with this difficulty, we must restrict $\psi \in \mathcal{H}$ in the following manner. Let $\psi = \sqrt{D}\exp(iS)$ with $D \in \operatorname{Den}(\mathfrak{M})$ and $S \in C^\infty(\mathfrak{M})$ then

$$\varrho = D + \operatorname{div}\left(\theta \lrcorner \mathbb{J} D\right) + \{D, S\} = D + \operatorname{div}(D[\theta - dS] \lrcorner \mathbb{J})$$

This expression may be interpreted as the momentum map as follows. In this interpretation, $(D\,dS, D)$ is an element of a Lie algebra generating the infinite-dimensional Lie group of strict contact diffeomorphisms acting on $\mathcal{H}$. Recall that there $\mathcal{H}$ is a space of half densities, thus the action of this Lie group follows from the canonical action of the group of diffeomorphisms. For further details on this construction, see [GT20]. The divergence term is, in general, nonzero because $dS$ is an exact differential while $\theta$ is not[1]. Thus, to ensure that $\varrho$ is positive definite, we must impose conditions on the phase of $\psi$. However, notice that this condition is sensitive to gauge transformations. Therefore, to preserve gauge invariance, we must enlarge (3.9) to

$$\theta \to \theta + d\alpha \text{ and } \psi \to e^{i\alpha}\psi \tag{3.12}$$

The same result is obtained directly from (3.6).

### 3.2.3 Dynamics

The Liouville equation (3.1) in this framework is provided by [BGT19]

$$i\frac{d\psi}{dt} = \mathbb{P}(H)\psi \tag{3.13}$$

In terms of the decomposition $\psi = \sqrt{D}\exp(iS)$, this may be written

$$\frac{d}{dt}D = -\{D, H\} \tag{3.14}$$

$$\frac{d}{dt}S = -\{S, H\} - H + \theta(X_H) \tag{3.15}$$

The term $\mathcal{L}_H := \theta(X_H) - H$ may be interpreted as the Lagrangian of the theory when $\theta$ is the Liouville one-form. This dynamics does not spoil the sign of (3.11), thus we may only impose the conditions over $S$ to render it positive as an initial condition.

---

[1] We may cancel this term by imposing the equality $dS \simeq \theta$ in a generalized formalism using von Neumann operators [GT22a].

## 3.3 Adapting Classical Field Theory to the Statistical Framework

In this section, we will adapt the geometrical description of a classical field theory to the statistical framework of KvH. We will focus on the real scalar field. In this case, the classical field theory phase space will be given by the cotangent bundle to a real line bundle $\pi_{\Sigma L} : L \to \Sigma$. For simplicity, we will consider the trivial case in which the space of sections $\Gamma(\Sigma, L) = \mathcal{N}$ is a space of smooth functions over $\Sigma$. In this thesis we will usually consider $\Sigma$ to be compact. Therefore, the NF space is $\mathcal{N} = C_c^\infty(\Sigma)$ as in *Example* 2.3. That is, $\varphi_{\mathbf{x}}$ will be a coordinate that represents a field. The simplest example will be the case of a Klein-Gordon field theory. In this work, we will consider only linear equations, and therefore the corresponding manifold of fields will admit a linear structure, which will significantly simplify the geometrical setting.

Other descriptions of gauge fields or higher-order couplings in the equation may be studied in terms of jet bundles, as in [Got&04, GIM04]. The statistical descriptions of this general setting are beyond the scope of this work. We point out that the measure-theoretical technicalities have been already treated to some extent in [Mas08, Str19].

### 3.3.1 Choosing the Space of Classical Fields and Their Momenta

If $\mathcal{N}$ represents the set of fields, the corresponding phase space should correspond to the cotangent bundle $T^*\mathcal{N} \simeq \mathcal{N} \times \mathcal{N}'$, where we used the linearity of $\mathcal{N}$ and the corresponding triviality of the cotangent bundle. The coordinate representation of an element of this space is given by $(\varphi_{\mathbf{x}}, \pi^{\mathbf{y}})$, with the coordinate systems described above. This space is endowed with a weakly symplectic form, which takes the local expression

$$\omega_{T^*\mathcal{N}} = d\pi^x \wedge d\varphi_x. \tag{3.16}$$

This situation is problematic in the sense that the model space for the fiber and the base of the cotangent bundle $T^*\Gamma(\Sigma, L)$ are different. To describe the statistical system, we need to express both coordinates over the same model space. To do so, we use the Riemannian structure $h$ of $\Sigma$ and use it to define a measure $d\mathrm{Vol}$ and the natural structure of rigged Hilbert space given by

$$\mathcal{N} \subset L^2(\Sigma, d\mathrm{Vol}) \subset \mathcal{N}' \tag{3.17}$$

As a result, the elements of $\mathcal{N}$ correspond to a linear subspace of its dual, and we can write the coordinate representation of the image of an element of $\mathcal{N}$ as $\delta^{\mathbf{x}y}\varphi_y$. In order to apply the theory of integration developed in chapter 2, the manifold of fields must be $\mathcal{M}_F = \mathcal{N}' \times \mathcal{N}'$, and coordinates $(\varphi^{\mathbf{x}}, \pi^{\mathbf{y}})$ are to be understood as Darboux coordinates for a densely defined weakly symplectic structure

$$\omega_{\mathcal{M}_F} = \delta_{xy}d\pi^x \wedge d\varphi^y = \int_\Sigma \frac{d^dx}{\sqrt{|h|}} \left[d\pi(x) \otimes d\varphi(x) - d\varphi(x) \otimes d\pi(x)\right], \tag{3.18}$$

where we interpret $\pi(x), \varphi(x)$ as densities of weight 1. Notice that this symplectic structure is defined over the dense subspace $L^2(\Sigma, d\text{Vol}) \subset \mathcal{N}'$.

If we consider a space of smooth functions such as Hida test functions, described in section 2.7, we can also consider a Poisson bracket[2]. That is, if $P, Q \in (\mathcal{N}) \otimes (\mathcal{N})$ are infinitely often differentiable functions, there is a well-defined operation of the form:

$$\{P, Q\}_{\mathcal{M}_F} = \delta^{xy} \left( \partial_{\varphi^x} P \partial_{\pi^y} Q - \partial_{\pi^x} P \partial_{\varphi^y} Q \right) :=$$
$$\int_\Sigma d\text{Vol}(x) \left( \frac{\partial P}{\partial \varphi(x)} \frac{\partial Q}{\partial \pi(x)} - \frac{\partial P}{\partial \pi(x)} \frac{\partial Q}{\partial \varphi(x)} \right), \quad (3.19)$$

Summarizing: our classical phase space shall be considered to be a linear subspace of $\mathcal{M}_F = \mathcal{N}' \times \mathcal{N}'$. This subspace corresponds to the injection of $\mathcal{N}$ by the rigged Hilbert space structure introduced on $L^2(\Sigma, d\text{Vol})$. This structure is considered to be canonical as $d\text{Vol}$ is the measure induced by the Riemannian structure of the Cauchy hypersurface $\Sigma$. On this space, we can consider two (almost) equivalent structures: a (weakly) symplectic structure $\omega_{\mathcal{M}_F}$ and a Poisson tensor $\{\cdot, \cdot\}_{\mathcal{M}_F}$. It is important to remark on a subtle aspect of the construction: while the symplectic form on $T^*\mathcal{N}$ (Equation (3.16)) is canonical, the one on $\mathcal{M}_F$ (Equation (3.18)) is not, since it depends on the choice of measure $d\text{Vol}$. The same is true for the Poisson bracket (3.19).

### 3.3.2 States and Observables: Introducing a Complex Structure

We will introduce Gaussian statistical states in such a way that the covariance of the Gaussian is a Riemannian form compatible with the existing structure of $(\mathcal{M}_F, \omega_{\mathcal{M}_F})$. To do so, we introduce an additional (almost) complex structure $J_{\mathcal{M}_F}$ satisfying:

- it is densely defined,

- it is such that $J_{\mathcal{M}_F}^2 = -\mathbb{1}$ in its domain $D(J_{\mathcal{M}_F})$,

- it leaves $D(J_{\mathcal{M}_F}) \subseteq L^2(\Sigma, d\text{Vol}) \subset \mathcal{N}'$ invariant,

- and $\omega_{\mathcal{M}_F}(\cdot, J_{\mathcal{M}_F} \cdot) = -\omega_{\mathcal{M}_F}(J_{\mathcal{M}_F} \cdot, \cdot)$. This condition can be written as $J_{\mathcal{M}_F}^\dagger = -J_{\mathcal{M}_F}$ and means that the complex structure is compatible with the symplectic form $\omega_{\mathcal{M}_F}$. This is the imaginary part of the Hermitian product of $L^2(\Sigma, d\text{Vol})$. We will use this fact below to construct a different Hermitian product describing our Gaussian state.

---

[2]This Poisson structure is often mistaken for a symplectic structure in QFT quantization schemes since, for linear spaces, the difference is subtle (see [Bru&15] chapter 5 for further discussion on that matter).

It is simple to verify that there are infinitely many different structures satisfying those conditions. To express the most general complex structure, let $(\varphi^{\mathbf{x}}, \pi^{\mathbf{y}})$ be Darboux coordinates for $\mathcal{M}_F$ and let us consider the conventions for matrix operations of matrices of the type (2.82). In this case we extend them to include

$$(AB)^{\mathbf{x}}_{\mathbf{y}} = A^{\mathbf{x}}_z B^z_{\mathbf{y}}, \; (A^t)^{\mathbf{x}}_{\mathbf{y}} = \delta_{\mathbf{y}u} A^u_v \delta^{v\mathbf{x}}, \; A_{\mathbf{xy}} = \delta_{\mathbf{x}z} A^z_{\mathbf{y}} \text{ and } A^{\mathbf{xy}} = A^{\mathbf{x}}_z \delta^{z\mathbf{y}}. \tag{3.20}$$

Notice that the definition of these variables already depends on the choice of the Hilbert space structure for the Gel'fand triple.

An arbitrary (almost) complex structure on $\mathcal{M}_F$ is locally expressed in the canonical coordinates of the scalar field as

$$-J_{\mathcal{M}_F} = (\partial_{\varphi^y}, \partial_{\pi^y}) \begin{pmatrix} A^y_x & \Delta^y_x \\ D^y_x & -(A^t)^y_x \end{pmatrix} \begin{pmatrix} d\varphi^x \\ d\pi^x \end{pmatrix} =$$
$$= \partial_{\varphi^y} \otimes [A^y_x d\varphi^x + \Delta^y_x d\pi^x] + \partial_{\pi^x} \otimes [D^y_x d\varphi^x - (A^t)^y_x d\pi^y] \tag{3.21}$$

With the multiplication convention of (3.20), the condition $J^2 = -\mathbb{1}$ is translated into $A^2 + \Delta D = -\mathbb{1}$, $\Delta^t = \Delta, D^t = D, A\Delta = \Delta A^t$ and $A^t D = DA$. Notice that $D$ is fixed once $\Delta$ and $A$ are known. Let $K^{\mathbf{x}}_{\mathbf{y}}$ be the inverse of $\Delta^{\mathbf{x}}_{\mathbf{y}}$, i.e., $\Delta^{\mathbf{x}}_z K^z_{\mathbf{y}} = \delta^{\mathbf{x}}_{\mathbf{y}}$, then $D = (iA^t + \mathbb{1})K(iA - \mathbb{1})$.

This complex structure, together with the symplectic form (3.18), induces a (pseudo)-Riemannian structure $\mu_{J,\mathcal{M}_F}(\cdot, \cdot) = \omega_{\mathcal{M}_F}(\cdot, -J\cdot)$, which, in the coordinates above, reads:

$$\mu_{\mathcal{M}_F} = (d\varphi^x, d\pi^x) \begin{pmatrix} -D_{xy} & A^t_{xy} \\ A_{xy} & \Delta_{xy} \end{pmatrix} \begin{pmatrix} d\varphi^y \\ d\pi^y \end{pmatrix} =$$
$$-D_{xy}d\varphi^x \otimes d\varphi^y + \Delta_{xy}d\pi^x \otimes d\pi^y + A_{xy}(d\varphi^y \otimes d\pi^x + d\pi^x \otimes d\varphi^y) \tag{3.22}$$

To obtain a Riemannian structure from here, we must also impose $\Delta_{\mathbf{xy}} > 0 > D_{\mathbf{xy}}$. This extra structure induces a Hermitian structure

$$h_{\mathcal{M}_C} = \frac{\mu_{\mathcal{M}_C} - i\omega_{\mathcal{M}_C}}{2},$$

and a Gaussian state is given by the measure $\mu_r$ whose characteristic functional is

$$C_g(\chi_{\mathbf{x}}, \eta_{\mathbf{x}}) = \int_{\mathcal{N}' \times \mathcal{N}'} D\mu_r(\varphi^{\mathbf{x}}, \pi^{\mathbf{x}}) e^{i(\chi_x \varphi^x + \eta_x \pi^x)} := \exp\left(\frac{1}{2}\mu^{-1}_{\mathcal{M}_F}[(\chi_{\mathbf{x}}, \eta_{\mathbf{x}}), (\chi_{\mathbf{x}}, \eta_{\mathbf{x}})]\right). \tag{3.23}$$

With

$$\mu^{-1}_{\mathcal{M}_F} = (\partial_{\varphi^x}, \partial_{\pi^x}) \begin{pmatrix} \Delta^{xy} & -A^{xy} \\ -(A^t)^{xy} & -D^{xy} \end{pmatrix} \begin{pmatrix} \partial_{\varphi^y} \\ \partial_{\pi^y} \end{pmatrix} = \tag{3.24}$$

## Complex coordinates

The description given above is expressed in simpler terms using complex (holomorphic and antiholomorphic) coordinates. These coordinates are built upon a change of coordinates, locally given by

$$d\tilde{\varphi}^x = d\varphi^x \qquad\qquad d\tilde{\pi}^x = A_y^x d\varphi^y + \Delta_y^x d\pi^y \qquad (3.25)$$

$$\partial_{\tilde{\varphi}^x} = \partial_{\varphi^x} - (KA)_x^y \partial_{\pi^y} \qquad\qquad \partial_{\tilde{\pi}^x} = K_x^y \partial_{\pi^y} \qquad (3.26)$$

Using these coordinates the complex structure is cast into the canonical form $J_{\mathcal{M}_F} = \partial_{\tilde{\pi}^x} \otimes d\tilde{\varphi}^x - \partial_{\tilde{\varphi}^x} \otimes d\tilde{\pi}^x$, $\mu_{\mathcal{M}_F} = K_{xy}(d\tilde{\varphi}^x \otimes d\tilde{\varphi}^y + d\tilde{\pi}^x \otimes d\tilde{\pi}^y)$, and $\omega_{\mathcal{M}_F} = K_{xy}d\tilde{\pi}^x \wedge d\tilde{\varphi}^y$. We can cast the structures into this canonical form with other changes of variables, but we choose this one because it leaves $\varphi^x$ (the manifold of fields) invariant. Notice that we are transforming the original cotangent bundle structure we began with, as the new coordinates combine base and fiber coordinates of the original bundle structure.

Having chosen a complex structure, we can define an isomorphism for $\mathcal{M}_F$ to become a complex vector space $\mathcal{M}_C = \mathcal{M}_F^{\mathbb{C}}$. Notice that while $\mathcal{M}_F$ depends on the choice of the Hilbert space structure for the Gel'fand triple (and the original $T^*\mathcal{N}$), the definition of $\mathcal{M}_C$ also depends on the choice of the complex structure. This fact will be important later. Notice, anyway, that the symplectic structures on $T^*\mathcal{N}'$, $\mathcal{M}_F$, and $\mathcal{M}_C$ are all diffeomorphic, but the diffeomorphisms depend on the metric structure $h$ on $\Sigma$ (for $\mathcal{M}_F$) and the complex structure $J_{\mathcal{M}_C}$ (for $\mathcal{M}_C$).

Having chosen the complex manifold $\mathcal{M}_C$, we consider holomorphic and antiholomorphic coordinates which are locally written as

$$(\underset{\sim}{\phi}^{\mathbf{x}}, \underset{\sim}{\bar{\phi}}^{\mathbf{y}}) = \frac{1}{\sqrt{2}}(\tilde{\varphi}^{\mathbf{x}} - i\tilde{\pi}^{\mathbf{x}}, \tilde{\varphi}^{\mathbf{y}} + i\tilde{\pi}^{\mathbf{y}}). \qquad (3.27)$$

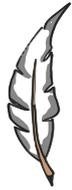

### Coordinates vs distributions

It is convenient to distinguish between $\phi^{\mathbf{x}} \in \mathcal{N}'_{\mathbb{C}}$ and the holomorphic coordinate of $\phi^{\mathbf{x}} \in \mathcal{M}_C$. On one hand, in this work, we may think of $\underset{\sim}{\phi}^{\mathbf{x}}$ as an object with physical meaning. That means that it is the solution of a classical equation of motion or is an object with a parametric dependence on time that has a definite meaning. On the other hand, $\phi^{\mathbf{x}}$ will often represent a placeholder for integration without any particular dependence on time. This distinction will be important later and in chapter 5 when we describe time evolution.

We use this sign convention for the holomorphic coordinates, in accordance with chapter 2, to ease the discussion on quantization in chapter 4. In this new manifold, the complex structure becomes $J_{\mathcal{M}_C}$, which is now written as:

$$J_{\mathcal{M}_C} = -i\partial_{\underset{\sim}{\phi^x}} \otimes d\underset{\sim}{\phi^x} + i\partial_{\underset{\sim}{\bar{\phi}^x}} \otimes d\underset{\sim}{\bar{\phi}^x}. \tag{3.28}$$

At the tangent space level, we can always write the projection operators $P_\pm = \frac{1}{2}(\mathbb{1} \pm iJ)$ such that $P_+\partial_{\phi^x} = \partial_{\underset{\sim}{\phi^x}}$ and $P_-\partial_{\bar{\phi}^x} = \partial_{\underset{\sim}{\bar{\phi}^x}}$.

As the triple $(\mu_C, \omega_C, J_C)$ defines a Kähler structure on $\mathcal{M}_C$, there exists a (non-unique) globally defined on a dense subspace (because the manifold is linear) Kähler potential $\mathcal{K}$

$$\mathcal{K}(\underset{\sim}{\bar{\phi}^x}, \underset{\sim}{\phi^x}) = \underset{\sim}{\bar{\phi}^x} K_{xy} \underset{\sim}{\phi^y}. \tag{3.29}$$

From the potential, we can define the corresponding Kähler form:

$$\omega_{\mathcal{M}_C} = i\partial\bar{\partial}\mathcal{K}, \tag{3.30}$$

where $\partial$ and $\bar{\partial}$ represent the Dolbeault operators. With the Kähler form and the expression of $J_{\mathcal{M}_C}$ from (3.28), we obtain the expression of the Riemannian form $\mu_{\mathcal{M}_C}$ and the Hermitian form:

$$h_{\mathcal{M}_C} = \frac{\mu_{\mathcal{M}_C} - i\omega_{\mathcal{M}_C}}{2} = K_{xy} d\underset{\sim}{\phi^x} \otimes d\underset{\sim}{\bar{\phi}^y},$$

with inverse $h_{\mathcal{M}_C}^{-1} = \Delta^{xy}\partial_{\underset{\sim}{\bar{\phi}^x}} \otimes \partial_{\underset{\sim}{\phi^y}}$.

Finally, the expression of the Poisson bracket (3.19) in holomorphic coordinates becomes:

$$\{P, Q\}_{\mathcal{M}_C} = i\Delta^{xy}\left(\partial_{\underset{\sim}{\phi^x}} P \partial_{\underset{\sim}{\bar{\phi}^y}} Q - \partial_{\underset{\sim}{\bar{\phi}^x}} P \partial_{\underset{\sim}{\phi^y}} Q\right). \tag{3.31}$$

In this case, we will rename the measure given by (3.23) as $\mu_c$, whose characteristic functional $C$ is obtained as

$$C(\rho_\mathbf{x}, \bar{\rho}_\mathbf{x}) = \int_{\mathcal{N}'_\mathbb{C}} D\mu_c(\phi^\mathbf{x}) e^{i(\overline{\rho_x\phi^x} + \rho_x\phi^x)} = e^{-\bar{\rho}_x\Delta^{xy}\rho_y}. \tag{3.32}$$

It is important to notice that the choice of the rigged Hilbert space made in (3.17), to express the system in Darboux coordinates, is different for holomorphic coordinates. In the first case, we built the triple from the real classical fields onto the real momentum fields. In holomorphic coordinates, the natural triple is defined from the complexified fields $\mathcal{N}_\mathbb{C}$ onto the complexified conjugate fields $\mathcal{N}'^*_\mathbb{C}$:

$$\mathcal{N}_\mathbb{C} \subset \mathcal{H}_\Delta \cong \mathcal{H}^*_K \subset \mathcal{N}'^*_\mathbb{C} \tag{3.33}$$

where $\mathcal{H}_\Delta = \overline{(\mathcal{N}_\mathbb{C}, \langle \cdot, \cdot \rangle_\Delta)}$ is the Hilbert space obtained by completion of $\mathcal{N}_\mathbb{C}$ with the scalar product induced by the tensor $\Delta$ as $\langle \chi_\mathbf{x}, \xi_\mathbf{y} \rangle_\Delta = \bar{\chi}_x \Delta^{xy} \xi_y$. This space is identified with the dense subset of $\mathcal{N}'^*_\mathbb{C}$ on which the Hermitian form $h_{\mathcal{M}_C}$ is finite.

This Hilbert space is the Cameron-Martin Hilbert space of $L^2_{Hol}(\mathcal{N}'_{\mathbb{C}}, D\mu_c)$. It has a physical interpretation for $\mathcal{H}_\Delta$ as the space corresponding to the *one-particle-state* structure. The space comprises the initial conditions for all the solutions of "positive frequency".

On the other hand, we can consider a similar construction from the dual perspective if we consider the completion of the dense subset of $\mathcal{N}'^*_{\mathbb{C}}$ on which the scalar product $\langle \cdot, \cdot \rangle_K$ induced by the Hermitian tensor $K$ is defined. Both spaces $\mathcal{H}_\Delta$ and $\mathcal{H}^*_K$ are identified by means of the Riesz representation theorem, realizing the duality $(\mathcal{N}_{\mathbb{C}}, \mathcal{N}'^*_{\mathbb{C}})$ with the scalar product $\langle \cdot, \cdot \rangle_\Delta$. In the following, we will consider them as the same space.

Summarizing: the classical phase space of fields, adapted to the statistical description, is described as a Kähler complex manifold $\mathcal{M}_C$, which has an associated rigged Hilbert space structure encoding the physical *one-particle-space* structure.

## 3.4 Koopman Van Hove theory for field theories

In general, we may describe any other statistical state with a Hida distribution $\varrho \in (\mathcal{N})' \otimes (\mathcal{N})'$ and any observable with a real Hida test function $O \in (\mathcal{N}) \otimes (\mathcal{N})$ such that

$$\langle O \rangle = \int_{\mathcal{N}' \times \mathcal{N}'} D\mu_r \varrho O \qquad (3.34)$$

In particular, we may restrict our study to physically relevant Gaussian states, and $\varrho \in L^1(\mathcal{N}' \times \mathcal{N}', \mu_r)$ will represent absolutely continuous states with respect to the Gaussian measure at hand. This may suffice if $\mu_r$ represents a stationary state of the dynamics.

In this section, we will assume the last formalism and study prequantization theory for these states. We will also discuss the evolution equation that governs the evolution. In order to keep the discussion simple, we will recur to heuristic arguments that will be made rigorous for the quantum evolution in chapter 5.

### 3.4.1 Prequantization I: The definition of the bundle and the states

To develop the theory of Koopman-van Hove (KvH) exposed in section 3.2 in the infinite-dimensional case, we start by defining the complex prequantum line bundle $B$ over the classical field phase space. This is a Hermitian line bundle $\pi_{\mathcal{M}_C, B} : B \to \mathcal{M}_C$, associated with a $U(1)$–principal bundle on $\mathcal{M}_C$. On this bundle, we define a principal connection that is required to have, as a local curvature form, the symplectic form of $\mathcal{M}_C$. This can be done if we consider the local connection one-form to be defined by the corresponding symplectic potential $\theta$ (as our field manifold is a linear space, this choice is well-defined globally). By using this connection, we can define a covariant derivative $\nabla$ on the associated bundle $B$. We use the geometric conventions of [Woo97].

The next step is to find a prequantized Hilbert space $\mathscr{H}_P$. To define this space, we focus on the case exposed in the previous section. We consider the set of square-integrable sections of $B$ with respect to the measure $\mu_c$ introduced in (3.32). By construction, this measure fulfills the important property of being invariant under symplectic transformations. With this choice, the scalar product of $\Psi^s, \Phi^s \in \Gamma(\mathcal{M}_C, B)$, sections of the prequantum bundle, can be written as:

$$\langle \Phi^s, \Psi^s \rangle = \int_{\mathcal{M}_C} D\mu_c \Phi^{s*} \Psi^s. \tag{3.35}$$

In this way, we have introduced a measure on the set of sections of the prequantum bundle, and we can define our pure quantum states to be the square-integrable sections with respect to the measure $\mu_c$, or equivalently, those sections with finite norm. Now we want to trivialize the bundle and factor out those sections with respect to a preferred one $\Psi_r$. This section will be our reference. This preference will be stated in terms of a particularly simple expression for its covariant derivative given by (3.38). We will ask this section not only to have norm 1 for this measure, but to have its local-on-$\phi^{\mathbf{x}}_{\sim}$ complex Hermitian product constant, i.e.,

$$\Psi_r^*(\bar{\phi}^{\mathbf{x}}_{\sim}, \phi^{\mathbf{x}}_{\sim}) \Psi_r(\bar{\phi}^{\mathbf{x}}_{\sim}, \phi^{\mathbf{x}}_{\sim}) = 1, \qquad \forall \ (\bar{\phi}^{\mathbf{x}}_{\sim}, \phi^{\mathbf{x}}_{\sim}) \in \mathcal{M}_C. \tag{3.36}$$

All other sections can then be represented as simple functions, which will be square integrable with respect to a measure defined by using $\Psi_r$. Notice that this situation is slightly different from the finite dimensional case. In that case we used the concept of (half-)density which are functions obtained using the symplectic musical isomorphism. This is, factoring out the symplectic measure $\Lambda$. This is impossible to reproduce in this case as we do not have a symplectic measure in the first place. For this reason we used as reference Gaussian states. The part of $\Psi$ that behaves as a density is encoded in $\mu_c$. As sections are also functions on $\mathcal{M}_C \cong \mathcal{N}'_{\mathbb{C}}$, we can write our prequantum Hilbert space as

$$\mathscr{H}_P = \left\{ \Phi : \mathcal{M}_C \to \mathbb{C} \text{ s.t. } \Phi \in L^2(\mathcal{N}'_{\mathbb{C}}, D\mu_c) \text{ as a function} \right\}. \tag{3.37}$$

The local expression of the reference section can be further determined by the covariant derivative of the bundle. We can define the reference section $\Psi_r$ adapted to the symplectic potential, hence, to the complex structure, requiring that

$$\nabla_X \Psi_r = -i\theta(X)\Psi_r. \tag{3.38}$$

Notice that while condition (3.36) fixes the modulus of the reference state, this condition fixes the phase. In any case, we can already remark that clearly the local expression of the state (or equivalently, the connection) depends on the choice of the complex structure on $\mathcal{M}_C$.

Using this section $\Psi_r$ as a trivialization of the bundle $B$, seen as a principal bundle of fiber $\mathbb{C}\backslash\{0\}$, any other section $\Phi^s$ is given by $\Phi^s = \Phi\Psi_r$. Here $\Phi : \mathcal{M}_C \to \mathbb{C}$ is identified with a regular smooth function of $(\mathcal{N}_{\mathbb{C}}) \otimes (\mathcal{N}_{\mathbb{C}})^* \subset L^2(\mathcal{N}'_{\mathbb{C}}, D\mu_c)$. Our notion of smooth function in this case is provided by Hida test functions. This notion also includes regularity properties under integration that we detailed in chapter 2. Hence, this decomposition is only valid for the subset of smooth sections in (3.37). However, this subset is dense, and as such, is enough to describe the prequantization procedure using it. The action of the covariant derivative is translated to the function $\Phi$ as:

$$\nabla_X(\Phi\Psi_r) = [X(\Phi) - i\theta(X)\Phi]\,\Psi_r. \tag{3.39}$$

Our prequantum Hilbert space is then identified with

$$\mathscr{H}_P \simeq L^2(\mathcal{N}'_{\mathbb{C}}, D\mu_c), \tag{3.40}$$

when referred to a reference section $\Psi_r$. The effect of the reference on (3.40) is that covariant derivatives incorporate an extra multiplicative factor $-i\theta(X)$ provided by the symplectic potential to the directional derivative.

Another key factor of this construction is that in the prequantum theory, every magnitude is obtained by integration of the classical degrees of freedom. This turns the classical manifold $\mathcal{M}_C$ into a rather auxiliary object.

### 3.4.2 Prequantization II: The prequantization of observables

The covariant differentiation is precisely the basic tool to build a prequantization mapping $\mathbb{P}$. We take as observables the real subset of Hida test functions that we identify as the set of smooth functions[3]

$$C_p^\infty(\mathcal{M}_C) \cong \mathrm{Re}\left[(\mathcal{N}_{\mathbb{C}}) \otimes (\mathcal{N}_{\mathbb{C}})^*\right].$$

For the classical functions $F \in C_p^\infty(\mathcal{M}_C)$ acting on the set of sections of $B$ the operator $\mathbb{P}(F)$ becomes

$$\mathbb{P}(F) = -i\hbar\nabla_{X_F} + F, \tag{3.41}$$

where $X_F$ is the Hamiltonian vector field associated with $F$. For the sake of simplicity, we will take $\hbar = 1$ in the rest of the paper.

When we consider the section $\Psi_r$, and the corresponding function $\Phi$, we can adapt $\mathbb{P}$ to the new setting described by (3.40). Thus, $\nabla_{X_F}$ must become a self-adjoint[4] operator on the Hilbert space of square-integrable functions, where $\Phi$ is contained. Notice that this condition of self-adjointness imposes further conditions on the symplectic potential $\theta$ that we will discuss below.

---

[3] Notice that this identification imposes also integrability properties and therefore is, in principle, more restrictive than the general notion of smooth function briefly discussed in section 2.8. According to (2.77), it is integrable under any $L^p$ norm, hence the subscript $p$.

[4] We assume that $F$ is such that this operator is indeed self-adjoint and not only formally self-adjoint.

The classical observables are chosen as functions on the classical phase space $F : \mathcal{M}_C \to \mathbb{R}$.

$$X_F \lrcorner \omega = -dF. \tag{3.42}$$

Remember that, being weakly symplectic, the equation above may not have a solution in general. Notice that if $F \in C_p^\infty(\mathcal{M}_F)$, using (2.77), the first Malliavin derivative is defined for them $F \in \mathbb{D}_\mu^{2,1}$. Therefore, the expression

$$dF = \partial_{\phi^x} F d\phi^x + \partial_{\bar\phi^x} F d\bar\phi^x,$$

ensures that $dF \in \mathcal{H}_\Delta \otimes L^2(\mathcal{N}'_\mathbb{C}, D\mu_c)$ and $\mathcal{H}_\Delta$ is the natural domain of $\omega$, so we can always find $X_F \in \mathcal{H}_\Delta \otimes L^2(\mathcal{N}'_\mathbb{C}, D\mu_c)$. Recall (2.50) and the discussion of section 2.6.4 for further details. Then, its prequantum counterpart is an operator $\mathbb{P}(F) : \mathscr{H}_P \to \mathscr{H}_P$ given by

$$\mathbb{P}(F)\Phi^s = F\Phi^s - i\nabla_{X_F}\Phi^s, \qquad \Phi^s \in \mathscr{H}_P. \tag{3.43}$$

It follows that this prequantization procedure $\mathbb{P}$ meets the Dirac prequantization conditions Q1)-Q3) of section 3.2. Factoring out the reference section $\Phi^s = \Phi\Psi^r$, we get that

$$\mathbb{P}(F)\Phi = F\Phi - iX_F\Phi - \theta(X_F)\Phi, \qquad \Phi^s \in \mathscr{H}_P. \tag{3.44}$$

Here $X_F\Phi$ is understood as a derivation. As pointed out below (3.27) in this work, we distinguish between coordinates $\phi^x \in \mathcal{M}_F$ and $\phi^x \in \mathcal{N}'_\mathbb{C}$. The latter is seen as a placeholder for integration without physical meaning; for this reason, when we express the mapping $\mathbb{P}$ in coordinates, for a function $F(\phi^x, \bar\phi^x)$, we write

$$\mathbb{P}[F(\phi^x, \bar\phi^x)] = F(\phi^x, \bar\phi^x) + \partial_{\bar\phi^x} F \Delta^{xy} \partial_{\phi^y} - \partial_{\phi^x} F \Delta^{xy} \partial_{\bar\phi^y} - \theta(X_F)(\phi^x, \bar\phi^x). \tag{3.45}$$

To render this operator self-adjoint, using Skorohod and Malliavin duality (2.62), we must impose that $\theta$ has an exact imaginary part:

$$\theta = \theta_r + \frac{i}{2}d\mathcal{K}(\phi^x, \bar\phi^x) = \theta_r + \frac{i}{2}d\left(\bar\phi^x K_{xy}\phi^y\right), \tag{3.46}$$

where $\theta_r$ is real. This is, the gauge freedom that we have in the choice of $\theta$ is only in the real part of the gauge potential.

### 3.4.3 Probability density and gauge freedom

In this Koopman-van Hove approach, a statistical field state is described by a section $\Psi^s \in \Gamma(\mathcal{M}_C, B)$ that can be factorized as $\Psi^s = \Psi\Psi^r$ with $\Psi \in L^2(\mathcal{N}'_\mathbb{C}, D\mu_c)$. However, the probability density is described by a Hida distribution that we assume to be integrable under the Gaussian state $\varrho \in L^1(\mathcal{N}'_{\mathbb{C}}, D\mu_c)$. To relate both of them, we postulate the analogous to (3.10):

$$\int_{\mathcal{N}'_{\mathbb{C}}} D\mu_c(\phi^{\mathbf{x}}) \varrho(\phi^{\mathbf{x}}, \bar{\phi}^{\mathbf{x}}) F(\phi^{\mathbf{x}}, \bar{\phi}^{\mathbf{x}}) = \int_{\mathcal{N}'_{\mathbb{C}}} D\mu_c \bar{\Psi} \mathbb{P}(F) \Psi. \tag{3.47}$$

Thus,

$$\varrho = |\Psi|^2 + \operatorname{div}\left(\theta \lrcorner \mathbb{J} |\Psi|^2\right) + i\{\Psi, \Psi^*\} + \Psi^*(iT + \theta(T))\Psi, \tag{3.48}$$

with $T = -i(\phi^x \partial_{\phi^x} - \bar{\phi}^x \partial_{\bar{\phi}^x})$, $\mathbb{J} = i\Delta^{xy}\partial_{\phi^x} \wedge \partial_{\bar{\phi}^y}$, and $\operatorname{div}(F^x \partial_{\phi^x} + G^x \partial_{\bar{\phi}^x}) = \partial_{\phi^x} F^x + \partial_{\bar{\phi}^x} G^x$. This expression, as in the finite-dimensional case, is invariant under gauge transformations:

$$\theta \to \theta + d\alpha \text{ and } \Psi \to e^{i\alpha}\Psi \text{ with } \alpha \in C^\infty_p(\mathcal{M}_F). \tag{3.49}$$

This gauge transformation is translated into a different connection on $B$ and may be interpreted as a phase choice on the reference section $\tilde{\Psi}_r$, which is the only freedom left when Equation (3.36) is fixed, along with the measure $\mu_c$.

### 3.4.4 Evolution

The non-existence of a uniform Lebesgue measure leads to some difficulties that must be treated with care in the evolution. Let's focus on the evolution equation (3.14). The first step is to describe the system in time-independent coordinates. Usually, we consider the canonical coordinates $(\varphi_{\mathbf{x}}, \pi^{\mathbf{x}})$ as time-independent. Thus, when the evolution of the dynamical system is generated by a Hamiltonian $H$, we get the equations of motion

$$\left(\frac{d}{dt} + \Gamma_c\right)\begin{pmatrix} \varphi^x \\ \pi^x \end{pmatrix} = \left\{\begin{pmatrix} \varphi^x \\ \pi^x \end{pmatrix}, H\right\} := F\begin{pmatrix} \varphi^x \\ \pi^x \end{pmatrix} \tag{3.50}$$

where $\Gamma_c = \begin{pmatrix} \delta_{xz}\dot{\delta}^{zy} & 0 \\ 0 & 0 \end{pmatrix}$ is regarded as a matrix term[5] that corrects the induced time dependence of $\varphi^x$. Notice that $\delta_{xz}\dot{\delta}^{zy}\varphi^x = \frac{\dot{h}}{2h}\varphi^y$. To lift this evolution to the Gaussian statistical state, we must consider the dependence of $\mu_c$ on the rest of the parameters. If we could write the uniform Lebesgue measure, we could refer the Gaussian measure $\mu_c$ to it as

$$D\mu_c \simeq D\phi^{\mathbf{x}} \frac{\exp\left(-\mu_{\mathcal{M}_C}(\bar{\phi}^x, \phi^x)\right)}{Z(t)} \tag{3.51}$$

---

[5] We will come back to this kind of expression in chapter 5 where we will interpret these kinds of terms as connections on a vector bundle.

where $Z(t) \propto \det(\mu_{\mathcal{M}_C})$ would be an infinite normalization factor. This kind of expression is common in the path integral formulation of the Schrödinger picture of QFT, see for example [LS96, LS98, CCQ04, HS15, HS17, HSU19, EHS21].

To extend the evolution equation (3.36) obtained in the finite-dimensional case, we must assume that the Gaussian part is factored out from the density operator

$$D \simeq \frac{\exp\left(-\mu_{\mathcal{M}_C}\left(\bar{\phi}^x, \phi^x\right)\right)}{Z}|\Psi|^2 \tag{3.52}$$

We want the Gaussian part of the evolution to be unaffected by the Hamiltonian; thus, we must choose $J_{\mathcal{M}_c}$ such that

$$\{\mu_{\mathcal{M}_C}\left(\bar{\phi}^x, \phi^x\right), H\} = 0 \tag{3.53}$$

This condition is met by choosing

$$J_{\mathcal{M}_C} = |F|^{-1}F, \tag{3.54}$$

in this way we get $\mathcal{L}_{X_H}J_{\mathcal{M}_C} = [F, J_{\mathcal{M}_C}] = 0$. Notice that this coincides with the prescription in [CCQ04] derived for QFT from a quite different set of physical requirements [AMA75]. Therefore the Liouville equation is

$$(\partial_t + 2\Gamma)D = -\{D, H\} \tag{3.55}$$

Thus the Koopman-van Hove equation is

$$i\left(\frac{d}{dt} + \Gamma\right)\Psi = \mathbb{P}(H)\Psi \tag{3.56}$$

With $\Gamma = -\frac{1}{2}\frac{d\mu_{\mathcal{M}_C}\left(\bar{\phi}^x, \phi^x\right)}{dt} - \frac{1}{2}\frac{d\log[\det(\mu_{\mathcal{M}_C})]}{dt} = -\frac{1}{2}\frac{d\mu_{\mathcal{M}_C}\left(\bar{\phi}^x, \phi^x\right)}{dt} - \frac{1}{2}\frac{d\,\text{Tr}[\log(\mu_{\mathcal{M}_C})]}{dt}$. The appearance of these kinds of functional determinants is common in QFT and is usually dealt with in perturbation theory as in [Wei96]. This derivation of the connection is the path proposed in [KM23] for the quantum case.

In any case, in this work, we are interested in the quantum evolution. In that case, in chapter 5, we will derive the appearance of an analogous connection with different, and mathematically rigorous, methods. Therefore, we will not discuss further the classical evolution equation of KvH and leave its study open for future works.

## 3.5 Kähler Structure for the Classical Klein-Gordon Theory

In this section, we will study the complex structure on the space of fields $J_{\mathcal{M}_C}$ that stems from KvH theory. As we will see, the prequantum operators will be hard to describe in the general case, and thus we will focus on their discussion in the quantum case in chapter 7.

For the spacetime structure in this chapter, we consider a $d + 1$ globally hyperbolic spacetime $(\mathcal{M}, \mathbf{g})$ endowed with a pseudo-Riemannian structure $\mathbf{g}$ of Lorentzian signature, with sign convention $(-, +, \cdots, +)$, and the Levi-Civita connection $\nabla_g$. Because of its condition of globally hyperbolic spacetime, $\mathcal{M}$ is diffeomorphic to $\mathbb{R} \times \Sigma$, where $\Sigma$ is a $d$-dimensional manifold diffeomorphic to every space-like Cauchy hypersurface of $\mathcal{M}$, and $\mathbb{R}$ represents the time parameter $t$. For our purposes in this work, we take $\Sigma$ compact and denote $\mathbf{h}_t$ as the $t$-parametric Riemannian metric induced from $\mathbf{g}$ by projection on $\Sigma$. In this chapter, we will treat the gravitational degrees of freedom as a parameter; that is, we will deal with a parametric theory of gravity [Kie12]. Nonetheless, this kind of projection is the basis of the Arnowitt-Deser-Misner (ADM) formalism of general relativity. Further details on this formalism may be found in [Gou07, Pad10]. In chapter 6 and chapter 7, we will use the results derived in this section to couple the theory to gravity and describe a QFT with back reaction.

To start with, we consider a one-parameter family of embeddings $\Sigma \to \mathcal{M}$. They are determined by a scalar function $\hat{t} : \mathcal{M} \to \mathbb{R}$ such that $\Sigma_t = \{\hat{t}(x) = t \text{ with } x \in \mathcal{M}\}$; i.e., $\Sigma_t$ are the level sets of $\hat{t}$. The structure of the whole spacetime projects into $(\Sigma_t, \mathbf{h}_t, \mathbf{D}_h)$ where $\mathbf{h}$ is a Riemannian structure and $\mathbf{D}$ is the corresponding Levi-Civita connection $\mathbf{D}\mathbf{h} = 0$.

We can also define the dynamics directly on $(\Sigma, \mathbf{h}, \mathbf{D})$. To this end, we take a coordinate system $\mathbf{x} = (x^i)$, $i = 1, \cdots, d$, over $\Sigma$ that induces a coordinate system over $\mathcal{M}$ of $(t, \mathbf{x})$. In this coordinate system, we obtain the relation

$$\partial_t = N\mathbf{n} + N^i \frac{\partial}{\partial x^i}, \tag{3.57}$$

where $N\mathbf{n} = \vec{\nabla}\hat{t}$ and $N = [-\mathbf{g}(\vec{\nabla}\hat{t}, \vec{\nabla}\hat{t})]^{-\frac{1}{2}}$ is the lapse function, while $\vec{N} = N^i \frac{\partial}{\partial x^i}$ is the shift vector. This foliation is depicted in Figure 3.1. Thus, if we provide as data $(N, \vec{N})$ over $\Sigma$, the embedding is fully determined. In chapter 6, we will use this fact to identify the action of the groupoid of hypersurface deformations of $\Sigma$.

In the Klein-Gordon classical theory, the dynamics of a scalar field can be defined as a parametric curve $\mathbb{R} \to \mathcal{N}$, where $\mathcal{N}$ is a nuclear space of test functions. We denote this solution by $\varphi_{\mathbf{x}}(t)$. Associated with the space of solutions, we build a Gel'fand triple with respect to the Hilbert space $L^2(\Sigma, d\mathrm{Vol})$, where $d\mathrm{Vol}$ represents the measure associated with the 3-metric tensor $\mathbf{h}$.

With these ingredients, we can introduce a Lagrangian formalism, as it is usually done in QFT. From the usual Lagrangian density in four dimensions, we can write the following bilinear form:

$$\mathcal{L}(t) = \int_\Sigma d\mathrm{Vol}\frac{1}{2}\Big[\frac{(\dot{\varphi})^2}{N} - \frac{N^i}{N}2\dot{\varphi}D_i\varphi + \Big(\frac{N^iN^j}{N} - Nh^{ij}\Big)(D_i\varphi)(D_j\varphi) - m^2(\varphi)^2\Big] \tag{3.58}$$

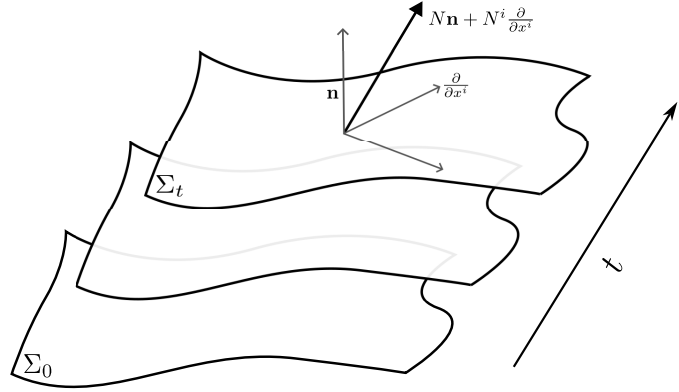

Figure 3.1: Foliation of a spacetime manifold provided by the family of Cauchy hypersurfaces $\Sigma_t$ with defined lapse function $N$ and shift vector $\vec{N}$.

where $N$, $N^i$, and $h^{ij}$ are functions on $\Sigma$.

In order to define the canonical momentum, we must choose between different options. One possible path is to define $\pi^{\mathsf{x}}$ as a density and therefore to build it as an element of $\mathcal{N}'$, which is the choice made in [CCQ04]. Another option is to consider $\varphi_{\mathsf{x}}$ as an element of a Rigged Hilbert space and profit from this structure to let $\pi_{\mathsf{x}}$ be an element of the same space as the field $\varphi_{\mathsf{x}}$. This option is the one we will choose to describe Fock quantization in section 4.5.2. If we derive the momenta in this prescription, with respect to the Gel'fand triple of $\mathcal{N} \subset L^2(\Sigma, dVol) \subset \mathcal{N}'$, we obtain as canonical momentum

$$\pi_{\mathsf{x}} = \delta_{\mathsf{x}\mathsf{y}} \frac{\partial \mathcal{L}}{\partial \dot{\varphi}_y} = \frac{1}{N}\Big[\dot{\varphi}_x - N^i D_i \varphi_x\Big] \tag{3.59}$$

The Hamiltonian is defined using the same rigging structure $\mathcal{H} = \delta^{xy}\pi_x\dot{\varphi}_y - \mathcal{L}$ and can be written as

$$H = \int_{\Sigma} d^d x [N\mathscr{H} + N^i \mathscr{H}_i] \tag{3.60}$$

where $\mathscr{H}^{\mathsf{x}}$ is a density called super Hamiltonian and $\mathscr{H}_i^{\mathsf{x}}$ are densities called supermomenta given by

$$\mathscr{H}^x = \frac{\sqrt{h}(x)}{2}[\pi_x^2 + h^{ij}D_i\varphi_x D_j\varphi_x + m^2\varphi_x^2] \tag{3.61}$$

$$\mathscr{H}_i^x = \sqrt{h}\pi_x D_i\varphi_x \tag{3.62}$$

In this chapter, we are interested in the choice $(\varphi^{\mathsf{x}}, \pi^{\mathsf{x}})$ as coordinates; thus, we obtain

$$\mathscr{H}^{\mathsf{x}} = \frac{1}{2\sqrt{h}}[(\pi^2)^{\mathsf{x}} + h^{ij}D_i\varphi^{\mathsf{x}}D_j\varphi^{\mathsf{x}} + m^2(\varphi^2)^{\mathsf{x}}] \tag{3.63}$$

$$\mathscr{H}_i^x = \sqrt{h}\pi^{\mathsf{x}}D_i\varphi^{\mathsf{x}} \tag{3.64}$$

As we can see, the definition of the super Hamiltonian involves products of the same distribution at the same space point **x**. This indicates, following [TW87], that point-splitting regularization procedures are also needed in the CSFT. As we indicated below (2.43), we can use Wick monomials to regularize the theory; see *Remark* 4.2. In fact, this is precisely the discussion that we will encounter for the quantum case in section 4.4.1.

The generator of (3.50) is

$$F = \left( \begin{array}{cc} N^i D_i & N \\ -\Theta & \Lambda \end{array} \right)$$

where $-\Theta = N D^i D_i + (D^i N) D_i - N m^2$ and $\Lambda = N^i D_i + D_i N^i$. This implies that $\Theta^x_y$ is a symmetric operator.

We can write down an explicit expression for $J_{\mathcal{M}_C}$ using the prescription (3.54), studying first two limiting cases where $J_{\mathcal{M}_F}$ can be written down straightforwardly.

**Null Shift**  In this case, we set $N^i = 0$, and we get

$$J_0 = \left( \begin{array}{cc} 0 & \Theta^{-\frac{1}{2}} N^{\frac{1}{2}} \\ -N^{-\frac{1}{2}} \Theta^{\frac{1}{2}} & 0 \end{array} \right) \tag{3.65}$$

**Huge Shift**  In this case, we consider $N^i N_i \gg N^2$, and therefore we set $N \simeq 0$. This is, of course, a limiting approximation because the lapse function can never vanish. Then we get

$$J_\infty = \left( \begin{array}{cc} (-\alpha^2)^{-\frac{1}{2}} N^i D_i & 0 \\ 0 & (-\Lambda^2)^{-\frac{1}{2}} \Lambda \end{array} \right) \tag{3.66}$$

with $\alpha = N^i D_i$.

**Interpolating Cases**  The other cases can be obtained as

$$J_{\mathcal{M}_c} = A_0 J_0 + A_\infty J_\infty \tag{3.67}$$

where $A_0 = |F|^{-1} |F_0|$ and $A_\infty = |F|^{-1} |F_\infty|$ are operators, being

$$|F_0| = \left( \begin{array}{cc} (-N\Theta)^{\frac{1}{2}} & 0 \\ 0 & (-\Theta N)^{\frac{1}{2}} \end{array} \right), \quad |F_\infty| = \left( \begin{array}{cc} (-\alpha^2)^{\frac{1}{2}} & 0 \\ 0 & (-\Lambda^2)^{\frac{1}{2}} \end{array} \right), \tag{3.68}$$

$$|F|^2 = |F_0|^2 + |F_\infty|^2 - \left( |F_0| J_0 J_\infty |F_\infty| + |F_\infty| J_\infty J_0 |F_0| \right) \tag{3.69}$$

In general, this expression is difficult to work out explicitly because of the need to compute $|F|^{-1}$. The derivation of the general complex structure can be found in [CCQ04] with the corresponding changes due to $\varphi$ being considered a scalar instead of a density.

To complete the KvH picture, we must choose a gauge for the covariant derivative. In these coordinates, the natural choice is the Liouville gauge, corrected by (3.46). This is, the symplectic potential is

$$\theta = \pi^x \delta_{xy} d\varphi^y + \frac{i}{2} d\Big(\pi^x \Delta_{xy} \pi^y - \varphi^x D_{xy}\varphi^x + \varphi^x A_{xy}\pi^y + \varphi^x A_{xy}^t \pi^y\Big). \tag{3.70}$$

However, since we do not have an explicit expression for $J_{\mathcal{M}_C}$, the particular form of the prequantum operators is very difficult to write down. In this work, we will focus our efforts on the quantum case. For this matter, in chapter 4, we will discuss different gauges; in chapter 5, we will understand the quantum evolution; and in chapter 7, we will expand this example in the quantum case.

Nonetheless, we can draw some conclusions from this analysis. As we advanced in the introduction, the time dependence of $J_{\mathcal{M}_c}$ and, in turn, the whole Kähler structure, is through the geometric content. We can encapsulate this fact into a slightly more general context of parametric theories on a curved spacetime.

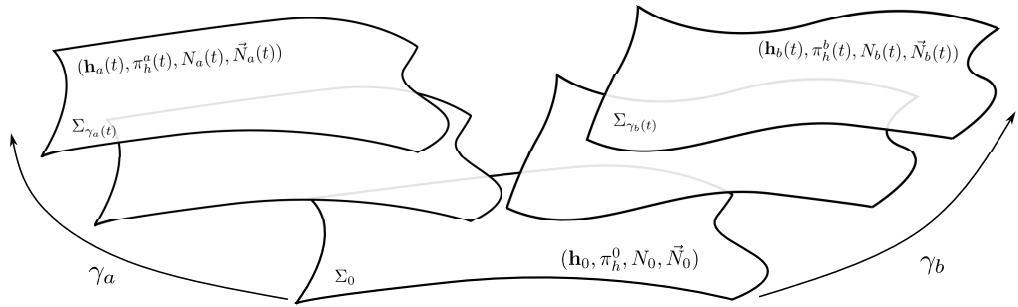

Figure 3.2: Two different paths $\gamma_a$ and $\gamma_b$ that parametrize the space-time content of classical gravitational theories. These curves are defined by $\gamma_a(0) = \gamma_b(0) = (\mathbf{h}_0, \pi_h^0, N_0, \vec{N}_0)$ and are given by $\gamma_{a/b}(t) = (\mathbf{h}_{a/b}(t), \pi_h^{a/b}(t), N_{a/b}(t), \vec{N}_{a/b}(t))$

**Remark 3.1. *(Parametric Theory on a Curved Spacetime)***

*For a given background spacetime, the geometric content on each hypersurface is the one given by the choice of the foliation, and thus, we have a description of the spacetime in terms of a curve $\gamma : \mathbb{R} \to T^* \mathrm{Riem}(\Sigma) \times \mathcal{C}\Gamma(\mathcal{A}_\Sigma)$. Here, the space $\mathrm{Riem}(\Sigma)$ contains the Riemannian metrics $\mathbf{h}$ over $\Sigma$, and we should also consider its cotangent bundle $T^* \mathrm{Riem}(\Sigma)$ with coordinates $(\mathbf{h}, \pi_h)$. On the other hand, $\mathcal{C}\Gamma(\mathcal{A}_\Sigma)$ will be introduced in (6.12) of section 6.2.1 and is the space of lapse functions and shift vectors. These spaces and their relation with General Relativity will be studied in detail in chapter 6.*

*A theory in which gravity is introduced as a background by selecting a curve*
$\gamma(t) = (\mathbf{h}(t), \pi_h(t), N(t), \vec{N}(t))$ *is what we call a parametric theory on a curved space-time or, when it is clear by context, simply parametric theory. In this case the geometrical content at the hypersurface $\Sigma_t$ is known beforehand for $t \geq 0$. In these theories, the dependence of the geometric variables manifests itself as an additional dependence on the Hamiltonian time. Figure 3.2 depicts an example of two different parametrizations of this kind.*

We can clearly see that the Hamiltonian $H(t)$ and the derived Kähler structure of the Klein-Gordon theory represent a parametric theory.

On the other hand, in a theory where, instead of being a background, the metric evolves coupled to the quantum matter, and therefore exhibits *backreaction* from the matter evolution, this geometric curve is unknown beforehand. In this case, one must make the Kähler structure dependent on the gravitational degrees of freedom instead of time. This is the subject of study of chapter 6.

# Chapter 4

# Quantization: Representations and Orderings

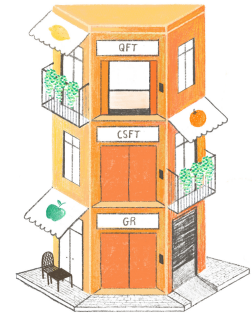

## 4.1 Introduction

Our goal in this chapter is to describe a non-covariant quantization program of a classical theory. As our aim is a geometric description of the resulting quantum model, we will consider a geometric quantization framework adapted to our infinite-dimensional manifold. However, this program is only valid over the excessively restrictive set of linear functions. Thus, to complete this quantization program, we must choose an ordering prescription.

The goal of this chapter is to describe different quantization mappings $Q$. In principle, the quantization mapping $Q$ is supposed to meet the Dirac quantization conditions [Hal13]

Q1) $Q(F + G) = Q(F) + Q(G)$

Q2) $Q(F)Q(G) - Q(G)Q(F) = -i\hbar Q(\{F, G\})$

Q3) $Q(F) = F\mathbb{1}$ if $F$ is constant

Q4) **(Irreducibility condition)** For a given set of classical observables $\{f_i\}_\mathcal{I}$ such that $\{f_i, g\} = 0 \ \forall i \in \mathcal{I}$ implies $g$ is constant; then if an operator $A$ commutes with every $Q(f_i)$, then $A$ is a multiple of the identity.

As we have seen in the previous chapter, $Q1) - Q3)$ describe a purely classical theory. The irreducibility condition, however, is not met by $\mathbb{P}$. This is, in the last instance, a reflection of the gauge freedom of the KvH theory. Moreover, it is known that no quantization satisfying Q1-Q4 exists. To fulfill those equations, we quantize only a subset of classical observables, such as the linear ones in Darboux coordinates $(f_x \varphi^x + g_x \pi^x)$.

Nonetheless, we want to describe the quantization of functions of arbitrary degree. For that matter, higher-order functions are allowed to break Q2) in such a way that

Q2') $Q(F)Q(G) - Q(G)Q(F) = -i\hbar Q(\{F, G\}) + o(\hbar^2)$.

The geometric quantization procedure, roughly speaking, is similar to the usual construction on finite-dimensional symplectic manifolds (see [Woo97, Hal13, Tuy16]). The main novelty of this approach is the extensive use of tools from Gaussian analysis described in chapter 2. This stage comes in two steps:

- Firstly, a pre-quantization is defined, which builds an initial Hilbert space. This part is common to the statistical description of a classical system in KvH described in the previous section, thus

$$Q(F) = -i\hbar \nabla_{X_F} + F, \qquad (4.1)$$

- Secondly, the definition of a polarization completes the quantization procedure for linear operators. This is the choice of a Lagrangian submanifold on the phase space of fields. The quantum Hilbert space is obtained by restriction of prequantum sections to this submanifold. Quantum operators are also restricted to those preserving the polarization in a suitable way.

The definition of the polarization is the novelty of the quantum procedure that breaks the gauge freedom of the KvH approach. Each gauge represents a different representation of the quantum theory, as we will show in section 4.2. The relation among every representation will be implemented through unitary operators and will be described in section 4.3. To complete the quantization program, in section 4.4 we add

- Thirdly, we expand the quantization procedure with an ordering prescription. This is a predetermined order to write monomials of any order from the quantization prescription of linear ones. In particular, we will use Weyl and Wick ordering prescriptions.

As in the previous chapter, we will present the construction adapted to $\mathcal{M}_F$ and we will assume that this manifold is linear and modeled over the DNF space $\mathcal{N}'$. We will make use of the Kähler structure $(\omega, \mu, J)_{\mathcal{M}_F}$ introduced in chapter 2. Following that study of the classical KvH formalism, we can isolate the purely quantum features of the theory from the technicalities already present in the classical case.

The first important difference is interpretative. In the quantum case, the Gaussian state $\mu_c$ includes parts of the vacuum of the QFT. In the definition of the measure, the expression of the modulus of the reference $\Psi_r$ usually appears in the literature as the modulus of the vacuum of the theory [LS96, LS98, CCQ04, HS15, HS17, HSU19, EHS21]. In those works, the reference is referred to the informally defined Lebesgue measure $D\phi$. In this way, $\Psi'_0$ is an informally defined function representing the vacuum, and its modulus is such that $D\phi|\Psi'_0|^2 \simeq D\mu_c$ in a heuristic way. However,

Equation (3.36) leaves room to add a phase to the reference state to describe the vacuum of the theory that we denote $\Psi_0$. The vacuum, as usual, will be defined as the state annihilated by the annihilation operator.

The second main difference that we will encounter is related to the loss of gauge freedom present in KvH. The condition of an irreducible representation introduces ordering problems in the quantization prescription. However, other aspects, such as two-point splitting regularization, usually regarded as QFT quirks, were already present in the classical case.

Throughout the chapter, we will develop different representations of a QFT that appear from different equally valid choices of polarizations. The relation among them, as pointed out above, is briefly discussed in section 4.3, but a question remains open. What is precisely the relation among representations of the same QFT? Focusing on the $C^*$-algebra of bounded observables, the answer to this question is provided by Algebraic QFT in section 4.5. In particular, we will illustrate how to relate every representation derived in this chapter with Fock quantization.

## 4.2 Geometric quantization for linear functions: types of quantization

Our aim in this section is to describe the different representations of a QFT that will suit our study. Our presentation is similar to others in the literature. Examples of this construction for field theories can be found in Oeckl's work [Oec12b, Oec12a, Oec12c], while, for finite-dimensional systems, [Woo97, Hal13] contain detailed discussions of the aspects discussed below. Some representations, such as the field-momenta and antiholomorphic representations, to our knowledge, have not been studied before.

The prequantization provides a Hilbert space $\mathscr{H}_P$ too big to realize the quantum states since, in principle, $\Phi \in \mathscr{H}_P$ are functions with a domain on the whole manifold, while, in regular quantum theories, the domain must have half the degrees of freedom of $\mathcal{M}_C$. To deal with this discrepancy, we must restrict the domain of functions $\Phi$ to a Lagrangian submanifold of $\mathcal{M}_C$. This is a maximal isotropic submanifold with respect to the symplectic product induced by $\omega_{\mathcal{M}_F}$.

Notice, though, that in this setting the concept of Lagrangian submanifold is not as straightforward as in the finite-dimensional case treated in [Woo97]. First, we must remember that $\omega_{\mathcal{M}_C}$ is defined, at most, on a dense subset of the vector fields of $\mathcal{M}_C \cong \mathcal{N}'_{\mathbb{C}}$. Therefore, the condition of (co)isotropy must take into account this fact. To be more precise, let $\Psi$ be a pre-quantum state, and consider a translation by $\Psi(\phi^{\mathsf{x}}) \mapsto \Psi(\phi^{\mathsf{x}} + h^{\mathsf{x}})$. Only when $h^{\mathsf{x}} \in \mathcal{H}_K \simeq \mathcal{H}_\Delta$, the Cameron-Martin Hilbert space, of

$$\mathcal{N}_{\mathbb{C}} \subset \mathcal{H}_\Delta \cong \mathcal{H}_K^* \subset \mathcal{N}'^*_{\mathbb{C}} \tag{4.2}$$

can we reabsorb this into the Gaussian measure (3.32) with an integrable Radon-Nikodym factor [Hid&93, NØP09]. Hence, we can consider $\mathcal{H}_\Delta$ as the only well-defined tangent vectors on the linear manifold $\mathcal{M}_C \simeq \mathcal{N}'_\mathbb{C}$ for the quantum counterpart, since they are the only ones that leave the Hilbert space invariant. Moreover, this space represents the allowed directions of translation that leave invariant the spaces $L^p(\mathcal{N}'_\mathbb{C}, D\mu_c)^1$ and has the physical interpretation of a *one-particle state structure*. It is also the main ingredient of the Wiener-Ito decomposition theorem that provides the particle interpretation of a QFT written as a Gaussian integral.

To bypass this difficulty on the definition of Lagrangian submanifolds, we will introduce a polarization; this is a distribution of vector fields spanning a foliation of $\mathcal{M}_C$ in Lagrangian submanifolds. We will introduce it at the level of the *one-particle state structure* introduced in the complex description of the classical phase space (3.33). Thus, we will choose a Lagrangian subspace $P$ of the complex Hilbert space $P \subset \mathcal{H}_\Delta$ on which the symplectic form $\omega_{\mathcal{M}_C}$ is well defined. The quantum Hilbert space is given by:

$$\mathscr{H} = \overline{\{\Psi^s \in \mathscr{H}_{\overline{P}} \text{ s.t. } \nabla_{\bar{X}}\Psi^s = 0 \ \forall X \in P\}} \tag{4.3}$$

As, again, the space is linear, we can choose a symplectic potential $-d\Theta = \omega_{\mathcal{M}_C}$ adapted to the polarization such that

$$\Theta(\bar{X}) = 0 \ \forall \ X \in P \tag{4.4}$$

Under this assumption, the definition of the polarization simplifies enormously since, due to (3.39), we can identify the polarized Hilbert space as:

$$\mathscr{H} = \overline{\{\Psi \in C^\infty(\mathcal{M}_C) \text{ s.t. } d\Psi(\bar{X}) = 0 \ \forall X \in P\}} \tag{4.5}$$

Roughly speaking, we can consider two kinds of polarizations. We will describe them through their symplectic potentials:

- Real polarizations where the linear space $P$ is characterized by $P = P^*$ and lead to:

  - Schrödinger quantization, described by functions $\Psi(\varphi^\mathbf{x})$.

    $$\theta_S = -i(\tilde{\varphi}^x + i\tilde{\pi}^x)K_{xy}d\tilde{\varphi}^y = \pi^x\delta_{xy}d\varphi^y - i\varphi^x K_{xy}d\varphi^y + \varphi^x(KA)_{xy}d\varphi^y.$$

  - Field-Momentum quantization, described by functions $\Psi(\pi^\mathbf{x})$.

    $$\breve{\theta}_M = i(\breve{\pi}^x - i\breve{\varphi}^x)D_{xy}^{-1}d\breve{\pi}^y = -\varphi^x\delta_{xy}d\pi^y + i\pi^x D_{xy}^{-1}d\pi^y + \pi^x(AD^{-1})_{xy}d\pi^y.$$

    Coordinates $(\breve{\varphi}^\mathbf{x}, \breve{\pi}^\mathbf{x})$ are described in (4.22).

---

[1] See Proposition 6.9 of [Hu16].

- (Anti)holomorphic quantizations for which $P \cap P^* = 0$:
    - Holomorphic quantization, described by functions $\Psi(\phi^{\mathsf{x}})$.

    $$\Theta = -i\bar{\phi}^x K_{xy} d\phi^y$$

    - Antiholomorphic quantization, described by functions $\Psi(\bar{\phi}^{\mathsf{x}})$.

    $$\bar{\Theta} = i\phi^x D_{xy}^{-1} d\bar{\phi}^y$$

    With $D_{xy}^{-1} D^{yz} = \delta_x^z$.

Intermediate situations are combinations of those two extreme cases, at least in the finite-dimensional setting; see [Woo97] for further details. In this section, we will describe the geometric quantization procedures that lead to each of these representations. We denote these quantization mappings $\mathcal{Q}, \mathcal{Q}_s, \overline{\mathcal{Q}}_m$ and $\overline{\mathcal{Q}}$ holomorphic, Schrödinger, field-momentum, and antiholomorphic, respectively.

## 4.2.1 Holomorphic quantization

As the holomorphic case is easier to handle in many ways and possesses better analytical properties described in chapter 2, we will begin our discussion with it.

For the holomorphic case, the complex structure $J_{\mathcal{M}_C}$ provides a projection $X \to \frac{1}{2}(\mathbb{1} + iJ_{\mathcal{M}_C})X$ for the elements of the Cameron Martin tangent space $X \in \mathcal{H}_\Delta$. The set of coordinates adapted to this quantization procedure is $(\phi^{\mathsf{x}}, \bar{\phi}^{\mathsf{x}})$ as described in (3.27), and the measure of the prequantum space is described by $\mu_c$, defined in (3.32) by

$$C(\rho_{\mathsf{x}}, \bar{\rho}_{\mathsf{x}}) = \int_{\mathcal{N}'_{\mathbb{C}}} D\mu_c(\phi^{\mathsf{x}}) e^{i(\overline{\rho_x \bar{\phi}^x} + \rho_x \phi^x)} = e^{-\bar{\rho}_x \Delta^{xy} \rho_y}. \tag{4.6}$$

As $\mathcal{M}_C$ is a linear space, the polarization is taken to be $P_{Hol} = \frac{1}{2}(\mathbb{1} + iJ_{\mathcal{M}_C})\mathcal{H}_\Delta$. Identifying coordinates $\phi^{\mathsf{x}} \in \mathcal{N}'_{\mathbb{C}}$ with holomorphic coordinates of $\phi^{\mathsf{x}} \in \mathcal{M}_C$, the adapted symplectic potential is obtained from the (non-unique) Kähler potential $\mathcal{K} = \bar{\phi}^x K_{xy} \phi^y$:

$$\Theta = -i\partial_{\phi^x} \mathcal{K} d\phi^x. \tag{4.7}$$

The set of states of the quantum Hilbert space given by (4.5) is therefore the space of holomorphic (square integrable) functions $\Phi : \mathcal{N}'_{\mathbb{C}} \to \mathbb{C}$. In this representation, we identify the reference and the vacuum sections $\Psi_r^H = \Psi_0$, because, as we will see, it will be annihilated by the annihilation operator. These functions $\Phi$ represent the excitations with respect to the vacuum state $\Psi_0$ (3.38), in such a way that the physical state corresponds to the section of the complex line bundle $\Phi\Psi_0$, which is square integrable with respect to the measure $\mu_c$, or, equivalently, to the set of square integrable holomorphic functions $\mathcal{Q}$ if we take the section $\Psi_0$ as a reference:

$$\mathscr{H}_{Hol} = L^2_{Hol}(\mathcal{N}'_{\mathbb{C}}, D\mu_c). \tag{4.8}$$

This is, by construction, a well-defined Hilbert space, but it is not the most common quantum model in QFT. Some examples of the study of this representation of QFT can be found in [Oec12c, Oec12b, Oec12a].

Instead, it is more common to represent the states as elements of a Bosonic (or fermionic) Fock space, see for instance [Wal94, Bru&15]. From the properties of Gaussian integration, we know that the Segal isomorphism **Definition** 2.10 allows us to identify the vectors in $L^2(\mathcal{N}'_\mathbb{C}, D\mu_c)$ with the vectors of the symmetric Fock space constructed on the space of one-particle states $\mathcal{H}_\Delta$. As the quantum states correspond to holomorphic functions $\Psi \in L^2_{Hol}(\mathcal{N}'_\mathbb{C}, D\mu_c)$, we can write:

$$\mathcal{I}(\Psi) = \left(\sqrt{n!}\psi^n_{\vec{x}_n}\right)^\infty_{n=0},\tag{4.9}$$

where $\psi^n_{\vec{x}_n}$ correspond to the coefficients of the state $\Psi$ with respect to the space of Wick complex monomials described in (2.38). Thus, the Segal isomorphism will be the connection with respect to the regular Fock space description. We will develop the latter further in section 4.5.

The physical interpretation of the *one-particle-state* structure $\mathcal{H}_\Delta$ is the space that comprises the initial conditions for all the solutions of "positive frequency." In the quantum part, these are to be interpreted as particles, in contrast with the elements of the conjugate $\mathcal{H}^*_\Delta$, which represent the corresponding antiparticles.

**Quantizing linear operators**

At the same time, as the connection is also holomorphic, the quantization mapping $\mathcal{Q}$ is simple to compute and defines holomorphic linear operators on $\mathscr{H}_{Hol}$. In particular, using (3.43) for (4.7), we verify that

$$\begin{aligned}
\mathcal{Q}(\phi^\mathbf{y})\Psi(\phi^\mathbf{x}) &= \phi^\mathbf{y}\Psi(\phi^\mathbf{x}),\\
\mathcal{Q}(\bar{\phi}^\mathbf{y})\Psi(\phi^\mathbf{x}) &= \Delta^{\mathbf{y}z}\partial_{\phi^z}\Psi(\phi^\mathbf{x}), &\forall\,\Psi \in \mathscr{H}_{Hol}.
\end{aligned}\tag{4.10}$$

Here we distinguish again $\phi^\mathbf{x} \in \mathcal{N}'_\mathbb{C}$ and the holomorphic coordinate $\phi^\mathbf{x} \in \mathcal{M}_C$. In this work, we will consider bare coordinates $\phi^\mathbf{x}$ as simple placeholders of integration. In particular, they do not depend on time or any other structure built over the Cauchy hypersurface $\Sigma$. This is done to highlight the physical meaning of the coordinate $\phi^\mathbf{x} \in \mathcal{M}_C$. We will choose different systems of holomorphic coordinates of $\mathcal{M}_C$ with different interpretations that we will represent over the same $\mathcal{N}'_\mathbb{C}$ after quantization. This is crucial to relate different quantizations of the theory, as we will see at the end of the section.

Notice that these operators correspond to the creation and annihilation operators, as was to be expected: $a^{\dagger\mathbf{x}} = \mathcal{Q}(\phi^\mathbf{x})$ while $a^\mathbf{x} = \mathcal{Q}(\bar{\phi}^\mathbf{x})$. Thus, we obtain that annihilation and creation operators correspond to the Malliavin derivative and Skorohod integral for holomorphic functions. This interpretation that we obtain from geometric quantization was already noted in the stochastic calculus literature from their algebraic relations [Hen09].

Notice also that (4.10) implies that $a^{\mathbf{x}}1 = 0$. We identify $\Psi_r^H = 1$ as the representation of the holomorphic function $\Psi_r^H \in L_{Hol}^2(\mathcal{N}_r', D\mu_c)$ of the reference section $\Psi_r^H$ in $\mathscr{H}_P$. Thus, we conclude that $a^{\mathbf{x}}\Psi_r^H = 0$. This implies that the reference section is annihilated by the annihilation operator, and therefore we can identify the reference and the vacuum $\Psi_r^H = \Psi_0$ as we did above.

### 4.2.2 Schrödinger quantization

Let us consider now the other most common example, the case of Schrödinger quantization, as we can see in [LS96, LS98, CCQ04, HS15, HS17, HSU19, EHS21]. In the Schrödinger representation, the quantum states are represented by the wave functional $\Psi(\varphi^{\mathbf{x}})$, i.e., they must depend only on the field variables $\varphi^{\mathbf{x}}$, and not on the field momenta $\pi^{\mathbf{x}}$. This is derived from (3.40) by applying (4.3) with a real polarization spanned by the momentum directions in $\mathcal{M}_F$. See [Oec12c] for a similar construction and [CCQ04] for an algebraic derivation of the representation.

#### The measure and the connection

The change from complex to real coordinates does not really affect the structure of the line bundle $B$ required to define the prequantization: we construct identical bundles on the two identical base manifolds $\mathcal{M}_C$ and $\mathcal{M}_F$ with different coordinates; and for each bundle, we can define a connection. These two connections need not be identical (since we can consider several different connections for each case), but they must have a curvature proportional to the symplectic form.

Identifying $\mathcal{M}_F$ and $\mathcal{M}_C$ also means that the measure on the space of fields can also be considered associated with the functional

$$C_S(\xi_{\mathbf{x}}, \eta_{\mathbf{x}}) = \int_{\mathcal{N}' \times \mathcal{N}'} D\mu(\varphi^{\mathbf{x}}, \pi^{\mathbf{x}}) e^{i(\xi_x \varphi^x + \eta_x \pi^x)}. \tag{4.11}$$

Real polarizations will be understood as taking only one copy of $\mathcal{N}'$, which is to restrict to the coordinate $\varphi^{\mathbf{x}}$. For this reason, in order to avoid the rescaling factor $\frac{1}{\sqrt{2}}$ in the quantization prescription, we must rescale the covariance of the measure adapted to the holomorphic prescription $\mu_c$. Taking $\rho_{\mathbf{x}} = \frac{1}{\sqrt{2}}(\xi_{\mathbf{x}} + i\eta_{\mathbf{x}})$ and $\phi^{\mathbf{x}} = \frac{1}{\sqrt{2}}(\varphi^{\mathbf{x}} - i\pi^{\mathbf{x}})$, we must have that $\mu$ is defined by the characteristic functional

$$C_S(\xi_{\mathbf{x}}, \eta_{\mathbf{x}}) = C_H\left(\frac{\bar{\rho}_{\mathbf{x}}}{\sqrt{2}}, \frac{\rho_{\mathbf{x}}}{\sqrt{2}}\right) = C_H\left(\frac{\xi_{\mathbf{x}} - i\eta_{\mathbf{x}}}{2}, \frac{\xi_{\mathbf{x}} + i\eta_{\mathbf{x}}}{2}\right). \tag{4.12}$$

Again, we will consider a pre-quantum Hilbert space of sections of the bundle $B$, square-integrable with respect to the measure $\mu$. We can also identify a trivializing section $\Psi_r^S$ compatible with the connection in the sense of Equation (3.38). With this trivialization, the space of states becomes the set of square-integrable functions on $\mathcal{N}'$ with respect to the measure $\mu$.

From the isomorphism discussed above between $\mathcal{M}_C$ and $\mathcal{M}_F$, we can also conclude the equivalence of the connections on both spaces and the corresponding covariant derivatives. Strictly speaking, we should use different notation for the vector fields with respect to which we differentiate, since they are isomorphic and not identical, but we will use the same notation to make the manuscript easier to read. We shall use then the notation of Equation (3.39) to represent the covariant derivative in the real polarization case, keeping in mind the fact that the only relevant vector fields are those tangent to the submanifold spanned by the field directions, as we will see now.

### The Schrödinger polarization

In the Schrödinger case, the polarization is spanned by the momentum directions, i.e., the real subspace $P_S = \mathrm{Im}(H_\Delta)$. We can consider an adapted symplectic potential with respect to $P_S$ satisfying

$$\theta_S(X) = 0 \qquad \forall X \in P_S. \tag{4.13}$$

This means that $\theta_S$ must be a linear combination of the set $\{d\tilde{\varphi}^{\mathsf{x}}\}$ of "basical" forms. The coefficients are read from the identification of the Gel'fand triple and the canonical Liouville 1-form of $T^*\mathcal{N}'$. Writing it in terms of the variables we introduced in the previous section, we obtain:

$$\theta_S = -i(\tilde{\varphi}^x + i\tilde{\pi}^x)K_{xy}d\tilde{\varphi}^y = \pi^x\delta_{xy}d\varphi^y - i\varphi^x K_{xy}d\varphi^y + \varphi^x(KA)_{xy}d\varphi^y. \tag{4.14}$$

Notice that we added the term $-i\tilde{\varphi}^x K_{xy}d\tilde{\varphi}^y$ (which is just an exact one-form equal to the differential of the norm of the $\bar{\varphi}^x$) to the usual conventions because in this way we ensure that the quantization prescription provides Hermitian operators in $L^2(\mathcal{N}', D\mu)$, as we will see below, and $2\Theta = \theta_S + \theta_M$ with

$$\theta_M = -(\tilde{\varphi}^x + i\tilde{\pi}^x)K_{xy}d\tilde{\pi}^y. \tag{4.15}$$

From this point of view, $\theta_S$ is the restriction of $\Theta$ to $P_S$ directions with a scale factor of 2 to keep up with the conventions in changes of covariances. In this way, we can establish the Schrödinger picture as a restriction of the holomorphic case to real directions. Nonetheless, this restriction must be considered with care.

The first conclusion of this analysis on the polarization is that functions on $\mathscr{H}_S$ (i.e., the quantum states) must be functions depending on the submanifold $\mathcal{N}'$ with coordinates $\varphi^{\mathbf{x}}$ rather than on the whole $\mathcal{M}_F$, as was to be expected. Indeed, for any $X \in P_S$, $\theta_S(X) = 0$ implies that

$$\nabla_X \Psi = 0 \Rightarrow d\Psi(X) = 0, \forall X \in P. \tag{4.16}$$

As the reference section $\Psi^S_r$ must satisfy this condition, any other section in the set of square-integrable ones will be obtained as a product by a function that is also annihilated by the momentum directions. Nonetheless, the definition of a measure for this set of functions is not immediate, since the measure in the original scheme above was defined for functions depending on both the fields and their momenta. However, we can repeat the construction for a different Gel'fand triple defined as

$$\mathcal{N} \subset \mathcal{H}^S_\Delta \subset \mathcal{N}', \tag{4.17}$$

where $\mathcal{H}^S_\Delta$ is the subspace of $\mathcal{H}_\Delta$ defined by the field states $\xi_{\mathbf{x}}$. This triple defines a corresponding dual product $\langle \xi_{\mathbf{x}}, \varphi^{\mathbf{x}} \rangle = \xi_x \varphi^x$ which allows us to define a measure over $\mathcal{N}'$ that, with a slight abuse of notation, we denote with the same symbol $\mu$ by the functional

$$C(\xi_{\mathbf{x}}) = \int_{\mathcal{N}'} D\mu(\varphi^{\mathbf{x}}) e^{i\xi_x \varphi^x}. \tag{4.18}$$

Notice that, by construction,

$$C(\xi_{\mathbf{x}}) = C_S(\xi_{\mathbf{x}}, 0) = e^{-\frac{1}{4}\xi_x \Delta^{xy} \xi_y}, \tag{4.19}$$

for $C_S$ the functional defined in Equation (4.11).

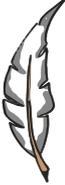

**Notational changes in the mathematical discussion**

The real measure $\mu$ defined above uses a different convention than the one in chapter 2. This is because this definition is adapted to the Schrödinger representation while (2.7) is designed to ease the mathematical discussion between real and complex measures. The difference amounts to a scale factor of 2 in the covariance.

Again, the Bochner-Minlos theorem ensures that there exists a unique measure satisfying this condition, and therefore we can define the set of (polarized) quantum states to be the space of square-integrable functions (with respect to the reference section) $\mathscr{H}_S = L^2(\mathcal{N}', D\mu)$.

**Quantizing linear operators**

We can readily quantize linear operators, such as the field operator $\varphi^x$ or the field momentum $\pi^x$, whose action on the states $\Phi \in L^2(\mathcal{N}', D\mu)$ will be:

$$\mathcal{Q}_s(\varphi^{\mathbf{y}})\Phi(\varphi^{\mathbf{x}}) = \varphi^{\mathbf{y}}\Phi(\varphi^{\mathbf{x}}),$$
$$\mathcal{Q}_s(\pi^x\delta_{x\mathbf{y}})\Phi(\varphi^{\mathbf{x}}) = (-i\partial_{\varphi^{\mathbf{y}}} + i\varphi^x K_{x\mathbf{y}} - \varphi^x(KA)_{x\mathbf{y}})\,\Phi(\varphi^{\mathbf{x}}). \tag{4.20}$$

Analogously, we can also introduce creation and annihilation operators from the field and momentum operators as:

$$a^{\mathbf{x}} = \frac{\Delta^{\mathbf{x}y}\partial_{\varphi^y} - iA_y^{\mathbf{x}}\varphi^y}{\sqrt{2}}, \qquad a^{\dagger\mathbf{x}} = \sqrt{2}\varphi^{\mathbf{x}} - \frac{\Delta^{\mathbf{x}y}\partial_{\varphi^y} - iA_y^{\mathbf{x}}\varphi^y}{\sqrt{2}}. \tag{4.21}$$

These operators satisfy that $\mathcal{Q}_s(\varphi^{\mathbf{x}}) = \frac{a^{\mathbf{x}} + a^{\dagger\mathbf{x}}}{\sqrt{2}}$ and $\mathcal{Q}_s(\pi^{\mathbf{x}}) = -iK_y^{\mathbf{x}}\frac{a^y - a^{\dagger y}}{\sqrt{2}}$. Notice that the complex exact term of (4.14) is essential for the quantized momentum to be self-adjoint, and it provides the right prescription for quantization obtained in [CCQ04] from an algebraic rather than geometric quantization procedure.

The discussion about the vacuum representation for this quantization is more involved than in the holomorphic case. Notice that, for $A_{\mathbf{y}}^{\mathbf{x}} \neq 0$, the constant function 1 is not annihilated by $a^x$. This is because the reference section $\Psi_r^S$ is not yet the vacuum of the theory. To get the correct vacuum section, we profit from the freedom of choice explored in section 3.4.3 and multiply it by a phase factor $\Psi_0 = \Psi_0^{Sch}\Psi_r^S$ with $\Psi_0^{Sch} \in L^2(\mathcal{N}', D\mu)$. We will delve further into this issue in section 4.3.1.

### 4.2.3 Quantization in the Antiholomorphic and Field-Momenta representations

Another real polarization is $P_M = Re(\mathcal{H}_\Delta)$. In this case, we could proceed as we have done so far with the Hilbert space associated with the measure $\mu_c$ and the symplectic potential (4.15), but it soon leads to cumbersome expressions for the quantized operators. This is because the change of coordinates (3.25) is adapted to deal with the Schrödinger representation in the fields space. Here we will develop antiholomorphic quantization, modifying that change of coordinate prescription to obtain a different measure $\mu_c$. We will then develop the Schrödinger picture in the field-momentum space. From now on, we will denote it simply as the *Field-Momentum representation*. In order to quantize this theory in a way akin to the one adapted to the Schrödinger picture, we should respect the momentum in the change of coordinates. We modify (3.25) to

$$d\breve{\varphi}^x = (A^t)_y^x d\pi^y - D_y^x d\varphi^x \qquad d\breve{\pi}^x = d\pi^x, \tag{4.22}$$

with $\phi^{\mathbf{x}} = \frac{1}{\sqrt{2}}(\breve{\pi}^{\mathbf{x}} - i\breve{\varphi}^{\mathbf{x}})$. Notice that this prescription is conjugate to $\phi^{\mathbf{x}}$, hence we denote it antiholomorphic coordinate. This treatment, following the arguments below

(3.25), modifies (3.32), and the measure considered for this case is $\nu_c$, defined by

$$\check{C}(\rho_{\mathsf{x}}, \bar{\rho}_{\mathsf{x}}) = \int_{\mathcal{M}_C} D\nu_c(\phi^{\mathsf{x}}) e^{i(\overline{\rho_x \phi^x} + \rho_x \phi^x)} = e^{\bar{\rho}_x D^{xy} \rho_y}. \tag{4.23}$$

As in the holomorphic case, $\phi^{\mathsf{x}}$ represents a coordinate of $\mathcal{M}_C$, while $\phi^{\mathsf{x}} \in \mathcal{N}'_{\mathbb{C}}$ is a placeholder for integration. This also leads to an antiholomorphic representation in $\mathscr{H}_{\overline{Hol}} = L^2_{\overline{Hol}}(\mathcal{N}'_{\mathbb{C}}, D\nu_c)$. The Kähler potential in these coordinates, with $D^{-1}_{xy} D^{yz} = \delta^z_x$, is[2] $\mathcal{D} = -\bar{\phi}^x D^{-1}_{xy} \phi^y$, and the symplectic one form is $\bar{\Theta} = -i\partial_{\underset{\smile}{\bar{\phi}^x}} \mathcal{D} d\underset{\smile}{\bar{\phi}^x}$. Thus, the quantization mapping becomes

$$\overline{\mathcal{Q}}(\underset{\smile}{\bar{\phi}^{\mathsf{x}}}) = \bar{\phi}^{\mathsf{x}}, \qquad \overline{\mathcal{Q}}(\phi^{\mathsf{x}}) = -D^{xy}\partial_{\bar{\phi}^y}. \tag{4.24}$$

These are interpreted as creation $\overline{\mathcal{Q}}(\bar{\phi}^{\mathsf{x}}) = b^{\dagger,\mathsf{x}}$ and annihilation $b^{\mathsf{x}} = \overline{\mathcal{Q}}(\phi^{\mathsf{x}})$ operators of the antiholomorphic representation. Notice that the vacuum of the theory in the antiholomorphic representation coincides with the reference section $\Psi_0^{\overline{H}} = \Psi_r$.

Following our steps with the Schrödinger picture, we define the Hilbert space of the field-momentum representation $\mathscr{H}_M = L^2(\mathcal{N}', D\nu)$, where $\nu$ is the Gaussian measure whose characteristic functional is

$$\check{C}_M(\xi_{\mathsf{x}}) = e^{\frac{1}{4}\xi_x D^{xy}\xi_y}. \tag{4.25}$$

The field-momentum polarization is therefore adapted to this Cameron-Martin Hilbert space, $P_M = Re(\mathcal{H}_{-D})$, and we choose the adapted symplectic potential

$$\check{\theta}_M = i(\check{\pi}^x - i\check{\varphi}^x)D^{-1}_{xy}d\check{\pi}^y = -\varphi^x\delta_{xy}d\pi^y + i\pi^x D^{-1}_{xy}d\pi^y + \pi^x(AD^{-1})_{xy}d\pi^y. \tag{4.26}$$

It follows from this prescription that

$$\overline{\mathcal{Q}}_m(\pi^{\mathsf{y}})\Phi(\pi^{\mathsf{x}}) = \pi^{\mathsf{y}}\Phi(\pi^{\mathsf{x}}),$$
$$\overline{\mathcal{Q}}_m(\varphi^x\delta_{x\mathsf{y}})\Phi(\pi^{\mathsf{x}}) = \left(i\partial_{\pi^{\mathsf{y}}} + i\pi^x D^{-1}_{x\mathsf{y}} + \pi^x(AD^{-1})_{x\mathsf{y}}\right)\Phi(\pi^{\mathsf{x}}). \tag{4.27}$$

In the momentum field space, creation and annihilation operators are

$$b^{\mathsf{x}} = -\frac{D^{\mathsf{x}y}\partial_{\pi^y} - i(A^t)^{\mathsf{x}}_y\pi^y}{\sqrt{2}}, \qquad b^{\dagger\mathsf{x}} = \sqrt{2}\pi^{\mathsf{x}} + \frac{D^{\mathsf{x}y}\partial_{\pi^y} - i(A^t)^{\mathsf{x}}_y\pi^y}{\sqrt{2}}. \tag{4.28}$$

These operate dually to the creation and annihilation operators in the field space, satisfying that $\mathcal{Q}(\pi^{\mathsf{x}}) = \frac{b^x + b^{\dagger x}}{\sqrt{2}}$ and $\mathcal{Q}(\varphi^x) = -i(D^{-1})^{\mathsf{x}}_y\frac{b^y - b^{\dagger y}}{\sqrt{2}}$. The discussion about the vacuum section of this theory is analogous to that in the Schrödinger representation. We will deal with it in section 4.3.1.

---

[2] To see this result, we use (4.36) and (A.10).

## 4.3 Integral transforms among quantization procedures. Consistency and Unitary equivalence.

Once we have shown different quantization procedures of a quantum field theory over different spaces $\mathscr{H}_{Hol}$, $\mathscr{H}_{\overline{Hol}}$, $\mathscr{H}_S$, and $\mathscr{H}_M$, we must study the relations of the quantization procedures under changes of coordinates and representations. This step is crucial to understanding the consistency of the quantization procedure of a classical theory. Firstly, as we will explain further in section 4.5, a unitary isomorphism of Hilbert spaces is not enough to claim the equivalence between two theories. Two descriptions of the same theory must be related by a unitary isomorphism of the algebra of observables. In this sense, we will relate $\mathscr{H}_{Hol}$ with $\mathscr{H}_S$ and $\mathscr{H}_{\overline{Hol}}$ with $\mathscr{H}_M$ through unitary isomorphisms respecting the algebra of creation and annihilation operators in each case. These relations are provided by a modification of the Segal-Bargmann transform **Definition** 2.13 that we will denote

$$\tilde{\mathcal{B}}_{Sch} : L^2(\mathcal{N}', D\mu) \to L^2_{Hol}(\mathcal{N}'_{\mathbb{C}}, D\mu_c),$$

$$\tilde{\mathcal{B}}_{Mom} : L^2(\mathcal{N}', D\nu) \to L^2_{\overline{Hol}}(\mathcal{N}'_{\mathbb{C}}, D\nu_c).$$

However, the main question that we want to answer is: To what extent do different quantization procedures depend on the choice of coordinates? Notice that if we preserve the algebra of observables, the quantization procedure is, in general, spoiled. Recall that we denote the quantization mappings of the previous section as $\mathcal{Q}$ Holomorphic, $\mathcal{Q}_s$ Schrödinger, $\overline{\mathcal{Q}}_m$ Field-Momentum, and $\overline{\mathcal{Q}}$ antiholomorphic. As we explained above, an algebra-preserving mapping will be $\tilde{\mathcal{B}}_{Sch}$. If we apply this property to the quantization of the momenta, we get

$$\tilde{\mathcal{B}}_{Sch}\mathcal{Q}_s(\pi^{\mathsf{x}})\tilde{\mathcal{B}}_{Sch}^{-1} = \tilde{\mathcal{B}}_{Sch}iK_y^{\mathsf{x}}\frac{a^{\dagger y} - a^y}{\sqrt{2}}\tilde{\mathcal{B}}_{Sch}^{-1} = iK_y^{\mathsf{x}}\mathcal{Q}\Big(\frac{\phi^y - \bar{\phi}^y}{\sqrt{2}}\Big) = \mathcal{Q}(K_y^{\mathsf{x}}\tilde{\pi}^y).$$

In general, $K_y^{\mathsf{x}}\tilde{\pi}^y \neq \pi^{\mathsf{x}}$ according to (3.25). Thus, quantization is not preserved. To answer the question posed at the beginning of the paragraph, we will provide unitary isomorphisms that relate every picture with each other via quantization-preserving integral transforms that we display in this diagram

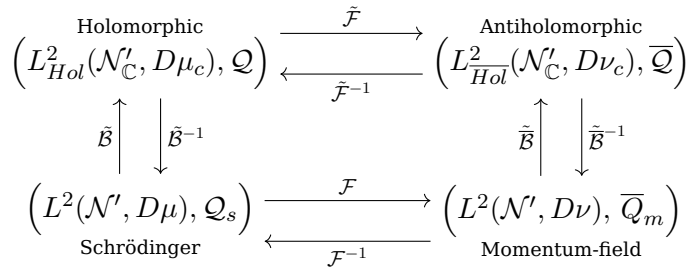

Figure 4.1: Quantization preserving mappings.

Here, $\tilde{\mathcal{B}}$ will be the regular Segal-Bargmann transform of **Definition** **2.13** with a normalization change to keep up with our conventions. We will also introduce a notion of Fourier transform $\mathcal{F}$ in infinite-dimensional spaces.

In summary, in this section we will see that, if we choose a function $F(\varphi^{\mathsf{x}}, \pi^{\mathsf{x}})$, we can express it in different sets of coordinates, (4.67) for holomorphic $F(\phi^{\mathsf{x}}, \bar{\phi}^{\mathsf{x}})$ or (4.22) for antiholomorphic $F(\underset{\smile}{\phi}^{\mathsf{x}}, \underset{\smile}{\bar{\phi}}^{\mathsf{x}})$, and the different quantization mappings are related by

$$\mathcal{Q}_s[F(\varphi^{\mathsf{x}}, \pi^{\mathsf{x}})] = \mathcal{F}^{-1}\overline{\mathcal{Q}}_m[F(\varphi^{\mathsf{x}}, \pi^{\mathsf{x}})]\mathcal{F} =$$
$$\tilde{\mathcal{B}}^{-1}\mathcal{Q}[F(\phi^{\mathsf{x}}, \bar{\phi}^{\mathsf{x}})]\tilde{\mathcal{B}} = \tilde{\mathcal{B}}^{-1}\tilde{\mathcal{F}}^{-1}\overline{\mathcal{Q}}[F(\underset{\smile}{\phi}^{\mathsf{x}}, \underset{\smile}{\bar{\phi}}^{\mathsf{x}})]\tilde{\mathcal{F}}\tilde{\mathcal{B}}. \quad (4.29)$$

### 4.3.1 Algebra-preserving isomorphisms: Segal-Bargmann modified transforms

Slightly modifying the Segal-Bargmann transform of **Definition** **2.13**, we can establish a unitary isomorphism

$$\tilde{\mathcal{B}}_{Sch} : L^2(\mathcal{N}', D\mu) \to L^2_{Hol}(\mathcal{N}'_{\mathbb{C}}, D\mu_c)$$

that preserves the algebra spanned by $a^{\mathsf{x}}, a^{\dagger\mathsf{x}}$. In order to define this modified Segal-Bargmann transform, we should deal with the extra $1/2$ factor appearing in the characteristic functional $C$ (4.19). To do so, let us first define $\Psi_{Hol} = \tilde{\mathcal{B}}_{Sch}(\Psi_{Sch})$. Then, we choose

$$\Psi_{Sch}(\varphi^{\mathsf{x}}) = e^{if(\varphi^{\mathsf{x}})} \int D\mu_S(\pi^{\mathsf{x}}) \Psi_{Hol}\big(\sqrt{2}[\varphi^{\mathsf{x}} - i\pi^{\mathsf{x}}]\big), \quad (4.30)$$

where $e^{if(\varphi^{\mathsf{x}})}$ is a phase factor to be determined. This is similar to the definition in [Oec12c], with the addition of the phase factor. For instance, if we take $f = 0$, and we denote the corresponding transformation as $\tilde{\mathcal{B}}$, it reads:

$$\tilde{\mathcal{B}}^{-1}\phi^{\mathsf{x}}\tilde{\mathcal{B}} = \sqrt{2}\varphi^{\mathsf{x}} - \frac{\Delta^{\mathsf{x}y}\partial_{\varphi^y}}{\sqrt{2}}, \text{ and } \tilde{\mathcal{B}}^{-1}\partial_{\phi^{\mathsf{x}}}\tilde{\mathcal{B}} = \frac{1}{\sqrt{2}}\partial_{\varphi^{\mathsf{x}}}. \quad (4.31)$$

These relations are derived explicitly in **A.1.1**. Thus, departing from the expressions of $a^x$ and $a^{\dagger x}$ shown under (4.10), we cannot recover the expressions of (4.21). To solve this problem, we choose, up to a constant,

$$f(\varphi^{\mathsf{x}}) = \frac{1}{2}(KA)_{xy} : \varphi^2 : |^{xy}_{\frac{\Delta}{2}}.$$

We choose the Wick ordered monomial to get a well-defined chaos decomposition, even though the pointwise product must be dealt with carefully, as we show in

 Representing $\tilde{\mathcal{B}}_{Sch} = \tilde{\mathcal{B}}e^{-if(\varphi^{\mathsf{x}})}$ and $\tilde{\mathcal{B}}_{Sch}^{-1} = e^{if(\varphi^{\mathsf{x}})}\tilde{\mathcal{B}}^{-1}$, we obtain

$$\tilde{\mathcal{B}}_{Sch}^{-1}\phi^{\mathsf{x}}\tilde{\mathcal{B}}_{Sch} = e^{if}\tilde{\mathcal{B}}^{-1}\phi^{\mathsf{x}}\tilde{\mathcal{B}}e^{-if} = \sqrt{2}\varphi^{\mathsf{x}} - \frac{\Delta^{\mathsf{xy}}\partial_{\varphi^y} - iA_y^{\mathsf{x}}\varphi^y}{\sqrt{2}},$$

$$\tilde{\mathcal{B}}_{Sch}^{-1}\partial_{\phi^{\mathsf{x}}}\tilde{\mathcal{B}}_{Sch} = e^{if}\tilde{\mathcal{B}}^{-1}\partial_{\phi^{\mathsf{x}}}\tilde{\mathcal{B}}e^{-if} = \frac{\partial_{\varphi^x} - i(KA)_{\mathsf{xy}}\varphi^y}{\sqrt{2}}. \tag{4.32}$$

With this choice, we preserve the form of the creation and annihilation operators for each picture. This may be interpreted as a nontrivial phase in the relation of the Schrödinger and holomorphic vacua, as is explained at the end of subsubsection 4.2.2. Indeed, $\Psi_0^{Hol} = 1$ represents the vacuum of the Fock space in the holomorphic representation because it is annihilated by the operator $a^{\mathsf{x}} = \Delta^{\mathsf{xy}}\partial_{\phi^y}$, as noted at the end of subsubsection 4.2.1. Computing the Schrödinger vacuum with (4.30), we obtain $\tilde{\mathcal{B}}_{Sch}^{-1}(1) = \Psi_0^{Sch} = \exp\left[\frac{i}{2}(KA)_{xy} : \varphi^2 : \left|\begin{smallmatrix}xy\\ \frac{\Delta}{2}\end{smallmatrix}\right.\right]$. This is indeed the true representation of the vacuum because it is annihilated by the annihilation operator in (4.21).

Similarly, for the momentum space, we can establish a unitary isomorphism with the antiholomorphic representation

$$\tilde{\mathcal{B}}_{Mom} : L^2(\mathcal{N}', D\nu) \to L^2_{\overline{Hol}}(\mathcal{N}'_{\mathbb{C}}, D\nu_c).$$

Let $\hat{\Psi}_{\overline{Hol}} = \tilde{\mathcal{B}}_{Mom}(\hat{\Psi}_{Mom})$. Then, its inverse is provided by the expression

$$\hat{\Psi}_{Mom}(\pi^{\mathsf{x}}) = e^{ig}\int D\nu(\varphi^{\mathsf{x}})\hat{\Psi}_{\overline{Hol}}(\sqrt{2}[\pi^{\mathsf{x}} + i\varphi^{\mathsf{x}}]). \tag{4.33}$$

Thus, with $g = \frac{i}{2}(AD^{-1})_{xy} : \pi^2 : \left|\begin{smallmatrix}xy\\ -\frac{D}{2}\end{smallmatrix}\right.$ we show in A.1.2 that

$$\tilde{\mathcal{B}}_{Mom}^{-1}\bar{\phi}^{\mathsf{x}}\tilde{\mathcal{B}}_{Mom} = b^{\dagger\mathsf{x}} \text{ and } -\tilde{\mathcal{B}}_{Mom}^{-1}D^{\mathsf{xy}}\partial_{\bar{\phi}^y}\tilde{\mathcal{B}}_{Mom} = b^{\mathsf{x}}. \tag{4.34}$$

This implies that the creation and annihilation operators of the antiholomorphic representations are transformed into (4.28), the corresponding operators of the momentum-field representation. Also, we see that the vacuum of the theory, the section annihilated by the annihilation operator $b^{\mathsf{x}}$, is represented by the phase function $\Psi_0^{Mom} = \exp\left[\frac{i}{2}(AD^{-1})_{xy} : \pi^2 : \left|\begin{smallmatrix}xy\\ -\frac{D}{2}\end{smallmatrix}\right.\right]$.

### 4.3.2 Preservation of the quantization mapping: Segal-Bargmann and Fourier transforms.

As we explained at the beginning of this section, the isomorphisms of the previous section spoil the quantization procedure. However, as we prove in (A.3) and (A.6), the Bargmann-Segal transforms preserve the quantization mappings such that

$$\begin{aligned}
\tilde{\mathcal{B}}^{-1}\mathcal{Q}(\varphi^{\mathsf{x}})\tilde{\mathcal{B}} &= \mathcal{Q}_s(\varphi^{\mathsf{x}}), & \tilde{\mathcal{B}}^{-1}\mathcal{Q}(\pi^{\mathsf{x}})\tilde{\mathcal{B}} &= \mathcal{Q}_s(\pi^{\mathsf{x}}), \\
\tilde{\mathcal{B}}^{-1}\overline{\mathcal{Q}}(\varphi^{\mathsf{x}})\tilde{\mathcal{B}} &= \overline{\mathcal{Q}}_m(\varphi^{\mathsf{x}}), & \tilde{\mathcal{B}}^{-1}\overline{\mathcal{Q}}(\pi^{\mathsf{x}})\tilde{\mathcal{B}} &= \overline{\mathcal{Q}}_m(\pi^{\mathsf{x}}).
\end{aligned} \tag{4.35}$$

Recall that $\tilde{\mathcal{B}}^{-1}$ is defined in (4.30) for $f = 0$ and $\tilde{\overline{\mathcal{B}}}^{-1}$ is defined in (4.33) for $g = 0$. In chapter 5, using the tools of integral kernels, we will also write explicit expressions for $\tilde{\mathcal{B}}$ in (5.56) and $\tilde{\overline{\mathcal{B}}}$ in (5.60). These are the regular Segal-Bargmann transforms introduced in **Definition** 2.13 with a $\sqrt{2}$ factor multiplying the domain. This fact motivates the definition of the Fourier transform as the quantization-preserving isometries that close the diagram of Figure 4.1. (Anti)holomorphic coordinates adapted to field (3.25) or momentum-field (4.22) representations are related via an antiholomorphic transformation

$$d\bar{\phi}^{\mathsf{x}} = i(K + iKA)^{\mathsf{x}}_y d\phi^y, \qquad -i(D^{-1} + iAD^{-1})^{\mathsf{x}}_y d\phi^y = d\bar{\phi}^{\mathsf{x}}. \tag{4.36}$$

This result can be proved using the relation $D = (iA^t + \mathbb{1})K(iA - \mathbb{1})$ as described in (A.10). Thus, we define the Fourier transform $\tilde{\mathcal{F}}$ as a relation between holomorphic and antiholomorphic representations defined by the unitary isomorphism

$$\tilde{\mathcal{F}}: \begin{array}{ccc} L^2_{Hol}(\mathcal{N}'_{\mathbb{C}}, D\mu_c) & \to & L^2_{\overline{Hol}}(\mathcal{N}'_{\mathbb{C}}, D\nu_c) \\ \Psi_{Hol}(\phi^{\mathsf{x}}) & \mapsto & \hat{\Psi}_{\overline{Hol}}(\bar{\phi}^{\mathsf{x}}) = \Psi_{Hol}[i(D^{-1} - iAD^{-1})^{\mathsf{x}}_y \bar{\phi}^y], \end{array} \tag{4.37}$$

with inverse

$$\tilde{\mathcal{F}}^{-1}: \begin{array}{ccc} L^2_{\overline{Hol}}(\mathcal{N}'_{\mathbb{C}}, D\nu_c) & \to & L^2_{Hol}(\mathcal{N}'_{\mathbb{C}}, D\mu_c) \\ \hat{\Psi}_{\overline{Hol}}(\bar{\phi}^{\mathsf{x}}) & \mapsto & \Psi_{Hol}(\phi^{\mathsf{x}}) = \hat{\Psi}_{\overline{Hol}}[i(K + iKA)^{\mathsf{x}}_y \phi^y]. \end{array} \tag{4.38}$$

For a proof of unitarity see A.1.3. Notice that the transformation acts non-trivially over creation and annihilation operators because

$$\tilde{\mathcal{F}}\phi^{\mathsf{x}}\tilde{\mathcal{F}}^{-1} = i(D^{-1} - iAD^{-1})^{\mathsf{x}}_y \bar{\phi}^y, \quad \tilde{\mathcal{F}}\partial_{\phi^{\mathsf{x}}}\tilde{\mathcal{F}}^{-1} = i(K + iKA)^y_{\mathsf{x}}\partial_{\bar{\phi}^y} \tag{4.39}$$

as we prove in A.1.3. However, these are precisely the transformations required to preserve the quantization mappings

$$\tilde{\mathcal{F}}\mathcal{Q}(\phi^{\mathsf{x}})\tilde{\mathcal{F}}^{-1} = -i(K - iKA)^{\mathsf{x}}_y \tilde{\mathcal{F}}\mathcal{Q}(\bar{\phi}^y)\tilde{\mathcal{F}}^{-1} = \overline{\mathcal{Q}}(\phi^{\mathsf{x}})$$
$$\tilde{\mathcal{F}}\mathcal{Q}(\bar{\phi}^{\mathsf{x}})\tilde{\mathcal{F}}^{-1} = -i(D^{-1} + iAD^{-1})^{\mathsf{x}}_y \tilde{\mathcal{F}}\mathcal{Q}(\phi^y)\tilde{\mathcal{F}}^{-1} = \overline{\mathcal{Q}}(\bar{\phi}^{\mathsf{x}})$$

Finally, to close the diagram of Figure 4.1, the remaining ingredient is the Fourier transform $\mathcal{F}$ defined by

$$\mathcal{F}: \begin{array}{ccc} L^2(\mathcal{N}', D\mu_S) & \to & L^2(\mathcal{N}', D\nu_M) \\ \Psi(\varphi^{\mathsf{x}}) & \mapsto & \hat{\Psi}(\pi^{\mathsf{x}}) = \tilde{\overline{\mathcal{B}}}^{-1}\tilde{\mathcal{F}}\tilde{\mathcal{B}}\,\Psi, \end{array} \tag{4.40}$$

Then, as we show explicitly in A.1.4, the quantization mappings are preserved because

$$\mathcal{F}\varphi^y\delta_{y\mathsf{x}}\mathcal{F}^{-1} = i\partial_{\pi^{\mathsf{x}}} + i\pi^y D^{-1}_{y\mathsf{x}} + \pi^y(AD^{-1})_{y\mathsf{x}}$$
$$\mathcal{F}^{-1}\pi^y\delta_{y\mathsf{x}}\mathcal{F} = -i\partial_{\varphi^{\mathsf{x}}} + i\varphi^y K_{y\mathsf{x}} - \varphi^y(KA)_{y\mathsf{x}}. \tag{4.41}$$

Thus, the quantization prescription for linear operators in the Schrödinger picture (4.20) is transformed into the analogous quantization prescription in the field-momentum picture (4.27) and vice versa. Hence, the quantization mappings are respected by the Fourier transform in these representations.

The discussion above is carried on over linear functions, but we can straightforwardly generalize it for arbitrary functions after the choice of one of the ordering prescriptions discussed in section 4.4.

These transformations do not respect, in general, the algebra of observables. We showed in the previous section that the isomorphisms of algebras are achieved by adding the corresponding phase factors to (4.30) and (4.33), which are lacking in this section. For this reason, the transforms of Figure 4.1 are not suited for discussions about the vacuum of every representation of the theory.

For completeness, we will show here how this representation fails to preserve the representation of the algebra of observables. Let's denote $a^{\mathbf{x}}, a^{\dagger,\mathbf{x}}$ the creation and annihilation operators of the Schrödinger picture provided by (4.21). Let also $b^{\mathbf{x}}, b^{\dagger,\mathbf{x}}$ represent the creation and annihilation operators of the momentum-field picture given by (4.28). Writing in terms of these operators the expression of the quantized operators $\hat{\varphi}^{\mathbf{x}}$ and $\hat{\pi}^{\mathbf{x}}$, the relation above can be rewritten as

$$\mathcal{F}\frac{a^{\dagger \mathbf{x}} + a^{\mathbf{x}}}{\sqrt{2}}\mathcal{F}^{-1} = i(D^{-1})^{\mathbf{x}}_y\frac{b^{\dagger y} - b^y}{\sqrt{2}}, \qquad iK^{\mathbf{x}}_y\mathcal{F}\frac{a^{\dagger y} - a^y}{\sqrt{2}}\mathcal{F}^{-1} = \frac{b^{\dagger \mathbf{x}} + b^{\mathbf{x}}}{\sqrt{2}}.$$

Thus, we can write

$$\mathcal{F}a^{\mathbf{x}}\mathcal{F}^{-1} = i(\Delta - D^{-1})^{\mathbf{x}}_y\frac{b^y}{2} + i(\Delta + D^{-1})^{\mathbf{x}}_y\frac{b^{\dagger y}}{2}$$

$$\mathcal{F}a^{\dagger \mathbf{x}}\mathcal{F}^{-1} = -i(\Delta + D^{-1})^{\mathbf{x}}_y\frac{b^y}{2} - i(\Delta - D^{-1})^{\mathbf{x}}_y\frac{b^{\dagger y}}{2} \qquad (4.42)$$

This implies that creation and annihilation operators mix, in general, in a nontrivial manner under $\mathcal{F}$. Notice that, from this expression, it is immediate to conclude that the mixing depends on the properties of the complex structure. Indeed, whenever $A^{\mathbf{x}}_{\mathbf{y}} = 0$ we get $D^{-1} = -\Delta$. Hence, in those cases, there will be no mixing between creation and annihilation operators under the Fourier transform. Under this assumption, $\tilde{\mathcal{B}}_{Sch}$ and $\tilde{\bar{\mathcal{B}}}_{Mom}$ reduce to $\tilde{\mathcal{B}}$ and $\tilde{\bar{\mathcal{B}}}$ respectively, and they form a unitary isomorphism of the algebra of observables. The Fourier transforms $\mathcal{F}$ and $\tilde{\mathcal{F}}$ also produce an isomorphism that respects the canonical commutation relations. To see this result recall that $[a^{\mathbf{x}}, a^{\dagger \mathbf{y}}] = \Delta^{\mathbf{xy}}$ and $[b^{\mathbf{x}}, b^{\dagger \mathbf{y}}] = -D^{\mathbf{xy}}$ and $-D^{\mathbf{xy}} = K^{\mathbf{xy}}$ for this particular case. In a nutshell, the diagram Figure 4.1 only represents unitary isomorphisms of the algebra of observables if $A^{\mathbf{x}}_{\mathbf{y}} = 0$.

## 4.4 Wick and Weyl orderings and the star product: Quantization of arbitrary functions

The last ingredient to conclude our quantization program is to enlarge the prescription to higher-order functions, as we explained at the beginning of the chapter. For this matter, we will focus on the holomorphic representation. This representation is particularly well-behaved for describing a QFT over a Cauchy hypersurface $\Sigma$. This is because the space of pure states for this theory is considered to be

$$L^2_{Hol}(\mathcal{N}'_\mathbb{C}, D\mu_c),$$

for a suited choice of Gaussian measure $D\mu_c$. In this space, the Wiener-Ito decomposition is particularly simple, and the Skorokhod integral amounts to simple multiplication by $\phi^\mathbf{x}$, which is also reflected in the simple form of creation and annihilation operators of (4.10). Moreover, this space possesses nice analytical properties briefly explained in section 2.5.3. In the sections above, we provided a detailed account of this representation as well as its relation to several other representations of QFT for linear operators. The goal of this section will be to present the extension of this quantization procedure to the nonlinear cases. For that matter, we will use the new tool of reproducing kernels presented in section 2.5.3 to deal with ordering problems. These ordering problems arise from the ambiguity in the prescription for nonlinear operators coming from the nontrivial commutation relation $[a^\mathbf{x}, a^{\dagger\mathbf{y}}] = \Delta^{\mathbf{x}\mathbf{y}}$.

In this section, we will present the Weyl ordering prescription and the corresponding Moyal product using tools of Gaussian analysis. These tools ease the study of the analytical properties of the classical functions that allow for well-behaved definitions of $C^*$-algebras of observables. After that, we will briefly discuss the technology transfer to the Wick ordering prescription.

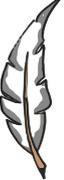

**Notation for Holomorphic quantization**

In the remainder of the chapter, we will only discuss the Holomorphic quantization mapping $\mathcal{Q}$. For simplicity, in the following sections, we identify the element $\phi^\mathbf{x} \in \mathcal{N}'_\mathbb{C}$ with $\underset{\sim}{\phi^\mathbf{x}} \in \mathcal{M}_C$ unless otherwise stated.

### 4.4.1 Weyl quantization

The first choice that we make is called Weyl quantization and we denote it $\mathcal{Q}_{Weyl}$. It is defined as follows. Firstly, any holomorphic quantization map must coincide with linear functions of $\mathcal{M}_C$, that is,

$$a^\mathbf{x} = \mathcal{Q}_{Weyl}(\bar{\phi}^\mathbf{x}) = \Delta^{\mathbf{x}y}\partial_{\phi^y}, \quad a^{\dagger\mathbf{x}} = \mathcal{Q}_{Weyl}(\phi^\mathbf{x}) = \phi^\mathbf{x}. \tag{4.43}$$

Then the Weyl quantization mapping assigns to each monomial $\phi^n \bar{\phi}^m$ an averaged assignment of every possible order. We can then define the Weyl quantization for any polynomial classical functional inductively. For linear operators, the quantization procedure must coincide with holomorphic geometric quantization: $\mathcal{Q}_{Weyl}[\phi^{\mathbf{x}}] = a^{\dagger \mathbf{x}}$, $\mathcal{Q}_{Weyl}[\bar{\phi}^{\mathbf{x}}] = a^{\mathbf{x}}$. For higher-order monomials, we proceed as

$$\mathcal{Q}_{Weyl}[(\phi^n \bar{\phi}^m)^{\mathbf{u}\vec{\mathbf{x}}, \vec{\mathbf{y}}\mathbf{v}}] =$$

$$a^{\dagger \mathbf{u}} \mathcal{Q}_{Weyl}[(\phi^{n-1}\bar{\phi}^m)^{\vec{\mathbf{x}}, \vec{\mathbf{y}}\mathbf{v}}] + \frac{[a^{\mathbf{u}}, a^{\dagger, z}]}{2} \mathcal{Q}_{Weyl}[\partial_{\bar{\phi}^z}(\phi^{n-1}\bar{\phi}^m)^{\vec{\mathbf{x}}, \vec{\mathbf{y}}\mathbf{v}}]$$

$$= \mathcal{Q}_{Weyl}[(\phi^n \bar{\phi}^{m-1})^{\mathbf{u}\vec{\mathbf{x}}, \vec{\mathbf{y}}}] a^{\mathbf{v}} + \mathcal{Q}_{Weyl}[\partial_{\phi^z}(\phi^n \bar{\phi}^{m-1})^{\mathbf{u}\vec{\mathbf{x}}, \vec{\mathbf{y}}}] \frac{[a^{\mathbf{u}}, a^{\dagger, z}]}{2}. \quad (4.44)$$

Notice that we can do this because $[a^{\mathbf{u}}, a^{\dagger, \mathbf{v}}] = \Delta^{\mathbf{uv}}$ commutes with $a^{\mathbf{x}}$ and $a^{\dagger, \mathbf{y}}$. From this expression, it follows that the action of Skorokhod integrals with covariance $\Delta^{\mathbf{xy}}/2$ is cast into

$$\mathcal{Q}_{Weyl}[\partial^{*\phi^x} F] = \mathcal{Q}_{Weyl}[(\phi^x - \frac{\Delta^{xy}}{2}\partial_{\bar{\phi}^y}) F] = a^{\dagger x} \mathcal{Q}_{Weyl}[F]$$

$$\mathcal{Q}_{Weyl}[\partial^{*\bar{\phi}^x} F] = \mathcal{Q}_{Weyl}[(\bar{\phi}^x - \frac{\Delta^{xy}}{2}\partial_{\phi^y}) F] = \mathcal{Q}_{Weyl}[F] a^x. \quad (4.45)$$

Thus, this quantization procedure is well-suited to quantize classical functions into a Hilbert space $\mathcal{O}_{cl}$ defined by

$$\mathcal{O}_{cl} = L^2(\mathcal{N}'_{\mathbb{C}}, D\mathcal{W}) \text{ with } \int D\mathcal{W}(\phi^{\mathbf{x}}) e^{i(\overline{\rho_x \bar{\phi}^x} + \rho_x \phi^x)} = e^{-\frac{\bar{\rho}_x \Delta^{xy} \rho_y}{2}}. \quad (4.46)$$

Indeed, for functions in this Hilbert space, we can consider a chaos decomposition

$$F[\phi, \bar{\phi}] = \sum F^{(n, \bar{m})}_{\vec{x}\vec{y}} \mathcal{W}_{\frac{\Delta}{2}}(\phi^n \bar{\phi}^m)^{\vec{x}\vec{y}}. \quad (4.47)$$

Wick monomials $\mathcal{W}_{\frac{\Delta}{2}}(\phi^n \bar{\phi}^m)^{\vec{x}\vec{y}}$ defined in **Definition** 2.11 in this context may be interpreted as a necessary point-splitting regularization procedure in the product of distributions, see *Remark* 4.2 below and [BF82]. Notice that, because $\mathcal{M}_F$ is modeled over $\mathcal{N}'_{\mathbb{C}}$, this is needed already at the classical level, that is, for $\mathcal{O}_{cl}$.

**Remark 4.1.** (*$\hbar$ dependence of $\mathcal{W}$*)
*Notice that the covariance $\Delta^{\mathbf{xy}}$ of this Gaussian measure $\mathcal{W}$ comes from the commutation relations $[a^{\mathbf{u}}, a^{\dagger, \mathbf{v}}] = \Delta^{\mathbf{uv}}$. In this commutation relations we assume $\hbar = 1$. If we drop this assumption, we get $[a^{\mathbf{u}}, a^{\dagger, \mathbf{v}}] = \hbar \Delta^{\mathbf{uv}}$. This implies that the $\mathcal{W}$ depends on $\hbar$ in such a way that that its characteristic functional is $\exp\left(-\hbar \frac{\bar{\rho}_x \Delta^{xy} \rho_y}{2}\right)$. Thus, the regularization of $\mathcal{O}_{cl}$ is dependent on $\hbar$.*

Now, quantizing the function $F$ means constructing the operator with the same chaos coefficients. For one such monomial, we obtain

$$\mathcal{Q}_{Weyl}\left[F^{(n, \bar{m})}_{\vec{x}\vec{y}} \mathcal{W}_{\frac{\Delta}{2}}(\phi^n \bar{\phi}^m)^{\vec{x}\vec{y}}\right] = F^{(n, \bar{m})}_{\vec{x}\vec{y}} (a^{\dagger n})^{\vec{x}} (a^m)^{\vec{y}}. \quad (4.48)$$

It is straightforward to obtain the action on the total classical function by linearity, although we must proceed with care because the resulting operators are, in general, unbounded.

An important property of Weyl quantization is the transformation of complex conjugation into involution:

$$\mathcal{Q}_{Weyl}(F^*) = \mathcal{Q}_{Weyl}(F)^\dagger. \tag{4.49}$$

Therefore, real classical functions of $\mathcal{O}_{cl}$ are quantized into Hermitian operators. As we will see, the machinery of integral kernels of section 2.7.1 is better suited to deal with the finer details of this quantization, which we explore in the next section.

**Weyl quantization from integral kernels**

We can re-express the Weyl prescription, acting over the dense subset of Hida test functions $(\mathcal{N}_{\mathbb{C}}) \subset L^2_{Hol}(\mathcal{N}'_{\mathbb{C}}, D\mu_c)$, using integral kernel representations as explained in section 2.7.1. That is, we will use (2.79) to rewrite the quantization prescription. This quantization prescription is particularly simple for the algebra of trigonometric exponentials $\mathcal{T}(\mathcal{N}'_{\mathbb{C}})$. For $\mathcal{E}_\chi = \exp\left(i\overline{\chi_x \phi^x} + i\chi_x \phi^x\right)$, we have

$$\mathcal{Q}_{Weyl}\left[\mathcal{E}_\chi\right]\Psi(\sigma^{\mathbf{x}}) =$$
$$\int_{\mathcal{N}'_{\mathbb{C}}} D\mu_c(\phi^{\mathbf{x}}) \exp\left(i\chi_x \sigma^x + i\overline{\chi_x \phi^x} - \frac{\bar\chi_x \Delta^{xy}\chi_y}{2} + \bar\phi^x K_{xy}\sigma^y\right)\Psi(\phi^{\mathbf{x}}). \tag{4.50}$$

We know from chapter 2 that $\mathcal{T}(\mathcal{N}'_{\mathbb{C}})$ is dense in $\mathcal{O}_{cl}$, and therefore we extend Weyl quantization to the whole space by linearity. In chapter 5, we will explore further this map in (5.37) and, for the antiholomorphic picture in (5.50).

**Moyal product**

To complete the picture of Weyl quantization, we need to know how the composition of operators is translated as a multiplicative operation in the space of classical functions. For that matter, we define the Moyal star product, denoting it $\star_m$, as the one that preserves operator composition after quantization. This is for an appropriate subset of classical functions $F, G \in W \subset \mathcal{O}_{cl}$[3]

$$\mathcal{Q}_{Weyl}(F)\,\mathcal{Q}_{Weyl}(G) = \mathcal{Q}_{Weyl}(F \star_m G).$$

The space $(W, \star_m)$ must be an algebra, and we denote it the algebra of classical Weyl-quantizable functions. We will characterize this algebra below, endowing it with a norm. Using (4.50) together with (2.79), it follows that $\mathcal{E}_\rho$ with the Moyal product is a representation of Weyl relations

---

[3]We will prove that $W$ is indeed a subset of $\mathcal{O}_{cl}$ in the next section.

$$\mathcal{E}_\rho \star_m \mathcal{E}_\alpha = \exp\left(\frac{\bar{\rho}_x \Delta^{xy}\alpha_y - \bar{\alpha}_x \Delta^{xy}\rho_y}{2}\right)\mathcal{E}_{\rho+\alpha}. \tag{4.51}$$

These Weyl relations will play a crucial role in our discussion about canonical quantization since they represent the canonical commutation relations in an exponential form. Using the identity $e^{\sigma^x \partial_{\phi^x}} F(\phi^\mathsf{x}, \bar{\phi}^\mathsf{x}) = F(\phi^\mathsf{x} + \sigma^\mathsf{x}, \bar{\phi}^\mathsf{x})$ over trigonometric exponentials $\mathcal{E}_\rho$, and because of the density of $\mathcal{T}(\mathcal{N}'_\mathbb{C})$, we can rewrite Weyl relations and extend the Moyal star product of any two functions $F, G \in W \subset \mathcal{O}_{cl}$ with an integral representation:

$$F \star_m G = \int D\mathcal{W}(\sigma^\mathsf{x}) \int D\mathcal{W}(\zeta^\mathsf{x}) F(\phi^\mathsf{x}+\sigma^\mathsf{x}, \bar{\phi}^\mathsf{x}+\bar{\zeta}^\mathsf{x}) G(\phi^\mathsf{x}+\zeta^\mathsf{x}, \bar{\phi}^\mathsf{x}-\bar{\sigma}^\mathsf{x}). \tag{4.52}$$

This expression is tightly related to the one in the covariant formalism of QFT shown in [Dit90]. Notice that this product acquires a dependency on $\hbar$ in the way discussed in *Remark* 4.1. As a consequence, we can express this star product as a power series in terms of $\hbar$.

### The algebra of Weyl classical quantizable functions

Our goal in this section is to study the algebra $W$ and to prove that it is indeed a subset of $\mathcal{O}_{cl}$. To do so, we start by noticing that a dense subalgebra should be $(\mathcal{T}(\mathcal{N}'_\mathbb{C}), \star_m)$. Then we must topologize this algebra in such a way that the $\star_m$ product is continuous. The natural way to do this is, since $\star_m$ is a reflection of the operator product under the map $\mathcal{Q}_{Weyl}$, the topology is the coarsest one making this map continuous. Thus, the topology is the one induced by $\mathcal{Q}_{Weyl}$ with the operator norm in the image. In more practical terms, we endow $W$ with a norm such that for $F \in W$,

$$\|F\|_{Weyl} = \|\mathcal{Q}_{Weyl}(F)\| = \sup_{\|\Psi\|_{\mu_c}=1} \|\mathcal{Q}_{Weyl}(F)\Psi\|_{\mu_c}. \tag{4.53}$$

It is indeed a norm because $\mathcal{Q}_{Weyl}$ is linear and respects involution. We start by studying the operator norm of the generators of the algebra, i.e., the trigonometric exponentials.

**Lemma 4.1.** *Trigonometric exponentials* $\mathcal{E}_\chi = \exp\left(i\overline{\chi_x \phi^x} + i\chi_x \phi^x\right)$ *are quantized into unitary operators of* $L^2_{Hol}(\mathcal{N}'_\mathbb{C}, D\mu_c)$.

*Proof.* Equation (4.49) allows us to compute the operator norm in the following way:

$$\|\mathcal{E}_\chi\|^2_{Weyl} = \sup_{\|\Psi\|=1} \langle\Psi|\mathcal{Q}_{Weyl}(\overline{\mathcal{E}_\chi} \star_m \mathcal{E}_\chi)\Psi\rangle = \sup_{\|\Psi\|=1} \langle\Psi|\Psi\rangle = 1. \tag{4.54}$$

Here we have used the relation $\overline{\mathcal{E}_\chi} = \mathcal{E}_{-\chi}$ and the Weyl relations $\overline{\mathcal{E}_\chi} \star_m \mathcal{E}_\chi = \mathcal{E}_0 = 1$. $\quad\square$

In order to compute the operator norm of a generic element of $\mathcal{T}(\mathcal{N}'_{\mathbb{C}})$, we must understand the action of $\mathcal{Q}_{Weyl}(\mathcal{E}_\chi)$ over a vector $\Psi$. For that matter, it is convenient to decompose trigonometric exponentials in terms of holomorphic (and antiholomorphic) coherent states $\mathcal{K}_\chi = \exp(\chi_x \phi^x)$ in the following way:

$$\mathcal{E}_\chi = \exp\left(\frac{\bar{\chi}_x \Delta^{xy} \chi_y}{2}\right) \overline{\mathcal{K}_{-i\chi}} \star_m \mathcal{K}_{i\chi}. \tag{4.55}$$

Quantized coherent states act in a straightforward way over the vector $\Psi$:

$$\mathcal{Q}_{Weyl}(\overline{\mathcal{K}_\chi})\Psi(\phi^{\mathbf{x}}) = \Psi(\phi^{\mathbf{x}} + \Delta^{\mathbf{x}y}\bar{\chi}_y),$$
$$\mathcal{Q}_{Weyl}(\mathcal{K}_\chi)\Psi(\phi^{\mathbf{x}}) = e^{\chi_x \phi^x}\Psi(\phi^{\mathbf{x}}). \tag{4.56}$$

Then we can compute for later use:

$$\langle\Psi|\mathcal{Q}_{Weyl}(\mathcal{E}_\chi)\Psi\rangle = e^{\frac{\bar{\chi}_x \Delta^{xy}\chi_y}{2}}\langle\mathcal{Q}_{Weyl}(\mathcal{K}_{-i\chi})\Psi|\mathcal{Q}_{Weyl}(\mathcal{K}_{i\chi})\Psi\rangle$$
$$= e^{\frac{\bar{\chi}_x \Delta^{xy}\chi_y}{2}}\int D\mu_c(\phi^{\mathbf{x}})e^{i(\overline{\chi_x \phi^x} + \chi_x\phi^x)}\overline{\Psi(\phi^{\mathbf{x}})}\Psi(\phi^{\mathbf{x}}). \tag{4.57}$$

This expression allows us to compute the Weyl-induced norm of a generic element $F \in \mathcal{T}(\mathcal{N}'_{\mathbb{C}})$. By definition, it can be written, in general in a non-unique manner, as $F(\phi^{\mathbf{x}}, \bar\phi^{\mathbf{x}}) = \sum_{n=1}^{N} F_n \mathcal{E}_{\chi^n}$. It is easy to see that the norm of $F$ in $\mathcal{O}_{cl}$ is simply computed by means of the characteristic functional (4.46). Then, using (4.57) and the characteristic functional of $D\mu_c$ defined in (4.6), the following equality holds:

$$\|F\|^2_{cl} = \sum_{n,m=0}^{N} F_n^* F_m e^{-\frac{\overline{(\chi^m - \chi^n)}_x \Delta^{xy}(\chi^m - \chi^n)_y}{2}} = \langle 1|\mathcal{Q}_{Weyl}(\bar{F} \star_m F)1\rangle_{\mu_c}. \tag{4.58}$$

Here the state 1 is simply the constant function 1. This expression and (4.53) provide us with the bound $\|F\|_{cl} \le \|F\|_{Weyl}$ and, by density, this is valid for every $F \in W$. The equality does not hold in general. From this fact, we have that $W \hookrightarrow \mathcal{O}_{cl}$ is a dense subspace such that the inclusion is continuous. We also have that

$$(W, \star_m, \|\cdot\|_{Weyl})$$

is a $C^*$ algebra with complex conjugation as involution. In this way, $\mathcal{Q}_{Weyl}$ is a $C^*$-isomorphism between $W$ and $\hat{W} = \mathcal{Q}_{Weyl}(W)$. The latter is the von Neumann algebra of quantum observables, which is a representation of Weyl relations (4.51) as a closed subalgebra of the algebra of bounded operators $\hat{W} \subset B\left(L^2_{Hol}(\mathcal{N}'_{\mathbb{C}}, D\mu_c)\right)$.

An important remark to notice is that in previous sections we obtained an expression for the quantization of linear functions of $\mathcal{O}_{cl}$. The result is the set of linear combinations of creation and annihilation operators, which may be unbounded. This means that, even though we can enlarge $\mathcal{Q}_{Weyl}$ to make sense over the whole space of classical functions $\mathcal{O}_{cl}$, the norm $\|\cdot\|_{cl}$ just provides a lower bound on the operator norm, and the Moyal star product $\star_m$ is not well defined outside $W$. In simpler terms, even though we could quantize the whole $\mathcal{O}_{cl}$, we cannot treat it as an algebra. This fact, even though it may seem abstract *a priori*, is the source of the renormalization program. In most cases, we must deal with Hamiltonians containing interaction terms such that $\hat{H} \in \mathcal{O}_{cl}$ but not in $W$. In the computation of the propagator of the theory, we need to compute products of $\hat{H}$ with itself. The naive treatment of this kind of expression leads to divergent integrals that must be treated with care in a renormalization program. We will not deal with this feature in this work and refer elsewhere [Nai05, AA15] for further information on the standard renormalization program.

Finally, let us provide an upper bound to $\|F\|_{Weyl}$. In general, it is difficult to obtain an explicit expression for it. Nonetheless, by Hölder's inequality, we obtain

$$|\langle\Psi|\mathcal{Q}_{Weyl}(\mathcal{E}_\chi)\Psi\rangle| \leq e^{\frac{\bar{\chi}_x \Delta^{xy}\chi_y}{2}}|\langle\Psi|\Psi\rangle|,$$

that for an element written as $F(\phi^{\mathsf{x}}, \bar{\phi}^{\mathsf{x}}) = \sum_{n=1}^{\infty} F_n \mathcal{E}_{\chi^n}$ leads to

$$\|F\|_{Weyl}^2 = \sup_{\|\Psi\|=1} \sum_{n,m=1}^{\infty} F_n^* F_m \langle\Psi|\mathcal{Q}_{Weyl}(\mathcal{E}_{\chi^m - \chi^n})\Psi\rangle_{\mu_c} \leq$$
$$\sum_{n,m=1}^{\infty} |F_n^* F_m| e^{\frac{\overline{(\chi^n - \chi^m)}_x \Delta^{xy}(\chi^n - \chi^m)_y}{2}}.$$

This is the best that we can do; in turn, it leads to a criterion of convergence for a function of $\mathcal{O}_{cl}$ to be in $W$ that may be used in a regularization program.

### 4.4.2 Wick holomorphic quantization

The most common ordering in QFT is not Weyl quantization, but Wick quantization, which assigns to regular monomials the so-called *normal order* prescription. This prescription is designed to quantize regular monomials into normal ordered products of creation and annihilation operators that guarantee null vacuum expectation values.

In order to find an explicit expression for this operation, we use the Wick operator $\mathcal{W}_{\frac{\Delta}{2}}$ for general functionals introduced in *Example* 2.6. Using this, Wick quantization is defined from Weyl quantization as $\mathcal{Q}_{Wick} = \mathcal{Q}_{Weyl} \circ \mathcal{W}_{\frac{\Delta}{2}}$.

**Remark 4.2. (Regularizator of the classical theory)** *In Weyl quantization, the operator $\mathcal{W}_{\frac{\Delta}{2}}$ plays the role of a point-splitting regulator of the classical theory. As an example, we get*

$$\mathcal{W}_{\frac{\Delta}{2}}(C_{xy}\phi^x\bar{\phi}^y) = C_{xy}\phi^x\bar{\phi}^y - C_{xy}\frac{\Delta^{xy}}{2}$$

thus the ill-defined product of distributions $\phi^x \bar{\phi}^y$, see [BF82], is regularized in such a way that it is integrable under the measure $D\mathcal{W}$.

We won't cover this case in full detail; the results of the previous section can be related to this case just by studying the properties of $\mathcal{W}_{\frac{\Delta}{2}}$. Our main interest in this section will be to find a correct star product $\star_w$ to find a representation of the Wick relations. Writing $\mathcal{Q}_{Weyl}$ in an integral kernel representation

$$\mathcal{Q}_{Wick}\left[\mathcal{E}_\chi\right]\Psi(\sigma^{\mathbf{x}}) =$$
$$\int_{\mathcal{N}'_{\mathbb{C}}} D\mu_c(\phi^{\mathbf{x}}) \exp\left(i\chi_x\sigma^x + i\overline{\chi_x\phi^x} + \bar{\phi}^x K_{xy}\sigma^y\right)\Psi(\phi^{\mathbf{x}}). \quad (4.59)$$

Wick star product follows from its definition $\mathcal{Q}_{Wick}(F)\,\mathcal{Q}_{Wick}(G) = \mathcal{Q}_{Wick}(F \star_w G)$ as

$$F \star_w G = \int D\mathcal{W}(\zeta^{\mathbf{x}}) F(\phi^{\mathbf{x}}, \bar{\phi}^{\mathbf{x}} + \sqrt{2}\zeta^{\mathbf{x}}) G(\phi^{\mathbf{x}} + \sqrt{2}\zeta^{\mathbf{x}}, \bar{\phi}^{\mathbf{x}}). \quad (4.60)$$

Then a representation of the Weyl algebra is obtained by the Wick-ordered trigonometric exponentials:

$$:\mathcal{E}_\chi := \exp\left(i\overline{\chi_x\phi^x} + i\chi_x\phi^x - \frac{\bar{\chi}_x \Delta^{xy}\chi_y}{2}\right),$$

$$:\mathcal{E}_\rho : \star_w : \mathcal{E}_\alpha := \exp\left(\frac{\bar{\rho}_x \Delta^{xy}\alpha_y - \bar{\alpha}_x\Delta^{xy}\rho_y}{2}\right) : \mathcal{E}_{\rho+\alpha} : . \quad (4.61)$$

This procedure is often used in quantum field theory because classical functions representing Hamiltonians are expressed in terms of regular monomials. Nonetheless, the space of classical functions possesses better analytical properties in Weyl quantization, and the difference between a Hamiltonian written in terms of regular or Wick monomials often amounts to an infinite constant that is physically interpreted as a trivial shift of the ground energy without physical meaning.

## 4.5 Algebraic Quantum Field Theory: Fock Quantization

To complete our study on quantization, we must show the connection of the representations described above with canonical quantization through the program of Algebraic Quantum Field Theory (AQFT). We present here a quick summary of its ingredients in order to introduce the most common QFT representation found in introductory books on the subject, i.e., Fock quantization. We refer to [Wal94, Bru&15] for a thorough analysis of the subject. Roughly speaking, the starting point of this construction relies on the abstract characterization of the $C^*$-algebra of quantum observables. It turns out that the predictions of a QFT can be derived in different representations as long as their algebras are unitarily isomorphic. Thus, we consider two representations equivalent if they are related by this class of isomorphism.

So far, we have used geometric quantization to find a particular class of representations of a QFT. In those cases, we obtain a Hilbert space as a representation of the class of pure states and the algebra of observables as the image of a quantization mapping from a class of classical observables. In those cases, though, geometric quantization comes with the ambiguity associated with the choice of polarization, and we are forced to study separately (anti)holomorphic, Schrödinger, and field-momentum representations. In this section, we show that, using the Gelfand-Naimark-Segal (GNS) construction, the necessary condition to achieve equivalence is to represent Weyl relations and an abstract equivalent state.

We already know that the vacuum state is represented in each of the representations displayed in this chapter. Thus, using the isomorphisms of section 4.3.1, we can see that the Schrödinger and holomorphic representations are equivalent, while field-momentum is equivalent to the antiholomorphic representation. In the previous section, we showed how we obtain Weyl relations from the different ordering prescriptions. Extending this analysis to the antiholomorphic representation, we would also obtain Weyl relations in a different set of coordinates, finally showing the equivalence of all representations. This last step will not be covered in detail in this work.

In summary, our goal in this section is to find Weyl relations that correspond to the abstract axioms that characterize the $C^*$-algebra of quantum observables from the geometric narrative that has been our guideline in this work. Then, we find how AQFT recovers the space of (not necessarily pure) states from this abstract setting. Finally, we will use it to establish the relation of holomorphic quantization, under an ordering prescription, with Fock space quantization commonly found in the literature [Wal94]. We claim that this analysis also leads to the equivalence of all representations exposed in this chapter, but we skip this part as it is just a paraphrase of the arguments given throughout the chapter.

### 4.5.1 The $C^*$ Algebra of Quantum Observables and the GNS Construction

The first problem that we encountered in order to quantize a classical theory was how to model our phase space $T^*\mathcal{N}$, where we know that there is a canonical symplectic form. So far, our approach was to model this space treating both base and fiber as distributions and then consider $\mathcal{M}_F = \mathcal{N}' \times \mathcal{N}'$. This choice, even though it complicates the treatment of the classical theory itself, suits the needs of KvH theory and geometric quantization for QFT. In this section, we are interested in AQFT, thus we start with a different choice of model space for our classical theory. In our model, the manifold of fields is $\mathcal{M}_A = \mathcal{N} \times \mathcal{N}$ with coordinates $(\varphi_{\mathbf{x}}, \pi_{\mathbf{x}})$, and it is endowed with a symplectic form

$$\omega_{\mathcal{M}_A} = \delta^{xy} d\pi_x \wedge d\varphi_y = \int_\Sigma d^d x \sqrt{|h|} \left[ d\pi(x) \otimes d\varphi(x) - d\varphi(x) \otimes d\pi(x) \right]. \qquad (4.62)$$

Notice the difference with (3.18). This is because here the coordinates are functions and not densities of weight one. At this stage, we can write Weyl relations in a coordinate-free manner. Let $\rho, \alpha \in T\mathcal{M}_A$ be vector fields; then the Weyl algebra is a $C^*$ algebra with generators $\mathcal{R}(\cdot)$ that fulfill Weyl relations

$$\mathcal{R}(\rho)^* = \mathcal{R}(-\rho), \quad \mathcal{R}(\rho)\mathcal{R}(\alpha) = e^{\frac{i}{2}\omega_{\mathcal{M}_A}(\rho,\alpha)}\mathcal{R}(\rho + \alpha). \tag{4.63}$$

Notice that particular representations of this algebra are given by (4.51) and (4.61). These are particular examples of a representation of Weyl relations in the algebra of bounded operators[4] of a particular Hilbert space.

Once the algebra is known, we need to recover the space of states of the theory. To carry on with that study, we use the GNS construction. Our starting point is a particular state of the algebra, i.e., a linear functional $\varpi : \mathscr{A} \to \mathbb{C}$ which is positive definite (i.e., $\varpi(aa^*) \geq 0$) with $\varpi(1) = 1$. With this state, we can define the GNS representation of the $C^*$ algebra:

**Theorem 4.2. (GNS Construction)** *Let $\mathscr{A}$ be a $C^*$-algebra with unit and let $\varpi : \mathscr{A} \to \mathbb{C}$ be a state. Then there exists a Hilbert space $\mathscr{H}$, a representation $\pi : \mathscr{A} \to B(\mathscr{H})$, and a vector $\Psi \in \mathscr{H}$ such that,*

$$\varpi(A) = \langle \Psi, \pi(A)\Psi \rangle_{\mathscr{H}}.$$

*Furthermore, the vector $\Psi$ is cyclic. The triplet $(\mathscr{H}, \pi, \Psi)$ with these properties is unique (up to unitary equivalence).*

In this fashion, we recover the usual prescription of quantum theories where we look for a representation of the algebra of observables as a subalgebra of the space of bounded operators acting on a separable Hilbert space. In this way, the GNS construction ensures that it is enough to restrict the study to the physically meaningful states $\varpi : \mathscr{A} \to \mathbb{C}$. That is, we can study the representation $\pi(\mathscr{A}) \subset B(\mathscr{H})$ in our physically relevant setting instead of every possible abstract $C^*$-algebra.

We can study many physical aspects of quantum theories by asking properties over $\varpi$. For example, the quasi-free or Gaussian condition ensures that the states are completely determined by two-point correlation functions, and the Hadamard condition ensures that causality is preserved [Bru&15]. In our case, we are interested in the theory in a Hamiltonian form, and therefore such features are part of the particular Hamiltonian that describes the theory.

Our goal in this section was more modest than the general aim of AQFT as a whole [Bru&15]. We only intend to set the groundwork to understand the equivalence of the same theory described in different terms. So far, we studied quantization mappings $Q$ to establish particular representations of Weyl relations. This is our $C^*$-algebra of quantum observables, which always acts over $L^2_{Hol}(\mathcal{N}', D\mu_c)$ or similar. From GNS, we now know that this Hilbert space, which seemed the central object in geometric quantization, is, in fact, irrelevant since it can be reconstructed by means of a particular state $\varpi$. Thus, to recover holomorphic quantization, we set $\hat{\mathcal{R}}(\alpha) = Q_{Weyl}(\mathcal{E}_\alpha)$ as our generators of the $C^*$-algebra of observables. Then, we can reconstruct the Hilbert space knowing that the Hilbert space vector representing the vacuum state is given by the constant function 1. To recover the state of the algebra, we look at the GNS construction, and by (4.57), we obtain

$$\varpi_{Hol}(\hat{\mathcal{R}}(\alpha)) = e^{\frac{\bar{\alpha}_x \Delta^{xy} \alpha_y}{2}} \int D\mu_c(\phi^{\mathbf{x}}) e^{i(\overline{\alpha_x \phi^x} + \alpha_x \phi^x)} = e^{-\frac{\bar{\alpha}_x \Delta^{xy} \alpha_y}{2}}. \qquad (4.64)$$

---

[4] Here, for simplicity, we identify the classical function with the $\star$ product with its image under the corresponding quantization mapping, as we know that they are isomorphic.

This construction may seem too involved for our purposes so far, but it is crucial to understand the equivalence with Fock quantization.

### 4.5.2 Fock Space Representation

Now we will consider how the representation of the $C^*$ algebra can be chosen for the Hilbert space to be the usual Fock space of QFT. Besides, we will identify the role that the different geometric structures introduced in the manifold of classical fields play in the representation.

In a Fock space representation, we start by introducing on the manifold of classical fields a complex structure compatible with (4.62). For simplicity, we choose the same entries of (3.21) adapted to the new setting on which $\mathcal{M}_A = \mathcal{N} \times \mathcal{N}$:

$$-J_{\mathcal{M}_A} = (d\varphi_y, d\pi_y) \begin{pmatrix} A^y_x & \Delta^y_x \\ D^y_x & -(A^t)^y_x \end{pmatrix} \begin{pmatrix} \partial_{\varphi_x} \\ \partial_{\pi_x} \end{pmatrix}. \qquad (4.65)$$

This leads to the positive definite Riemannian metric $\mu_{\mathcal{M}_A}(\cdot, \cdot) = \omega_{\mathcal{M}_A}(J\cdot, \cdot)$. We can use it to define a Hermitian metric on $\mathcal{M}_A$:

$$h_{\mathcal{M}_A} = \frac{\mu_{\mathcal{M}_A} - i\omega_{\mathcal{M}_A}}{2}. \qquad (4.66)$$

Moreover, by linearity, $(\mathcal{M}_A, h_{\mathcal{M}_A})$ is completed to a complex Hilbert space that we denote $\mathcal{H}_{1ps}$, the Hilbert space of one-particle states. We will use this Hilbert space as a starting point to represent our quantum field theory. The next step is to notice that we must allow states with an arbitrary number of bosonic particles; thus, we represent our theory over $\mathcal{H}_{Fock} = \Gamma \mathcal{H}_{1ps}$, the bosonic, or symmetric, Fock space. This space is defined in section D.2.

With this structure at hand, we use a change of variables dual to (3.25):

$$d\tilde{\varphi}_x = d\varphi_x + (KA)^y_x d\pi_y, \quad d\tilde{\pi}_x = K^y_x d\pi_y,$$

$$(\psi_{\mathbf{x}}, \bar{\psi}_{\mathbf{y}}) = \frac{1}{\sqrt{2}}(\tilde{\varphi}_{\mathbf{x}} + i\tilde{\pi}_{\mathbf{x}}, \tilde{\varphi}_{\mathbf{y}} - i\tilde{\pi}_{\mathbf{y}}), \qquad (4.67)$$

to obtain, in coordinates, $h_{\mathcal{M}_A} = \Delta^{xy} d\bar{\psi}_x \otimes d\psi_y$. In those coordinates, it is clear that $\mathcal{H}_{1ps} = \mathcal{H}_\Delta$, where the latter is the completion of $\mathcal{N}_{\mathbb{C}}$ with the Hermitian form $h_{\mathcal{M}_A}$. An element of the Fock space is written in coordinates in a very straightforward manner as

$$\Psi = (\psi^{(0)}, \psi^{(1)}_x, \psi^{(2)}_{x_1 x_2}, \cdots, \psi^{(n)}_{\vec{x}_n}, \cdots), \tag{4.68}$$

with $\psi^{(n)}_{(\vec{x}_n)} = \psi^{(n)}_{\vec{x}_n}$. Annihilation and creation operators are also easily written as

$$\bar{\chi}_x a^x \Psi = (\bar{\chi}_x \Delta^{xy} \psi^{(1)}_y, \cdots, \sqrt{n}\ \bar{\chi}_x \Delta^{xy} \psi^{(n)}_{y\vec{x}_{n-1}}, \cdots),$$
$$\chi_x a^{\dagger,x} \Psi = (0, \psi^{(0)} \chi_x, \cdots, \sqrt{n+1}\ \psi^{(n)}_{(\vec{x}_n} \chi_{x_{n+1})}, \cdots). \tag{4.69}$$

Among all the possible choices, we pick $1 \in \mathcal{H}^0_\Delta$ as the cyclic vector of the GNS construction because we want to interpret it as the vacuum state of our theory. The last ingredient is the state; a Fock generator of the Weyl algebra is [Wal94, CCQ04], in holomorphic coordinates:

$$\mathcal{R}_F(\alpha) = \exp(\bar{\alpha}_x a^x + \alpha_x a^{\dagger,x}). \tag{4.70}$$

Using the BCH formula with $[a^\mathbf{x}, a^{\dagger,\mathbf{y}}] = \Delta^{\mathbf{xy}}$, we put annihilation operators to the right of creation ones, then the Fock state acting on the generator, which is the vacuum expectation value of it, is easily computed by...

$$\varpi_{Fock}(\mathcal{R}_F(\alpha)) = e^{-\frac{1}{2}h_{\mathcal{M}_A}(\bar{\alpha},\alpha)} \tag{4.71}$$

Thus, $\varpi_{Hol}$ of (4.64) and $\varpi_{Fock}$ act in the same way over their respective generators of the Weyl algebra. Because of the GNS construction, they are related by a unitary isomorphism and, as such, they represent the same theory. The same is true for the Schrödinger representation [CCQ04]. As anticipated, a unitary isomorphism at the level of the Hilbert space is not enough to establish the equivalence of the theories.

Indeed, we already knew that Fock space coordinates (4.68) are the coefficients of the chaos decomposition of the holomorphic representation under the Segal isomorphism $\mathcal{I}: L^2_{Hol}(\mathcal{M}_C, D\mu_c) \to \mathcal{H}_{Fock}$. However, to establish the relation in full, we must also preserve the $C^*$-algebra of observables. In this case, both theories are unitarily equivalent representations of the same algebra of observables if we represent the $C^*$-algebra $(W, \star_m)$ with the quantization mapping $\mathcal{Q}_{Weyl}$ studied in section 4.4.

In any case, it is important to remark that Fock quantization is fundamentally different from that of holomorphic quantization since it relies upon the notion of one-particle space structure from which the Fock space is constructed. Kähler structures in this setting are also fundamentally different from those of holomorphic quantization. In holomorphic quantization, this structure is densely defined in the domain of the wave function $\Psi(\phi^\mathbf{x})$, while in Fock quantization, it is defined dually over the coefficients of the one-particle states.

The relation between both pictures is also more involved than previously expected. In this section, we made the design choice of writing Fock operators and vectors in holomorphic coordinates; this is because the explicit isomorphism given by the Segal isomorphism is only obvious under the change of coordinates (4.67). The real power of Fock construction relies on its coordinate-free nature, which just depends on the Kähler structure of $(\omega, \mu, J)_{\mathcal{M}_A}$. We will keep this section simple at the cost of blurring its real power, but this is a well-known feature in the literature, and we refer elsewhere for a thorough discussion [Wal94, Bru&15].

# Chapter 5

# Quantum Evolution: The Modified Schrödinger Equation

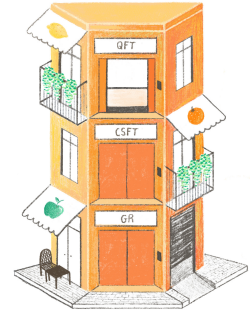

## 5.1 Introduction

The Hamiltonian approach that we present in this work presents some conceptual advantages over the relativistically covariant approaches. In the latter, the quantized quantities are solutions of some classical equations of motion. This approach, even though convenient for practical calculations [Bru&15], leaves no room for the distinction between kinematics and dynamics. In our approach, the quantized quantities are initial conditions or Cauchy data. This allows for the separation of both aforementioned aspects of the theory. In fact, in the previous chapter, we provide a careful depiction of the phase space and observables of a QFT. These are the fundamental kinematic aspects of the theory, as we pointed out in section 1.2. Now we must focus on the depiction of the dynamics of the theory.

First of all, in this chapter, we will focus on theories of QFT on curved spacetimes in which the latter is treated as a background. The general coupling to a dynamical spacetime with *backreaction* is more involved, and we skip its discussion to chapter 6. In this chapter, our point of view is that of a parametric theory, as described in *Remark* 3.1. In that case, the time parameter $t$ is treated as a label that marks the changes in the different geometric structures. We will elaborate more on this aspect in section 5.4.

In this chapter, we aim to postulate the dynamics of QFT using geometric arguments. In regular quantum mechanics, this evolution is just the Schrödinger equation. However, in general curved spacetimes, evolution will not preserve the Hilbert space of pure states $L^2_{Hol}(\mathcal{N}'_{\mathbb{C}}, D\mu_c)$ unless the spacetime is static or stationary [AMA75, CCQ04, AA15].

To gain a geometrical insight into how to study this change of Hilbert space structure, we should understand the kinematical aspects of the previous chapter in the language of differential geometry. For that matter, in section 5.2, we opt to rephrase the tools of the previous chapter using Kibble's program of geometrization of quantum mechanics [Kib79, AS99]. The procedure of geometrization is roughly as follows: given a quantum theory defined by a Hilbert space of pure states $\mathcal{H}$ and its algebra of observables $\hat{W}$, we describe each ingredient in geometrical terms. The Hilbert space is described as a complex manifold of pure states $\mathscr{P}$. We will say that we are working at the *second quantized (s.q.)* level when we model the space of pure states $\mathscr{P}$ as a manifold in this way. This manifold is endowed with a *s.q.* Kähler structure $(\mathcal{G}, \Omega, \mathcal{J})_{\mathscr{P}}$ that describes the Hilbert scalar product. Our objective is to describe this space in full generality without restricting the analysis to a particular measure $\mu_c$ determining the *s.q.* Kähler structure. For that matter, and in order to have good properties for differential geometry, we model $\mathscr{P}$ using Hida test functions $(\mathcal{N}_{\mathbb{C}})$.

This geometrization process opens the gate for a different kinematical description of the theory. From that study, we get the configuration space of all possible QFTs. This will be the space containing all allowed *s.q.* Kähler structures

$$(\mathcal{G}, \Omega, \mathcal{J})_{\mathscr{P}} \in \text{Kähl}(\mathscr{P}).$$

Together with this, we should restrict observables to be sesquilinear forms computed with this structure. This will be denoted $W_2(\mathscr{P})$.

Once again, Gaussian analysis tools from chapter 2 will be key to simplify the treatment of the theory in section 5.3. At this *s.q.* level, even the description of coordinates and functions over the manifold $\mathscr{P}$ will depend on the concept of Hida test function from section 2.7. These functions have been introduced with respect to the auxiliary concept of *white noise measure* $\beta$ (2.69) and, even though they are independent of this choice of auxiliary measure, are easier to handle in terms of it. This observation allows for a systematic use of white noise tools, such as integral kernels from section 2.7.1, to describe changes of coordinates, tensors, charts, etc. In particular, we are interested in the definition of real Darboux-like coordinates in this *s.q.* framework.

Once this description is understood, in section 5.4, we postulate dynamics as a Kähler-preserving flow. The situation is somewhat similar to the analysis shown in chapter 3. In that case, the dynamics of a Schrödinger-like equation for CSFT leads to some surprising aspects. There is the appearance of a connection $\Gamma_t$ in the time derivative and the need for the dependence of $\mu_c$ on the Hamiltonian of the classical theory. Nonetheless, that analysis was done with a heuristic treatment of Gaussian integration. In this case, we will recover similar structures to those in chapter 3 derived from the time dependence of the geometry of $\Sigma_t$ and its Riemannian structure. Using the previous analysis on the geometrization of quantum field theory, we derive a dynamical equation from first principles. This procedure will be consistent with our geometrical viewpoint and bound to reproduce the Schrödinger equation in the situations in which there is no ambiguity in its derivation [AMA75, CCQ04, AA15]. More precisely, we must match the unmodified Schrödinger equation in static and stationary spacetimes.

Finally, in section 5.5 we will explore the ambiguities in the choice of the connection $\Gamma_t$ for the time derivative $\nabla_t$. In particular, we will explore the dependence on the representation (holomorphic, antiholomorphic, Schrödinger and field-momentum) of a naive choice of $\nabla_t$. We will perform this analysis profiting from the discussion in section 4.3, summarized in Figure 4.1. That is, we will describe the integral transforms appearing in that diagram using the tools of the geometrization described above. Thus, we will start our discussion on the section describing the Fourier $\tilde{\mathcal{F}}$ and Segal-Bargmann $\tilde{\mathcal{B}}, \tilde{\bar{\mathcal{B}}}$ transforms using the techniques of integral kernels extensively studied throughout this chapter.

The central result of this chapter will be presented in section 5.5.4. This is, the choice of an unambiguous connection $\Gamma$ that respects the quantization mapping $\mathcal{Q}$ in compatible way with a coupling with gravity. The derivation of this condition is the main focus of chapter 6.

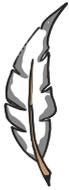

**Notation for Holomorphic and Weyl Quantization**

For reasons stated at the beginning of section 4.4, it is better to use the holomorphic representation to develop this study. For that matter, in this chapter, we identify $\phi^\mathsf{x}$ with $\phi^\mathsf{x}$ and the quantization mapping is extended to any function using the Weyl ordering prescription $\mathcal{Q} \simeq \mathcal{Q}_{Weyl}$ unless otherwise stated. However, we point out that the analysis can be straightforwardly extended to any other representation.

## 5.2 Kinematics of QFT: Geometric Quantum Field Theory

Kinematic aspects of a QFT as described in chapter 4 are the Hilbert space of pure states $\mathscr{H} = L^2_{Hol}(\mathcal{N}^r_\mathbb{C}, D\mu_c)$ and the algebra of observables $\hat{W} = \mathcal{Q}(W)$ that represent Weyl relations. Using the geometrization of QFT from [Kib79, AS99], kinematic aspects of a QFT are the description of the manifold of pure quantum states $\mathscr{P}$ with its additional spaces of *s.q.* Kähler structures

$$(\mathcal{G}, \Omega, \mathcal{J})_\mathscr{P} \in \text{Kähl}(\mathscr{P}) \tag{5.1}$$

and the algebra of observables as a representation of Weyl algebra $W_2(\mathscr{P})$, which is a subset of quadratic functions.

### 5.2.1 Geometrization of Quantum Field Theory I: The Manifold of Pure States

Due to the infinite-dimensional nature of the problem, the model space used to define the manifold structure is of crucial importance. For a well-posed geometric framework, we will choose to consider as reference a particular Fréchet-Nuclear space: the set of Hida test functions $\mathcal{N} = (\mathcal{N}_{\mathbb{C}})$. As we did in section 2.7, we will consider a suitable rigged Hilbert space

$$\mathcal{N} \subset L^2_{Hol}(\mathcal{M}_C, D\mu_c) \subset \mathcal{N}'^*. \tag{5.2}$$

As a result, the Hilbert space of the states of quantum fields $L^2_{Hol}(\mathcal{M}_C, D\mu_c)$ can be densely modeled as a $C^\infty$-manifold, considering $\mathcal{N}$ as the model space.

As the model space for our manifold is the space of test functions, we can introduce a system of coordinates associated with them on the space of quantum fields. In this case, we will denote test functions with a superscript $\Psi^\phi \in \mathcal{N}$, while $\Phi_\phi \in \mathcal{N}'^*$ stands for distributions. This convention is reversed from that of regular test functions and distributions (2.1), but turns out to be more convenient in this case because we stick to the geometric convention, and test functions will usually be treated as coordinates of the manifold $\mathscr{P}$. Of course, pointwise meaning holds for test functions, while it is lost in the distributional setting.

We endow the manifold with a Hermitian form trivially by using the scalar product of the Hilbert space. Since the topology of $\mathcal{N}$ is finer than that of $L^2_{Hol}(\mathcal{M}_C, D\mu_c)$, we have that its scalar product $\langle, \rangle_{\mu_c}$, inherited by the embedding, is a sesquilinear positive definite and continuous bilinear form on $\mathcal{N}$. Moreover, this form is weakly non-degenerate. This means that only the zero vector has zero norm also in $\mathcal{N}$, but the map $\langle, \rangle_{\mu_c} : \mathcal{N} \to \mathcal{N}'^*$ that maps every element $\Psi^\phi \in \mathcal{N}$ to $\langle \Psi^\phi, \cdot \rangle_{\mu_c} \in \mathcal{N}'^*$ is not surjective, although it is clearly injective.

The manifold $\mathscr{P}$ with this Hermitian structure, as anticipated, denotes the manifold of pure quantum states. As a linear complex manifold, it can be trivially treated as a real manifold endowed with a *s.q.* Kähler structure $(\mathcal{G}, \Omega, \mathcal{J})_{\mathscr{P}}$ related to the Hermitian tensor above. To recover the quantum theory, we follow [Kib79] and choose a preferred point in $|0\rangle \in \mathscr{P}$ that represents the vacuum and consider the tangent space of this point $T_0\mathscr{P} \simeq \mathcal{N}$. In linear spaces, which is the case that we are treating here, this distinction is blurred by the identification of $\mathscr{P}$ with $T_0\mathscr{P}$. In the following, we will identify both spaces, when it is possible, to simplify the discussion.

The description of $\mathscr{P}$ using directly the identification with Hida test functions $\Psi^\phi \in L^2_{Hol}(\mathcal{M}_C, D\mu_c)$ is equivalent to a description in holomorphic *s.q.* coordinates. In this case, there is a global chart for the linear manifold $\mathscr{P}$, and under the identification of both spaces, we can write

$$h_{\mathscr{P}}(\Phi^\phi, \Psi^\phi) = \frac{\mathcal{G}_{\mathscr{P}}(\Phi^\phi, \Psi^\phi) - i\Omega_{\mathscr{P}}(\Phi^\phi, \Psi^\phi)}{2} = \int_{\mathcal{N}'_\mathbb{C}} D\mu_c(\phi^{\mathsf{x}})\overline{\Phi(\phi^{\mathsf{x}})}\Psi(\phi^{\mathsf{x}}) := \overline{\Phi^\sigma}\Delta_{\bar\sigma\phi}\Psi^\phi.$$
(5.3)

Notice that in this case, the complex structure is completely determined by the definition of the space of quantum states, and we do not have the freedom to select it as in the case of the manifold of classical fields. This is because we are describing a given theory in geometrical terms. Below, in section 5.3, we will develop some important tools to express the geometrization of this section in concrete sets of coordinates and discuss to what extent we can freely choose the quantization procedure.

### 5.2.2 Geometrization of Quantum Field Theory II: The Algebra of Observables

Once we have described the space of pure states of the theory in the geometrized setting, the remaining task is to describe an algebra of observables that fulfills Weyl relations and is unitarily equivalent to the ones described so far. As before, we start in full generality and unveil several interesting properties in particular sets of coordinates using Fock quantization.

We must address first the issue of how to cast into this formalism a previously existing algebra. The description of the space of pure states $\mathscr{P}$ that we want to achieve is geometrical in nature, as a consequence, a description of the algebra should follow three conditions:

- The algebra of observables is a subset of the space of $C^\infty(\mathscr{P})$.

- It contains a dense representation of the Weyl algebra (4.63). This implies that the product of observables cannot be the pointwise product of functions.

- The Hamiltonian vector field generated from an element of the algebra $F$ given by $X_F = \{\cdot, F\}_{\mathscr{P}}$ is such that its flow preserves the Kähler structure. This means that $\mathcal{L}_{X_F} h_{\mathscr{P}} = 0$ and implies that quantum observables generate symmetries that cannot transform the structure $(\mathcal{G}, \Omega, \mathcal{J})_{\mathscr{P}}$ that defines the quantum phase space. If we want to change this structure with respect to a given evolution, it must be introduced as a change in the Kähler manifold from an external source such as an external gravitational field [Alo&24a].

To achieve this, we will be considering the sesquilinear forms defined in holomorphic s.q. coordinates as

$$f_{\hat{G}}(\bar\Psi, \Psi) = \int D\mu_c(\phi)\overline{\Psi(\phi)}\hat{G}\Psi(\phi),$$
(5.4)

for $G \in \mathcal{O}_{cl}$, $\hat{G} = \mathcal{Q}_{Weyl}(G)$. In formal terms, we will write

$$f_{\hat{G}}(\bar{\Psi}, \Psi) = \langle \Psi, \hat{G}\Psi \rangle_{\mathscr{P}} = \bar{\Psi}^{\bar{\phi}} \Delta_{\bar{\phi},\sigma} \hat{G}_{\gamma}^{\sigma} \Psi^{\gamma}, \tag{5.5}$$

where $\Delta_{\bar{\phi},\sigma}$ is introduced in (5.3) and $\hat{G}_{\gamma}^{\sigma}$ represents the integral kernel representation of the operator $\mathcal{Q}_{Weyl}(G)$. Integral kernel representations will be of huge importance in this *s.q.* description, and we will discuss them in depth in the next section. The product of sesquilinear forms follows directly from an extension of the Moyal product

$$f_{\hat{G}} *_m f_{\hat{H}} := f_{\mathcal{Q}_{Weyl}(G *_m H)} = f_{\hat{G}\hat{H}}. \tag{5.6}$$

With these quadratic functions, we can find a $C^*$-isomorphism from the algebra $(W, \star_m, \|\cdot\|_{Weyl})$, described in subsubsection 4.4.1, to the set

$$W_2(\mathscr{P}) = \left\{ f_{\hat{G}}(\bar{\Psi}, \Psi) | \ \hat{G} \in \hat{W} \right\}, \tag{5.7}$$

endowed with the product already described and a norm

$$\|f_{\hat{G}}\| := \sup \left\{ [f_{\hat{G}^{\dagger}} *_m f_{\hat{G}}(\Psi)]^{\frac{1}{2}}, \|\Psi\|_{h_{\mathscr{P}}} = 1 \right\} = \|G\|_{Weyl}. \tag{5.8}$$

Then it is clear that $W_2(\mathscr{P})$ is a representation of Weyl relations.

The most interesting property of this product is that it trivializes the Lie algebra representation of the commutator of quantum operators. It can be readily checked that

$$\{f_{\hat{G}}, f_{\hat{H}}\}_{\mathscr{P}} = -i(f_{\hat{G}} *_m f_{\hat{H}} - f_{\hat{H}} *_m f_{\hat{G}}) = f_{-i[\hat{G},\hat{H}]}. \tag{5.9}$$

The left-hand side of this expression only depends on the geometrical properties of the manifold, i.e., on its Kähler structure. In summary, a dense representation of Weyl relations $W_2(\mathscr{P})$ is contained in the algebra of sesquilinear forms endowed with the product $*_m$.

Due to the fact that quadratic functions in (5.5) do not mix holomorphic and antiholomorphic coordinates, this representation also fulfills the last condition. To see this, notice that $\mathcal{L}_{X_{f_A}} \Omega_{\mathscr{P}} = 0$ because the symplectic structure is always preserved by the Hamiltonian flow of $f_A$. We can explicitly write down the Hamiltonian vector field using (5.19) as $X_{f_A} = -i\left( \Psi^{\bar{\phi}} A_{\sigma}^{\phi} \partial_{\Psi^{\phi}} - \bar{\Psi}^{\sigma}(A^t)_{\bar{\sigma}}^{\bar{\phi}} \partial_{\bar{\Psi}^{\bar{\phi}}} \right)$. For the complex structure (5.14), we thus obtain

$$\mathcal{L}_{X_{f_A}} \left(\mathcal{J}_{\mathscr{P}}\right)_{\phi}^{\sigma} = \left(\mathcal{J}_{\mathscr{P}}\right)_{\phi}^{\alpha} \partial_{\Psi^{\alpha}} X_{f_A}^{\sigma} - \partial_{\Psi^{\sigma}} X_{f_A}^{\alpha} \left(\mathcal{J}_{\mathscr{P}}\right)_{\alpha}^{\phi} +$$
$$\left(\mathcal{J}_{\mathscr{P}}\right)_{\phi}^{\alpha} \partial_{\bar{\Psi}^{\alpha}} X_{f_A}^{\sigma} - \partial_{\bar{\Psi}^{\sigma}} X_{f_A}^{\alpha} \left(\mathcal{J}_{\mathscr{P}}\right)_{\alpha}^{\phi} = 0$$

Finally, due to the Kähler compatibility condition $\mathcal{G}_{\mathscr{P}} = \Omega_{\mathscr{P}}(\cdot, \mathcal{J}_{\mathscr{P}} \cdot)$, we get for (5.15) $\mathcal{L}_{X_{f_A}} h_{\mathscr{P}} = 0$.

In particular, we want the mean value of an observable (5.5) to be a real quantity. For that matter, we restrict observables to be associated with self-adjoint operators. To simplify the presentation, we skip the discussion on domain problems and restrict the definition to bounded operators, that is, operators $A \in W_2(\mathscr{P}) \subset \mathcal{O}_{cl}$. Then self-adjoint operators correspond to the set of quadratic polynomials on $\mathscr{P}$ defined as:

$$\mathcal{F}_2(\mathscr{P}) = \left\{ f_{\hat{A}}(\bar{\Psi}, \Psi) | \ \hat{A}^\dagger = \hat{A} \right\}. \tag{5.10}$$

This is equivalent, at the level of classical functions, to $A^* = A \in W_2(\mathscr{P}) \subset \mathcal{O}_{cl}$. We will explore how to express this condition in particular sets of *s.q.* coordinates in section 5.3.3. In particular, in holomorphic *s.q.* coordinates, the condition will be derived in (5.30). Thus, the algebra of quantum observables might be seen as the quantization of real classical functions. In addition, those were the legitimate observables of a classical theory in the first place.

In summary, $\mathcal{F}_2(\mathscr{P})$ is the algebra of observables of the quantum theory. Introducing the Jordan product $f_{\hat{A}} \circ f_{\hat{B}} = (f_{\hat{A}} *_m f_{\hat{B}} + f_{\hat{B}} *_m f_{\hat{A}})/2 = f_{(\hat{A}\hat{B}+\hat{B}\hat{A})/2}$, we can endow it with a product $*$ that

$$f_{\hat{A}} *_m f_{\hat{B}} = f_{\hat{A}} \circ f_{\hat{B}} + \frac{i}{2} \{ f_{\hat{A}}, f_{\hat{B}} \}_{\mathscr{P}} \tag{5.11}$$

This product is not internal to the algebra of observables. Instead it leaves $W_2(\mathscr{P})$ invariant.

## 5.3 Geometrized QFT in coordinates: Holomorphic and Darboux s.q. coordinates

The manifold $\mathscr{P}$ is modeled over the, rather cumbersome to deal with, set of Hida test functions $(\mathcal{N}_\mathbb{C})$. In the previous section, we already used holomorphic *s.q.* coordinates to describe $\mathscr{P}$. However, the main drawback of this choice of coordinates is similar in nature to (3.27). This is, holomorphic coordinates *s.q.* are adapted to a particular *s.q.* Kähler structure. In this section, we will discuss the implications of this statement, geometrizing the holomorphic representation of QFT. To do so, we look for an efficient way of describing charts, i.e., systems of coordinates, for $\mathscr{P}$.

The first step is to agree on a natural measure to use as a reference. This is, using the toolkit of chapter 2, to choose a canonical Gel'fand triple that generalizes (2.68). In that case, the integration with respect to $dVol_h$ is used as the natural choice. In this case, we follow [Hid&93] and choose for reference the white noise measure $\beta$ defined by the characteristic functional (2.69). We pick white noise for two reasons.

- First, it is the immediate generalization of (2.68), and the white noise Hilbert space $L^2(\mathcal{N}'_\mathbb{C}, D\beta)$ has as Cameron-Martin Hilbert space $L^2(\Sigma, dVol_h)$. This means that the white noise space is nothing but the Bosonic Fock space constructed from the canonical measure of $\Sigma$.

- Second, we do not have *a priori* knowledge of the measure $\mu_c$ that depends on the Hamiltonian of the particular QFT we are describing. As we discussed at the beginning of this chapter, it is important to distinguish the kinematical aspects of a theory from the dynamical ones. Kinematics, in this case, means the study of the configuration space of all possible QFTs. This will be the space containing all allowed *s.q.* Kähler structures

$$(\mathcal{G}, \Omega, \mathcal{J})_{\mathscr{P}} \in \text{Kähl}(\mathscr{P}) \tag{5.12}$$

Together with this, to complete the kinematical description, we need to describe the algebra of observables. In our case, we will focus on the quantization mapping $\mathcal{Q}$. Dynamics must be determined by the Hamiltonian of the theory that selects a particular curve in this space.

Thus, the natural triple identifying functions with distributions is

$$(\mathcal{N}_{\mathbb{C}}) \subset L^2_{Hol}(\mathcal{N}'_{\mathbb{C}}, D\beta) \subset (\mathcal{N}_{\mathbb{C}})'^*. \tag{5.13}$$

Proceeding as in section 2.7, Hida test functions are easily described with respect to an auxiliary Gaussian measure. In that section, we described several tools that will ease the description of coordinates, tensors, and operators that will represent the algebra of observables.

### 5.3.1 Holomorphic quantization in the geometrized setting I: The manifold of pure states

Let's focus on the geometrization of the space of pure states $\mathscr{P}$ of the holomorphic representations. In the rigging (5.13), we denote $\beta_{\bar{\phi},\sigma}$ as the bilinear form that identifies test functions $\Phi^\sigma$ with distributions $\Phi_\phi = (\beta_{\bar{\phi},\sigma}\Phi^\sigma)^*$. It also admits a weak inverse $\beta^{\phi,\bar{\sigma}}$ such that $\beta^{\gamma,\bar{\sigma}}\beta_{\bar{\sigma},\phi} = \delta^\gamma_\phi$ is the evaluation mapping, continuous for Hida test functions, as can be shown using the tools of reproducing kernels *Example* 2.10. In a nutshell, we will use $\beta_{\bar{\phi},\sigma}, \ \beta^{\gamma,\bar{\sigma}}$ as the auxiliary bilinears to raise and lower indices of our operators and tensors.

Let $\Psi^\phi$ be a test function representing a state of the manifold $\mathscr{P}$, the manifold of pure states. The notion of canonicity (5.13) suggests that we must write every bilinear form or operator as the action of an integral kernel with respect to the white noise. In this manner, for holomorphic quantization, we have a complex structure

$$\mathcal{J}_{\mathscr{P}} = i(d\Psi^\phi \otimes \partial_{\Psi^\phi} - d\bar{\Psi}^{\bar{\phi}} \otimes \partial_{\bar{\Psi}^{\bar{\phi}}}) \tag{5.14}$$

and the Hermitian structure in holomorphic coordinates

$$h_{\mathscr{P}} = \frac{\mathcal{G}_{\mathscr{P}} - i\Omega_{\mathscr{P}}}{2} = \Delta_{\bar{\sigma}\phi} d\bar{\Psi}^{\bar{\sigma}} \otimes d\Psi^\phi \tag{5.15}$$

is written using the bilinear $\Delta_{\bar{\sigma},\phi}$ expressed as a white noise kernel with

$$\tilde{\Upsilon}^{\bar{\sigma}}\Delta_{\bar{\sigma},\phi}\Psi^{\phi} = \iint D\beta(\sigma)D\beta(\phi)\overline{\Upsilon(\sigma)}e^{\sigma^x\Delta_{xy}\bar{\phi}^y}\Psi(\phi) = \int D\mu_c(\phi)\overline{\Upsilon(\phi)}\Psi(\phi) \qquad (5.16)$$

Following our discussion about reproducing kernels in *Example* 2.10, it is simple to understand that this expression, of course, does not have a pointwise meaning since the objects are bi-valued Hida distributions. We also obtain that the inverse bilinear $(\mathcal{K}\Delta)^{\sigma}_{\phi} = \delta^{\sigma}_{\phi}$ is $\mathcal{K}^{\phi,\bar{\sigma}} = \exp(\bar{\sigma}^x K_{xy}\phi^y)$.

It is convenient to describe the manifold with real coordinates. Let $\Psi^{\phi} = \frac{1}{\sqrt{2}}(\tilde{\Phi}^{\phi} + i\tilde{\Pi}^{\phi})$ such that the set of coordinates $\begin{pmatrix} \tilde{\Phi}^{\phi} \\ \tilde{\Pi}^{\phi} \end{pmatrix}$ are *real* in the sense that they are holomorphic functions with real coefficients in their chaos decomposition. This can be written as $\overline{\Phi(\bar{\phi}^{\mathsf{x}})} = \Phi(\bar{\phi}^{\mathsf{x}})$. In these coordinates

$$\mathcal{G}_{\mathscr{P}} = (d\tilde{\Phi}^{\bar{\phi}}\ d\tilde{\Pi}^{\bar{\phi}})\Delta_{\bar{\phi}\sigma}\begin{pmatrix} d\tilde{\Phi}^{\sigma} \\ d\tilde{\Pi}_{\sigma} \end{pmatrix}, \ \ \Omega_{\mathscr{P}} = (d\tilde{\Phi}^{\bar{\phi}}\ d\tilde{\Pi}^{\bar{\phi}})\Delta_{\bar{\phi}\sigma}\epsilon\begin{pmatrix} d\tilde{\Phi}^{\sigma} \\ d\tilde{\Pi}_{\sigma} \end{pmatrix},$$

$$\mathcal{J}_{\mathscr{P}} = (\partial_{\tilde{\Phi}^{\phi}}\ \partial_{\tilde{\Pi}^{\phi}})\delta^{\phi}_{\sigma}\epsilon\begin{pmatrix} d\tilde{\Phi}^{\sigma} \\ d\tilde{\Pi}_{\sigma} \end{pmatrix} \qquad \text{for } \epsilon = \begin{pmatrix} 0 & -1 \\ 1 & 0 \end{pmatrix}. \qquad (5.17)$$

To write down a dynamical system, it is also interesting to describe the Poisson bracket in this context. To do so, we define, associated with these coordinates, the Wirtinger derivatives

$$\partial_{\Psi^{\phi}} = \frac{1}{\sqrt{2}}(\partial_{\tilde{\Phi}^{\phi}} - i\partial_{\tilde{\Pi}^{\phi}}), \quad \partial_{\Psi^{\bar{\phi}}} = \frac{1}{\sqrt{2}}(\partial_{\tilde{\Phi}^{\bar{\phi}}} + i\partial_{\tilde{\Pi}^{\bar{\phi}}}), \qquad (5.18)$$

This fact can be proved by using the computational rule (2.83). Thus, the Poisson bracket is

$$\{\cdot,\cdot\}_{\mathscr{P}} = -i\mathcal{K}^{\phi,\bar{\sigma}}\partial_{\Psi^{\phi}} \wedge \partial_{\Psi^{\bar{\sigma}}}. \qquad (5.19)$$

Let $F(\tilde{\Phi}^{\phi},\tilde{\Pi}^{\phi})$ and $G(\tilde{\Phi}^{\phi},\tilde{\Pi}^{\phi})$ be well-behaved real functions over $\mathscr{N}$, then the bracket action is

$$\{F,G\}_{\mathscr{P}} = \iint D\beta(\phi)D\beta(\sigma)\exp(\bar{\sigma}^x K_{xy}\phi^y)\left(\frac{\partial F}{\partial\tilde{\Phi}(\sigma)}\frac{\partial\bar{G}}{\partial\tilde{\Pi}(\bar{\phi})} - \frac{\partial F}{\partial\tilde{\Pi}(\sigma)}\frac{\partial\bar{G}}{\partial\tilde{\Phi}(\bar{\phi})}\right), \quad (5.20)$$

The space of Hida test functions is nuclear and Fréchet, among other convenient properties suited for geometry discussed in section 2.8. This implies that the notions of Gateaux and Fréchet derivatives coincide and there is no ambiguity in the definition of this Poisson bracket.

### 5.3.2 Geometrization in Darboux second quantized (s.q.) coordinates

An important set of coordinates that will ease some of the discussion are Darboux-like coordinates that provide some sense of canonical coordinates in this involved framework. To craft a notion of Darboux *s.q.* coordinates, (5.13) suggests choosing as a canonical form for the symplectic structure the one corresponding to the white noise Hilbert space. In the spirit of the discussion at the beginning of this section, this is a direct generalization of Darboux coordinates in (4.62). All things considered, the canonical form is

$$\Omega_{\mathscr{P}} = \beta_{\bar{\phi},\sigma} d\Pi^{\bar{\phi}} \wedge d\Phi^{\sigma} = \int D\beta(\phi) \left[ d\Pi(\bar{\phi}) \otimes d\Phi(\phi) - d\Phi(\bar{\phi}) \otimes d\Pi(\phi) \right]. \tag{5.21}$$

We can simply find a system of Darboux *s.q.* coordinates with a change of *s.q.* variables $\begin{pmatrix} \tilde{\Phi}^{\phi} \\ \tilde{\Pi}^{\phi} \end{pmatrix} = \mathcal{C} \begin{pmatrix} \Phi^{\phi} \\ \Pi^{\phi} \end{pmatrix}$ that leads to an off diagonal form of the complex structure

$$\mathcal{C} = \begin{pmatrix} \mathcal{K} & 0 \\ 0 & \mathbb{1} \end{pmatrix}, \quad \mathcal{C}^{-1} = \begin{pmatrix} \Delta & 0 \\ 0 & \mathbb{1} \end{pmatrix}, \quad \mathcal{J} = \begin{pmatrix} 0 & \Delta \\ -\mathcal{K} & 0 \end{pmatrix}. \tag{5.22}$$

Finally, this symplectic structure induces a Poisson bracket over functions in Darboux s.q. coordinates

$$\{\cdot, \cdot\}_{\mathscr{P}} = \beta^{\phi,\bar{\sigma}} \partial_{\Phi^{\phi}} \wedge \partial_{\Pi^{\bar{\sigma}}}. \tag{5.23}$$

that over smooth real functions acts like

$$\{F, G\}_{\mathscr{P}} = \int D\beta(\phi) \left( \frac{\partial F}{\partial \Phi(\phi)} \frac{\partial \overline{G}}{\partial \Pi(\bar{\phi})} - \frac{\partial F}{\partial \Pi(\phi)} \frac{\partial \overline{G}}{\partial \Phi(\bar{\phi})} \right). \tag{5.24}$$

This set of coordinates will be important later to discuss the particular expression for the Hermitian of an operator.

### 5.3.3 Hermitian operators in real coordinates

In general, the Hermitian of an operator is given by the symplectic form, such that $\Omega(\cdot, A^{\dagger}) = \Omega(A, \cdot)$. For this reason, the explicit expression of the $\dagger$ operator in different systems of coordinates has different forms. Since it is determined by the symplectic form, the simplest expression of the $\dagger$ operation is in Darboux s.q. coordinates. In the case of interest of this section we doubled the degrees of freedom of the Hilbert space we want to describe by considering real coordinates. Considering the natural setting of complex holomorphic *s.q.* coordinates, the definition using the symplectic form would be enough to characterize the $\dagger$ operation. In the real case we must distinguish Hermitian operators $A^{\dagger}$ from adjoint operators. We call $A^{\dagger}$ the adjoint of $A$ only if $[A, \mathcal{J}] = 0$; otherwise, it will mix the holomorphic and antiholomorphic subspaces, not respecting the definition of the whole scalar product. Thus, in real coordinates

$$f_{\mathcal{O}}(\Phi, \Pi) = \frac{1}{2} \Omega_{\mathscr{P}} \left( \left( \begin{array}{c} \Phi^\phi \\ \Pi^\phi \end{array} \right), (\mathcal{J}\mathcal{O})^\sigma_\phi \left( \begin{array}{c} \Phi^\phi \\ \Pi^\phi \end{array} \right) \right). \tag{5.25}$$

Where $\mathcal{O}$ is a $2 \times 2$ matrix whose entries are operators acting on functions $\Phi^\sigma$ or $\Pi^\sigma$. This considered, (5.10) in Darboux *s.q.* coordinates is

$$\mathcal{F}_2(\mathscr{P}) = \left\{ f_{\hat{A}}(\Phi, \Pi) | \; \hat{A}^\dagger = \hat{A} \text{ and } [\hat{A}, \mathcal{J}] = 0 \right\}. \tag{5.26}$$

Recall that we assume $\hat{A}$ is bounded to avoid domain problems in this definition. Our goal in this section is then to describe the $\dagger$ operator. The symplectic form induced by the white noise measure possesses a particularly simple expression for the $\dagger$ of an operator. We will distinguish this case by denoting $\ddagger$ to the white noise Hermitian:

$$\mathcal{O}^\ddagger = \epsilon^t \mathcal{O}^t \epsilon. \tag{5.27}$$

In this operation, the $\epsilon$ matrix of (5.17) implements the anti-symmetry of the symplectic form, and we introduce the transpose of an operator as

$$\left( \begin{array}{cc} \mathcal{B} & \mathcal{C} \\ \mathcal{E} & \mathcal{L} \end{array} \right)^{t\sigma}_\phi = \left( \begin{array}{cc} \mathcal{B}^t & \mathcal{E}^t \\ \mathcal{C}^t & \mathcal{L}^t \end{array} \right)^\sigma_\phi, \text{ with } (\mathcal{B}^t)^\sigma_\phi = \beta^{\sigma\bar{\gamma}} \beta_{\bar{\delta}\phi} \mathcal{B}^{\bar{\delta}}_{\bar{\gamma}}. \tag{5.28}$$

Using the canonical expression of $\ddagger$, we can represent the adjoint operator in other systems of coordinates by specifying a change to Darboux *s.q.* coordinates. With a change of variables like (5.22), we get an expression for the holomorphic $\dagger_H$ and antiholomorphic $\dagger_{\bar{H}}$ cases

$$\mathcal{O}^{\dagger_H} = \mathcal{K} \mathcal{O}^\ddagger \Delta, \;\; \mathcal{O}^{\dagger_{\bar{H}}} = \mathcal{D}^{-1} \mathcal{O}^\ddagger \mathcal{D}. \tag{5.29}$$

where $(\mathcal{D}^{-1})^{\phi\bar{\sigma}} = \exp(-\bar{\sigma}^x D_{xy}^{-1} \phi^y)$. These simple expressions allow for a simpler analysis of some of the aspects discussed above.

The first example that we will write down is the condition (5.10) in holomorphic coordinates, which reads

$$\Delta_{\bar{\phi},\sigma} \hat{A}^\sigma_\rho \mathcal{K}^{\rho,\bar{\gamma}} = \left( \hat{A}^{\bar{\gamma}}_{\bar{\phi}} \right)^*. \tag{5.30}$$

Another interesting example is the Fourier transform $\tilde{\mathcal{F}}$. It is important to notice that $\tilde{\mathcal{F}}$ is not internal to either of the two systems of holomorphic or antiholomorphic coordinates, as shown in Figure 4.1, and therefore we obtain $\tilde{\mathcal{F}}^\dagger = \mathcal{K} \tilde{\mathcal{F}}^\ddagger \mathcal{D}$. Using an integral representation, as we will introduce in (5.47), it is immediate to see

$$\tilde{\mathcal{F}}^\dagger = \tilde{\mathcal{F}}^{-1}. \tag{5.31}$$

### 5.3.4  The algebra of observables of Fock quantization in the geometrized setting

The remaining task is to describe the representation of the Weyl algebra $W_2(\mathscr{P})$ (5.7) and of observables $\mathcal{F}_2(\mathscr{P})$ (5.10) in real coordinates. To do so, we describe the quantization mapping $\mathcal{Q}$ in realified *s.q.* holomorphic coordinates $\begin{pmatrix} \tilde{\Phi}^\phi \\ \tilde{\Pi}^\phi \end{pmatrix}$ and briefly discuss other approaches. We will follow the spirit of Fock quantization; this is, the representation of Weyl relations stems from the definition of creation and annihilation operators fulfilling the canonical commutation relations. Creation and annihilation operators in this setting are represented by the Skorokhod integral and its adjoint (4.10), based on the Malliavin derivative. To represent them, we introduce the operators

$$\tilde{\mathfrak{D}}^x = \Delta^{xy}\partial_{\phi^y} \quad \text{and} \quad \tilde{\mathfrak{D}}^{\dagger,x} = \phi^x. \tag{5.32}$$

In a generic system of coordinates, we denote this pair of adjoint operators $\mathfrak{D}^{\dagger,x}$ and $\mathfrak{D}^x$. They are indeed adjoint and not only Hermitian according to the distinction made in section 5.3.3 because the complex structure in realified holomorphic *s.q.* coordinates is equivalent to multiplying by the constant matrix $\mathcal{J} \simeq \epsilon$.

The main difference between this quantization and the ones studied in chapter 4 is that the role of the imaginary unit $i$ multiplying a function is given by $\mathcal{J}$. Thus, in general coordinates, our quantization mapping maps the imaginary unit to the *s.q.* complex structure $\mathcal{Q}(i) = \mathcal{J}$.[1] To see this condition in place, we must contract the operators with a direction of $\chi \in T_0\mathcal{M}_A \simeq \mathcal{M}_A$, the tangent space of the manifold on which the Hermitian structure (4.66) is defined. Recall that this manifold $\mathcal{M}_A$ is the one that spans the coefficients in the chaos decomposition and generates Fock quantization in section 4.5. In holomorphic coordinates for $\mathcal{M}_A$, according to this quantization rule, it must be expressed as $\tilde{\chi}_\mathbf{x} = \frac{1}{\sqrt{2}}(\tilde{\lambda}_\mathbf{x} + \mathcal{J}\tilde{\eta}_\mathbf{x})$. Therefore, we have

$$[\mathfrak{D}(\bar{\chi}_1), \mathfrak{D}^\dagger(\chi_2)] = \frac{1}{2}\Big(\mathbb{1}\mu_{\mathcal{M}_A}(\bar{\chi}_1, \chi_2) - \mathcal{J}\omega_{\mathcal{M}_A}(\bar{\chi}_1, \chi_2)\Big). \tag{5.33}$$

where the bilinears are those of the Hermitian structure of $\mathcal{M}_A$ (4.66). Then we can treat them as the creation and annihilation operators. Notice, however, that the holomorphic part of the state, in generic real coordinates, is the eigenspace of the $2 \times 2$ matrix $\mathcal{J}$ of eigenvalue $i$. This is, the holomorphic part will be given by the projection

$$\frac{1}{2}(\delta - i\mathcal{J})^\sigma_\phi \begin{pmatrix} \Phi^\phi \\ \Pi^\phi \end{pmatrix} = \begin{pmatrix} \Phi^\sigma \\ \Pi^\sigma \end{pmatrix}_{Hol}. \tag{5.34}$$

The operators $\mathfrak{D}^\dagger(\chi), \mathfrak{D}(\chi)$ act on both the holomorphic and antiholomorphic parts of the state.

---

[1]Notice that, what we are saying in this expression, is that the multiplication by the complex scalar $i$ is mapped into a tensor $\mathcal{J}$ acting over the real two dimensional vector $\begin{pmatrix} \tilde{\Phi}^\phi \\ \tilde{\Pi}^\phi \end{pmatrix}$.

As we see in this section, the relevant information of this kind of representation, as is the case of Fock quantization, relies on a change of description of $\mathcal{M}_A$. For this reason, it is interesting to express the direction $\chi \in \mathcal{M}_A$ in canonical coordinates $(\lambda_x, \eta_x)$ related to the holomorphic coordinates by the change of variable (4.67). Turning back to the case of holomorphic *s.q.* coordinates, operators $\tilde{\mathfrak{D}}, \tilde{\mathfrak{D}}^{\dagger_H}$ contracted with a direction $(\lambda_x, \eta_x)$ (and its conjugate, respectively) are given by

$$\tilde{\mathfrak{D}}(\bar{\chi}) = \tilde{\mathfrak{D}}(\lambda_x, \eta_x) = \frac{\lambda_x}{\sqrt{2}} \Delta^{xy} \partial_{\phi^y} - \frac{\eta_x}{\sqrt{2}} \begin{pmatrix} A^t & -\delta \\ \delta & A^t \end{pmatrix}^{xy} \partial_{\phi^y}$$

$$\tilde{\mathfrak{D}}^{\dagger_H}(\chi) = \tilde{\mathfrak{D}}(\lambda_x, \eta_x)^{\dagger_H} = \frac{\lambda_x \phi^x}{\sqrt{2}} - \frac{(\eta K)_x}{\sqrt{2}} \begin{pmatrix} A & \delta \\ -\delta & A \end{pmatrix}^x_y \phi^y. \tag{5.35}$$

Looking at (3.25), we can see that the first coordinate selects directions of the $\varphi^{\mathsf{x}}$ canonical coordinate, while the second selects the direction $\pi^{\mathsf{x}}$. Thus, in this system of coordinates, we get

$$\lambda_x \hat{\varphi}^x = \tilde{\mathfrak{D}}(\lambda_x, 0) + \tilde{\mathfrak{D}}(\lambda_x, 0)^{\dagger_H}, \quad \eta_x \hat{\pi}^x = \tilde{\mathfrak{D}}(0, \eta_x) + \tilde{\mathfrak{D}}(0, \eta_x)^{\dagger_H}. \tag{5.36}$$

In this way, we get $[\lambda_x \hat{\varphi}^x, \eta_x \hat{\pi}^x] = \epsilon \lambda_x \delta^{xy} \eta_y$, which are the canonical commutation relations. To complete the quantization procedure, the only remaining ingredient is the choice of ordering. As above, our choice is Weyl ordering. The remaining task is therefore to adapt (4.50) to the spirit of this chapter. This is, express it in terms of an integral kernel with respect to white noise. In this fashion, recalling (2.80), we get, in real holomorphic *s.q.* coordinates, the following result

$$\mathcal{Q}\left[\exp(i\lambda_x \varphi^x)\right]^{\bar{\phi}\sigma} = e^{\bar{\phi}^x \delta_{xy} \sigma^y - \frac{\lambda_x \Delta^{xy} \lambda_y}{2} + \epsilon \frac{\lambda_x (\Delta_y^x \bar{\phi}^y + \sigma^x)}{\sqrt{2}}}$$

$$\mathcal{Q}\left[\exp(i\eta_x \pi^x)\right]^{\bar{\phi}\sigma} = e^{\bar{\phi}^x \delta_{xy} \sigma^y + \frac{\eta_x D^{xy} \eta_y}{2} - \epsilon \frac{\eta_x (\bar{\phi}^x - K_y^x \sigma^y)}{\sqrt{2}}} e^{-\frac{\eta_x A^t{}_y^x (\bar{\phi}^y + K_z^y \sigma^z)}{\sqrt{2}}} \tag{5.37}$$

Once this is considered, the algebra of quadratic observables (5.7) is just given by (5.25).

These kinds of expressions for the kernels and creation and annihilation operators do not simplify under a change to the set of Darboux *s.q.* coordinates given by (5.22). For this reason, we skip the discussion in other systems of coordinates. Nonetheless, we leave open for future works the question: Is there any change of Darboux *s.q.* coordinates that simplifies (5.35) and coincides with (4.67) at the first quantized level?

## 5.4 Time Evolution: The (modified) Schrödinger equation

To postulate the equations of motion, we must first understand how our quantum structures, built on $\mathscr{P}$, depend on the time parameter. Recall that in this chapter we describe a parametric theory as in *Remark* 3.1. In that case, the parameter $t$ is treated as a label. This label describes a foliation of spacetime into leaves generated by a different embedding of the Cauchy hypersurface $\Sigma$ at each value of $t$. In our case, the *s.q.* Kähler structure depends on those objects, and therefore it will acquire a parametric dependence $(\Omega(t), \mathcal{G}(t), \mathcal{J}(t))_{\mathscr{P}}$. In last instance, as depicted in Figure 5.1, this dependence is inherited from the dependence of the Gaussian measure that defines the space of pure states that we geometrized in the previous sections. .

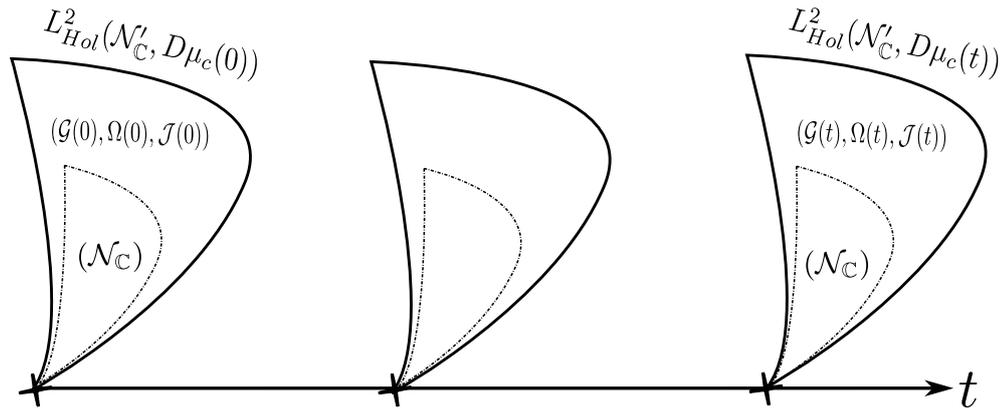

Figure 5.1: Dependence of the space of pure states $L^2_{Hol}(\mathcal{N}'_{\mathbb{C}}, D\mu_c)$ on the time parameter $t$ inherited by the Kähler *s.q.* structure $(\Omega(t), \mathcal{G}(t), \mathcal{J}(t))_{\mathscr{P}}$ over $\mathscr{P} = (\mathcal{N}_{\mathbb{C}})$ obtained from the geometrization procedure.

This is because the objects defining the geometric structure at each leaf have parametric dependence on $t$. These are the Riemannian metric $h^{ij}(t)$, its associated momenta $\pi_{ij}(t)$, and the lapse function and shift vector field $N(t), N^i(t)$. The particular dependence of $(\Omega(t), \mathcal{G}(t), \mathcal{J}(t))_{\mathscr{P}}$ on the geometry of $\Sigma$ is irrelevant for our discussion in this section, and we skip it to chapter 7. We also devote chapter 6 to understanding the source and nature of the dependence going beyond the parametric theory. Using Geometrodynamics, we study the structure of admissible $h^{ij}$ and $\pi_{ij}$, providing the Hamiltonian structure of General Relativity. This is, in the hybrid quantum-classical case, the study of the general coupling to gravity with *backreaction*.

### 5.4.1 Covariance of the time derivative

The first aspect to take into account is that, because the full Kähler structure $(\Omega(t), \mathcal{G}(t), \mathcal{J}(t))_{\mathscr{P}}$ depends on time, as well as the quantization procedure itself $\mathcal{Q}(t)$, we cannot assume explicit time independence for every set of s.q. coordinates, and as such, we must define a covariant derivative for the time parameter $\nabla_t$ that preserves the structure. This kind of connection is similar in nature to the one obtained in chapter 3 for CSFT.

Recall that we know from geometric quantization that the points of the manifold of pure states $\Psi$ represent different sections $\Psi_s$ of a Hermitian line bundle $\pi_{\mathcal{M}_C,B}: B \to \mathcal{M}_C$, associated with a $U(1)$-principal bundle on the manifold of classical fields $\phi^{\mathbf{x}} \in \mathcal{M}_C$.

In this scenario, the manifold of classical fields must be enlarged to accommodate time in the base of the bundle. In concordance with the picture in which $\mathcal{M}_C$ accounts for the Cauchy data of a theory in a globally hyperbolic spacetime, the bundle incorporating time in this theory is $\pi_{\mathcal{M}_C,B}: B_t \to \mathcal{M}_C \times \mathbb{R}$.

Consider thus the connection defined on the product $\mathcal{M}_C \times \mathbb{R}$ by the addition of the pullbacks of the connection one-forms with respect to the canonical projections. We will use $\nabla$ to represent the covariant derivative with respect to this new connection. The covariant derivative in the tangent directions of $\mathcal{M}_C$ has been discussed in [chapter 3](#) and [chapter 4](#) in regards to geometric quantization. In this section, we must study the covariant derivative in the tangent directions to the time parameter that we denote $\nabla_t$.

The covariant derivative $\nabla_t$ will be described by a connection $\Gamma^\phi_\sigma$, and we choose it such that

$$\nabla_t \Omega_{\sigma,\phi} = \partial_t \Omega_{\sigma,\phi} - \Omega_{\sigma,\gamma} \Gamma^\gamma_\phi - \Gamma^\gamma_\sigma \Omega_{\gamma,\phi} = 0,$$
$$\nabla_t \mathcal{G}_{\sigma,\phi} = \partial_t \mathcal{G}_{\sigma,\phi} - \mathcal{G}_{\sigma,\gamma} \Gamma^\gamma_\phi - \Gamma^\gamma_\sigma \mathcal{G}_{\gamma,\phi} = 0,$$
$$\nabla_t \mathcal{J}^\phi_\sigma = \partial_t \mathcal{J}^\phi_\sigma - \mathcal{J}^\phi_\gamma \Gamma^\gamma_\sigma + \Gamma^\sigma_\gamma \mathcal{J}^\phi_\gamma = 0. \tag{5.38}$$

Actually, because of the Kähler compatibility condition, any two of the conditions will imply the third. Using the discussion in [section 5.3.3](#), we can write the equivalent conditions

$$\nabla_t \mathcal{J} = \partial_t \mathcal{J} + [\Gamma, \mathcal{J}] = 0$$
$$\nabla_t O = \partial_t O + [\Gamma, O] \text{ such that } (\nabla_t O)^\dagger = \nabla_t(O^\dagger) \ \forall \ O \in \hat{W} \tag{5.39}$$

Notice that in that discussion, the $\dagger$ operator in different coordinates will introduce nontrivial dependence on the time parameter $t$.

These conditions will not be enough to select a unique $\Gamma$ to respect the dynamics. This selection requires careful study that we will develop in [section 5.5](#).

Another aspect that is affected by the time derivative is the quantization procedure itself. Recall from geometric quantization that we describe general sections as functions. This is because we factor out a special section $\Psi_r$ such that $\Psi_s = \Psi_r \Psi$, and the nontrivial structure of the bundle is rephrased in terms of the quantization procedure $\mathcal{Q}$. In that construction, we use the twofold nature of line bundles. On one hand, they are vector bundles and, as such, they are represented as vectors; but, on the other hand, they are also principal bundles using the multiplication of the fiber[2] $\mathbb{C}\backslash\{0\}$.

---

[2]Excluding the zero section, which is irrelevant in our construction.

When we treat the line bundle as a real manifold, the twofold nature of the line bundle is lost. In that context, we substitute the fiber of the complex line bundle $B_t$ with a two-dimensional real space, keeping only the vector bundle structure. We denote this bundle $\pi_{\mathcal{M}_C, \tilde{B}_t} : \tilde{B}_t \to \mathcal{M}_C \times \mathbb{R}$. This has implications in the decomposition of the section $\Psi_s = \Psi_r \Psi$. We choose to describe functions $\Psi$ with vectors $\begin{pmatrix} \Phi^\phi \\ \Pi^r \end{pmatrix}$. As is implied by (3.36) and the discussion about vacua in chapter 4, we must introduce a vacuum section $\Psi_0$, which is regarded as the multiplication by a phase with respect to the reference section $\Psi_r$. This should be recovered from a principal bundle structure. Then $\Psi_0$ is part of the associated $O(2)$-principal bundle. As a consistency criterion, we ask for the vacuum section to be parallel transported by this connection:

$$\nabla_t \tilde{\Psi}_0 = \partial_t \tilde{\Psi}_0 + [\Gamma, \tilde{\Psi}_0] = 0. \tag{5.40}$$

Sections $\Psi_s$ without dynamical evolution, which we will discuss later on, must also be parallel transported and therefore

$$\nabla_t \begin{pmatrix} \Phi^\phi \\ \Pi^\phi \end{pmatrix} = \frac{\partial}{\partial t} \begin{pmatrix} \Phi^\phi \\ \Pi^\phi \end{pmatrix} + \Gamma^\phi_\sigma \begin{pmatrix} \Phi^\sigma \\ \Pi^\sigma \end{pmatrix} = 0 \tag{5.41}$$

Notice that states in this representation are sections on a bundle $\pi : \tilde{B}_t \to \mathbb{R}$. Here the fibers are represented as functions of $(\mathcal{N}) \times (\mathcal{N})$, even though this equation, in general, will require a larger space to be solved, as we will discuss below.

### 5.4.2 Dynamics: The Modified Schrödinger equation

With the discussion we presented above, we are in a position to postulate the dynamics from geometrical principles. Our guideline is to modify the evolution, i.e., the Schrödinger equation, to respect the Kähler structure at the *second quantized* (*s.q.*) level $(\mathcal{G}, \Omega, \mathcal{J})_{\mathscr{P}}$.

Since we want it to preserve the symplectic structure, we opt for a Hamiltonian flow generated by a Hermitian operator $\hat{H}$ to postulate evolution. Then, for any function $F(t, \Psi, \bar{\Psi})$, its evolution is generated by some $f_{\hat{H}}$:

$$\frac{d}{dt} F = \partial_t F + \{F, f_{\hat{H}}\}_{\mathscr{P}}. \tag{5.42}$$

where $\hat{H}$ is a self-adjoint operator. The form of $f_{\hat{H}}$ is (5.25). With the Poisson Bracket $\{\cdot, \cdot\}_{\mathscr{P}}$, we can readily write down the modified Schrödinger equation in any set of real coordinates:

$$\nabla_t \begin{pmatrix} \Phi^\phi \\ \Pi^\phi \end{pmatrix} = -\mathcal{J} \hat{H}^\phi_\sigma \begin{pmatrix} \Phi^\sigma \\ \Pi^\sigma \end{pmatrix}. \tag{5.43}$$

In static and stationary spacetimes, we may find a time-independent expression of the complex structure $J_{\mathcal{M}_F}$ of (3.21). In this limiting case, the connection term will be null. In this case, $\nabla_t$ is simply $\partial_t$, and this equation becomes the regular Schrödinger equation.

It is clear that this equation does not have a solution on $(\mathcal{N}) \times (\mathcal{N})$. In quantum theory, it is desirable to solve this equation over a Hilbert space. Unfortunately, for quantum field theory, this is not possible either; to solve this equation, we must look for solutions on $(\mathcal{N})' \times (\mathcal{N})'$. This is our way out to deal with different instantaneous Hilbert spaces described in [AA15, Alo&24a]. A similar equation is postulated in [KM23] to deal with this problem.

## 5.5 Other representations: The connection ambiguity

The choice of the connection for the time derivative $\nabla_t$ is subject to a great deal of ambiguity. Let us showcase this with a simple example that we will explore in detail in this section. Since we are interested in the relation of the quantization mapping $\mathcal{Q}$ with this covariant time derivative $\nabla_t$ we will focus on its action on operators. Let $\mathcal{O}_\sigma^\phi$ be an operator. In order to preserve the symplectic structure, as it is discussed below, the simplest choice is to require

$$\nabla_t \mathcal{O} = \frac{1}{2} \left[ \frac{\partial \mathcal{O}}{\partial t} + \left( \frac{\partial \mathcal{O}^\dagger}{\partial t} \right)^\dagger \right]. \tag{5.44}$$

However, the explicit dependence of $\dagger$ on time is different in each representation, as we discussed in section 5.3.3. This changes (5.39) in nontrivial ways. In order to understand this ambiguity, in this section, we will study this choice of connection in different representations.

In the previous sections all we did was to geometrize the phase space $L^2_{Hol}(\mathcal{N}'_{\mathbb{C}}, D\mu_c)$, which is the quantum phase space of the holomorphic representation. However, in chapter 4 we identified four representations of the same quantum field theory. In principle, the geometrization of the previous sections should be repeated in every representation. Fortunately, thanks to the discussion in section 4.3, summarized in Figure 4.1, we only need to describe the integral transforms appearing in that diagram to complete the geometrized picture. Thus, we will start our discussion on the section describing the Fourier $\tilde{\mathcal{F}}$ and Segal-Bargmann $\tilde{\mathcal{B}}, \overline{\mathcal{B}}$ transforms using the techniques of integral kernels extensively studied throughout this chapter.

As a result of this study in section 5.5.4 we present the central result of this section. We will select a connection $\Gamma$ that respects the quantization mapping $\mathcal{Q}$ from physical criteria. This is, we will make this connection compatible with a coupling with gravity. However, the particular source of this condition will be explored in chapter 6.

### 5.5.1  The Fourier transform: Antiholomorphic quantization in the geometrized setting

The tool at hand to express the antiholomorphic picture is the Fourier transform $\tilde{\mathcal{F}}$ (4.37). To simplify this discussion, we will make the rather counterintuitive design choice of representing the space of antiholomorphic functions $(\mathcal{N}_{\mathbb{C}})^* \subset L^2_{\overline{Hol}}(\mathcal{N}_{\mathbb{C}}, D\nu_c)$ with holomorphic coordinates $(\mathcal{N}_{\mathbb{C}})$. To do this, we rephrase the appropriate sign conventions in the *s.q.* quantized structure.

$$\bar{h}_{\mathscr{P}} = \frac{\bar{\mathcal{G}}_{\mathscr{P}} + i\bar{\Omega}_{\mathscr{P}}}{2} = \mathcal{D}_{\bar{\sigma}\phi} d\bar{\Psi}^{\bar{\sigma}} \otimes d\Psi^{\phi}. \tag{5.45}$$

With the opposite sign convention to (5.45) and

$$\bar{\Upsilon}^{\bar{\sigma}}\mathcal{D}_{\bar{\sigma},\phi}\Psi^{\phi} = \iint D\beta(\sigma) D\beta(\phi)\overline{\Upsilon(\sigma)}e^{-\sigma^x D_{xy}\bar{\phi}^y}\Psi(\phi) = \int D\nu_c(\phi)\overline{\Upsilon(\phi)}\Psi(\phi). \tag{5.46}$$

We denote the real set of coordinates $\begin{pmatrix} \hat{\Phi}^{\phi} \\ \hat{\Pi}^{\phi} \end{pmatrix}$ such that $\Psi^{\phi} = \frac{1}{\sqrt{2}}(\hat{\Phi}^{\phi} + i\hat{\Pi}^{\phi})$. As anticipated, we will not discuss this representation directly, but rather we will use the Fourier transform $\tilde{\mathcal{F}}$ of Figure 4.1 as a white noise integral kernel such that $\begin{pmatrix} \hat{\Phi}^{\sigma} \\ \hat{\Pi}^{\sigma} \end{pmatrix} = \tilde{\mathcal{F}}^{\phi}_{\sigma}\begin{pmatrix} \tilde{\Phi}^{\sigma} \\ \tilde{\Pi}^{\sigma} \end{pmatrix}$. Thus, we express the Fourier transform with an integral kernel as

$$\tilde{\mathcal{F}}^{\phi}_{\sigma}\begin{pmatrix} \tilde{\Phi}^{\sigma} \\ \tilde{\Pi}^{\sigma} \end{pmatrix} = \int D\beta(\sigma) \exp\left[\phi^x\begin{pmatrix} AD^{-1} & -D^{-1} \\ D^{-1} & AD^{-1} \end{pmatrix}_{xy}\bar{\sigma}^y\right]\begin{pmatrix} \tilde{\Phi}(\sigma) \\ \tilde{\Pi}(\sigma) \end{pmatrix}. \tag{5.47}$$

We can also express the inverse as

$$(\tilde{\mathcal{F}}^{\phi\bar{\sigma}})^{-1} = \exp\left[-\phi^x\begin{pmatrix} KA & K \\ -K & KA \end{pmatrix}_{xy}\bar{\sigma}^y\right] =$$
$$e^{-\phi^x(KA)_{xy}\bar{\sigma}^x}\left[\cos\left(\phi^x K_{xy}\bar{\sigma}^y\right) + \epsilon \sin\left(\phi^x K_{xy}\bar{\sigma}^y\right)\right]. \tag{5.48}$$

Notice that the choice of holomorphic coordinates made at the beginning of this section pays off now, making this transformation an internal operator to the holomorphic functions, even if the analytic properties of $(\mathcal{N}_{\mathbb{C}})$ may be spoiled. Also, we can rewrite these kernels to use the composition table Table 2.1 of section 2.7.1. This allows for efficient calculus that will be of great importance in section 5.5. As an example, following these rules, it is easy to see that $\Delta^{\phi}_{\sigma} = (\tilde{\mathcal{F}}^t\mathcal{D}\tilde{\mathcal{F}})^{\phi}_{\sigma}$.

The algebra of observables is obtained by Fock quantization from (5.35). Let $\hat{\mathfrak{D}}(\lambda_x, \eta_x) = \tilde{\mathcal{F}}\widehat{\mathfrak{D}}(\lambda_x, \eta_x)\tilde{\mathcal{F}}^{-1}$ and $\hat{\mathfrak{D}}(\lambda_x, \eta_x)^{\dagger_{\bar{H}}} = \tilde{\mathcal{F}}\widehat{\mathfrak{D}}(\lambda_x, \eta_x)^{\dagger_H}\tilde{\mathcal{F}}^{-1}$. Using the tools of

section 2.7.1, we get

$$\hat{\mathfrak{D}}(\lambda_x, \eta_x) = -\frac{\eta_x D^{xy}}{\sqrt{2}} \partial_{\phi^y} - \frac{\lambda_x}{\sqrt{2}} \begin{pmatrix} A & \delta \\ -\delta & A \end{pmatrix}^{xy} \partial_{\phi^y},$$

$$\hat{\mathfrak{D}}(\lambda_x, \eta_x)^{\dagger_{\bar{H}}} = \frac{\eta_x \phi^x}{\sqrt{2}} + \frac{(\lambda D^{-1})x}{\sqrt{2}} \begin{pmatrix} A^t & -\delta \\ \delta & A^t \end{pmatrix}^x_y \phi^y. \tag{5.49}$$

This is, if we switch from the holomorphic to the antiholomorphic picture with the Fourier transform, we see how the roles of $\lambda$ and $\eta$ exchange. The duality is also adapted to the covariance of each picture, as described in Section 4.2.3. This is a mere reflection of the preservation of quantization by $\tilde{\mathcal{F}}$ shown in the diagram of Figure 4.1 and translates to

$$\overline{\mathcal{Q}}\big[\exp(i\lambda_x \varphi^x)\big]^{\bar{\phi}\sigma} = e^{\bar{\phi}^x \delta_{xy}\sigma^y - \frac{\lambda_x \Delta^{xy}\lambda_y}{2}} - \epsilon^{\frac{\lambda_x(\bar{\phi}^x + (D^{-1})^x_y \sigma^y)}{\sqrt{2}}} e^{-\frac{\lambda_x A^x_y(\bar{\phi}^y - (D^{-1})^y_z \sigma^z)}{\sqrt{2}}}$$

$$\overline{\mathcal{Q}}\big[\exp(i\eta_x \pi^x)\big]^{\bar{\phi}\sigma} = e^{\bar{\phi}^x \delta_{xy}\sigma^y + \frac{\eta_x D^{xy}\eta_y}{2}} - \epsilon^{\frac{\eta_x(D_{xy}\bar{\phi}^y - \sigma^x)}{\sqrt{2}}} \tag{5.50}$$

## 5.5.2 The Segal-Bargmann transform: Schrödinger and Field-momentum in the geometrized setting

In order to have an efficient description of the Schrödinger and Field-momentum representations in the geometrized setting, we need to enlarge the machinery of integral kernels in section 2.7.1 from holomorphic Hida test functions ($\mathcal{N}_{\mathbb{C}}$) to Hida test functions of real domain ($\mathcal{N}$). This process can be done without further conceptual complications. However, the reason why the reproducing kernel of *Example* 2.10 is so simple to write down relies on the simplicity of the holomorphic chaos decomposition (2.21), which is independent of any particular Gaussian measure used as reference. In the real case, this simplicity is lost because in the real chaos decomposition (2.41), we should use Wick-ordered monomials : $\varphi^n$ : $|^{\vec{x_n}}_\delta$ to express the white noise reproducing kernel:

$$\mathcal{RK}(\varphi^{\mathbf{x}}, \pi^{\mathbf{x}}) = \sum_{n=0}^{\infty} \frac{: \varphi^n : |^{\vec{x}_n}_\delta \delta^n_{\vec{x}_n, \vec{y}_n} : \pi^n : |^{\vec{y}_n}_\delta}{n!},$$

$$\int_{\mathcal{N}'} D\beta(\pi^{\mathbf{x}}) \mathcal{RK}(\varphi^{\mathbf{x}}, \pi^{\mathbf{x}}) \Psi(\pi^{\mathbf{x}}) = \Psi(\varphi^{\mathbf{x}}). \tag{5.51}$$

This necessity spoils the efficiency that motivated the use of integral kernels in the first place. In order to avoid cumbersome expressions, it is enough to study the white noise integral kernel representations of the Segal-Bargmann transforms $\tilde{\mathcal{B}}, \tilde{\tilde{\mathcal{B}}}$.

**Schrödinger representation**

Starting with $\tilde{\mathcal{B}}$, we can write the integral kernel of the expression (4.30) as

$$\left(\tilde{\mathcal{B}}^{-1}\right)^{\varphi\bar{\phi}} = \exp\left(\sqrt{2}\varphi^x \delta_{xy}\bar{\phi}^y - \frac{\bar{\phi}^x \Delta_{xy}\bar{\phi}^y}{2}\right) =: \exp\left(\sqrt{2}\varphi^x \delta_{xy}\bar{\phi}^y\right) :|_{\frac{\Delta}{2}}, \tag{5.52}$$

where we used the Wick order operator of *Example* 2.5. As we see, real coordinates are adapted to the covariance of $\Delta^s_{\varphi\pi}$ provided by

$$\tilde{\Upsilon}^\varphi \Delta^s_{\varphi,\pi} \Psi^\pi = \int_{\mathcal{N}'} D\mu(\varphi^{\mathbf{x}}) \overline{\Upsilon(\varphi^{\mathbf{x}})} \Psi(\varphi^{\mathbf{x}}) \tag{5.53}$$

where $\mu$ is the real measure adapted to the Schrödinger picture (4.19). Taking the inverse $\left(\mathcal{K}^s\right)^{\varphi,\pi}$, we write

$$\left(\mathcal{K}^s\right)^{\varphi,\pi} \tilde{\mathcal{B}}^\phi_\pi = \exp\left(\sqrt{2}\phi^x K_{xy}\varphi^y - \frac{\phi^x K_{xy}\phi^y}{2}\right) =: \exp\left(\sqrt{2}\phi^x K_{xy}\varphi^y\right) :|_{\frac{\Delta}{2}}, \tag{5.54}$$

Notice that in this context we can write the † operator in this representation

$$\mathcal{O}^{\dagger s} = \mathcal{K}^s \mathcal{O}^\dagger_{\Delta^s}. \tag{5.55}$$

Using the expressions above, we see that $\left(\mathcal{K}^s\right)^{\varphi,\pi} \tilde{\mathcal{B}}^\phi_\pi \Delta^\sigma_\phi = \left[\left(\tilde{\mathcal{B}}^{-1}\right)^{\varphi\bar{\sigma}}\right]^*$, which implies $\tilde{\mathcal{B}}^\dagger = \tilde{\mathcal{B}}^{-1}$. Notice that $\tilde{\mathcal{B}}$ is not internal to any space, similarly to the discussion about the Fourier transform at the end of section 5.3.3.

Contrary to the holomorphic case, (5.53) and its inverse do not admit simple expressions in terms of white noise reproducing kernels. However, $\tilde{\mathcal{B}}$ and its inverse admit a closed form adapted to the white noise bilinear $\beta_{\varphi,\pi}$ that we derive below. Notice that we can write

$$\left(\tilde{\mathcal{B}}^{-1}\right)^{\varphi\bar{\phi}} = \exp\left[\bar{\phi}^x\left(\delta - \frac{\Delta}{2}\right)_{xy}\bar{\phi}^y\right] : \exp\left(\sqrt{2}\varphi^x \delta_{xy}\bar{\phi}^y\right) :|_\delta,$$

The inverse transform would be provided by

$$\left(\tilde{\mathcal{B}}\right)^{\phi\varphi} = \exp\left[\bar{\phi}^x\left(\frac{\Delta}{2} - \delta\right)_{xy}\bar{\phi}^y\right] : \exp\left(\frac{\phi^x \delta_{xy}\varphi^y}{\sqrt{2}}\right) :|_\delta,$$

But we must eliminate the $\bar{\phi}^{\mathbf{x}}$ terms with Skorokhod-Malliavin duality; then we write

$$\exp\left[\bar{\phi}^x\left(\frac{\Delta}{2} - \delta\right)_{xy}\bar{\phi}^y\right] \mapsto \exp\left[\left(\frac{\Delta}{2} - \delta\right)^{xy}\partial_{\phi^x}\partial_{\phi^y}\right].$$

This operator commutes with the Wick order operator of *Example* 2.5, thus the integral kernel of this transformation is provided by

$$\left(\tilde{\mathcal{B}}\right)^{\phi\varphi} =: \exp\left[\frac{\phi^x \delta_{xy}\varphi^y}{\sqrt{2}} + \varphi^x\left(\frac{\Delta}{4} - \frac{\delta}{2}\right)_{xy}\varphi^y\right] :|_\delta, \tag{5.56}$$

**Field-momentum representation**

The analysis is formally equal for $\tilde{\bar{\mathcal{B}}}$. In this case, we use the bilinear $\mathcal{D}^m_{\varphi\pi}$ provided by

$$\tilde{\Upsilon}^\varphi \mathcal{D}^m_{\varphi,\pi} \Psi^\pi = \int_{\mathcal{N}'} D\nu(\varphi^{\mathsf{x}}) \overline{\Upsilon(\varphi^{\mathsf{x}})} \Psi(\varphi^{\mathsf{x}}) \tag{5.57}$$

where $\nu$ is the real measure adapted to the Field-momentum picture (4.25). We denote its inverse as $\left(\mathcal{D}^{m,-1}\right)^{\varphi,\pi}$. The † operator in this representation is

$$\mathcal{O}^{\dagger m} = \mathcal{D}^{m,-1} \mathcal{O}^\dagger_{\mathcal{D}^m}. \tag{5.58}$$

We write the integral kernels of (4.33) with

$$\begin{aligned}
\left(\tilde{\bar{\mathcal{B}}}^{-1}\right)^{\pi\bar{\phi}} &= \exp\left(\sqrt{2}\pi^x \delta_{xy} \bar{\phi}^y + \frac{\bar{\phi}^x D_{xy} \bar{\phi}^y}{2}\right) =: \exp\left(\sqrt{2}\pi^x \delta_{xy} \bar{\phi}^y\right) :|_{-\frac{D}{2}}, \\
\left(\mathcal{D}^{-1,m}\right)^{\pi,\varphi} \tilde{\bar{\mathcal{B}}}^\phi_\varphi &= \exp\left(-\sqrt{2}\phi^x D^{-1}_{xy} \pi^y + \frac{\phi^x D^{-1}_{xy} \phi^y}{2}\right) =: \exp\left(-\sqrt{2}\phi^x D^{-1}_{xy} \pi^y\right) :|_{-\frac{D}{2}},
\end{aligned} \tag{5.59}$$

Similarly, we obtain the white noise integral kernels

$$\begin{aligned}
\left(\tilde{\bar{\mathcal{B}}}^{-1}\right)^{\varphi\bar{\phi}} &= \exp\left[\bar{\phi}^x \left(\delta + \frac{D}{2}\right)_{xy} \bar{\phi}^y\right] : \exp\left(\sqrt{2}\varphi^x \delta_{xy} \bar{\phi}^y\right) :|_\delta, \\
\left(\tilde{\bar{\mathcal{B}}}\right)^{\phi\varphi} &=: \exp\left[\frac{\phi^x \delta_{xy} \varphi^y}{\sqrt{2}} - \varphi^x \left(\frac{D}{4} + \frac{\delta}{2}\right)_{xy} \varphi^y\right] :|_\delta.
\end{aligned} \tag{5.60}$$

### 5.5.3 Ambiguity in the choice of the connection

In this section, we will study the ambiguity that (5.38) leaves in choosing a connection. We are interested in the relation of the quantization mapping $\mathcal{Q}$ with this covariant time derivative, for this reason we will focus on its action on operators. Let $\mathcal{O}^\phi_\sigma$ be an operator acting on the holomorphic representation and written in holomorphic *s.q.* coordinates. As we mentioned above, the $\dagger_H$ operator has a particular dependence on the time parameter; thus, the simplest choice is to require

$$\nabla^0_t \mathcal{O} = \frac{1}{2}\left[\frac{\partial \mathcal{O}}{\partial t} + \left(\frac{\partial \mathcal{O}^{\dagger_H}}{\partial t}\right)^{\dagger_H}\right]. \tag{5.61}$$

This choice indeed defines a connection $_0\Gamma$ that fulfills the equivalent (5.39) because, in this case, the *s.q.* complex structure $\mathcal{J}$ is simply represented by the constant matrix $\epsilon$. To see the particular form of $_0\Gamma$, we rewrite (5.29)

$$\left(O^{\dagger_H}\right)^{\bar{\sigma}}_{\bar{\phi}} = \epsilon^{\mathsf{T}} \Delta_{\bar{\phi}\gamma} (O^{\mathsf{T}})^\gamma_\tau \mathcal{K}^{\tau\bar{\sigma}} \epsilon, \tag{5.62}$$

where T indicates the transpose as a matrix. We use this notion instead of (5.28) to ensure that the time independence of the transposition operation is maintained regardless of the time dependence of $\beta_{\bar{\phi}\sigma}$. Tracing the corresponding time dependences, we see that

$$\left(\frac{d\mathcal{O}^{\dagger H}}{dt}\right)^{\dagger H} = \dot{\mathcal{O}} + [\mathcal{K}^{\phi\bar{\gamma}}\dot{\Delta}_{\bar{\gamma}\sigma}, \mathcal{O}]. \tag{5.63}$$

where $\cdot$ represents the time derivative. We use here the identity $\dot{\mathcal{K}}^{\phi\bar{\gamma}}\Delta_{\bar{\gamma}\sigma} = -\mathcal{K}^{\phi\bar{\gamma}}\dot{\Delta}_{\bar{\gamma}\sigma}$, which is a consequence of one being the inverse of the other, and thus its contraction is $\delta^\phi_\sigma$, which is time independent. As a consequence, and using the explicit expressions of the white noise bilinears in (5.16) and below, we derive

$$_0\Gamma^\phi_\sigma = \frac{1}{2}\mathcal{K}^{\phi\bar{\gamma}}\dot{\Delta}_{\bar{\gamma}\sigma} = \delta^\phi_\sigma \frac{1}{2}\phi^x K_{xy}\dot{\Delta}^{yz}\partial_{\sigma^z}, \tag{5.64}$$

To show that this object indeed behaves as a connection, consider a change of s.q. coordinates $\Psi^\phi \to \mathcal{C}^\phi_\sigma \Psi^\sigma$ such as (5.22). Then, using (5.63), the covariant derivative gives $\mathcal{C}\nabla^0_t(\mathcal{O})\mathcal{C}^{-1} = \nabla^0_t(\mathcal{C}\mathcal{O}\mathcal{C}^{-1}) + [\mathcal{C}\dot{\mathcal{C}}^{-1}, \mathcal{C}\mathcal{O}\mathcal{C}^{-1}]$. Thus, it fulfills the inhomogeneous transformation law of a connection.

$$_0\Gamma \to \mathcal{C}\,_0\Gamma\mathcal{C}^{-1} + \mathcal{C}\frac{d\mathcal{C}^{-1}}{dt}.$$

This example shows a connection that fulfills

$$(\nabla^0_t\mathcal{O})^{\dagger H} = \nabla^0_t(\mathcal{O}^{\dagger H}) \text{ and } \nabla^0_t\mathcal{J} = 0.$$

Below, we extend this way of deriving the connection using each of the unitary transformations represented in the diagram of Figure 4.1. In order to simplify the discussion, the technical details will be dealt with in section A.2. Using these connections, we will study the choice of a unique connection from physical principles. Our guideline will be that the connection of the modified Schrödinger equation must be blind to the particular representation of QFT chosen to derive it; that is, in the last instance, arbitrary. For this reason, we want to respect the diagram of Figure 4.1. This is why we look for a connection equivariant with respect to the integral transformations described there.

As a final remark, notice that any connection $\Gamma$ that fulfills (5.39), in holomorphic *s.q.* coordinates, can be written as

$$(\nabla_t\mathcal{O})^{\dagger H} = \left(\partial_t\mathcal{O} + \left[\Gamma, \mathcal{O}\right]\right)^{\dagger H} = \partial_t\mathcal{O}^{\dagger H} - \left[\Gamma^{\dagger H} - \mathcal{K}^{\phi\bar{\gamma}}\dot{\Delta}_{\bar{\gamma}\sigma}, \mathcal{O}^{\dagger H}\right].$$

Thus

$$\Gamma = -\Gamma^{\dagger H} + \phi^x K_{xy}\dot{\Delta}^{yz}\partial_{\phi^z} = -\Gamma^{\dagger H} - \mathfrak{D}^{\dagger,x}\dot{K}_{xy}\mathfrak{D}^y. \tag{5.65}$$

As a consequence of this result, every connection considered in this section will be written in holomorphic *s.q.* coordinates

$$\Gamma = \frac{1}{2}\phi^x K_{xy}\dot{\Delta}^{yz}\partial_{\phi^z} + \gamma, \tag{5.66}$$

where $\gamma^\dagger = -\gamma$.

A particular limiting case, common to this and all the connections exposed below, is the case when the complex structure $J_{\mathcal{M}_F}$ of (3.21) is time independent. In that case, we obtain $\Gamma = 0$ and the modified Schrödinger equation (5.43) becomes the usual one.

**Antiholomorphic connection**

We could have chosen to derive the connection $_0\Gamma$ in the antiholomorphic representation. For this reason, we use the Fourier transform $\tilde{\mathcal{F}}$ to describe it. Let $\mathcal{O}_\sigma^\phi$ be an operator and $\hat{\mathcal{O}} = \mathcal{F}\mathcal{O}\mathcal{F}^{-1}$; thus, we postulate that the covariant derivative acts on this operator in a way such that, in *s.q.* holomorphic coordinates, it is written as

$$\tilde{\nabla}_t \mathcal{O} = \frac{1}{2}\tilde{\mathcal{F}}^{-1}\left[\frac{\partial\hat{\mathcal{O}}}{\partial t} + \left(\frac{\partial\hat{\mathcal{O}}^{\dagger_{\bar{H}}}}{\partial t}\right)^{\dagger_{\bar{H}}}\right]\tilde{\mathcal{F}}. \tag{5.67}$$

Repeating the procedure shown at the beginning of the section, we can write this expression with a connection such that

$$\tilde{\nabla}_t \mathcal{O} = \partial_t \mathcal{O} + [\tilde{\Gamma}, \mathcal{O}].$$

We skip the technical details to section A.2, where we write the connection in holomorphic *s.q.* coordinates

$$\tilde{\Gamma}_\sigma^\phi = \frac{1}{2}\left[\left(\tilde{\mathcal{F}}^{-1}\dot{\tilde{\mathcal{F}}}\right)_\sigma^\phi + (\mathcal{D}\tilde{\mathcal{F}})_{\sigma\bar{\gamma}}^{-1}\frac{d(\mathcal{D}\tilde{\mathcal{F}})^{\bar{\gamma}\phi}}{dt}\right]. \tag{5.68}$$

In section A.2, we compute the derivative of creation and annihilation operators and derive the explicit expression of this connection:

$$\tilde{\Gamma} = \frac{1}{2}\phi^x K_{xy}\dot{\Delta}^{yz}\partial_{\phi^z} + \frac{1}{2}\phi^x K_{xy}(\dot{A} - \epsilon\dot{\delta})^{yv}(AD^{-1} + \epsilon D^{-1})_v^u\partial_{\phi^u}$$
$$- \frac{1}{2}\phi^x[KD^{-1}(A^t - \epsilon\delta)]_{xy}(\dot{A}^t + \epsilon\dot{\delta})^{yv}\partial_{\phi^v}. \tag{5.69}$$

Notice, however, that this connection is of the form $\Gamma = a^{\dagger,x}\mathcal{K}_{xy}a^y$ for some kernel $\mathcal{K}$. This means it does not possess quadratic terms with repeated creation or annihilation operators. For this type of connection, it is impossible to obtain mixing between creation and annihilation operators with the parallel transport equation. In other terms, this prescription of the connection does not allow for Bogoliubov transformations to be generated by parallel transport, as is the case in [KM23]. For instance, particle production in an expanding universe is not recovered with this connection.

**Schrödinger connection**

Now we will use the connection derived in the Schrödinger representation. As in the previous case described, we describe it with the Segal-Bargmann transform $\mathcal{B}$. Let us denote $\mathcal{Q} = \tilde{\mathcal{B}}^{-1}\mathcal{O}\tilde{\mathcal{B}}$; thus, we postulate that the covariant derivative in Holomorphic *s.q.* coordinates is given by

$$\nabla_t \mathcal{O} = \frac{1}{2}\tilde{\mathcal{B}}\left[\frac{\partial \mathcal{Q}}{\partial t} + \left(\frac{\partial \mathcal{Q}^{\dagger_s}}{\partial t}\right)^{\dagger_s}\right]\tilde{\mathcal{B}}^{-1}. \tag{5.70}$$

Similarly to the previous cases, we get

$$\underset{\smile}{\Gamma}{}^{\phi}_{\sigma} = \frac{1}{2}\left[(\tilde{\mathcal{B}}\dot{\tilde{\mathcal{B}}}^{-1})^{\phi}_{\sigma} + (\tilde{\mathcal{B}}\mathcal{K}^s)_{\sigma\varphi}\frac{d(\varDelta^s\tilde{\mathcal{B}}^{-1})^{\varphi\phi}}{dt}\right]. \tag{5.71}$$

In [section A.2], we derive its explicit expression:

$$\underset{\smile}{\Gamma} = \frac{1}{2}\phi^x K_{xy}\dot{\varDelta}^{yz}\partial_{\phi^z} - \frac{1}{4}\left(\dot{\varDelta}^{xy}\partial_{\phi^x}\partial_{\phi^y} + \dot{K}_{xy}\phi^x\phi^y\right). \tag{5.72}$$

The anti-self-adjoint term in this case possesses two creation and two annihilation operators; thus, when computing the parallel transported creation and annihilation operators, Bogoliubov transformations will occur in this prescription. See the discussion in [chapter 7] for an explanation of these transformations.

**Field-Momentum connection**

The remaining connection is the one defined in the Field-momentum representation. To achieve this, the transform that we must use is the diagonal line in [Figure 4.1] given by $\overline{\tilde{\mathcal{B}}}^{-1}\tilde{\mathcal{F}} = \mathcal{F}\tilde{\mathcal{B}}^{-1}$. As before, we denote $\underset{\smile}{\mathcal{Q}} = \overline{\tilde{\mathcal{B}}}^{-1}\tilde{\mathcal{F}}\mathcal{O}\tilde{\mathcal{F}}^{-1}\overline{\tilde{\mathcal{B}}}$. This defines a connection through

$$\nabla_t \mathcal{O} = \frac{1}{2}\tilde{\mathcal{F}}^{-1}\overline{\tilde{\mathcal{B}}}\left[\frac{\partial \mathcal{Q}}{\partial t} + \left(\frac{\partial \mathcal{Q}^{\dagger_m}}{\partial t}\right)^{\dagger_m}\right]\overline{\tilde{\mathcal{B}}}^{-1}\tilde{\mathcal{F}}. \tag{5.73}$$

Similarly to the previous cases, we get

$$\underset{\smile}{\Gamma}{}^{\phi}_{\sigma} = \frac{1}{4}\left[\mathcal{K}^{\phi\gamma}\dot{\varDelta}_{\gamma\sigma} + \left[\tilde{\mathcal{F}}^{-1}\overline{\tilde{\mathcal{B}}}\frac{d\,\overline{\tilde{\mathcal{B}}}^{-1}\tilde{\mathcal{F}}}{dt}\right]^{\phi}_{\sigma} + (\tilde{\mathcal{F}}^{-1}\overline{\tilde{\mathcal{B}}}\mathcal{D}^{m,-1})_{\sigma\varphi}\frac{d(\mathcal{D}^m\overline{\tilde{\mathcal{B}}}^{-1}\tilde{\mathcal{F}})^{\varphi\phi}}{dt}\right], \tag{5.74}$$

In this case, in [section A.2], we decompose

$$\underset{\smile}{\Gamma} = \tilde{\Gamma} +$$
$$\frac{1}{4}\left[(A - \epsilon\delta)^{xu}(\dot{D}^{-1})_{uv}(A^t - \epsilon\delta)^{vy}(K\phi)_x(K\phi)_y - (A^t + \epsilon\delta)^{xu}(\dot{D}^{-1})_{uv}(A + \epsilon\delta)^{vy}\partial_{\phi^x}\partial_{\phi^y}\right]$$

We identify in this connection features of the two previous connections $\tilde{\Gamma}$ and $\underset{\smile}{\Gamma}$. This was to be expected as it is based on the composition of two transformations of the same type.

**Connections over the field and field-momentum operators**

The connections discussed above are cumbersome and difficult to treat in their current form. However, acting on operators $\hat{\varphi}^{\mathbf{x}}$ and $\hat{\pi}^{\mathbf{x}}$, they simplify enormously. Recall from [chapter 4](#) that

$$\hat{\varphi}^{\mathbf{x}} = \frac{\phi^{\mathbf{x}} + \Delta^{\mathbf{x}y}\partial_{\phi^y}}{\sqrt{2}},$$

$$\hat{\pi}_{\mathbf{x}} = -iK_{\mathbf{x}y}\frac{\phi^y - \Delta^{yz}\partial_{\phi^z}}{\sqrt{2}} - (KA)_{\mathbf{x}y}\frac{\phi^y + \Delta^{yz}\partial_{\phi^z}}{\sqrt{2}}.$$

We can simplify computations with the previous connections by using [Table 5.1](#).

| Holomorphic connection: | Antiholomorphic connection: |
|---|---|
| $\nabla_t^0 \hat{\varphi}^{\mathbf{x}} = \frac{1}{2}\dot{\Delta}^{\mathbf{x}y}K_{yz}\hat{\varphi}^z$ | $\bar{\nabla}_t \hat{\varphi}_{\mathbf{x}} = -\frac{1}{2}\dot{D}^{\mathbf{x}y}D_{yz}^{-1}\hat{\varphi}^z + \frac{(AD^{-1})_{\mathbf{x}y}}{dt}\hat{\pi}^y$ |
| $\bar{\nabla}_t^0 \hat{\pi}_{\mathbf{x}} = -\frac{1}{2}\hat{\pi}_z\dot{\Delta}^{zy}K_{y\mathbf{x}} - \frac{(KA)_{\mathbf{x}y}}{dt}\hat{\varphi}^y$ | $\bar{\nabla}_t \hat{\pi}^{\mathbf{x}} = \frac{1}{2}\dot{D}^{\mathbf{x}y}D_{yz}^{-1}\hat{\pi}^z$ |

| Schrödinger connection: | Momentum connection: |
|---|---|
| $\nabla_t \hat{\varphi}^{\mathbf{x}} = 0$ | $\nabla_t \hat{\varphi}_{\mathbf{x}} = \frac{(AD^{-1})_{\mathbf{x}y}}{dt}\hat{\pi}^y$ |
| $\nabla_t \hat{\pi}_{\mathbf{x}} = -\frac{(KA)_{\mathbf{x}y}}{dt}\hat{\varphi}^y.$ | $\nabla_t \hat{\pi}^{\mathbf{x}} = 0$ |

Table 5.1: Action of the different covariant derivatives over the operators $\hat{\varphi}_{\mathbf{x}}$ and $\hat{\pi}_{\mathbf{x}}$.

### 5.5.4 A connection motivated by Physics

The ambiguity in the choice of the connection provides incompatible evolution equations depending on this choice. This is inadmissible for the description of a physical theory. At this stage, we do not have a criterion to select one of the connections studied in the previous chapter over the others or to derive a new one. In fact, the derivation of our criterion will be motivated by the coupling with gravity and will be the subject of study in [chapter 6](#). In this chapter, we will borrow the condition of **Postulate** 6 that, in our case, imposes

$$\nabla_t \mathcal{Q}(f) = \mathcal{Q}(\partial_t f) \ \ \forall f(t) \in \mathcal{O}_{cl}. \tag{5.75}$$

Recall that $\mathcal{O}_{cl}$ is the Hilbert space of Weyl-quantizable functions [(4.46)](#), and we allow for a parametric dependence on time $t$.

We can give an interpretation to this condition by introducing the curvature tensor $R(\partial_t, X) = \nabla_t \nabla_X - \nabla_X \nabla_t$. Using this and the geometric quantization prescription [(4.1)](#), we get that

$$[\nabla_t, \mathcal{Q}(F)] = \mathcal{Q}(\partial_t F) - \mathcal{J}R(\partial_t, X_F) \tag{5.76}$$

As a result, condition (5.75) has the geometric interpretation of a curvatureless connection on the time direction, at least acting on holomorphic functions.

We argued in chapter 3 around (3.50) that it is convenient to consider the pair of classical variables $(\varphi_{\mathbf{x}}, \underline{\pi}^{\mathbf{x}})$ time-independent. As we anticipated, the covariant derivatives of the previous section act differently over the self-adjoint operators $\hat{\varphi}^{\mathbf{x}}$ and $\hat{\pi}^{\mathbf{x}}$.

Notice that none of these connections fulfill (5.75) for $(\varphi_{\mathbf{x}}, \underline{\pi}^{\mathbf{x}})$ time-independent. However, a slight modification of the momentum connection, using the canonical commutation relations, provides the desired result. Consider

$$\Gamma^{\phi}_{\sigma} = \underline{\Gamma}^{\phi}_{\sigma} - \frac{\epsilon}{2}\hat{\pi}^x\hat{\pi}^y\frac{(AD^{-1})_{xy}}{dt} \tag{5.77}$$

That is still of the form (5.66). Then, the associated connection fulfills

$$\nabla_t\hat{\pi}^{\mathbf{x}} = 0, \qquad \nabla_t\hat{\varphi}_{\mathbf{x}} = \frac{(AD^{-1})_{\mathbf{x}y}}{dt}\hat{\pi}^y - \left[\frac{\epsilon}{2}\hat{\pi}^x\hat{\pi}^y\frac{(AD^{-1})_{xy}}{dt}, \hat{\varphi}_{\mathbf{x}}\right] = 0 \tag{5.78}$$

Where we used the canonical commutation relations $[\hat{\varphi}^x, \hat{\pi}_y] = \epsilon\delta^x_y$. Using this analysis, we obtained the desired connection.

This connection is cumbersome to write down in the general case. However, in chapter 7, we show that in the cosmological case, the connection simplifies enormously and we can draw conclusions in a physically relevant scenario.

**Remark 5.1.** *(Ordering and time dependence)*

*Notice that the condition*

$$\nabla_t Q(f) = Q(\partial_t f)$$

*has been tested for the linear operators $\hat{\varphi}^{\mathbf{x}}$ and $\hat{\pi}^{\mathbf{x}}$. For higher-order operators, notice that the regulator $\mathcal{W}_{\Delta/2}$ of Remark 4.2 introduces additional dependencies on $t$ through the covariance $\Delta^{\mathbf{xy}}$. As a result, the extension to nonlinear operators is only possible for an ordering that produces non-regularized products of $\hat{\varphi}^{\mathbf{x}}$ and $\hat{\pi}^{\mathbf{x}}$. This is the case in Weyl ordering in which*

$$\mathcal{Q}(\varphi^{\mathbf{x}}\pi^{\mathbf{y}}) = \frac{1}{2}(\hat{\varphi}^{\mathbf{x}}\hat{\pi}^{\mathbf{y}} + \hat{\pi}^{\mathbf{y}}\hat{\varphi}^{\mathbf{x}})$$

# Chapter 6

# The Hybrid Coupling to Gravity

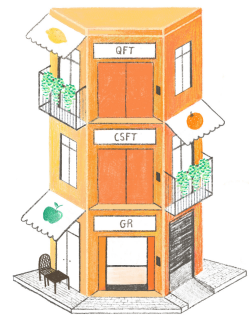

## 6.1 Introduction

So far, the gravitational content of the theory has been treated as an external parameter or *background*. This is the kind of theory that we introduced in section 3.5. Recall from that discussion that we label the geometrical content with a real parameter $s$. That parameter labels a family of embeddings of codimension 1 denoted by $\epsilon_s(\Sigma) = \Sigma_s \subset \mathcal{M}$ of the Cauchy hypersurface $\Sigma$ in $\mathcal{M}$, which is the Lorentzian manifold representing the globally hyperbolic spacetime. Some of the results of this chapter are only valid for $\Sigma$ of dimension $d = 3$, and therefore $\mathcal{M}$ of dimension $d + 1 = 4$.

The fact that the embedding is of codimension 1 simplifies the discussion enormously. Because of this fact, the embeddings are generated by a vector field provided by (3.57) in the whole spacetime $\mathcal{M}$. This vector field is completely specified on $\Sigma_s$ by the pair $(N_s, \vec{N}_s)$, denoted the lapse function and shift vector. For a fixed point in the parameter $s = s_0$, we have that $N_{s_0} \in C^\infty(\Sigma)$ and $\vec{N}_{s_0} \in \mathfrak{X}(\Sigma)$ is a vector field. As a consequence, we describe the parametric theory, as in *Remark* 3.1, specifying the curve $(\mathbf{h}_s, \pi_{h,s}) \in T^*\mathrm{Riem}(\Sigma)$ and the embedding $(N_s, N_s^i)$.

This description of a parametric theory is, in the last instance, independent of whether we are discussing CSFT in chapter 3 or QFT in chapter 4 and chapter 5. The fact that the evolution of the gravitational part is provided by a definite curve comes out as a consequence of the classical nature of gravity in our treatment.

In this chapter, we aim to turn the geometrical content into a dynamical variable without spoiling the classical nature of gravity. For this reason, we will derive the coupling of matter and gravity from geometrical arguments blind to the classical or quantum nature of gravity. The coupling of the different theories of gravity with CSFT is, of course, well known in the literature. However, the full coupling of QFT and gravity, describing the *backreaction* that the quantum degrees of freedom exert over the classical geometry of spacetime, is a novelty of our approach that we first discussed in [Alo&24a].

To be more precise, in this chapter, we will couple the quantization procedures $\mathcal{Q}$ of the previous chapters to gravity. We emphasize that, throughout this thesis, gravity will remain a classical theory described by General Relativity (GR). We will describe it in canonical formalism and investigate the implications of the coupling on the theory.

The first difficulty that we must address is that the canonical description of gravity admits at least two different types of variables to describe the configuration space of the theory.

On one hand, we may focus on the Riemannian structure $\mathbf{h}$. This description of the phase space of gravity will be called geometrodynamics. The configuration space of geometrodynamics is the superspace[1]

$$\mathcal{S}(\Sigma) = \mathrm{Riem}(\Sigma)/\mathrm{Diff}(\Sigma), \tag{6.1}$$

where $\mathrm{Riem}(\Sigma)$ is the space of Riemannian metrics defined on the manifold $\Sigma$ and $\mathrm{Diff}(\Sigma)$ is the diffeomorphism group. Our objective is to describe the theory with a non-reduced phase space in which the kinematical variables are the Riemannian 3-metric $\mathbf{h} \in \mathrm{Riem}(\Sigma)$ and its associated momenta $\pi_h \in T^*(\mathrm{Riem}(\Sigma))|_h$, defined in terms of the extrinsic curvature and seen as a tensor density of weight 1. We will sometimes identify $(\mathbf{h}, \pi_h)$ with their matrix components $(h_{ij}, \pi_h^{ij})$. The geometrodynamical phase space is given by a submanifold $\mathcal{M}_G$ that contains both variables $(\mathbf{h}, \pi_h) \in \mathcal{M}_G$, constructed as a submanifold of the cotangent bundle:

$$\mathcal{M}_G \subset T^*\mathrm{Riem}(\Sigma) \sim \mathrm{Riem}(\Sigma) \times \mathrm{Riem}'(\Sigma) \tag{6.2}$$

locally isomorphic to the Cartesian product of the Riemannian 3-metrics and symmetric (0,2)-tensor densities of weight 1 that we identify as distributions (which we have denoted by $\mathrm{Riem}'(\Sigma)$). To obtain the phase space of geometrodynamics, not every point in $T^*\mathrm{Riem}(\Sigma)$ is allowed, but only the ones that fulfill the geometrodynamical constraints that we will derive in section 6.2. For further details, we refer to [Kie12].

On the other hand, we can describe the theory using connection dynamics. In this kind of description, the connection one-form of the spacetime structure is the primary variable to describe the theory. The first example of this kind of theory is the Palatini description of gravity. However, this description in Hamiltonian form possesses the same structure as the geometrodynamical description [Rom93]. To overcome this difficulty and simplify the resolution of the constraints, Ashtekar introduced in [Ash87] a complex connection one-form that recovers gravity in its self-dual part. The simplification of the constraints to a polynomial form is, in turn, the basis of the Loop Quantum Gravity program. Therefore, the presentation of the hybrid quantum-classical coupling using these variables could be pursued to describe Wentzel-Kramers-Brillouin (WKB) limits of Loop Quantum Gravity. We refer again to [Kie12] for further details on this topic.

---

[1]This quotient is ill-defined when $\Sigma$ is a compact manifold, but it is possible to cure those problems by restricting ourselves to a smaller class of $\mathrm{Diff}(\Sigma)$. We will not treat those problems here and refer to [Kie12] and references therein for further detail.

In this chapter, we will describe General Relativity (GR) with the geometrodynamical description and leave the connection dynamical part open for future investigations. In this description, we will focus our attention on the so-called *seventh route to geometrodynamics* exposed in [HKT76]. In this approach, we interpret the geometrodynamical constraints as a requirement of the foliation independence of the whole (globally hyperbolic) spacetime $\mathcal{M}$ into spatial leaves $\Sigma_s$. This can be seen as a reflection of the $\mathcal{M}$-diffeomorphism invariance of GR, but the relation between both symmetries is not as straightforward as naively expected, this is shown in [BK72].

The *seventh route to geometrodynamics* derives GR as the only theory (in 3+1 dimensions) arising from a series of postulates, among the most important ones are the *Hamiltonian evolution rule* and the *path independence* principle. In this chapter, we will reinterpret these two principles using new tools that will suit our geometric taste. In more detail, following early ideas by Teitelboim [Tei73], we can see that the family of deformations of the leaf $\Sigma$ embedded in the space of codimension 1 $\mathcal{M}$ forms a Lie groupoid defined through its Lie algebroid [BFW13, Boj&16]. Using this understanding, we implement these two postulates and derive GR in the vacuum in section 6.2.

In section section 6.3, we add the *geometrodynamical principle of equivalence* that selects a unique coupling of matter with GR from purely geometric arguments. Finally, we use this same geometric structure to identify the interaction of quantum matter in section 6.4.

This analysis allows us to explore conditions over the quantization procedure. As a consequence, we obtain that compatibility with geometrodynamics leads to (5.75).

## 6.2 Geometrodynamics in the vacuum: The path independence principle

In this section, we will derive General Relativity (GR) in the vacuum as a symmetric theory under the Lie groupoid of deformations of $\Sigma$ embedded in $\mathcal{M}$. This Lie groupoid is generated by the Lie algebroid $\mathcal{A}_\Sigma$, which we describe below.

In this work, we will make several simplifying assumptions that will ease the technical mathematical details of the construction. We would like to emphasize that some of the mathematical details are a matter of current research, and thus the arguments made in this section must be treated as compelling arguments with geometric flavour rather than rigorous derivations.

The first simplifying assumption is that the embeddings should be treated at the level of the superspaces $\mathcal{S}(\Sigma) \hookrightarrow \mathcal{S}(\mathcal{M})$ introduced in[2] (6.1), thus accounting for the possible isometries of both manifolds. In this work, we will obviate this symmetry and work directly over $\mathrm{Riem}(\Sigma) \hookrightarrow \mathrm{pseudo\,Riem}(\mathcal{M})$. The derivation of the algebroid in the correct setting is given in [BFW13].

Conceptually, *the seventh route to geometrodynamics* studies the action of this algebroid $\mathcal{A}_\Sigma$ on the unreduced phase space of the gravitational theory $T^* \mathrm{Riem}(\Sigma)$. This takes us to the second simplifying assumption. The theory of Hamiltonian actions of Lie algebroids is a matter of current research [Ike21, BRW23, BW24]. Alas, the application of this formalism in the present case is still unclear [BRW23]. Instead of trying to apply it in its full rigor, we will follow [HKT76] and introduce the Hamilton action through postulates rather than deriving it from the theory. This is, we will postulate an action generated by a momentum map-like object initially proposed by Teitelboim [Tei73]. This object will be dubbed the supermomentum section for our Lie algebroid $\mathcal{A}_\Sigma$.

### 6.2.1 The Lie algebroid of hypersurface deformations and the supermomentum section.

If we consider a globally hyperbolic $3 + 1$-dimensional spacetime $\mathcal{M}$ with coordinates $X^\mu$, there exists a one-parameter family of spatial leaves, $\mathcal{M} \simeq \{\Sigma_s \mid s \in \mathbb{R}\}$, foliating $\mathcal{M}$. Each spatial leaf $\Sigma_s$ is the image under the $s$-dependent embedding $\varepsilon_s$ of a fixed reference hypersurface $\Sigma$, and it is characterized by a hyperplane equation $\tau(X^\mu) = s$, such that each $\varepsilon_s(\Sigma) = \Sigma_s \subset \mathcal{M}$ is a Cauchy surface $\forall s \in \mathbb{R}$. From the hyperplane equation, taking 3 embedding coordinates $x^i$ for $\Sigma$, one derives for $\Sigma_s$ the parametric hyperplane equations

$$X^\mu = X^\mu(x, s).$$

The Lorentzian metric $g$ of $\mathcal{M}$ induces a Riemannian 3-metric $h_s = \varepsilon_s^\star g$ on each $\Sigma_s$, whose coordinates can be defined as:

$$h_{ij}(x, s) = {}^4 g_{\mu\nu}(X^\alpha(x, s))\partial_{x^i} X^\mu(x, s)\partial_{x^j} X^\nu(x, s). \tag{6.3}$$

Each spatial leaf can also be characterized by a time-like normal vector field $\hat{n}_s$ of norm 1, in terms of which we define the second fundamental form

$$K_{ij}(s) = -\frac{1}{2}\pounds_{\hat{n}_s} h_{ij}(s), \tag{6.4}$$

where $\pounds_{\hat{n}_s}$ represents the Lie derivative with respect to the vector field $\hat{n}_s$. With these elements, we define an extrinsic geometry on $\Sigma_s$ describing how the 3-geometry evolves along the 4-foliation. Thus, we relate the initial coordinates in a 3-surface $\Sigma_s$ and the coordinates on $\Sigma_{s+\delta_s}$ as a transformation in the normal direction $\hat{n}_s$ to the original hypersurface and a tangential one $\vec{N}$.

---

[2]In the second case, the metrics are pseudo-Riemannian

In terms of this set of coordinates, as depicted in [Figure 3.1](), the foliation is described as:

$$\frac{d}{ds}X^\mu(s) = N_s(x)n_s^\mu(x) + N_s^i(x)\partial_{x^i}X^\mu(x,s). \tag{6.5}$$

where the expression in coordinates of the lapse function (obviating the omnipresent dependence on $x$) reads:

$$N_s := g_{\mu\nu}(X^\alpha(s))\left(\frac{d}{ds}X^\mu(s)\right)n_s^\nu, \tag{6.6}$$

measuring the size of the change in the normal direction. Analogously, the coordinates of the shift vector are:

$$N_s^i := g_{\mu\nu}(X^\alpha(s))\left(\frac{d}{ds}X^\mu(s)\right)\partial_{x^j}X^\mu(s)h_s^{ji}, \tag{6.7}$$

which weights the change in each of the tangential directions associated with $\partial_{x^i}$.

Both objects can be combined to define an evolution-generating vector field $E$ at each hypersurface, which encodes the foliation as:

$$E|_{\Sigma_s} = \vec{N}_s + N_s\hat{n}_s. \tag{6.8}$$

Now, considering an initial data Cauchy surface $\Sigma_0 = \varepsilon_0(\Sigma)$ embedded in $\mathcal{M}$ and $E$ (or lapse and shift $\forall s$), one can span the whole foliation $\{\Sigma_s\}$ equivalent to $\mathcal{M}$ (and all tensors defined at each leaf) when considering its flow $\alpha_E^s$, so that $\Sigma_s = \alpha_E^s\Sigma_0$. Thus, $E$ characterizes the foliation and relates the embeddings through $\alpha_E^s \circ \varepsilon_0 = \varepsilon_s$. Now we will focus on the infinitesimal action of deformations of the Cauchy hypersurface $\Sigma_0 = \Sigma$ and drop the subindex $s$ to indicate $(N, \vec{N}) \in C^\infty(\Sigma) \oplus \mathfrak{X}(\Sigma)$.

In this way, we can think about $\mathcal{L}_E$ as the infinitesimal deformation of the hypersurface $\Sigma$. As we advanced above, the Lie algebra structure spanned by the Lie derivatives $\mathcal{L}_E$ defined on a neighborhood of $\Sigma \subset \mathcal{M}$ is not projected over a Lie algebra structure defined on a bundle with base $\Sigma$ but to a Lie algebroid $\mathcal{A}_\Sigma$. To see this, we express the tangential and normal deformations using functional derivatives. The generators of such deformations $D(N, \vec{N})$ are

$$D(N, \vec{N}) := \int_\Sigma d^3x\left(N(x)n^\mu(x)\frac{\delta}{\delta X^\mu(x)} + N^i(x)\partial_{x^i}X^\mu(x)\frac{\delta}{\delta X^\mu(x)}\right), \tag{6.9}$$

Consider $(N, \vec{N}), (M, \vec{M}) \in C^\infty(\Sigma) \oplus \mathfrak{X}(\Sigma)$. If we compute their Lie bracket closing relations as in [[Tei73](), [Boj&16]()], we obtain

$$[D(N, \vec{N}), D(M, \vec{M})] =$$
$$D\big(\vec{N} \cdot M - \vec{M} \cdot N + \mathrm{div}_h(\vec{N})M - N\,\mathrm{div}_h(\vec{M})\,, [\vec{N}, \vec{M}] + N\,\mathrm{grad}_h M - M\,\mathrm{grad}_h N\big) \tag{6.10}$$

where $\vec{M} \cdot N$ indicates the action of the derivation $\vec{M}$ on the function $N$ and $\mathrm{grad}_h\, N$ the gradient with respect to the metric $h$, in coordinates $(\mathrm{grad}_h\, N)^i = h^{ij}\partial_{x^j} N$ and $\mathrm{div}_h(\vec{N}) = \frac{1}{\sqrt{h}}\partial_{x^i}(\sqrt{h}N^i)$.

Once these relations have been proved, we notice that the deformation of the initial manifold is completely encoded in the closing relations over the space of lapse and shifts. However, the presence of $\mathrm{grad}_h$ and $\mathrm{div}_h$ in (6.10) does not define a Lie algebra structure because the closing relations have structure functions instead of constants. Instead, this structure can be understood in terms of Lie algebroids [BFW13, Boj&16]. For completeness, we expose below the description of this structure. The uninterested reader can skip the description of this structure.

**Lie algebroid structure**

To understand this structure in full, this algebroid should be described over the space of embeddings $\Sigma \hookrightarrow \mathcal{M}$ under isometry equivalence classes. This is the route pursued in [BFW13]. In this work, working under the simplifying assumptions exposed at the beginning of the section, we will follow [Boj&16] and describe the algebroid over the space $\mathrm{Riem}(\Sigma)$. This approach presents some drawbacks that we will discuss below. Nonetheless, these do not affect the discussion of the rest of the chapter.

**Definition 6.1.** *(Lie Algebroid)*

*A Lie algebroid is a vector bundle $B$ over a smooth base manifold $M$ together with a Lie bracket $[\cdot,\cdot]_M$ on the set $\Gamma(B)$ of sections of $B$ and a bundle map $\rho\colon \Gamma(B) \to \Gamma(TM)$, called the anchor, provided that $\rho\colon (\Gamma(B), [\cdot,\cdot]_B) \to (\mathfrak{X}(M), [\cdot,\cdot])$ is a homomorphism of Lie algebras. This is, for any $\alpha,\ \beta \in \Gamma(B)$ and*

$$\rho\left([\alpha,\beta]_B\right) = [\rho(\alpha),\rho(\beta)]\ ,$$

*where $[\cdot,\cdot]$ is the commutator of vector fields. The anchor map also fulfills the Leibniz identity. This is for any $f \in C^\infty(M)$*

$$[\alpha, f\beta]_B = f[\alpha,\beta]_B + (\rho(\alpha)f)\,\beta$$

For a deeper understanding of the theory of Lie algebroids and groupoids, we refer elsewhere [Mac87]. To describe the algebroid of hypersurface deformations, we consider the trivial vector bundle

$$\mathcal{A}_\Sigma = \mathrm{Riem}(\Sigma) \times C^\infty(\Sigma) \oplus \mathfrak{X}(\Sigma) \text{ with sections } \Gamma(\mathcal{A}_\Sigma) \simeq C^\infty(\Sigma) \oplus \mathfrak{X}(\Sigma) \qquad (6.11)$$

For the applications we want to address, we introduce the space of constant sections

$$\mathcal{C}\Gamma(\mathcal{A}_\Sigma) = C^\infty(\Sigma) \oplus \mathfrak{X}(\Sigma) \qquad (6.12)$$

endowed with a Lie bracket defined by

$$(N, \vec{N}), (M, \vec{M}) \in C^\infty(\Sigma) \oplus \mathfrak{X}(\Sigma)$$
$$[\![(N, \vec{N}), (M, \vec{M})]\!] :=$$
$$\left( \vec{N} \cdot M - \vec{M} \cdot N + \operatorname{div}_h(\vec{N}) M - N \operatorname{div}_h(\vec{M}) \,, \, [N, \vec{M}] + N \operatorname{grad}_h M - M \operatorname{grad}_h N \right) \tag{6.13}$$

To complete the structure, we must provide the anchor $\rho : \mathcal{A}_\Sigma \to T \operatorname{Riem}(\Sigma)$. This anchor is

$$\rho(N, \vec{N}, \mathbf{h}) f(\mathbf{h}) = -\pounds_{\vec{N}} f(\mathbf{h}) + 2 K_{ij} \frac{\partial f(\mathbf{h})}{\partial h_{ij}} \tag{6.14}$$

where $K_{ij}$ is the second fundamental form of (6.4). With the anchor, we can extend the Lie bracket $[\![\cdot, \cdot]\!]$ to any section.

**Remark 6.1.** *To complete the description of the anchor, we must provide a closed form for $K_{ij}$. In the phase space description that we will describe below, we write [Pad10]*

$$K_{ij} = -\frac{2\kappa}{\sqrt{h}} \left( \pi_h^{kl} h_{ki} h_{lj} - \frac{1}{2} h_{kl} \pi_h^{kl} h_{ij} \right) \tag{6.15}$$

*where $\kappa$ is a constant that we will fix in (6.22). Thus, in the last instance, the complete description of the Lie algebroid structure would depend on the equations of motion that we want to derive, as is the case in [Boj&16].*

*In order to avoid this dynamical dependence on the description, another possibility is to consider the space on which the generators (6.9) act. This is, the space of embeddings of $\Sigma \hookrightarrow \mathcal{M}$ reduced by the equivalence classes of isometries. This space is called the space of $\Sigma$-universes denoted $\mathcal{U}\Sigma$ and is the case treated in [BFW13]. As we mentioned above, this case will not be treated in this work.*

### Hamiltonian actions

In our case, we do not need to understand the full structure of $\mathcal{A}_\Sigma$. Instead, following [HKT76], the physically relevant subset is the space of constant sections $\mathcal{C}\Gamma(\mathcal{A}_\Sigma) = C^\infty(\Sigma) \oplus \mathfrak{X}(\Sigma)$. This is the space containing the lapse functions and shift vectors $(N, \vec{N})$. Notice that the bracket $[\![\cdot, \cdot]\!]_x$ does not respect the space of constant sections as it introduces dependencies on $\mathbf{h}$. Nonetheless, with $\mathcal{C}\Gamma(\mathcal{A}_\Sigma)$, we generate all possible hypersurface deformations that span all possible foliations of $\mathcal{M}$. Following [HKT76], GR can be derived as a theory invariant under foliations, and as such, we will use the action of the algebroid $\mathcal{A}_\Sigma$ to derive it.

The theory of Hamiltonian actions of Lie algebroids [Ike21, BRW23, BW24] introduces a momentum section that plays the role of a momentum map for regular Lie algebras. Alas, the application of this formalism in the present case is still unclear [BRW23]. Instead of using this formalism, we will follow [HKT76] and postulate that the action of the Lie algebroid is provided by this momentum map-like structure, see [BFW13].

**Postulate 1. *(Hamiltonian action with the supermomentum section)***

Let $\mathcal{M}_F$ be the phase space of a theory representing a field $(\varsigma, \pi_\varsigma) \in \mathcal{M}_F$ with a canonical Poisson bracket $\{\cdot, \cdot\}_{\mathcal{M}_F}$. The action of the hypersurface deformations spanned by $\mathcal{C}\Gamma(\mathcal{A}_\Sigma)$ is provided by a supermomentum section that mimics the behavior of a momentum map for Lie algebra actions. This is, there is a function

$$(\mathscr{H}, \vec{\mathscr{H}}) : \quad \mathcal{M}_F \quad \longrightarrow \quad \mathcal{C}\Gamma(\mathcal{A}_\Sigma)' = C^\infty(\Sigma)' \oplus \mathfrak{X}(\Sigma)'$$

where $(\mathscr{H}, \vec{\mathscr{H}})$ are also denoted supermagnitudes. Concretely, $\mathscr{H}(\varsigma, \pi_\varsigma, x)$ is a density dubbed superhamiltonian and $\vec{\mathscr{H}}(\varsigma, \pi_\varsigma, x)$ is a one-form density called supermomentum.

Extending the notation of (2.1), we can express

$$(N, \vec{N})_x \cdot (\mathscr{H}, \vec{\mathscr{H}})^x := N_x \mathscr{H}^x(\varsigma, \pi_\varsigma) + \vec{N}_x \cdot \vec{\mathscr{H}}^x(\varsigma, \pi_\varsigma) = N_x \mathscr{H}^x(\varsigma, \pi_\varsigma) + \vec{N}_{x;i} \mathscr{H}^{x;i}(\varsigma, \pi_\varsigma)$$

The Hamiltonian action is defined in the following way. Let $(N_s, \vec{N}_s)_x$ generate a curve on the groupoid. For any function $F(\varsigma, \pi_\varsigma)$, we define the action as

$$\frac{d}{ds} F = \{F, (N_s, \vec{N}_s)_x \cdot (\mathscr{H}, \vec{\mathscr{H}})^x\}_{\mathcal{M}_F}$$

This action should be equivariant; this implies that the supermagnitudes should fulfill the closing relations

$$\{(N, \vec{N})_x \cdot (\mathscr{H}, \vec{\mathscr{H}})^x, (M, \vec{M})_x \cdot (\mathscr{H}, \vec{\mathscr{H}})^x\}_{\mathcal{M}_F} = [\![(N, \vec{N}) \, , \, (M, \vec{M})]\!]_x \cdot (\mathscr{H}, \vec{\mathscr{H}})^x \quad (6.16)$$

That imposes closing relations among the supermagnitudes. Notice that we can extend this action to any section $\Gamma(\mathcal{A}_\Sigma)$. In fact, this last expression only makes sense in this general context.

General Relativity (GR) will be the theory that describes the evolution of phase space variables in a foliation-independent manner. For this reason, the generators of the evolution of the system will be provided by generators of hypersurface deformations presented above and will fulfill the following postulate.

**Postulate 2. *(Path independence principle)***

The generator of the evolution in $\mathcal{M}_F$ must be a generator of the action of hypersurface deformations described in **Postulate 1**. The foliation $\epsilon_s(\Sigma) = \Sigma_s$ is generated by a curve $(N_{s,\mathbf{x}}, \vec{N}_{s,\mathbf{x}}) : \mathbb{R} \to \mathcal{C}\Gamma(\mathcal{A}_\Sigma)$. Thus, the Hamiltonian of the theory is

$$H(\varsigma, \pi_\varsigma, N_{s,\mathbf{x}}, \vec{N}_{s,\mathbf{x}}) = (N_s, \vec{N}_s)_x \cdot (\mathscr{H}, \vec{\mathscr{H}})^x. \quad (6.17)$$

We also impose that the evolution from $\Sigma_{t_i}$ to $\Sigma_{t_f}$ must be independent of the path $(N_{s,x}, \vec{N}_{s,x}) : [t_i, t_f] \to \mathcal{C}\Gamma(\mathcal{A}_\Sigma)$ chosen to connect both hypersurfaces. Two different paths of this kind are depicted in *Figure 6.1*. In order to fulfill this principle, we must restrict the phase space $\mathcal{M}_F$, introduced in **Postulate 1**, to a submanifold $(\varsigma, \pi_\varsigma) \in \mathcal{M}_{FG} \subset \mathcal{M}_F$ defined through the constraints

$$\left( \mathscr{H}(\varsigma, \pi_\varsigma), \vec{\mathscr{H}}(\varsigma, \pi_\varsigma) \right) = (0, 0). \tag{6.18}$$

*Because of the closure relations imposed in* **Postulate 1**, *we describe a system of first-class constraints. This means their Poisson brackets can be expressed as linear combinations of the constraints themselves, and the submanifold* $\mathcal{M}_{LG}$ *is well defined. Moreover, the constraints are conserved by the dynamics; thus, it is enough to impose them over the initial conditions.*

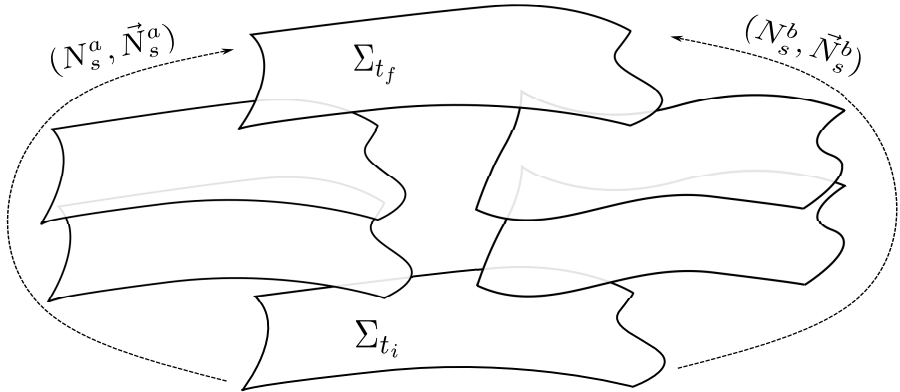

Figure 6.1: Two paths of $\mathcal{C}\Gamma(\mathcal{A}_\Sigma)$ marking a closed curve in the space space of embeddings of codimension one of a hypersurface $\Sigma$. The paths are provided by $(N_{\mathbf{x}}^{a/b}(s), \vec{N}_{\mathbf{x}}^{a/b}(s))$. Notice that the closed curve do not include elements of $T^*\mathrm{Riem}(\Sigma)$ thus we do not include the content of $(\mathbf{h}, \pi_h)$ in this picture as we did in Figure 3.2.

These postulates are the central axioms of the so-called *seventh route to geometro-dynamics* [HKT76, Kie12]. In order to describe any path-independent theory, we only need to specify the phase space $\mathcal{M}_F$, find its supermomentum section, and ask for (6.18) to hold. In the next sections, we will look for these ingredients to couple matter theories to GR.

### 6.2.2 General Relativity in the vacuum

Our objective in this section is to derive, or regain in the language of [HKT76], the supermagnitudes of GR as the only possible theory fulfilling **Postulate 1** and other physically motivated requirements. The main motivation of this presentation is to provide a set of ingredients that presents GR as inevitable from **Postulate** 1 and **Postulate** 2. We will limit our attention to these postulates and refer to [HKT76] for concrete proofs. We also note that the proofs presented in that work are only valid in a 3+1 spacetime.

## Phase space

Firstly, one must choose a phase space $\mathcal{M}_F$. As we introduced at the beginning of the chapter around (6.2), the phase space of gravity $\mathcal{M}_G$ is a subset of $T^\star\mathrm{Riem}(\Sigma) \sim \mathrm{Riem}(\Sigma) \times \mathrm{Riem}'(\Sigma)$. That is, the kinematical variables are the Riemannian 3-metric $\mathbf{h} \in \mathrm{Riem}(\Sigma)$ and its associated momenta $\pi_h \in T^\star(\mathrm{Riem}(\Sigma))|_h$, which are tensor densities of weight 1. In accordance with the notation of previous chapters, we identify the entries of the Riemannian structure with functions and denote them by $h_{ij;\mathbf{x}}$, while $\pi_h^{ij;x}$ is identified with a distribution. This cotangent bundle $T^\star\mathrm{Riem}(\Sigma)$ is endowed with a canonical symplectic form $\omega_G := -dh_{ij;x} \wedge d\pi_h^{ij;x}$ that can be associated with a Poisson bracket over given, for $f, g \in C^\infty\big(T^\star\mathrm{Riem}(\Sigma)\big)$, by

$$\{f,g\}_G = \partial_{\left[h_{ij;x}f\partial_{\pi_h^{ij;x}}\right]}g :=$$
$$\int_\Sigma d^3x \left( \frac{\partial f}{\partial h_{ij}(x)} \frac{\partial g}{\partial \pi_h^{ij}(x)} - \frac{\partial g}{\partial h_{ij}(x)} \frac{\partial f}{\partial \pi_h^{ij}(x)} \right). \quad (6.19)$$

At an operational level, the discussion about the definition of the differential structure is pretty much the same as in chapter 2. Thus, simply adding the appropriate Kronecker deltas, we get

$$\frac{\partial \pi^{kl;y}}{\partial \pi^{ij;x}} = \delta_{ij}^{kl}\delta_x^y, \qquad \frac{\partial h_{kl;y}}{\partial h_{ij;x}} = \delta_{kl}^{ij}\delta_y^x, \qquad \frac{\partial h_{kl;y}}{\partial \pi^{ij;x}} = \frac{\partial \pi^{kl;y}}{\partial h_{ij;x}} = 0 \,.$$

## Supermagnitudes

We are now equipped with the kinematical tools needed to describe the gravitational supermagnitudes $(\mathscr{H}_G, \vec{\mathscr{H}_G})^{\mathbf{x}}$ of **Postulate** 1. We will use them to elucidate the particular shape of these supermagnitudes.

Let's take $(0, \vec{N}) \in \mathcal{CT}(\mathcal{A}_\Sigma)$. This transformation represents a tangential deformation in $\Sigma$ and can be seen as only an *initial data reshuffling*. Therefore, let $f(h_{ij;\mathbf{x}}, \pi^{ij;\mathbf{x}}) \in C^\infty\big(T^\star\mathrm{Riem}(\Sigma)\big)$, thus

$$\mathcal{L}_{\vec{N}}F = \{F, \vec{\mathscr{H}_x}\}_G \cdot \vec{N}^x. \quad (6.20)$$

This relation must also apply to smeared versions of the phase space coordinates $f^x h_{ij;x} + g_x \pi^{ij;x}$. Using this, the explicit expression of the Lie derivative $\mathcal{L}_{\vec{N}}$ of these objects, and the closure relations (6.16) among *stretchings* $(0, \vec{N}), (0, \vec{M}) \in \mathcal{C}\Gamma(\mathcal{A}_\Sigma)$, it is shown in [HKT76] that the only possible supermomentum is

$$\mathscr{H}_G^{i;x}(\mathbf{h}, \pi_h) = -2D_j \pi^{ji;x}, \tag{6.21}$$

where $D$ represents the Levi-Civita connection of $(\Sigma, \mathbf{h})$.

To obtain a unique superhamiltonian, we must add two postulates.

**Postulate 3. (Time reversal symmetry [HKT76])**

*The time reversal operation is introduced in $\mathcal{C}\Gamma(\mathcal{A}_\Sigma) \times T^\star Riem(\Sigma)$ by the transformations*

$$(N, \vec{N})_\mathbf{x} \to (N, -\vec{N})_\mathbf{x} \quad and \quad (h_{ij;\mathbf{x}}, \pi^{ij;\mathbf{x}}) \to (h_{ij;\mathbf{x}}, -\pi^{ij;\mathbf{x}}).$$

*We postulate that the supermomentum section of the gravitational phase space $(N, \vec{N})_x \cdot (\mathscr{H}_G, \vec{\mathscr{H}}_G)^x$ is invariant under time reversal. This is*

$$\mathscr{H}(h_{ij;\mathbf{x}}, \pi^{ij;\mathbf{x}}) = \mathscr{H}(h_{ij;\mathbf{x}}, -\pi^{ij;\mathbf{x}}) \quad and \quad \vec{\mathscr{H}}(h_{ij;\mathbf{x}}, \pi^{ij;\mathbf{x}}) = -\vec{\mathscr{H}}(h_{ij;\mathbf{x}}, -\pi^{ij;\mathbf{x}}).$$

In order to state the next postulate, we introduce the following definition:

**Definition 6.2. (Ultralocality)** *A function $f(h_{ij;\mathbf{x}}, \pi^{ij;\mathbf{x}})$ is said to be ultralocal on the metric $h_{ij;\mathbf{x}}$ if it does not depend on its derivatives $\partial_\mathbf{y} h_{ij;\mathbf{x}}$ or derivatives of higher order.*

*The function $f$ is called ultralocal on gravitational momenta if it does not depend on derivatives of $\pi^{ij;\mathbf{x}}$.*

**Postulate 4. (Ultralocality on gravitational momenta [HKT76])**

*To ensure compatibility with (6.4)*

$$\{h_{ij;x}, N_y \mathscr{H}_G^y\}_G = -2K(h_{ij;\mathbf{x}}, \pi^{ij;\mathbf{x}})N_x,$$

*where $K(h_{ij;\mathbf{x}}, \pi^{ij;\mathbf{x}})$ is ultralocal on gravitational momenta.*

From these two postulates, again referring to [HKT76], the GR superhamiltonian is regained:

$$\mathscr{H}_G^\mathbf{x}(\mathbf{h}, \pi_h) = \frac{1}{2} \frac{(2\kappa)}{\sqrt{h}} G_{ijkl} \pi_h^{ij} \pi_h^{kl} - (2\kappa)^{-1} \sqrt{h}(R(\mathbf{h}) - 2\Lambda), \tag{6.22}$$

where $G_{ijkl} := \frac{1}{2\sqrt{h}} \left(h_{ik}h_{jl} + h_{il}h_{jk} - h_{ij}h_{kl}\right)$ is De Witt's metric and $R(\mathbf{h})$ is the Ricci scalar. Constants $\Lambda$ and $\kappa$ are arbitrary. We will choose them such that $\Lambda$ is the cosmological constant and $\kappa = 8\pi G/c^4$, where $G$ is Newton's constant of gravitation and $c$ is the speed of light. Recurring to the Lagrangian approach, we can go further and drop **Postulate** 3 as shown in [Kuc74].

### Evolution

General Relativity (GR) in the vacuum describes the evolution of $(h_{ij;\mathbf{x}}, \pi^{ij;\mathbf{x}})$ over the foliation $\Sigma_s$ specified by a curve $(N_{s,\mathbf{x}}, \vec{N}_{s,\mathbf{x}}) : \mathbb{R} \to \mathcal{C}\Gamma(\mathcal{A}_\Sigma)$. According to **Postulate** 3, the Hamiltonian of the theory is

$$H(\mathbf{h}, \pi_h, N_{s,\mathbf{x}}, \vec{N}_{s,\mathbf{x}}) = (N_s, \vec{N}_s)_x \cdot (\mathcal{H}_G, \vec{\mathcal{H}}_G)^x. \tag{6.23}$$

As we discussed in **Postulate** 3, the evolution equation generated by this Hamiltonian is ill-defined on the whole space $T^* \operatorname{Riem}(\Sigma)$. The choice of foliation, that is, the path $(N_s, \vec{N}_s)_{\mathbf{x}}$ chosen, is arbitrary, and we may describe the same theory with different foliations with common intermediate hypersurfaces. In order to achieve this, as we anticipated in (6.2), we must reduce the phase space of the theory to the constrained subspace

$$\mathcal{M}_G = \Big\{ (\mathbf{h}, \pi_h) \in T^* \operatorname{Riem}(\Sigma) \text{ s.t } \mathcal{H}_G^{\mathbf{x}}(\mathbf{h}, \pi_h) = 0 \text{ and } \vec{\mathcal{H}}_G^{\mathbf{x}}(\mathbf{h}, \pi_h) = 0 \Big\}. \tag{6.24}$$

The procedure described above, the *seventh route to geometrodynamics*, is essentially different from that of the ADM formalism. In that case, the Hamiltonian formalism is obtained by projecting the Einstein-Hilbert action over $\Sigma_s$. In that case, the symmetry associated with GR is the invariance under diffeomorphisms of $\mathcal{M}$ instead of the path independence principle. The relation between both symmetries is that they represent subsets of a larger group of transformations [BK72]. As a consequence, in the ADM formalism, the constraints arise as a result of the Gauss-Codazzi equations; see, for example, [Gou07, Pad10].

## 6.3 Geometrodynamics with matter

In order to complete our gravitational picture, we must introduce matter into the mix. That is, we must describe how to couple a field described through its phase space $\mathcal{M}_F$ with canonical coordinates $(\varsigma_{\mathbf{x}}, \pi_\varsigma^{\mathbf{x}}) \in \mathcal{M}_F$. The pathway to follow at this stage is just a rephrasing of the previous section.

### Phase space

The phase space of this theory is jointly described as a submanifold of $\mathcal{M}_F \times T^* \operatorname{Riem}(\Sigma)$ where the Poisson bracket is provided by the sum

$$\{\cdot, \cdot\} = \{\cdot, \cdot\}_G + \{\cdot, \cdot\}_{\mathcal{M}_F}. \tag{6.25}$$

An important step at this stage is to ensure that this is indeed a Poisson Bracket. The only remaining property to check is that, if we find a dependence of $\{\cdot, \cdot\}_{\mathcal{M}_F}$ on $(\mathbf{h}, \pi_h)$, the Jacobi identity holds.

**Supermagnitudes**

Firstly, imposing **Postulate** 1, this is the Hamiltonian action of the algebroid of hypersurface deformations $\mathcal{A}_\Sigma$, we should find supermagnitudes $(\mathscr{H}^{\times}, \vec{\mathscr{H}}^{\times})$ that describe the theory.

Our aim is to introduce a new postulate that fixes the form of the supermagnitudes for the theory with matter. For simplicity, we will restrict our attention to the minimal coupling case as in [HKT76].

**Postulate 5.** *(Geometrodynamical equivalence principle)*

*A matter theory minimally coupled to gravity fulfills the geometrodynamical principle of equivalence when the supermagnitudes of* **Postulate** 1 *are written as*

$$\mathscr{H}^{\times} = \mathscr{H}_G^{\times}(\mathbf{h}, \pi_h) + \mathscr{H}_M^{\times}(\mathbf{h}, \varsigma, \pi_\varsigma), \tag{6.26}$$

$$\vec{\mathscr{H}}^{\times} = \vec{\mathscr{H}}_G^{\times}(\mathbf{h}, \pi_h) + \vec{\mathscr{H}}_M^{\times}(\mathbf{h}, \varsigma, \pi_\varsigma). \tag{6.27}$$

*where* $(\mathscr{H}_M^{\times}, \vec{\mathscr{H}}_M^{\times}) : T^* \operatorname{Riem}(\Sigma) \times \mathcal{M}_F \to \mathcal{A}_\Sigma'$ *are the matter supermagnitudes. They depend ultralocally on* $\mathbf{h}$ *and do not depend on* $\pi_h$. *In the absence of the matter part, the equivalence principle recovers GR in the vacuum with the supermagnitudes* (6.21) *and* (6.21).

It is convenient to rephrase the closing relations (6.16) for the matter and gravity supermagnitudes separately. We already know that $(\mathscr{H}_G, \vec{\mathscr{H}}_G)^{\times}$ alone fulfills those closing relations for the full Poisson bracket $\{\cdot, \cdot\}$ because $\{\cdot, \cdot\}_M$ does not derive the gravitational variables. Let us explore the Poisson brackets of crossed supermagnitudes involving only the gravitational Poisson bracket. On the one hand, we obtain the cancellation

$$N_x M_y \{\mathscr{H}_G^x, \mathscr{H}_M^y\}_G + N_x M_y \{\mathscr{H}_M^x, \mathscr{H}_G^y\}_G = 0 \tag{6.28}$$

because of ultralocality (on $\pi$ and $h$ for both derivatives) and antisymmetry. On the other hand, as a result of (6.20), the relation

$$\pounds_{\vec{N}}(\vec{M}_y \cdot \vec{\mathscr{H}}_G^y) = \{\vec{M}_y \cdot \vec{\mathscr{H}}_G^y, \vec{N}_x \cdot (\vec{\mathscr{H}}_M + \vec{\mathscr{H}}_G)^x\} = \{\vec{M}_y \cdot \vec{\mathscr{H}}_G^y, \vec{N}_x \cdot \vec{\mathscr{H}}_G^x\}_G$$

must hold. As a consequence of the independence of $\vec{\mathscr{H}}_G$ on matter variables, we obtain

$$\{\vec{M}_y \cdot \vec{\mathscr{H}}_G^y, \vec{N}_x \cdot \vec{\mathscr{H}}_M^x\}_G = 0. \tag{6.29}$$

Using (6.21), this can only happen if $\vec{\mathscr{H}}_M^{\times}(\varsigma, \pi_\varsigma)$ does not depend on $\mathbf{h}$ at all. This, in turn, also implies

$$\{M_y \mathscr{H}_G^y, \vec{N}_x \cdot \vec{\mathscr{H}}_M^x\}_G = 0. \tag{6.30}$$

Lastly, the remaining crossed term does not cancel. Instead, we have

$$\{M_y \mathscr{H}_M^y, \vec{N}_x \cdot \vec{\mathscr{H}}_G^x\}_G = 2 \frac{\partial \mathscr{H}_M^y}{\partial h_{ij;x}} M_y D_i N_{j;x}. \tag{6.31}$$

After this analysis, we can rewrite the geometrodynamical principle of equivalence, introducing the matter closing relations. This is a modification of the relations (6.16) that should hold for the matter supermagnitudes.

**Postulate 5.** *(Matter closing relations)*

*A matter theory minimally coupled to gravity fulfills the geometrodynamical principle of equivalence when the supermagnitudes of* **Postulate 1** *are written as*

$$\mathscr{H}^{\mathbf{x}} = \mathscr{H}_G^{\mathbf{x}}(\mathbf{h}, \pi_h) + \mathscr{H}_M^{\mathbf{x}}(\mathbf{h}, \varsigma, \pi_\varsigma), \tag{6.32}$$

$$\vec{\mathscr{H}}^{\mathbf{x}} = \vec{\mathscr{H}}_G^{\mathbf{x}}(\mathbf{h}, \pi_h) + \vec{\mathscr{H}}_M^{\mathbf{x}}(\varsigma, \pi_\varsigma). \tag{6.33}$$

*where* $(\mathscr{H}_M^{\mathbf{x}}, \vec{\mathscr{H}}_M^{\mathbf{x}}) : T^* \operatorname{Riem}(\Sigma) \times \mathcal{M}_F \to \mathcal{A}_\Sigma'$ *are the matter supermagnitudes. They do not depend on* $\pi_h$, *and the superhamiltonian depends ultralocally on* $\mathbf{h}$. *In the absence of the matter part, the equivalence principle recovers GR in the vacuum with the supermagnitudes (6.21) and (6.21), and the matter supermagnitudes fulfill the matter closing relations*

$$\{(N, \vec{N})_x \cdot (\mathscr{H}_M, \vec{\mathscr{H}}_M)^x, (M, \vec{M})_x \cdot (\mathscr{H}_M, \vec{\mathscr{H}}_M)^x\}_{\mathcal{M}_F} =$$

$$[\![(N, \vec{N}) \,, \, (M, \vec{M})]\!]_x \cdot (\mathscr{H}_M, \vec{\mathscr{H}}_M)^x + 2\big(M_y D_i N_{j;x} - N_y D_i M_{j;x}\big)\frac{\partial \mathscr{H}_M^y}{\partial h_{ij;x}}. \tag{6.34}$$

**Example 6.1.** *(Supermagnitudes of Klein-Gordon theory)*

*The minimally coupled classical Klein-Gordon theory described in* section 3.5 *fulfills the matter closing relations for the supermagnitudes*

$$\mathscr{H}^{\mathbf{x}} = \frac{1}{2\sqrt{h}}[(\pi^2)^{\mathbf{x}} + h^{ij} D_i \varphi^{\mathbf{x}} D_j \varphi^{\mathbf{x}} + m^2 (\varphi^2)^{\mathbf{x}}], \tag{6.35}$$

$$\mathscr{H}_i^x = \sqrt{h}\pi^{\mathbf{x}} D_i \varphi^{\mathbf{x}}. \tag{6.36}$$

*Defined over the phase space* $\mathcal{M}_F$ *with coordinates* $(\varphi^{\mathbf{x}}, \pi^{\mathbf{x}})$ *with the Poisson bracket*

$$\{P, Q\}_{\mathcal{M}_F} = \delta^{xy} \left(\partial_{\varphi^x} P \partial_{\pi^y} Q - \partial_{\pi^x} P \partial_{\varphi^y} Q\right)$$

**Evolution**

Requiring the path independence **Postulate** 2, the generator of the evolution is the Hamiltonian $H = (N_s, \vec{N}_s)_x \cdot (\mathscr{H}, \vec{\mathscr{H}})^x$ for some curve $(N_s, \vec{N}_s)_{\mathbf{x}} : \mathbb{R} \to \mathcal{A}_\Sigma$, and the phase space must be constrained by

$$\big(\mathscr{H}(\mathbf{h}, \pi_h, \varsigma, \pi_\varsigma), \vec{\mathscr{H}}(\mathbf{h}, \pi_h, \varsigma, \pi_\varsigma)\big) = (0, 0).$$

In general, for any function $f(s, N_s, \vec{N}_s, \mathbf{h}, \pi_h, \varsigma, \pi_\varsigma)$, we can write

$$\frac{dF}{ds} = \{F, H\} + \partial_s(N_x^i)\frac{\partial}{\partial N^i x} F + \partial_s F. \tag{6.37}$$

## 6.4 Hybrid Geometrodynamics: Hybrid closing relations

The key aspect of our proposal to couple quantum matter to classical gravity relies on the fact that the analysis of the previous section is blind to the quantum or classical nature of matter. The main ingredient needed to describe the quantum-classical hybrid geometrodynamics is the phase space of the quantum theory. For that matter, we slightly adapt the discussion of chapter 5.

### 6.4.1 Phase space

A priori, the phase space of a quantum theory coupled with gravity is

$$T^* \operatorname{Riem}(\Sigma) \times \mathscr{P} \simeq T^* \operatorname{Riem}(\Sigma) \times (\mathcal{N}_\mathbb{C}).$$

The phase space above possesses a hybrid Poisson bracket constructed from (5.19)

$$\{\cdot, \cdot\} = \{\cdot, \cdot\}_G + \{\cdot, \cdot\}_\mathscr{P}. \tag{6.38}$$

Notice that the expression $\{\cdot, \cdot\}_\mathscr{P}$ has a complicated dependence on $\mathbf{h}$ through the bilinear $\mathcal{K}^{\phi,\sigma}$. However, using a set of Darboux *s.q.* coordinates $(\Phi^\phi, \Pi_\phi)$, as in (5.24), we can write $\{F, G\}_\mathscr{P} = \partial_{\Phi_\phi} F \partial_{\Pi^\phi} G - \partial_{\Pi^\phi} F \partial_{\Phi_\phi} G$, we can eliminate this dependence and trivially show that $\{\cdot, \cdot\}$ fulfills Jacobi. This procedure, in turn, selects $(\Phi^\phi, \Pi_\phi)$ as a set of canonical variables independent of any external parameter, similarly to $(\varphi_\mathbf{x}, \pi^\mathbf{x})$ in (3.50).

Moreover, this bracket comes from the Kähler structure (5.1) defined on $\mathscr{P}$. The discussion of chapter 5 about the description of this structure does not really depend on the parametric nature of the gravity in that context. Pictorically, we should understand the dependence on the real parameter $t$ of Figure 5.1 as a choice of path of a parametric theory depicted in Figure 3.2. As a result in this chapter we will understand this dependence as in Figure 6.2. This is, we conclude that the structures of Kähl($\mathscr{P}$) should allow for a dependence on $\mathcal{C}\Gamma(\mathcal{A}_\Sigma) \times T^* \operatorname{Riem}(\Sigma)$ as explicitly exemplified in (3.54).

$$\left(\mathcal{G}(N_\mathbf{x}, \vec{N}_\mathbf{x}, \mathbf{h}, \pi_h), \Omega(N_\mathbf{x}, \vec{N}_\mathbf{x}, \mathbf{h}, \pi_h), \mathcal{J}(N_\mathbf{x}, \vec{N}_\mathbf{x}, \mathbf{h}, \pi_h)\right)_\mathscr{P}$$

In order to accommodate this dependence into the description, a quantum state will be described by a section on a trivial bundle

$$\mathcal{C}\Gamma(\mathcal{A}_\Sigma) \times T^* \operatorname{Riem}(\Sigma) \times (\mathcal{N}_\mathbb{C}).$$

This is, a quantum state is described by a function $\Psi\left(\phi^\mathbf{x}, N_\mathbf{x}, \vec{N}_\mathbf{x}, \mathbf{h}, \pi_h\right)$.

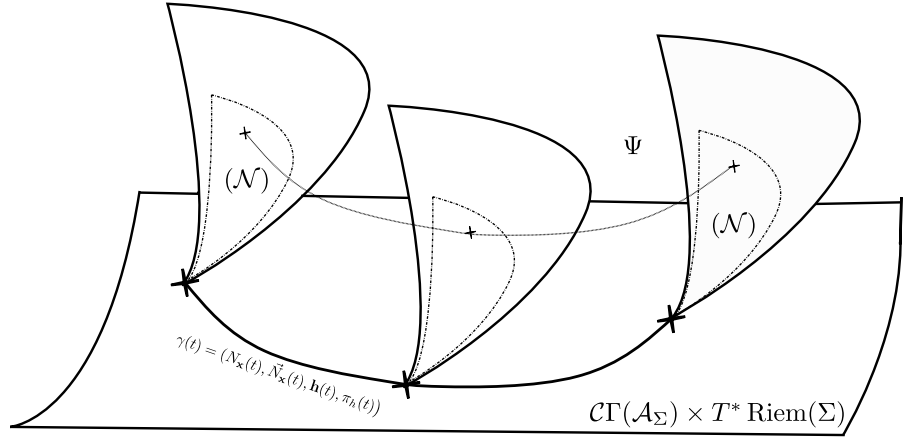

Figure 6.2: Parametric dependence of a quantum state $\Psi$ through the choice of a parametric curve $\gamma$ in the space $\mathcal{C}\Gamma(\mathcal{A}_\Sigma) \times T^* \operatorname{Riem}(\Sigma)$. This is the structure behind the dependence on a real parameter $t$ of Figure 5.1.

### 6.4.2 Supermagnitudes

The complicated dependence of the quantum states on lapse and shift complicates the discussion. This fact, however, is a reflection of the complicated relation between kinematics and dynamics that arises in QFT, which we explored in chapter 4 and chapter 5. Lapse and shift $(N_\mathbf{x}, \vec{N}_\mathbf{x})$ are fundamentally dynamical features of the theory according to **Postulate** 2. Therefore, their introduction in the definition of the state must be an intermediate step that should not spoil the path independence principle.

Similarly to the discussion below section 5.4.1, once the vacuum section is factored out, we get a Hermitian line bundle $\pi : B \to \mathcal{C}\Gamma(\mathcal{A}_\Sigma) \times T^* \operatorname{Riem}(\Sigma)$ in which we will introduce a connection $\Gamma$. Notice that this structure is the analogue of the one discussed in (5.41). To recover that structure, we must choose a curve $\gamma : \mathbb{R} \to \mathcal{C}\Gamma(\mathcal{A}_\Sigma) \times T^* \operatorname{Riem}(\Sigma)$. In fact, this curve is the one that parametrizes the gravitational theory that we mentioned in the introduction.

The base of the bundle $\mathcal{C}\Gamma(\mathcal{A}_\Sigma) \times T^* \operatorname{Riem}(\Sigma)$ is a linear space; therefore, we will identify it with its tangent space using the notation $\vec{\zeta} = (\zeta_{N;\mathbf{x}}, \zeta_{N^i;\mathbf{x}}, \zeta_{h_{ij};\mathbf{x}}, \zeta_{\pi^{ij}}^{\mathbf{x}}) = \left(N_\mathbf{x}, \vec{N}_\mathbf{x}, \mathbf{h}, \pi_h\right)$ to identify a direction in this tangent space. The covariant derivative $\nabla_{\vec{\zeta}}$ in components for the section $\Psi$ is

$$\nabla_{\vec{\zeta}}\Psi = \vec{\zeta}_x \cdot \left(\partial_{N_x}, \partial_{N_{ix}}, \partial_{h_{ij;x}}, \partial_{\pi^{ij;x}}\right)\Psi + \zeta_x^i \cdot \Gamma_{\zeta_x^i}\Psi \tag{6.39}$$

The remaining ingredient to complete the quantum picture is the quantization mapping $\mathcal{Q}$. We continue with the conventions of chapter 4 and choose the holomorphic Weyl quantization mapping. The quantization mapping also depends on the base

manifold

$$\mathcal{Q}\big(N_{\mathbf{x}}, \vec{N}_{\mathbf{x}}, \mathbf{h}, \pi_h\big).$$

With this, we define a compatibility condition

**Definition 6.3. (Curvature-free quantization mapping)** *The covariant derivative* $\nabla$ *and the quantization mapping* $\mathcal{Q}$ *define a curvature-free quantization mapping, in the sense of* (5.76), *if they fulfill the compatibility condition*

$$\nabla_{\vec{\zeta}}\mathcal{Q}(f) = \zeta_x^i \partial_{\zeta_x^i}\mathcal{Q}(f) + [\zeta_x^i \cdot \Gamma_{\zeta_x^i}, \mathcal{Q}(f)] = \mathcal{Q}(\zeta_x^i \partial_{\zeta_x^i} f) \qquad (6.40)$$

**Postulate 6. (Hybrid supermagnitudes)**

*The matter supermagnitudes of the quantum theory are obtained from the supermagnitudes* $\big(\mathscr{H}_M(\mathbf{h}, \pi_h, \varphi^{\mathbf{x}}, \pi^{\mathbf{x}}), \vec{\mathscr{H}}_M(\mathbf{h}, \pi_h, \varphi^{\mathbf{x}}, \pi^{\mathbf{x}})\big)$ *of its corresponding classical theory such that*

1. $(\mathscr{H}_M^{\mathbf{x}}, \vec{\mathscr{H}}_M^{\mathbf{x}})$ *depend on* $\pi^{\mathbf{x}}$ *at most quadratically. That is, we cannot allow for kinetic terms in the Hamiltonian of order greater than two in the classical theory to be quantized.*

2. *The classical supermagnitudes fulfill the matter closing relations of* **Postulate 5**.

3. *The quantization mapping* $\mathcal{Q}$ *is curvature-free.*

4. *The space of quantum states is constrained to*

$$\mathcal{M}_Q = \{\nabla_{\vec{\zeta}}\Psi^\phi = 0 \qquad \forall\, \vec{\zeta} \in T[\mathcal{C}\Gamma(\mathcal{A}_\Sigma) \times T^* \operatorname{Riem}(\Sigma)] \text{ s.t. } \Psi^\phi \in \Gamma(B)\}$$

*This space is also known as the horizontal bundle* $\operatorname{Hor}^\nabla$. *Heuristically, we interpret this condition in the following way: the only dependence of the quantum state comes from the factorization of the vacuum to be included in the Gaussian measure. This is the kind of factorization that we discussed in* (3.52). *As we discussed in* chapter 3, *this argument cannot be made rigorous. This is because it implies the existence of a Lebesgue-like measure* $D\phi$. *We refer to* [Alo&24a] *for further details on this constraint.*

*Then the generators of the quantum evolution are*

$$\mathscr{H}_Q^{\mathbf{x}} = \langle \Psi, Q(\mathscr{H}_M^{\mathbf{x}})\Psi\rangle, \qquad \vec{\mathscr{H}}_Q^{\mathbf{x}} = \langle \Psi, Q(\vec{\mathscr{H}}_M^{\mathbf{x}})\Psi\rangle \qquad (6.41)$$

*where* $\langle,\rangle$ *denotes the Hermitian product* (5.15) *and we drop the superindex of* $\Psi^\phi$ *to alleviate notation.*

We still have to check that this definition fulfills the matter closing relations. On one hand, notice that as a consequence of conditions 3 and 4, we have for any function $f$

$$\partial_{\zeta^i}\langle\Psi, Q(f)\Psi\rangle = \langle\Psi, \nabla_{\zeta^i}Q(f)\Psi\rangle = \langle\Psi, Q(\partial_{\zeta^i}f)\Psi\rangle.$$

This important relation implies that the dependence introduced in the section $\Psi$ and the quantization mapping $Q$ on the geometric data $(N_{\mathbf{x}}, \vec{N}_{\mathbf{x}}, \mathbf{h}, \pi_h)$ is such that the quantum matter supermagnitudes possess the same dependence as the classical ones, protecting the path independence principle.

On the other hand, notice that (5.9) implies

$$\{\langle\Psi, Q(f)\Psi\rangle, \langle\Psi, Q(g)\Psi\rangle\}_{\mathscr{P}} = \langle\Psi, \, -iQ(f\star_m g - g\star_m f)\Psi\rangle \quad (6.42)$$

where $\star_m$ is the Moyal product of (4.52). Because of condition 2, we can use the well-known result that if $f$ and $g$ are at most quadratic in momenta $\pi^{\mathbf{x}}$, we get $f\star_m g - g\star_m f = i\{f, g\}_{\mathcal{M}_F}$. Then we obtain

$$\{(N, \vec{N})_x \cdot (\mathscr{H}_Q, \vec{\mathscr{H}}_Q)^x, (M, \vec{M})_x \cdot (\mathscr{H}, \vec{\mathscr{H}})^x\}_{\mathscr{P}} =$$
$$\langle\Psi, Q\big[\{(N, \vec{N})_x \cdot (\mathscr{H}_M, \vec{\mathscr{H}}_M)^x, (M, \vec{M})_x \cdot (\mathscr{H}_M, \vec{\mathscr{H}}_M)^x\}_{\mathcal{M}_C}\big]\Psi\rangle \quad (6.43)$$

Then, using condition 1 and the previous considerations, the quantum generators of **Postulate** 6 fulfill the matter closing relations

$$\{(N, \vec{N})_x \cdot (\mathscr{H}_M, \vec{\mathscr{H}}_Q)^x, (M, \vec{M})_x \cdot (\mathscr{H}_M, \vec{\mathscr{H}}_Q)^x\}_{\mathcal{M}_F} =$$
$$[\![(N, \vec{N})\, ,\, (M, \vec{M})]\!]_x \cdot (\mathscr{H}_Q, \vec{\mathscr{H}}_Q)^x + 2\big(M_y D_i N_{j;x} - N_y D_i M_{j;x}\big)\frac{\partial\mathscr{H}_Q^y}{\partial h_{ij;x}} \quad (6.44)$$

A more detailed proof is developed in [Alo&24a]. In that work, it is shown that the matter closing relations on the quantum part imply a weaker version of the curvature-free condition for the quantization mapping.

### 6.4.3 Evolution

Finally, according to **Postulate** 2, the generator of the evolution is

$$H = (N_{x,s}, \vec{N}_{x,s}) \cdot (\mathscr{H}_G^x + \langle\Psi, Q(\mathscr{H}_M^x)\Psi\rangle, \vec{\mathscr{H}}_G^x + \langle\Psi, Q(\vec{\mathscr{H}}_M^x)\Psi\rangle), \quad (6.45)$$

constrained to fulfill

$$(\mathscr{H}_G^x + \langle\Psi, Q(\mathscr{H}_M^x)\Psi\rangle, \vec{\mathscr{H}}_G^x + \langle\Psi, Q(\vec{\mathscr{H}}_M^x)\Psi\rangle) = (0, 0).$$

For simplicity, we denote $H_G = (N_{x,s}, \vec{N}_{x,s}) \cdot (\mathscr{H}_G^x, \vec{\mathscr{H}}_G^x)$ and $H_M = (N_{x,s}, \vec{N}_{x,s}) \cdot (\mathscr{H}_M^x, \vec{\mathscr{H}}_M^x)$. Then our analysis leads to a system of equations

$$\frac{d}{ds}\Psi = \left( \frac{-i}{\hbar}\mathcal{Q}(H_M) - \{h_{ij;x}, H\}_G \Gamma_{h_{ij;x}} + \{\pi_h^{ij;x}, H\}_G \Gamma_{\pi_h^{ij;x}} - \dot{N}_x \Gamma_{N_x} - \dot{N}_x^i \Gamma_{N_x^i} \right) \Psi,$$

$$\frac{d}{ds}h_{ij} = 2NG_{ijkl}\pi_h^{kl} + 2(D_i N_s^k)h_{kj},$$

$$\frac{d}{ds}\pi_h^{ij} = -\partial_{h_{ij}}H_G - \langle \Psi \mid Q(\partial_{h^{ij}}H_M)\Psi \rangle. \tag{6.46}$$

In this case, the connection (5.77) can be adapted to this context by substituting the time parameter $t$ with the geometric variables $(N_x, \vec{N}_x, \mathbf{h}, \pi_h)$. Notice that the backreaction term $-\langle \Psi \mid Q(\partial_{h^{ij}}H_M)\Psi \rangle$ is simply the expected value of the quantized version of the classical backreaction term; thus, the interesting part of the dynamics lies in the quantum part.

# Chapter 7

# Applications to Cosmology: Klein-Gordon Theory on Curved Spacetimes

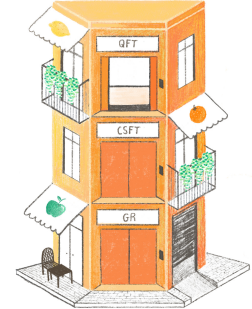

## 7.1 Introduction

In this chapter, we will apply the formalism derived in the previous chapters to the quantum Klein-Gordon theory over a fixed curved geometry. That is, we will deal with the parametric theory of *Remark* 3.1 and leave the *backreaction* terms of (6.46) for future work. This study is a continuation of the classical study in the KvH formalism of section 3.5; thus, we recommend going through that section before reading this chapter.

The first issue that we must address is the choice of complex structure $J_{\mathcal{M}_c}$ that we introduce in the classical phase space $(\mathcal{M}_c, \omega_{\mathcal{M}_c})$ to quantize the theory. In principle, the prescription made in (3.54) was chosen to be adapted to the classical evolution. Nonetheless, this prescription coincides with those studied in the context of QFT in [CCQ04], derived from a quite different set of physical requirements in the covariant formalism by [AMA75]. Moreover, the latter conditions are studied only in the static and stationary cases, while our classical prescription is valid in a general framework. As a matter of fact, how to choose this prescription in the quantum case is still an open problem [MO21]. For this reason, we stick to the prescription (3.54) also in the quantum case.

In this section, we will be interested in cosmological models for our spacetimes. The reason for choosing these kinds of solutions is twofold. First, they represent physically relevant models that constitute a simple benchmark where the modifications proposed to the quantum evolution can be tested. Second, the cosmological principle states that the universe at cosmological scales must be homogeneous and isotropic. This assumption imposes huge restrictions on the form of the spacetime metric $\mathbf{g}$, regardless of the quantum or classical nature of matter. In particular, the universes allowed by this principle are the so-called Friedman-Lemaître-Robertson-Walker (FLRW) spacetimes. We refer to [Pad10] for further details. For these universes, there is a system of coordinates $(t, r, \theta, \phi)$ for the spacetime $\mathcal{M}$ in which the metric is

written

$$\mathbf{g} = -dt \otimes dt + a^2(t)\Big[\frac{dr \otimes dr}{1 - kr^2} + r^2\big(d\theta + \sin^2\theta d\phi\big)\Big], \tag{7.1}$$

with $k = 0, \pm 1$, $(r, \theta, \phi) \in [0, \infty) \times [0, \pi] \times [0, 2\pi]$, and $a(t)$ is the positive scale factor. These coordinates suggest a foliation of spacetime into Cauchy hypersurfaces $\mathbb{R}^3 \simeq \Sigma^1$ with induced geometric data

$$N = 1, \vec{N} = 0,$$
$$\mathbf{h} = a^2(t)\Big[\frac{dr \otimes dr}{1 - kr^2} + r^2\big(d\theta + \sin^2\theta d\phi\big)\Big],$$
$$K_{ij} = -2\kappa G_{ijkl}\pi_h^{kl} = -\frac{\dot{a}(t)}{a(t)}h_{ij} \tag{7.2}$$

where $K_{ij}$ represents the second fundamental form and $G_{ijkl}$ is De Witt's metric with inverse $G^{ijkl} = \frac{\sqrt{h}}{2}\big(h^{ik}h^{jl} + h^{il}h^{jk} - 2h^{ij}h^{kl}\big)$. Remember that we introduced $\kappa$ in (6.22). Notice that this describes the curve needed to describe a parametric theory introduced in *Remark* 3.1.

As a consequence, these spacetimes possess the property of having a null shift $\vec{N}$, and therefore the complex structure reduces to the limiting case of (3.65)

$$J_{\mathcal{M}_F} = \begin{pmatrix} 0 & \Theta^{-\frac{1}{2}}N^{\frac{1}{2}} \\ -N^{-\frac{1}{2}}\Theta^{\frac{1}{2}} & 0 \end{pmatrix} \tag{7.3}$$

with $-\Theta = ND^iD_i + (D^iN)D_i - Nm^2$. Thus, we will study the quantization of this case separately.

## 7.2 Quantum Models in the Limiting Cases

In this section, we will focus on the case of a parametric theory as in *Remark* 3.1. That is, we treat the gravitational content as a background. In previous chapters, we have seen how the choice of a complex structure on the phase space of classical fields

$$-J_{\mathcal{M}_F} = (\partial_{\varphi^y}, \partial_{\pi^y})\begin{pmatrix} A_x^y & \Delta_x^y \\ D_x^y & -(A^t)_x^y \end{pmatrix}\begin{pmatrix} d\varphi^x \\ d\pi^x \end{pmatrix} =$$
$$= \partial_{\varphi^y} \otimes [A_x^y d\varphi^x + \Delta_x^y d\pi^x] + \partial_{\pi^x} \otimes [D_x^y d\varphi^x - (A^t)_x^y d\pi^y] \tag{7.4}$$

---

[1] Notice that we allow for $\Sigma$ to be non-compact in this example. This can be done without further complications just by imposing suitable boundary conditions at infinity

is a crucial ingredient of the quantization procedure. Recall from (3.21) the conditions over the matrix entries associated with $J^2 = -\mathbb{1}$. In this section, we will discuss the limiting cases of null and huge shifts discussed in section 3.5.

### 7.2.1 A Universe with Huge Shift

Let's first briefly discuss the limiting case $N \simeq 0$, whose complex structure (3.66) is

$$J_\infty = \begin{pmatrix} (-\alpha^2)^{-\frac{1}{2}} N^i D_i & 0 \\ 0 & (-\Lambda^2)^{-\frac{1}{2}} \Lambda \end{pmatrix} \tag{7.5}$$

with $\alpha = N^i D_i$ and $\Lambda = N^i D_i + D_i N^i$. This case may be physically interpreted as the appearance of a horizon in our spacetime. This limiting case breaks the quantization formalism in the following sense. The covariance of the Gaussian measure defined in (4.6) is $\Delta^{\mathbf{xy}} = 0$. This means that the measure represents a Dirac delta with an infinite-dimensional domain around $\phi^{\mathbf{x}} = 0$. This fact might actually be reconciled with a physical interpretation. However, the momentum operator of the Schrödinger picture (4.20) needs the definition of the inverse $K_{\mathbf{xy}}$. In this case, the momentum operator diverges with unclear consequences. This, in turn, may be interpreted as the splitting of the underlying Cauchy hypersurface $\Sigma$ into two regions, inside and outside the horizon. In each of these regions, the formalism exposed so far can be developed without further complications. However, once the spacetime develops one of these horizons, the purity of the quantum state in each of these regions may be compromised. We may, for example, trace out the degrees of freedom corresponding to wavefunctions completely supported inside the horizon. See e.g. [Wal94] for this kind of partial trace in spacetimes with bifurcating Killing Horizons.

### 7.2.2 A Universe with Null Shift

As we advanced in the introduction, the relevant case for cosmological applications is $\vec{N} = 0$. In that situation, the entries of (7.4) are

$$\varphi_x \Delta^{xy} \xi_y = \int_\Sigma d^d x \sqrt{h} \varphi(x) \Theta^{-\frac{1}{2}} N^{\frac{1}{2}} \xi(x) \tag{7.6}$$

with $A^{\mathbf{x}}_{\mathbf{y}} = (A^t)^{\mathbf{x}}_{\mathbf{y}} = 0$ and $D^{\mathbf{xy}} = -K^{\mathbf{xy}}$ with

$$\varphi_x K^{xy} \xi_y = \int_\Sigma d^d x \sqrt{h} \varphi(x) N^{-\frac{1}{2}} \Theta^{\frac{1}{2}} \xi(x) \tag{7.7}$$

Also, the Hamiltonian of the Klein-Gordon theory (3.60) in holomorphic coordinates is written $H = \bar{\phi}^x \Theta_{xy} \phi^y$. As we pointed out in (5.1), the quantum Hamiltonian acting over this space is obtained with Weyl quantization. However, this procedure produces a vacuum energy in a similar way to the constant appearing in *Remark* 4.2. This constant is important to describe the coupling with gravity in (6.46) but can be set to zero in the parametric case. In order to simplify the discussion, we pick Wick ordering and obtain

$$\hat{H} = \Theta_{xy}(a^\dagger)^x a^y \tag{7.8}$$

In holomorphic s.q. coordinates, this expression is just $\phi^y(\sqrt{\Theta N})^x_y \partial_{\phi^x}$ where $N^x_y \varphi_x = N\varphi_y$.

In order to write down the modified Schrödinger equation, we need to compute the connection term $\Gamma$ of (5.77). In this context, due to the fact that the diagonal terms of the complex structure are null, its expression simplifies enormously. To alleviate notation we denote $(\dot{\delta}{}^\circ\delta_\circ)^x_z := \dot{\delta}{}^{xz}\delta_{zy} = \frac{1}{2}\partial_t \log h \delta^x_y$, the connection (5.77) reduces to

$$
\begin{aligned}
\Gamma = &\frac{1}{2}\phi^x K_{xy}\dot{\Delta}^{yz}\partial_{\phi^z} - \frac{1}{4}\Big(\dot{\Delta}^{xy}\partial_{\phi^x}\partial_{\phi^y} + \dot{K}_{xy}\phi^x\phi^y\Big) + \\
&\frac{1}{2}\Big(\phi^x(\dot{\delta}{}^\circ\delta_\circ)^z_x\partial_{\phi^z} - \phi^x K_{xy}(\dot{\delta}{}^\circ\delta_\circ)^y_u\Delta^{uz}\partial_{\phi^z}\Big) + \\
&\frac{1}{4}\Big([\Delta^\circ,\dot{\delta}{}^\circ\delta_\circ]^{xy}_+\partial_{\phi^x}\partial_{\phi^y} - [K_\circ,\dot{\delta}{}^\circ\delta_\circ]^+_{xy}\phi^x\phi^y\Big).
\end{aligned}
\tag{7.9}
$$

Where we denoted separately $[\Delta^\circ,\dot{\delta}{}^\circ\delta_\circ]^{\mathbf{xy}}_+ = \Delta^{\mathbf{x}u}(\dot{\delta}{}^\circ\delta_\circ)^{\mathbf{y}}_u + (\dot{\delta}{}^\circ\delta_\circ)^{\mathbf{x}}_u\Delta^{u\mathbf{y}}$ and $[K_\circ,\dot{\delta}{}^\circ\delta_\circ]^+_{\mathbf{xy}} = K_{\mathbf{x}u}(\dot{\delta}{}^\circ\delta_\circ)^u_{\mathbf{y}} + (\dot{\delta}{}^\circ\delta_\circ)^u_{\mathbf{x}}K_{u\mathbf{y}}$. In this notation, we can see how the prescription of this connection is just a correction to the Schrödinger connection (5.72).

**(Modified) Schrödinger Equation**

In the holomorphic setting, the (modified) Schrödinger equation is

$$i\left[\frac{\partial}{\partial t} + \Gamma\right]\Psi = \phi^y(\sqrt{\Theta N})^x_y\partial_{\phi^x}\Psi \tag{7.10}$$

There are new terms in the Schrödinger equation that involve the time derivatives $\dot{\Delta}^{\mathbf{xy}}$, $K_{\mathbf{xy}}$, and $\dot{\delta}_{\mathbf{xy}}$. The new left-hand side term $i\Gamma$ has two parts:

- The factor $\frac{i}{2}\phi^x K_{xy}\dot{\Delta}^{yz}\partial_{\phi^z}$ is an anti-adjoint correction to the time derivative needed in the evolution to preserve the probabilistic nature of the quantum theory. In the absence of this term, the time derivative breaks the Kähler structure $(\mathcal{G},\Omega,\mathcal{J})_{\mathscr{P}}$, and norm conservation is not guaranteed. This is an already known fact for the Schrödinger equation in curved spacetimes [HS15, HS17, HSU19, EHS21].

- The remainder of the term is a self-adjoint operator. As such, this term can be interpreted as a correction to the quantum part of the dynamics of the system induced by the curvature of the spacetime. We can characterize this effect as the result of an unstable vacuum.

**Particle Creation in a Dynamical Spacetime**

One of the main novelties of our approach, suggested also in similar terms by the recent work [KM23], is that it provides a dynamical equation that describes particle creation in an expanding universe. Particle creation in expanding universes has been traditionally studied by the following arguments, see e.g. [BD82, Wal94, PT09]. First, we must assume that we can describe in our system the asymptotic past $\mathscr{I}^-$ and future $\mathscr{I}^+$ in which the vacuum can be described without ambiguities. In these asymptotic limits, we take a basis of solutions of the classical equations of motion with definite positive energy that serve as a basis for the quantum Hilbert spaces $\mathcal{H}_{\mathscr{I}^-}$ and $\mathcal{H}_{\mathscr{I}^+}$. The time dependence of the spacetime metric induces a parametric dependence on the Hamiltonian of the classical theory, as is the usual case in a parametric theory, see *Remark* 3.1. This implies that the basis of $\mathcal{H}_{\mathscr{I}^-}$ has, formally, a nontrivial overlap with the basis of $\mathcal{H}_{\mathscr{I}^+}$, and this translates into a nontrivial overlap of creation and annihilation operators

$$a^{\mathbf{x}}_{\mathscr{I}^-} = \alpha(x) a^{\mathbf{x}}_{\mathscr{I}^+} + \beta^*(x) a^{\dagger,\mathbf{x}}_{\mathscr{I}^+} \tag{7.11}$$

with $|\alpha(x)|^2 - |\beta(x)|^2 = 1$. This kind of relation is named a Bogoliubov transformation. This transformation preserves the commutation relations $[a^{\dagger,\mathbf{x}}, a^{\mathbf{y}}] = \Delta^{\mathbf{xy}}$ but spoils the stability of the vacuum. That is, if $a^{\mathbf{x}}_{\mathscr{I}^-}|0\rangle_{\mathscr{I}^-} = 0$, then $a^{\mathbf{x}}_{\mathscr{I}^+}|0\rangle_{\mathscr{I}^-} \neq 0$ and vice versa. Since there is a particle interpretation available in $\mathscr{I}^{\pm}$, this is easily interpreted as particle creation in the dynamical universe. This kind of analysis is also responsible for Hawking radiation in collapsing black holes and the Unruh effect for non-inertial observers in flat Minkowski spacetimes, see [Wal94].

In our case, the situation is different. The ultimate reason to recur to asymptotic states in the previous arguments is to keep under control the normalization of the states with well-established bases of the corresponding Hilbert spaces. In our case, this normalization is built-in and automatically introduced by the connection. Also, the relation of the classical basis with the asymptotic quantum Hilbert spaces is implemented in our formalism by the compatibility condition $\nabla_t \mathcal{Q}(f) = \mathcal{Q}(\partial_t f)$ detailed in (5.75).

In our case, we describe instantaneous vacuum states that fulfill the equation (5.40) or equivalently, are annihilated by the creation and annihilation operators parallel transported by $\nabla_t a^{\mathbf{x}}_t = 0$. To put this argument in equations, consider that at initial time $t = 0$, creation and annihilation operators are provided by $a^{\dagger,\mathbf{x}}_0 = \phi^{\mathbf{x}}$ and $a^{\mathbf{y}}_0 = \Delta^{\mathbf{y}x} \partial_{\phi^x}$. At this point in time, the vacuum phase $\Psi_0 = 1$, and thus the vacuum and reference $\Psi_r$ sections coincide[2]. However, the covariant derivatives of these operators are

$$\nabla_t \phi^{\mathbf{x}} = \frac{1}{2}\big(\phi^y K_{yz}\dot{\Delta}^{z\mathbf{x}} - \dot{\Delta}^{\mathbf{x}y}\partial_{\phi^y} + \phi^y(\dot{\delta}^{\circ}\delta_{\circ})^{\mathbf{x}}_y - \phi^y K_{yz}(\dot{\delta}^{\circ}\delta_{\circ})^z_u\Delta^{u\mathbf{x}} +$$
$$\Delta^{\mathbf{x}y}(\dot{\delta}^{\circ}\delta_{\circ})^z_y\partial_{\phi^z} + (\dot{\delta}^{\circ}\delta_{\circ})^{\mathbf{x}}_y\Delta^{yz}\partial_{\phi^z}\big)$$

$$\nabla_t(\Delta^{\mathbf{x}y}\partial_{\phi^y}) = \frac{1}{2}\big(\dot{\Delta}^{\mathbf{x}y}\partial_{\phi^y} - \phi^y K_{yz}\dot{\Delta}^{z\mathbf{x}} - \Delta^{\mathbf{x}y}(\dot{\delta}^{\circ}\delta_{\circ})^z_y\partial_{\phi^z} + (\dot{\delta}^{\circ}\delta_{\circ})^{\mathbf{x}}_y\Delta^{yz}\partial_{\phi^z} +$$
$$\phi^y(\dot{\delta}^{\circ}\delta_{\circ})^{\mathbf{x}}_y + \phi^y K_{yz}(\dot{\delta}^{\circ}\delta_{\circ})^z_u\Delta^{u\mathbf{x}}\big)$$

---

[2] For a discussion of the difference between these two sections, we refer to chapter 4, for example, the discussion below (4.21)

As we can see[3], the appearance of the Bogoliubov transformations also stems from the parametric dependence on time of the operators. In terms of creation and annihilation operators, this equation becomes

$$\partial_t a_t^{\mathbf{x}} = \frac{1}{2}(a_t^{\dagger,y} - a_t^y)\dot{K}^z_y\Delta^{\mathbf{x}}_z - \frac{1}{2}(\dot{\delta}^{\circ}\delta_{\circ})^{\mathbf{x}}_y(a_t^y + a_t^{\dagger,y}) \tag{7.12}$$

with $\partial_t a_t^{\dagger \mathbf{x}} = (\partial_t a_t^{\mathbf{x}})^{\dagger}$.

In summary, in the simplifying case of null shift, the formalism developed throughout this thesis provides a dynamical equation that mixes creation and annihilation operators during the evolution. Moreover, in spite of its simplicity, this case describes physically relevant cosmological models as we will see below. To our knowledge, the study of the phenomenon of particle creation in dynamical spacetimes has typically been studied using asymptotically flat spacetime states [BD82, Wal94, PT09]. Some authors have hinted at the need for a correction to the Schrödinger equation [AA15], and more recently, others have linked these modifications to a connection generating this effect [KM23]. However, the treatment of the ambiguities in selecting this connection and the specific dynamical equation that drives particle creation is a novel aspect of this framework.

## 7.3 Quantum Klein-Gordon theory in FLRW spacetimes

As we anticipated in the introduction, the analysis of the null shift case applies when we particularize our study to the Klein-Gordon Theory on a FLRW spacetime. For simplicity, we may express the metric (7.1) in terms of a new coordinate

$$\chi = \int_0^r \frac{d\tau}{\sqrt{1 - k\tau^2}} \tag{7.13}$$

---

[3]Using these expressions, it is straightforward to see that (5.78) follows using $\varphi^{\mathbf{x}} = \frac{\phi^{\mathbf{x}} + \Delta^{\mathbf{x}y}\partial_{\phi^y}}{\sqrt{2}}$ and $\pi_{\mathbf{x}} = -i\frac{K_{\mathbf{x}y}\phi^y - \partial_{\phi^{\mathbf{x}}}}{\sqrt{2}}$

Then the pseudo-Riemannian metric is written as $-dt^2 + \mathbf{h}$ and

$$\mathbf{h} = \begin{cases} a^2(t)\left[d\chi^2 + \sin^2\chi(d\theta^2 + \sin^2\theta d\phi^2)\right] & \text{with } \chi \in [0, 2\pi) \text{ if } k = 1 \\ a^2(t)\left[dx^2 + dy^2 + dz^2\right] & \text{with } x, y, z \in \mathbb{R}^3 \text{ if } k = 0 \\ a^2(t)\left[d\chi^2 + \sinh^2\chi(d\theta^2 + \sin^2\theta d\phi^2)\right] & \text{with } \chi \in [0, \infty) \text{ if } k = -1 \end{cases} \quad (7.14)$$

From these sets of coordinates, it follows that the spatial Cauchy hypersurfaces $\Sigma$ represent a spherical space for $k = 1$, flat for $k = 0$, and hyperbolic for $k = -1$. The complex structure $\mathcal{J}_{\mathcal{M}_F}$ in these cases is also studied in [CCQ04]. Recall that in these cases the lapse $N = 1$ while the shift vector $\vec{N} = 0$ is null. To complete the picture, we need to compute the covariant Laplacian $D^i D_i = \frac{1}{\sqrt{h}}\partial_i(h^{ij}\sqrt{h}\partial_j) = \frac{1}{a^2(t)}\nabla^2$ and the volume factor $\sqrt{h} = a^3(t)\mathsf{Vol}_k$ with

$$\mathsf{Vol}_k, \nabla^2 = \begin{cases} \sin^2\chi\sin\theta, & \frac{1}{\sin^2\chi}\left[\partial_\chi(\sin^2\chi\partial_\chi) + \frac{1}{\sin\theta}\left[\partial_\theta(\sin\theta\partial_\theta) + \partial_\phi^2\right]\right] & k = 1 \\ 1, & \partial_x^2 + \partial_y^2 + \partial_z^2 & k = 0 \\ \sinh^2\chi\sin\theta, & \frac{1}{\sinh^2\chi}\left[\partial_\chi(\sinh^2\chi\partial_\chi) + \frac{1}{\sin\theta}\left[\partial_\theta(\sin\theta\partial_\theta) + \partial_\phi^2\right]\right] & k = -1 \end{cases}$$
$$(7.15)$$

From these expressions, and using only the definition (2.6), we can compute the operator

$$(\dot{\delta}{\circ}\delta_\circ)^{\mathbf{x}}_{\mathbf{y}} = 3\frac{\dot{a}(t)}{a(t)}\delta^{\mathbf{x}}_{\mathbf{y}}$$

The factor $\frac{\dot{a}(t)}{a(t)}$ is a measurable parameter in cosmological models called the Hubble parameter [Pad10].

The entries of the complex structure are computed using the classical Klein-Gordon Hamiltonian (3.60) in this background

$$H_{KG} = \int_\Sigma \frac{a^3(t)}{2}\left(\pi^2(u) - \frac{\varphi(u)\nabla^2\varphi(u)}{a^2(t)} + m^2\varphi(u)^2\right)\mathsf{Vol}_k(u)du^3. \quad (7.16)$$

Defining $M(t) = a(t)m$, the parameters of the quantum theory are

$$\Theta^{\mathbf{x}}_{\mathbf{y}} = \frac{\delta^{\mathbf{x}}_{\mathbf{y}}}{a^2}(-\nabla^2 + M^2), \qquad \Delta^{\mathbf{x}}_{\mathbf{y}} = a\delta^{\mathbf{x}}_{\mathbf{y}}\left(\sqrt{-\nabla^2 + M^2}\right)^{-1},$$
$$K^{\mathbf{x}}_{\mathbf{y}} = \frac{1}{a}\delta^{\mathbf{x}}_{\mathbf{y}}\sqrt{-\nabla^2 + M^2} \qquad \hat{H} = \frac{1}{a}\phi^x\left(\sqrt{-\nabla^2 + M^2}\right)^y_x\partial_{\phi^y}. \quad (7.17)$$

Thus we can compute

$$\dot{\Delta}^{\mathbf{xy}} = \Delta^{\mathbf{xy}}\frac{\dot{a}}{a}\left(4 - \frac{M^2}{M^2 - \nabla^2}\right) = \left[\dot{a}a^3\mathsf{Vol}_k\left(4 - \frac{M^2}{M^2 - \nabla^2}\right)\frac{1}{\sqrt{M^2 - \nabla^2}}\right]^{\mathbf{xy}}$$
$$\dot{K}_{\mathbf{xy}} = -K_{\mathbf{xy}}\frac{\dot{a}}{a}\left(4 - \frac{M^2}{M^2 - \nabla^2}\right) = \left[-\frac{\dot{a}}{a^3\mathsf{Vol}_k}\left(4 - \frac{M^2}{M^2 - \nabla^2}\right)\sqrt{M^2 - \nabla^2}\right]_{\mathbf{xy}} \quad (7.18)$$

Gathering together all terms of the connection, we obtain the modified Schrödinger equation

$$i\left[\partial_t + \frac{\dot{a}}{a}\phi^y\left(2 - \frac{1}{2}\frac{M^2}{M^2 - \nabla^2}\right)^x_y\partial_{\phi^x}\right]\Psi =$$
$$\dot{a}a^3\Big[\text{Vol}_k\Big(\frac{1}{2} + \frac{1}{4}\frac{M^2}{M^2 - \nabla^2}\Big)\frac{1}{\sqrt{M^2 - \nabla^2}}\Big]^{xy}\partial_{\phi^x}\partial_{\phi^y}\Psi +$$
$$-\frac{\dot{a}}{a^3}\Big[\frac{1}{\text{Vol}_k}\Big(\frac{1}{2} + \frac{1}{4}\frac{M^2}{M^2 - \nabla^2}\Big)\sqrt{M^2 - \nabla^2}\Big]_{xy}\phi^x\phi^y\Psi +$$
$$\frac{1}{a}\phi^x\left(\sqrt{-\nabla^2 + M^2}\right)^y_x\partial_{\phi^y}\Psi \quad (7.19)$$

Hence, in this scenario, we obtain a modification of the propagator of the theory proportional to the Hubble parameter $\frac{\dot{a}}{a}$.

**Example 7.1.** *(Static FLRW spacetimes)*
*In the particular case of a constant $a$, the connection terms cancel, and we recover the usual Schrödinger equation. We can express the solution of that equation with an integral kernel*

$$\Psi(\phi, t) = \int D\beta(\sigma)\exp[\phi^x(\delta_{xy} - itK)_{xy}\bar{\sigma}^y]\Psi(\sigma, 0) \quad (7.20)$$

*In this solution, we interpret $K_{\mathbf{xy}}$ as the energy of the system.*

### 7.3.1 Particle production

Particle production is usually expressed in terms of creation and annihilation operators. In this case, equation (7.12) is written

$$\partial_t\begin{pmatrix} a^{\mathbf{x}} \\ a^{\dagger,\mathbf{x}} \end{pmatrix} = \frac{\dot{a}}{a}\left[\frac{1}{2}\begin{pmatrix} -1 & 1 \\ 1 & -1 \end{pmatrix}\frac{M^2}{M^2 - \nabla^2} + \begin{pmatrix} -1 & -2 \\ -2 & -1 \end{pmatrix}\right]\begin{pmatrix} a^{\mathbf{x}} \\ a^{\dagger,\mathbf{x}} \end{pmatrix} \quad (7.21)$$

If we focus on the flat case with $\Sigma = \mathbb{R}^3$, we can use the Fourier transform $f(k) = \int d^3x \exp(i\vec{x}\cdot\vec{k})f(x)$ as a change of coordinates of $\Sigma$. If we express creation and annihilation operators in the Fourier-transformed domain, the Laplacian becomes $-\nabla^2 \to \vec{k}^2$, a multiplicative factor. Using this, we can rewrite the equation in terms of the frequency $\omega_{\vec{k},t}^2 = \vec{k}^2/M^2(t) = \vec{k}^2/(m^2 a^2(t))$

$$\partial_t\begin{pmatrix} a^{\vec{k}} \\ a^{\dagger,\vec{k}} \end{pmatrix} = \frac{\dot{a}}{a}\left[\frac{1}{2}\begin{pmatrix} -1 & 1 \\ 1 & -1 \end{pmatrix}\frac{1}{1 + \omega_{\vec{k},t}^2} + \begin{pmatrix} -1 & -2 \\ -2 & -1 \end{pmatrix}\right]\begin{pmatrix} a^{\vec{k}} \\ a^{\dagger,\vec{k}} \end{pmatrix} \quad (7.22)$$

As we mentioned at the end of section 7.2.2, the construction of QFT in curved spacetimes is usually done resorting to a basis of the Hilbert space provided by classical solutions of the equations of motion. In that kind of analysis, there is a vast number of solutions for Bogoliubov coefficients in (7.11) for specific cosmological models in terms of the frequency $\omega_{\vec{k},t}$ [BD82, PT09, For21]. Using this equation, we just need to provide $a(t)$ and solve (7.22) to compare with these results. We plan to address this issue in future works.

# Chapter 8

# Conclusions and Future Work

> *Dicen que las flores se marchitan*
> *Pero yo tengo riegos árabes en el jardín,*
> *de mi mezquita*
>
> Javier Ibarra

## 8.1 Conclusions

In this thesis, we presented a detailed study of Hamiltonian QFTs of the scalar field on curved spacetimes. We developed a mathematical formalism to efficiently address the geometric aspects of these theories and studied the interplay between mathematical tools borrowed from Gaussian analysis and their physical interpretations.

More concisely, in chapter 2, we gathered together a set of results and methods to study the Hilbert spaces $L^2(\mathcal{N}'_{\mathbb{C}}, D\mu_c)$ of Gaussian integration theory in infinite-dimensional spaces. The main contribution of this thesis to that analysis is merely introducing a convenient notation to describe physical models later on. Also, in this work, the complex domain proved to be of utmost importance in the study of the quantum theory of the scalar field. For this reason, we presented a review of the aforementioned results in the complex domain, departing from the standard presentations of the field. See e.g. [Hid&93, Hu16].

Some of the features that are usually regarded as genuine quirks of QFT are already found in the mathematical theory of Gaussian analysis. For instance, the Wick theorem, which deals with vacuum expectation values in QFT, and Feynman diagrams, which are considered a computational tool for perturbative quantum theories, see e.g. [Nai05]. However, in section 2.6, we presented those ingredients as connatural tools to Gaussian analysis.

In chapter 3, we started a project to understand to what extent we can identify some of these ingredients already at the classical level. For that matter, we extended the Koopman-van Hove (KvH) formalism of classical mechanics to the case of Classical Statistical Field Theory (CSFT). That is, we used prequantization theory to represent the kinematics of the classical theory and studied the evolution in this representation.

The main conclusions that we derived from that analysis are the following:

- Let $(\mathcal{M}_F, \omega_F)$ be the phase space of a classical field theory of the scalar field. To properly describe (Gaussian) statistical states, $\mathcal{M}_F$ is modeled over strong Duals of Nuclear-Fréchet (DNF) spaces that we denoted $\mathcal{N}'$ in this thesis. Also, we described the classical Gaussian statistical state through the complex structure $J_F$ of (3.21), inducing a Kähler structure $(\omega_F, \mu_F, J_F)$.

  This kind of complex structure has been widely known as a tool in the field of QFT on curved spacetimes, see e.g. [AMA75, Wal94, CCQ04, MO21]. The novelty in this proposal is that we introduce it in a purely classical context. The need for its appearance is the consistency of the statistical description of classical states.

- The KvH dynamics imposes restrictions in the form of $J_F$. In section 3.4.4, we provided the prescription $J_F = |X_H|^{-1} X_H$, with $H$ being the Hamiltonian of the theory, already at the classical level. This coincides with the standard quantum prescription for static and stationary spacetimes [CCQ04] but is not limited to that case.

Built upon prequantum theory, in chapter 4, we developed a quantization program. This quantization program is based on geometric quantization [Woo97] and the choice of an ordering prescription. For that matter, we systematically used tools of Gaussian analysis to build quantization mappings $Q$. The novelties of this analysis may be summarized as follows:

- Geometric quantization prescriptions produce different representations of QFT. Among them, the most interesting ones we found are the holomorphic, Schrödinger, antiholomorphic, and field-momentum representations. The concrete relations among these quantization procedures can be described by integral transforms and are summarized by the diagram

$$
\begin{array}{ccc}
\overset{\text{holomorphic}}{\left(L^2_{Hol}(\mathcal{N}'_{\mathbb{C}}, D\mu_c), \mathcal{Q}\right)} & \underset{\tilde{\mathcal{F}}^{-1}}{\overset{\tilde{\mathcal{F}}}{\rightleftarrows}} & \overset{\text{Antiholomorphic}}{\left(L^2_{\overline{Hol}}(\mathcal{N}'_{\mathbb{C}}, D\nu_c), \overline{\mathcal{Q}}\right)} \\[2mm]
\tilde{\mathcal{B}} \Big\uparrow \Big\downarrow \tilde{\mathcal{B}}^{-1} & & \overline{\tilde{\mathcal{B}}} \Big\uparrow \Big\downarrow \overline{\tilde{\mathcal{B}}}^{-1} \\[2mm]
\underset{\text{Schrödinger}}{\left(L^2(\mathcal{N}', D\mu), \mathcal{Q}_s\right)} & \underset{\mathcal{F}^{-1}}{\overset{\mathcal{F}}{\rightleftarrows}} & \underset{\text{Momentum-field}}{\left(L^2(\mathcal{N}', D\nu), \overline{\mathcal{Q}}_m\right)}
\end{array}
$$

  This diagram, in turn, introduces the Fourier transform $\mathcal{F}$ of section 4.3.2 for the infinite-dimensional case. To our knowledge, this transform is a new feature of this formalism and is not related to other Fourier-like transforms, such as the one presented in [Hid&93].

- Wick and Weyl orders, the Moyal product, and other $\star$ products admit a straightforward characterization using tools of Gaussian analysis. Moreover, in sec-tion 4.4, we used Gaussian analysis tools to provide concrete analytical descriptions of the domain of the quantization mappings $Q$.

On the physical side of the discussion, this analysis shed some light on the regularization needs of the theories we are dealing with. To showcase this, take a classical observable $f : \mathcal{M}_F \to \mathbb{R}$ and Weyl Holomorphic quantization $\mathcal{Q}_{Weyl}$. As we pointed out in that section, the quantization procedure produces a bounded operator $\mathcal{Q}_{Weyl}(f)$ when $f$ is square integrable under some Gaussian measure $\mathcal{W}$ described in (4.46). This kind of argument imposes conditions on the classical observable $f$ and not on the quantum part of the theory. Moreover, in section 3.5, we already found a simple example that CSFT already needs a point-splitting regularization procedure in order to be described consistently.

The dynamical aspects of QFT were dealt with in chapter 5. A naive postulate of the Schrödinger equation on curved spacetimes leads to norm losses in the evolution, jeopardizing the probabilistic interpretation of Hamiltonian quantum theories. In order to avoid this effect, we followed [Kib79] and geometrized QFT. We presented this geometrization in different sets of *s.q.* coordinates that we were able to introduce using white noise analysis, a particular branch of Gaussian analysis. The compatibility of time evolution with the geometric structure described here led us to the central result of this thesis:

- From geometric considerations, we obtain a modification to the Schrödinger equation via a covariant time derivative $\nabla_t$ with connection $\Gamma_t$ such that

$$i\hbar(\partial_t + \Gamma_t)\Psi = \hat{H}\Psi.$$

  This connection $\Gamma_t$ has an anti-self-adjoint part that compensates the norm losses that the self-adjoint quantum Hamiltonian $\hat{H}$ is not able to compensate.

Geometric considerations alone are not able to uniquely fix the connection term $\Gamma_t$. In order to select a criterion that fixes this connection, in this thesis, we recurred to the hybrid quantum-classical coupling with gravity in chapter 6. In that chapter, through a series of postulates, we reviewed the content of [Alo&24a] and used Geometrodynamics to minimally couple a QFT of matter with classical General Relativity (GR). The scope of that chapter is slightly more general than the particular application to the main project of this thesis. For this reason, the conclusion that we extracted from that development was:

- In order to be compatible with the hybrid description of quantum matter coupled to classical gravity, the covariant time derivative $\nabla_t$ of the modified Schrödinger equation must be compatible with the quantization mapping $Q$ in such a way that

$$\nabla_t Q(f) = Q(\partial_t f).$$

This condition uniquely fixes the connection term $\Gamma_t$. In section 5.5, we derive this connection to be (5.77) for Weyl quantization in the holomorphic representation. This result straightforwardly translates to every other representation.

Finally, in chapter 7, we applied the theory derived in previous chapters to the case of quantum Klein-Gordon theory on a FLRW spacetime. In that scenario, we were able to discuss physically relevant features that we summarize as follows:

- The modified Schrödinger equation, obtained from the conservation of geometric structures of QFT, introduces new phenomenology in the system. In particular, for the case of Klein-Gordon theory on a FLRW spacetime, the phenomenon of particle creation in expanding universes is introduced through a dynamical differential equation (7.21). This example is of utmost importance in physics as these spacetimes are the only ones compatible with the cosmological principle and are the state-of-the-art model compatible with experiments.

To our knowledge, the study of the phenomenon of particle creation in dynamical spacetimes has been successfully addressed by always recurring to asymptotic flat states of the spacetime [BD82, Wal94, PT09]. Some authors hinted at the need for a correction to the Schrödinger equation [AA15] and, recently, others even identified it with a connection generating this effect [KM23]. Nonetheless, the treatment of the ambiguities present in the choice of the connection and the concrete dynamical equation that generates particle creation is a novelty of this thesis.

In a nutshell, the geometric path that we chose as a guideline to study Hamiltonian representations of QFT on curved spacetimes led us to the systematic application of Gaussian analysis tools to define and reinterpret aspects in both the mathematical and physical parts of the discussion. Firstly, we developed a mathematically coherent formalism that describes quantization. Secondly, we proposed modifications to the Schrödinger equation that preserve mathematical coherence beyond the initial time. Finally, we explored the phenomenology of these modifications in the case of Klein-Gordon theory in Friedman-Lemaître-Robertson-Walker (FLRW) spacetimes.

## 8.2 Future Lines of Research

This thesis lays the groundwork for further exploration of several topics that were not fully addressed here. Although significant effort was made to handle the mathematical challenges of integration theory in infinite-dimensional spaces, and various nontrivial results emerged from this analysis, the physical soundness of the matter sector was left largely unexamined. The focus here was primarily on the scalar field in curved backgrounds, with an emphasis on the minimally coupled Klein-Gordon theory in highly symmetric FLRW spacetimes. Future research could broaden the scope of this work by addressing these simplifications. Some potential directions for this expanded research include:

- **Quantum photons in cosmological spacetimes**. The first generalization that stems from this work is the treatment of Abelian gauge theories. More precisely, the modified Schrödinger equation of the scalar field in FLRW, given in (5.42), introduces non-trivial contributions to the propagator of the theory. What are the modifications in the propagator for a photon? Furthermore, photons coming from the Cosmic Microwave Background, which we are detecting at this moment, have been traveling through space since the recombination epoch. This is, these photons should feel the integrated effect of the corrected propagator for more than 13 billion years [Bau22]. What are the measurable implications of this correction?

- **Quantum non-Abelian Yang-Mills theories in the Schrödinger formulation**. In general, integration problems in infinite-dimensional spaces must be addressed to unveil non-perturbative effects of non-Abelian Yang-Mills theories. Usually, these problems are treated using lattice gauge theories, which involve discretizing the Euclidean version of space-time using a lattice, see e.g. [Nai05, Sch14]. In this procedure, the integrals become finite and tractable in computer simulations. Other searches for non-perturbative effects relying on integration theory can be found in [Str19]. Closely related, stochastic quantization also offers some insight into the formalization of path integral formulations [Mas08].

  Most of the approaches presented above are described in the covariant formalism of QFT. This is because the canonical Hamiltonian formulation possesses the so-called Gribov problem [Hei96]. That is, the configuration space of non-Abelian Yang-Mills theories admits configurations with a non-trivial first integral Chern class. This implies that there is a discrete global quantity, denoted the winding number, that generates infinitely many separated and discrete sectors [Str19, GR20]. Transitions among these sectors are not reachable by perturbative methods and are usually studied by semiclassical solutions denoted instantons, which are the source of the so-called Strong CP problem, see [Alo&22] and references therein.

  Gaussian analysis may offer a perspective in the study of some of these problems, already in the case of flat spacetime. Furthermore, confinement is a well-known feature that these theories should present, and particle interpretations are not valid in this context. For these reasons, we ask: What is the analogous Wiener-Ito decomposition in this confined case? What are the descriptions in the analogous holomorphic, Schrödinger, antiholomorphic, or field-momentum representations? Is there any analogous Fourier transform in this case?

- **Fermions in the canonical formalism**. The treatment of fermions in the path integral formulation is introduced using Grassmann variables. This problem can be addressed using stochastic quantization [Alb&20]. However, little is known about how to apply the tools of our formalism to fermions [Oec14].

- **QFT in richer curved backgrounds**. The cosmological examples presented here exhibit a lot of symmetries. An interesting problem to address is how to model quantum matter as a probe of classical black hole solutions. This can be understood by introducing the Schwarzschild metric as background. Other canonical approaches [HS15, Sch18] have found that quantum probes of matter are unable to reach the singularity; thus, the spacetime is quantum complete. Are we able to reproduce similar results in our case?

Other interesting examples are quantum probes in collapsing classical models, such as inhomogeneous but isotropic dust collapse models. In these cases, the background metric is called the Lemaître-Tolman-Bondi spacetime [BK95, VW11]. The question of the previous paragraph can be enlarged to the case of singularities generated dynamically. Moreover, these collapses also form Killing horizons [She24]; this is the arena where Hawking radiation arises [Wal94]. As we pointed out below (7.5), our formalism hits some limiting cases when the spacetime develops horizons. How can we explore these phenomena in our framework?

One step further would be to consider the hybrid coupling to gravity chapter 6 and investigate the quantum dust collapsing in a classical background and study singularity formation from quantum sources.

- **Semiclassical WKB limit of Quantum gravity coupled to matter**. The semiclassical WKB limits of quantum gravity have already been studied, see e.g. [Kie12]. Can we derive the formalism of chapter 6 from a semiclassical WKB limit of Wheeler-deWitt quantum gravity? Also, the description in that chapter can be rephrased in terms of Ashtekar self-dual variables for gravity, which are the basis of loop quantum gravity [Rov04, Kie12]. Can we describe the hybrid theory in Ashtekar variables as a WKB limit of loop quantum gravity?

## 8.3  Conclusiones

En esta tesis hemos presentado un estudio detallado de las Teorías Cuánticas de Campos (QFT) Hamiltonianas del campo escalar en espaciotiempos curvados. Hemos desarrollado un formalismo matemático para abordar de manera eficiente los aspectos geométricos de estas teorías y estudiado la relación entre las herramientas matemáticas del análisis gaussiano en dimension infinita y sus interpretaciones físicas.

Más concretamente, en el Capítulo 2, presentamos un conjunto de resultados y métodos para estudiar los espacios de Hilbert $L^2(\mathcal{N}'_{\mathbb{C}}, D\mu_c)$ de la teoría de integración gaussiana en espacios de dimensión infinita. La principal contribución de esta tesis a ese análisis se limita a la introducción de una notación conveniente para describir modelos físicos más adelante. Además, en este trabajo, las funciones de dominio complejo han demostrado ser de suma importancia en el estudio de la teoría cuántica del campo escalar. Por esta razón, hemos presentado una revisión de los resultados en dominio complejo, alejándonos del tratamiento estándar del campo de estudio. Véase e.g. [Hid&93, Hu16].

Algunas de las características que suelen considerarse como peculiaridades de la teoría cuántica ya se encuentran en el desarrollo matemático del análisis gaussiano. Por ejemplo, el teorema de Wick, que trata con valores esperados de vacío en QFT, y los diagramas de Feynman, que se consideran una herramienta de la teorías cuántica perturbativa, ver por ejemplo [Nai05]. Sin embargo, en la Sección 2.6, hemos visto esos ingredientes como herramientas connaturales al análisis gaussiano.

En el Capítulo 3, hemos iniciado un proyecto para entender en qué medida podemos identificar algunos de estos ingredientes ya a nivel clásico. Para ese propósito, ampliamos el formalismo de Koopman van-Hove (Koopman-van Hove (KvH)) de la mecánica clásica al caso de la Teoría estadística clásica de campos (Classical Statistical Field Theory (CSFT)). Es decir, usamos la teoría de pre-cuantización para representar la cinemática de la teoría clásica y estudiamos la evolución en esta representación. Las principales conclusiones que derivamos de ese análisis son las siguientes:

- Sea $(\mathcal{M}_F, \omega_F)$ el espacio de fases de una teoría clásica de campo escalar. Para describir adecuadamente los estados estadísticos (gaussianos), $\mathcal{M}_F$ se modela sobre espacios Duales a espacios Fréchet-Nuclear (strong Duals of Nuclear-Fréchet (DNF)) que denotamos $\mathcal{N}'$ en esta tesis. Además, describimos el estado estadístico clásico gaussiano a través de la estructura compleja $J_F$ de (3.21), induciendo una estructura de Kähler $(\omega_F, \mu_F, J_F)$.

  Este tipo de estructura compleja ha sido usualmente reconocida como una herramienta en el campo de la QFT en espacios-tiempo curvos, ver e.g. [AMA75, Wal94, CCQ04, MO21]. La novedad en esta propuesta es que la introducimos en un contexto puramente clásico. La necesidad de su aparición radica en la consistencia de la descripción estadística de estados clásicos.

- La dinámica de KvH impone restricciones en la forma de $J_F$. En la Sección 3.4.4, proponemos la prescripción $J_F = |X_H|^{-1}X_H$, donde $H$ es el Hamiltoniano de la teoría, ya a nivel clásico. Esto coincide con la prescripción cuántica estándar para espaciotiempos estáticos y estacionarios [CCQ04], pero no se limita a ese caso.

Sobre la base de la teoría de pre-cuantización, en el Capítulo 4, hemos desarrollado un programa de cuantización. Este programa de cuantización se basa en la cuantización geométrica [Woo97] y la elección de una prescripción de ordenamiento. Para ello, utilizamos sistemáticamente herramientas del análisis gaussiano para construir *mappings* de cuantización $Q$. Las novedades de este análisis se pueden resumir de la siguiente manera:

- Las prescripciones de cuantización geométrica producen diferentes representaciones de la QFT. Entre ellas, las más interesantes que encontramos son las representaciones holomorfa, de Schrödinger, anti-holomorfa y de campo-momento. Las relaciones concretas entre estos procedimientos de cuantización se pueden describir mediante transformaciones integrales y se resumen en el diagrama

$$
\begin{array}{ccc}
\overset{\text{Holomorfa}}{\left(L^2_{Hol}(\mathcal{N}'_\mathbb{C}, D\mu_c), \mathcal{Q}\right)} & \underset{\tilde{\mathcal{F}}^{-1}}{\overset{\tilde{\mathcal{F}}}{\rightleftarrows}} & \overset{\text{Antiholomorfa}}{\left(L^2_{\overline{Hol}}(\mathcal{N}'_\mathbb{C}, D\nu_c), \overline{\mathcal{Q}}\right)} \\
\tilde{\mathcal{B}}\uparrow\,\downarrow\tilde{\mathcal{B}}^{-1} & & \tilde{\overline{\mathcal{B}}}\uparrow\,\downarrow\tilde{\overline{\mathcal{B}}}^{-1} \\
\underset{\text{Schrödinger}}{\left(L^2(\mathcal{N}', D\mu), \mathcal{Q}_s\right)} & \underset{\mathcal{F}^{-1}}{\overset{\mathcal{F}}{\rightleftarrows}} & \underset{\text{Campo-Momento}}{\left(L^2(\mathcal{N}', D\nu), \overline{Q}_m\right)}
\end{array}
$$

  Este diagrama, a su vez, introduce la transformada de Fourier $\mathcal{F}$ de la Sección 4.3.2 para el caso de dimensión infinita. Hasta donde sabemos, esta transformación es una característica nueva de este formalismo y no está relacionada con otras transformadas del tipo Fourier, como la presentada en [Hid&93].

- Los órdenes de Wick y Weyl, el producto de Moyal y otros productos $\star$ admiten una caracterización directa usando herramientas del análisis gaussiano. Además, en la Sección 4.4, utilizamos herramientas del análisis gaussiano para proporcionar descripciones analíticas concretas del dominio de los *mappings* de cuantización $Q$.

  En la parte física del estudio, este análisis muestra aspectos interesantes de las necesidades de regularización de las teorías con las que estamos tratando. Por ejemplo, tomemos un observable clásico $f : \mathcal{M}_F \to \mathbb{R}$ y la cuantización holomorfa de Weyl $\mathcal{Q}_{Weyl}$. Como señalamos en esa sección, el procedimiento de cuantización produce un operador acotado $\mathcal{Q}_{Weyl}(f)$ cuando $f$ es de cuadrado integrable bajo cierta medida gaussiana $\mathcal{W}$ descrita en (4.46). Este tipo de argumento impone condiciones sobre el observable clásico $f$ y no sobre la parte cuántica de la teoría. Por otra parte, en la sección 3.5, ya encontramos un caso simple en el que la CSFT ya necesita un procedimiento de regularización de división puntual para ser descrita de manera coherente.

Los aspectos dinámicos de la QFT se han tratado en el Capítulo 5. Postular directamente la ecuación de Schrödinger en espacios-tiempo curvos lleva a pérdidas de norma en la evolución, poniendo en peligro la interpretación probabilística de las teorías cuánticas hamiltonianas. Para evitar este efecto, siguiendo[Kib79], geometrizamos la QFT. Presentamos esta geometrización en diferentes conjuntos de coordenadas en *segunda cuantización* (*s.q.*) que hemos introducido usando el análisis de ruido blanco, una rama particular del análisis gaussiano. La compatibilidad de la evolución temporal con la estructura geométrica descrita aquí nos ha llevado al resultado central de esta tesis:

- A partir de consideraciones geométricas, obtenemos una modificación a la ecuación de Schrödinger a través de una derivada temporal covariante $\nabla_t$ con conexión $\Gamma_t$ tal que

$$i\hbar(\partial_t + \Gamma_t)\Psi = \hat{H}\Psi.$$

  Esta conexión $\Gamma_t$ tiene una parte anti-autoadjunta que compensa las pérdidas de norma que el Hamiltoniano cuántico autoadjunto $\hat{H}$ no es capaz de compensar.

Las consideraciones geométricas por sí solas no son capaces de fijar de manera única el término de conexión $\Gamma_t$. Para seleccionar un criterio que fije esta conexión, en esta tesis recurrimos al acoplamiento híbrido cuántico-clásico con la gravedad en el Capítulo 6. En ese capítulo, a través de una serie de postulados, revisamos el contenido de [Alo&24a] y utilizamos la geometrodinámica para acoplar mínimamente una QFT de materia con la Relatividad General (General Relativity (GR)) clásica. El contenido de ese capítulo es ligeramente más general que la aplicación particular al proyecto principal de esta tesis. Por esta razón, la conclusión que hemos extraído de ese desarrollo es simplemente:

- Para ser compatible con la descripción híbrida de materia cuántica acoplada a la gravedad clásica, la derivada temporal covariante $\nabla_t$ de la ecuación de Schrödinger modificada debe ser compatible con el *mapping* de cuantización $Q$ de tal manera que

$$\nabla_t Q(f) = Q(\partial_t f).$$

  Esta condición fija de manera unívoca el término de conexión $\Gamma_t$. En la Sección 5.5, derivamos que esta conexión debe ser (5.77) para la cuantización de Weyl en la representación holomorfa. Este resultado se puede traducir directamente a cualquier otra representación.

Finalmente, en el Capítulo 7, aplicamos la teoría desarrollada en los capítulos anteriores, al caso de la teoría cuántica de Klein-Gordon en un espaciotiempo Friedman-Lemaître-Robertson-Walker (FLRW). En este escenario, pudimos discutir características físicamente relevantes que resumimos de la siguiente manera:

- La ecuación de Schrödinger modificada, obtenida de la conservación de estructuras geométricas de la QFT, introduce nueva fenomenología en el sistema. En particular, para el caso de la teoría de Klein-Gordon en un espaciotiempo FLRW, el fenómeno de creación de partículas en universos en expansión se introduce a través de una ecuación diferencial dinámica (7.21). Este ejemplo es de suma importancia en física, ya que estos espacios-tiempo son los únicos compatibles con el principio cosmológico y conforman el actual paradigma compatible con los experimentos.

Hasta donde sabemos, el estudio del fenómeno de creación de partículas en espacios-tiempo dinámicos ha sido abordado con éxito recurriendo siempre a estados asintóticamente planos del espaciotiempo [BD82, Wal94, PT09]. Algunos autores han insinuado la necesidad de una corrección a la ecuación de Schrödinger [AA15] y, recientemente, otros incluso la han identificado con una conexión que genera este efecto [KM23]. No obstante, el tratamiento de las ambigüedades presentes en la elección de la conexión y la ecuación dinámica concreta que genera la creación de partículas es una novedad de esta tesis.

En resumen, el camino geométrico que elegimos como guía para estudiar representaciones Hamiltonianas de la QFT en espacios-tiempo curvados nos ha llevado a la aplicación sistemática de herramientas de análisis gaussiano para definir y reinterpretar aspectos tanto en las partes matemáticas como físicas de la discusión. Primero, hemos desarrollado un formalismo matemáticamente coherente que describe la cuantización. En segundo lugar, hemos propuesto modificaciones a la ecuación de Schrödinger que preservan la coherencia matemática más allá del tiempo inicial. Finalmente, exploramos la fenomenología de estas modificaciones en el caso de la teoría de Klein-Gordon en espaciotiempos Friedman-Lemaître-Robertson-Walker (FLRW).

# Appendix A

# Explicit Calculations with Integral Transforms

## A.1 Segal-Bargmann and Fourier transforms

In this appendix, we will prove that the Segal-Bargmann transforms and the Fourier transform conserve the quantization mappings of section 4.3.2. We also provide proof for some relations exposed in section 4.3.1.

### A.1.1 Schrödinger picture

Let $\tilde{\mathcal{B}}(\Psi_S) = \Psi_H$, then we get the relation between both pictures with

$$\tilde{\mathcal{B}}^{-1}(\Psi_H)(\varphi^{\mathbf{x}}) = \int D\mu_S(\pi^{\mathbf{x}})\Psi_H(\sqrt{2}\phi^{\mathbf{x}}) =$$

$$\sum_{n=0}^{\infty} \int D\mu_S(\pi^{\mathbf{x}})2^{\frac{n}{2}}\psi_{\vec{x}_n}^{(n)}[(\varphi - i\pi)^n]^{\vec{x}_n} =$$

$$\sum_{n=0}^{\infty} \psi_{\vec{x}_n}^{(n)} \frac{\partial^n}{\partial \xi_{\vec{x}_n}^n} \int D\mu_S(\pi^{\mathbf{x}})e^{\sqrt{2}\xi_z(\varphi - i\pi)^z}\bigg|_{\xi_{\mathbf{x}}=0} =$$

$$\sum_{n=0}^{\infty} \psi_{\vec{x}_n}^{(n)} \frac{\partial^n}{\partial \xi_{\vec{x}_n}^n} \left[ e^{\sqrt{2}\xi_z\varphi^z} \int D\mu_S(\pi^{\mathbf{x}})e^{-i\sqrt{2}\xi_z\pi^z} \right]_{\xi_{\mathbf{x}}=0} =$$

$$\sum_{n=0}^{\infty} \psi_{\vec{x}_n}^{(n)} \frac{\partial^n}{\partial \xi_{\vec{x}_n}^n} e^{\sqrt{2}\xi_x\varphi^x - \frac{\xi_u \Delta^{uv}\xi_v}{2}}\bigg|_{\xi_{\mathbf{x}}=0} =$$

$$\sum_{n=0}^{\infty} 2^{\frac{n}{2}} \psi_{\vec{x}_n}^{(n)} \frac{\partial^n}{\partial \tilde{\xi}_{\vec{x}_n}^n} e^{\xi_x\varphi^x - \frac{\tilde{\xi}_u \Delta^{uv}\tilde{\xi}_v}{4}}\bigg|_{\tilde{\xi}_{\mathbf{x}}=0} =$$

$$\sum_{n=0}^{\infty} 2^{\frac{n}{2}} \psi_{\vec{x}_n}^{(n)} : \varphi^n :|_{\frac{\vec{x}_n}{\Delta}}^{\vec{x}_n} = \Psi_S(\varphi^{\mathbf{x}}).$$

To compute $\tilde{\mathcal{B}}^{-1}\phi^x\tilde{\mathcal{B}}$, let us compute the transform of $\chi_x\phi^x\Psi_H(\phi^{\mathbf{x}})$. We get

$$\tilde{\mathcal{B}}^{-1}(\chi_x \phi^x \Psi_H)(\varphi^{\mathbf{x}}) = \int D\mu_S(\pi^{\mathbf{x}}) \chi_y \sqrt{2} \phi^y \Psi_H(\sqrt{2}\phi^{\mathbf{x}}) =$$

$$\sum_{n=0}^{\infty} \sqrt{2} \chi_y \int D\mu_S(\pi^{\mathbf{x}}) 2^{\frac{n}{2}} (\varphi - i\pi)^y \psi_{\vec{x}_n}^{(n)} [(\varphi - i\pi)^n]^{\vec{x}_n} =$$

$$\sum_{n=0}^{\infty} \chi_y \psi_{\vec{x}_n}^{(n)} \frac{\partial^{n+1}}{\partial \xi_{\vec{x}_n,y}^{n+1}} \int D\mu_S(\pi^{\mathbf{x}}) e^{\sqrt{2}\xi_z(\varphi - i\pi)^z} \Big|_{\xi_{\mathbf{x}}=0} =$$

$$\sum_{n=0}^{\infty} \sqrt{2} \chi_y \frac{\partial}{\partial \xi_y} 2^{\frac{n}{2}} \psi_{\vec{x}_n}^{(n)} \frac{\partial^n}{\partial \xi_{\vec{x}_n}^n} e^{\xi_x \varphi^x - \frac{\xi_u \Delta^{uv} \xi_v}{4}} \Big|_{\xi_{\mathbf{x}}=0} =$$

$$= \chi_x \left( \sqrt{2}\varphi^x - \frac{\Delta^{xy}\partial_{\varphi^y}}{\sqrt{2}} \right) \Psi_S(\varphi^{\mathbf{x}}).$$

Here we used $\frac{\partial}{\partial \xi_y} \exp\left( \xi_x \varphi^x - \frac{\xi_u \Delta^{uv} \xi_v}{4} \right) = \varphi^y - \frac{\Delta^{yx}}{2} \partial_{\varphi^x}$. Then we proved

$$\tilde{\mathcal{B}}^{-1} \phi^x \tilde{\mathcal{B}} = \sqrt{2}\varphi^x - \frac{\Delta^{xy}\partial_y}{\sqrt{2}}. \tag{A.1}$$

Finally, for $\tilde{\mathcal{B}}^{-1}\partial_{\phi^y}\tilde{\mathcal{B}}$, we compute the transform of $\chi_x \Delta^{xy} \partial_{\phi^y} \Psi_H(\phi^{\mathbf{x}})$. Notice that

$$\tilde{\mathcal{B}}^{-1}(\chi_x \Delta^{xy} \partial_{\phi^y} \Psi_H)(\varphi^{\mathbf{x}}) = \tilde{\mathcal{B}}^{-1}\left( \sum_{n=0}^{\infty} \chi_y \Delta^{yz} \partial_{\phi^z} \psi_{\vec{x}_n}^{(n)} (\phi^n)^{\vec{x}_n} \right) =$$

$$\chi_y \Delta^{yz} \sum_{n=0}^{\infty} \tilde{\mathcal{B}}^{-1}\left( \partial_{\phi^z} \psi_{\vec{x}_n}^{(n)} (\phi^n)^{\vec{x}_n} \right) = \chi_y \Delta^{yz} \sum_{n=0}^{\infty} \tilde{\mathcal{B}}^{-1}\left( \psi_{z,\vec{x}_{n-1}}^{(n)} n(\phi^{n-1})^{\vec{x}_{n-1}} \right) =$$

$$\chi_y \Delta^{yz} \sum_{n=0}^{\infty} 2^{\frac{n-1}{2}} n\psi_{z,\vec{x}_n}^{(n)} : \varphi^{n-1} : \big|_{\frac{\Delta}{2}}^{\vec{x}_{n-1}} = \frac{\chi_y \Delta^{yz}\partial_{\varphi^z}}{\sqrt{2}} \Psi_S(\varphi^{\mathbf{x}}).$$

Thus we proved

$$\tilde{\mathcal{B}}^{-1}\partial_{\phi^x}\tilde{\mathcal{B}} = \frac{\partial_{\varphi^x}}{\sqrt{2}}. \tag{A.2}$$

Adding the corresponding phase factor, we also derive (4.31) from these expressions.

Now we will prove that $\tilde{\mathcal{B}}$ preserves the quantization mappings. According to (3.25), we can write $\varphi^x = \frac{\phi^x + \bar{\phi}^x}{\sqrt{2}}$ and $\pi^x = iK_y^x \frac{\phi^y - \bar{\phi}^y}{\sqrt{2}} - (KA)_y^x \frac{\phi^y + \bar{\phi}^y}{\sqrt{2}}$. Using the definition (4.10) and the relations proven above, we see that

$$\tilde{\mathcal{B}}^{-1}\mathcal{Q}(\varphi^x)\tilde{\mathcal{B}} = \varphi^x = \mathcal{Q}_s(\varphi^x),$$
$$\tilde{\mathcal{B}}^{-1}\mathcal{Q}(\pi^x \delta_{xy})\tilde{\mathcal{B}} = -i\partial_{\varphi^y} + i\varphi^z K_{zy} - \varphi^x(KA)_{xy} = \mathcal{Q}_s(\pi^x \delta_{xy}) \tag{A.3}$$

## A.1.2 Momentum field picture

Similarly, we define an isomorphism

$$\tilde{\overline{\mathcal{B}}} : L^2(\mathcal{N}', D\nu_M) \to L^2_{\overline{Hol}}(\mathcal{N}'_{\mathbb{C}}, D\nu_c).$$

such that if $\hat{\Psi}_{\overline{H}} = \tilde{\overline{\mathcal{B}}}(\hat{\Psi}_M)$, then the inverse is provided by (4.33) with phase factor $g = 0$. In terms of the chaos decomposition, this is

$$\tilde{\overline{\mathcal{B}}}^{-1}(\hat{\Psi}_{\bar{H}})(\pi^{\mathsf{x}}) = \int D\nu_M(\varphi^{\mathsf{x}}) \Psi_{\bar{H}}(\sqrt{2}\bar{\phi}^{\mathsf{x}}) =$$

$$\sum_{n=0}^{\infty} \int D\nu_M(\varphi^{\mathsf{x}}) 2^{\frac{n}{2}} \hat{\psi}_{\vec{x}_n}^{(n)} [(\pi + i\varphi)^n]^{\vec{x}_n} =$$

$$\sum_{n=0}^{\infty} \hat{\psi}_{\vec{x}_n}^{(n)} \frac{\partial^n}{\partial \xi_{\vec{x}_n}^n} \int D\nu_M(\varphi^{\mathsf{x}}) e^{\sqrt{2}\xi_z(\pi + i\varphi)^z} \Big|_{\xi_{\mathsf{x}}=0} =$$

$$\sum_{n=0}^{\infty} \hat{\psi}_{\vec{x}_n}^{(n)} \frac{\partial^n}{\partial \xi_{\vec{x}_n}^n} \left[ e^{\sqrt{2}\xi_z\pi^z} \int D\nu_M(\varphi^{\mathsf{x}}) e^{i\sqrt{2}\xi_z\varphi^z} \right]_{\xi_{\mathsf{x}}=0} =$$

$$\sum_{n=0}^{\infty} \hat{\psi}_{\vec{x}_n}^{(n)} \frac{\partial^n}{\partial \xi_{\vec{x}_n}^n} e^{\sqrt{2}\xi_x\pi^x + \frac{\xi_u D^{uv}\xi_v}{2}} \Big|_{\xi_{\mathsf{x}}=0} =$$

$$\sum_{n=0}^{\infty} 2^{\frac{n}{2}} \hat{\psi}_{\vec{x}_n}^{(n)} \frac{\partial^n}{\partial \tilde{\xi}_{\vec{x}_n}^n} e^{\xi_x\pi^x + \frac{\tilde{\xi}_u D^{uv}\tilde{\xi}_v}{4}} \Big|_{\tilde{\xi}_{\mathsf{x}}=0} =$$

$$\sum_{n=0}^{\infty} 2^{\frac{n}{2}} \hat{\psi}_{\vec{x}_n}^{(n)} : \pi^n :|_{-\frac{D}{2}}^{\vec{x}_n} = \Psi_M(\pi^{\mathsf{x}}).$$

To compute $\tilde{\overline{\mathcal{B}}}^{-1}\bar{\phi}^x\tilde{\overline{\mathcal{B}}}$, let us compute the transform of $\chi_x\bar{\phi}^x\hat{\Psi}_H(\bar{\phi}^{\mathsf{x}})$. We get

$$\tilde{\overline{\mathcal{B}}}^{-1}(\chi_x\bar{\phi}^x\hat{\Psi}_H)(\varphi^{\mathsf{x}}) = \int D\mu_S(\pi^{\mathsf{x}})\chi_y\sqrt{2}\bar{\phi}^y\hat{\Psi}_H(i\sqrt{2}\bar{\phi}^{\mathsf{x}}) =$$

$$\sum_{n=0}^{\infty} \sqrt{2}\chi_y \int D\nu_M(\varphi^{\mathsf{x}}) 2^{\frac{n}{2}} (\pi + i\varphi)^y \hat{\psi}_{\vec{x}_n}^{(n)} [(\pi + i\varphi)^n]^{\vec{x}_n} =$$

$$\sum_{n=0}^{\infty} \chi_y \hat{\psi}_{\vec{x}_n}^{(n)} \frac{\partial^{n+1}}{\partial \xi_{\vec{x}_n,y}^{n+1}} \int D\nu_M(\varphi^{\mathsf{x}}) e^{\sqrt{2}\xi_z(\pi + i\varphi)^z} \Big|_{\xi_{\mathsf{x}}=0} =$$

$$\sum_{n=0}^{\infty} \sqrt{2}\chi_y \frac{\partial}{\partial \xi_y} 2^{\frac{n}{2}} \hat{\psi}_{\vec{x}_n}^{(n)} \frac{\partial^n}{\partial \xi_{\vec{x}_n}^n} e^{\xi_x\pi^x + \frac{\xi_u D^{uv}\xi_v}{4}} \Big|_{\xi_{\mathsf{x}}=0}$$

$$= \chi_x \left( \sqrt{2}\pi^x + \frac{D^{xy}\partial_{\pi^y}}{\sqrt{2}} \right) \hat{\Psi}_M(\pi^{\mathsf{x}}),$$

where we used $\frac{\partial}{\partial \xi_y} \exp\left(\xi_x \pi^x + \frac{\xi_u D^{uv} \xi_v}{4}\right) = \pi^y + \frac{D^{yx}}{2} \partial_{\pi^x}$. Then

$$\tilde{\bar{\mathcal{B}}}^{-1} \bar{\phi}^x \tilde{\bar{\mathcal{B}}} = \sqrt{2}\pi^x + \frac{D^{xy}\partial_{\pi^y}}{\sqrt{2}}. \tag{A.4}$$

Finally, for $\tilde{\bar{\mathcal{B}}}^{-1} \partial_{\bar{\phi}^y} \tilde{\bar{\mathcal{B}}}$ we compute the transform of $-\chi_x D^{xy} \partial_{\bar{\phi}^y} \hat{\Psi}_H(\bar{\phi}^{\mathbf{x}})$.

$$\tilde{\bar{\mathcal{B}}}^{-1}(-\chi_x D^{xy}\partial_{\bar{\phi}^y}\hat{\Psi}_H)(\pi^{\mathbf{x}}) = \tilde{\bar{\mathcal{B}}}^{-1}\left(\sum_{n=0}^{\infty} -\chi_y D^{yz}\partial_{\bar{\phi}^z}\hat{\psi}^{(n)}_{\vec{x}_n}(\bar{\phi}^n)^{\vec{x}_n}\right) =$$

$$-\chi_y D^{yz}\sum_{n=0}^{\infty}\tilde{\bar{\mathcal{B}}}^{-1}\left(\partial_{\bar{\phi}^z}\hat{\psi}^{(n)}_{\vec{x}_n}(\bar{\phi}^n)^{\vec{x}_n}\right) = -\chi_y D^{yz}\sum_{n=0}^{\infty}\tilde{\bar{\mathcal{B}}}^{-1}\left(\hat{\psi}^{(n)}_{z,\vec{x}_{n-1}}n(\bar{\phi}^{n-1})^{\vec{x}_{n-1}}\right) =$$

$$-\chi_y D^{yz}\sum_{n=0}^{\infty}2^{\frac{n-1}{2}}n\hat{\psi}^{(n)}_{z,\vec{x}_n}:\pi^{n-1}:\Big|_{-\frac{D}{2}}^{\vec{x}_{n-1}} = -\frac{\chi_y D^{yz}\partial_{\pi^z}}{\sqrt{2}}\hat{\Psi}_M(\pi^{\mathbf{x}}).$$

Thus proving

$$\tilde{\bar{\mathcal{B}}}^{-1}\partial_{\bar{\phi}^y}\tilde{\bar{\mathcal{B}}} = \frac{1}{\sqrt{2}}\partial_{\pi^y} \tag{A.5}$$

Adding the phase factor of (4.33), we also prove the relations (4.34).

Now we will prove that $\tilde{\bar{\mathcal{B}}}$ preserves the quantization mappings. According to (4.22), we can write $\varphi^x = -i(D^{-1})^x_y \frac{\check{\phi}^y - \check{\bar{\phi}}^y}{\sqrt{2}} + (AD^{-1})^x_y \frac{\check{\phi}^y + \check{\bar{\phi}}^y}{\sqrt{2}}$ and $\pi^x = \frac{\check{\phi}^y + \check{\bar{\phi}}^y}{\sqrt{2}}$. Using the definition (4.24) and the relations proven above, we see that

$$\tilde{\bar{\mathcal{B}}}^{-1}\overline{\mathcal{Q}}(\varphi^x\delta_{xy})\tilde{\bar{\mathcal{B}}} = i\partial_{\pi^y} + i\pi^z D^{-1}_{zy} + \pi^x(AD^{-1})_{xy} = \overline{\mathcal{Q}}_m(\varphi^x\delta_{xy}),$$
$$\tilde{\bar{\mathcal{B}}}^{-1}\overline{\mathcal{Q}}(\pi^x)\tilde{\bar{\mathcal{B}}} = \pi^x = \overline{\mathcal{Q}}_m(\pi^x). \tag{A.6}$$

### A.1.3 Holomorphic and Antiholomorphic Fourier transform

Let $\hat{\Psi}(\bar{\phi}^{\mathbf{x}}) = \tilde{\mathcal{F}}[\Psi(\phi^{\mathbf{x}})]$ and $\hat{\Phi}(\bar{\phi}^{\mathbf{x}}) = \tilde{\mathcal{F}}[\Phi(\phi^{\mathbf{x}})]$. We denote the chaos decomposition of each function

$$\Psi(\phi^{\mathbf{x}}) = \sum_{n=0}^{\infty}\psi^{(n,0)}_{\vec{x}_n}(\phi^n)^{\vec{x}_n}, \qquad\qquad \Phi(\phi^{\mathbf{x}}) = \sum_{n=0}^{\infty}\varphi^{(n,0)}_{\vec{x}_n}(\phi^n)^{\vec{x}_n}, \tag{A.7}$$

$$\hat{\Psi}(\bar{\phi}^{\mathbf{x}}) = \sum_{n=0}^{\infty}\hat{\psi}^{(0,\bar{n})}_{\vec{x}_n}(\bar{\phi}^n)^{\vec{x}_n}, \qquad\qquad \hat{\Phi}(\bar{\phi}^{\mathbf{x}}) = \sum_{n=0}^{\infty}\hat{\varphi}^{(0,\bar{n})}_{\vec{x}_n}(\bar{\phi}^n)^{\vec{x}_n}. \tag{A.8}$$

By the definition of $\tilde{\mathcal{F}}$ in (4.37) in terms of the chaos decomposition, we get

$$\hat{\psi}^{(0,\bar{n})}_{\vec{x}_n} = (i)^n\psi^{(n,0)}_{\vec{y}_n}\left[(D^{-1} - iAD^{-1})^n\right]^{\vec{y}_n}_{\vec{x}_n}. \tag{A.9}$$

To prove that $\tilde{\mathcal{F}}$ is an isometric isomorphism, we derive from the relations displayed under (3.21) that

$$
\begin{aligned}
\Delta^{xy} &= -(D^{-1} + iAD^{-1})^x_u D^{uv}(D^{-1} - iAD^{-1})^x_v, \\
D^{xy} &= -(K + iKA)^x_u \Delta^{uv}(K - iKA^{-1})^x_v, \\
\delta^x_y &= -(K - iKA)^x_u(D^{-1} + iAD^{-1})^u_y, \\
\delta^x_y &= -(D^{-1} - iAD^{-1})^x_u(K + iKA)^u_y.
\end{aligned}
\tag{A.10}
$$

Recall from the Wiener-Ito decomposition theorem, Equation 2.32, with definitions of the measures given by (4.6) and (4.23), that

$$
\int D\nu_c(\phi^{\mathbf{x}})\overline{\hat{\Phi}(\bar{\phi}^{\mathbf{x}})}\hat{\Psi}(\bar{\phi}^{\mathbf{x}}) = \sum_{n=0}^{\infty} n!\, (-1)^n \overline{\hat{\varphi}^{(0,\bar{n})}_{\vec{x}_n}}[D^n]^{\vec{x}_n \vec{y}_n}\hat{\psi}^{(0,\bar{n})}_{\vec{y}_n} =
$$

$$
\sum_{n=0}^{\infty} n!\, (i^2)^n \overline{\varphi^{(n,0)}_{\vec{u}_n}}[(D^{-1} + iAD^{-1})^n]^{\vec{u}_n}_{\vec{x}_n}[D^n]^{\vec{x}_n \vec{y}_n}[(D^{-1} - iAD^{-1})^n]^{\vec{v}_n}_{\vec{y}_n}\psi^{(n,0)}_{\vec{v}_n} =
$$

$$
\sum_{n=0}^{\infty} n!\, \overline{\varphi^{(n,0)}_{\vec{u}_n}}[\Delta^n]^{\vec{u}_n \vec{v}_n}\psi^{(n,0)}_{\vec{v}_n} = \int D\mu_c(\phi^{\mathbf{x}})\overline{\Phi(\phi^{\mathbf{x}})}\Psi(\phi^{\mathbf{x}}).
\tag{A.11}
$$

This proves the isometry. The isomorphism is shown by inverting (A.9) with

$$
(i)^n \hat{\psi}^{(0,\bar{n})}_{\vec{x}_n}[(K + iKA)^n]^{\vec{x}_n}_{\vec{y}_n} = \psi^{(n,0)}_{\vec{y}_n}.
\tag{A.12}
$$

In turn, using Wiener-Ito decompositions and (A.12) , (A.9) it follows that

$$
\tilde{\mathcal{F}}(\phi^x \Psi) = i(D^{-1} - iAD^{-1})^x_y \bar{\phi}^y \hat{\Psi} \ \text{ and } \ \tilde{\mathcal{F}}(\partial_{\phi^x}\Psi) = i(K + iKA)^y_x \partial_{\bar{\phi}^y}\hat{\Psi}
\tag{A.13}
$$

proving (4.39).

### A.1.4 Fourier transform $\mathcal{F}$ of linear operators

Let us use the shorthand notation

$$
\begin{aligned}
a^x_0 &= \Delta^{xy}\frac{\partial_{\dot{\phi}^y}}{\sqrt{2}}, & a^{\dagger x}_0 &= \sqrt{2}\varphi^x - \Delta^{xy}\frac{\partial_{\dot{\phi}^y}}{\sqrt{2}}, \\
b^x_0 &= -D^{xy}\frac{\partial_{\pi^y}}{\sqrt{2}}, & b^{\dagger x}_0 &= \sqrt{2}\pi^x + \frac{D^{xy}\partial_{\pi^y}}{\sqrt{2}}.
\end{aligned}
\tag{A.14}
$$

We are interested in the action of $\mathcal{F} = \tilde{\mathcal{B}}^{-1}\tilde{\mathcal{F}}\tilde{\mathcal{B}}$ on linear operators. By virtue of (A.1),(A.2), (A.4), (A.5), and the action of the $\tilde{\mathcal{F}}$ operator (A.13), it is immediate to see that[1] we get

---

[1] Recalling (A.10) the relations displayed under (3.21).

$$\mathcal{F}a_0^x \mathcal{F}^{-1} = -i(D^{-1} + iAD^{-1})^x_y b_0^y,$$
$$\mathcal{F}a_0^{\dagger x} \mathcal{F}^{-1} = i(D^{-1} - iAD^{-1})^x_y b_0^{\dagger y},$$
$$\mathcal{F}^{-1} b_0^x \mathcal{F} = -i(K - iKA)^x_y a_0^y,$$
$$\mathcal{F}^{-1} b_0^{\dagger x} \mathcal{F} = i(K + iKA)^x_y a_0^{\dagger y}, \tag{A.15}$$

Notice that $\varphi^x = \frac{a_0^x + a_0^{\dagger x}}{\sqrt{2}}$ and $\pi^x = \frac{b_0^x + b_0^{\dagger x}}{\sqrt{2}}$, thus we can compute

$$\mathcal{F}\varphi^x \mathcal{F}^{-1} = \mathcal{F}\left(\frac{a_0^x + a_0^{\dagger x}}{\sqrt{2}}\right)\mathcal{F}^{-1} = i(D^{-1})^x_y \frac{b_0^{\dagger y} - b_0^y}{\sqrt{2}} + (AD^{-1})^x_y \frac{b_0^{\dagger y} + b_0^y}{\sqrt{2}}, \tag{A.16}$$

$$\mathcal{F}^{-1} \pi^x \mathcal{F} = \mathcal{F}^{-1}\left(\frac{b_0^x + b_0^{\dagger x}}{\sqrt{2}}\right)\mathcal{F} = iK^x_y \frac{a_0^{\dagger y} - a_0^y}{\sqrt{2}} + (KA)^x_y \frac{a_0^{\dagger y} + a_0^y}{\sqrt{2}}. \tag{A.17}$$

Plugging in the definitions (A.14), we obtain

$$\mathcal{F}\varphi^x \delta_{xy} \mathcal{F}^{-1} = i\partial_{\pi^y} + i\pi^z D^{-1}_{zy} + \pi^x (AD^{-1})_{xy},$$
$$\mathcal{F}^{-1} \pi^x \delta_{xy} \mathcal{F} = -i\partial_{\varphi^y} + i\varphi^z K_{zy} - \varphi^x (KA)_{xy}, \tag{A.18}$$

proving (4.41) and thus the preservation of the quantization mapping.

## A.2 Explicit expressions for the connections and the quantized operators in *second quantized* coordinates.

In this section, we will explicitly compute the connections introduced in section 5.5.

### A.2.1 Antiholomorphic connection

In this section, we will compute explicitly the expressions of the connection in (5.67) in holomorphic *s.q.* coordinates. Let $\mathcal{O}^{\bar{\phi}}_\sigma$ be an operator, then the expression of the covariant derivative is

$$\tilde{\nabla}_t \mathcal{O} = \frac{1}{2}\tilde{\mathcal{F}}^{-1}\left[\frac{\partial \hat{\mathcal{O}}}{\partial t} + \left(\frac{\partial \hat{\mathcal{O}}^{\dagger_{\bar{H}}}}{\partial t}\right)^{\dagger_{\bar{H}}}\right]\tilde{\mathcal{F}}. \tag{A.19}$$

To compute this term, recall the kernels with respect to white noise of the Fourier transform (5.47) and its inverse. Replicating the discussion at the beginning of section 5.5, using $\tilde{\mathcal{F}}^{-1}\dot{\tilde{\mathcal{F}}} = -\dot{\tilde{\mathcal{F}}}^{-1}\tilde{\mathcal{F}}$ and

$$\tilde{\mathcal{F}}^{-1}\frac{d\tilde{\mathcal{F}}\mathcal{O}\tilde{\mathcal{F}}^{-1}}{dt}\tilde{\mathcal{F}} = \dot{\mathcal{O}} + [\tilde{\mathcal{F}}^{-1}\dot{\tilde{\mathcal{F}}}, \mathcal{O}]. \tag{A.20}$$

The second term is a combination of the one above and the one at the beginning of [section 5.5](#).

$$(\tilde{\mathcal{F}}^{-1}) \left( \frac{\partial \tilde{\mathcal{F}} \mathcal{O}^{\dagger_H} \tilde{\mathcal{F}}^{-1}}{\partial t} \right)^{\dagger_{\bar{H}}} \tilde{\mathcal{F}} = \dot{\mathcal{O}} + \left[ (\mathcal{D}\tilde{\mathcal{F}}^{\phi\bar{\gamma}})^{-1} \frac{d(\mathcal{D}\tilde{\mathcal{F}})_{\bar{\gamma}\sigma}}{dt}, \mathcal{O} \right]. \tag{A.21}$$

Then, explicitly we get that in holomorphic second quantized coordinates

$$\tilde{\Gamma}_\sigma^\phi = \frac{1}{2} \left[ (\tilde{\mathcal{F}}^{-1} \dot{\tilde{\mathcal{F}}})_\sigma^\phi + (\mathcal{D}\tilde{\mathcal{F}}^{\phi\bar{\gamma}})^{-1} \frac{d(\mathcal{D}\tilde{\mathcal{F}})_{\bar{\gamma}\sigma}}{dt} \right]. \tag{A.22}$$

To get an explicit expression, notice that (A.22) leads always to a connection $\tilde{\Gamma} = \phi^x \tilde{\Gamma}_x^y \partial_{\phi^y}$. In this way, we get

$$\tilde{\nabla}_t(\phi^x) \partial_{\phi^x} = \tilde{\nabla}_t(\mathfrak{D}^{\dagger x}) K_{xy} \mathfrak{D}^y = \tilde{\Gamma}. \tag{A.23}$$

Using (5.65), we can rewrite

$$\tilde{\Gamma} = -\mathfrak{D}^{\dagger,x} K_{xy} \tilde{\nabla}_t(\mathfrak{D}^y) + \mathfrak{D}^{\dagger,x} \dot{K}_{xy} \mathfrak{D}^y = -\mathfrak{D}^{\dagger,x} \tilde{\nabla}_t(K_{xy} \mathfrak{D}^y). \tag{A.24}$$

To complete this expression, we need the explicit expressions of the covariant derivatives of (5.32) that we directly derive from (5.67). For simplicity, consider a real direction $\xi_x$, then we see that

$$\tilde{\mathcal{F}} \xi_x \tilde{\mathfrak{D}}^x \tilde{\mathcal{F}}^{-1} = -\xi_x \begin{pmatrix} A & \delta \\ -\delta & A \end{pmatrix}^{xy} \partial_{\phi^y}, \tag{A.25}$$

$$\tilde{\mathcal{F}} \, \xi_x \tilde{\mathfrak{D}}^{\dagger_H,x} \tilde{\mathcal{F}}^{-1} = \xi_x \begin{pmatrix} AD^{-1} & D^{-1} \\ -D^{-1} & AD^{-1} \end{pmatrix}^x_y \phi^y. \tag{A.26}$$

Let the time derivative be represented by a $\cdot$, consider also $\xi_x$ time-independent. We will consider that the coordinates $(\phi^x, \bar{\phi}^x)$ are just placeholders for integration and, as such, do not depend on the time parameter.

Let us consider the derivative of the operator $O = \xi_x \tilde{\mathfrak{D}}^x$. We can compute every term of the derivative using the rule (2.81) and the definitions of the $\dagger$ operators given by (5.29).

$$\tilde{\mathcal{F}}^{-1} \frac{d\tilde{\mathcal{F}} \xi_x \tilde{\mathfrak{D}}^x \tilde{\mathcal{F}}^{-1}}{dt} \tilde{\mathcal{F}} = -\xi_x \begin{pmatrix} \dot{A} & \dot{\delta} \\ -\dot{\delta} & \dot{A} \end{pmatrix}^{xy} \begin{pmatrix} AD^{-1} & -D^{-1} \\ D^{-1} & AD^{-1} \end{pmatrix}^z_y \partial_{\phi^z},$$

$$\tilde{\mathcal{F}}^{-1} \left( \frac{d\tilde{\mathcal{F}} \xi_x \tilde{\mathfrak{D}}^{\dagger_H,x} \tilde{\mathcal{F}}^{-1}}{dt} \right)^{\dagger_{\bar{H}}} \tilde{\mathcal{F}} = -\xi_x \left[ \frac{d}{dt} \begin{pmatrix} AD^{-1} & D^{-1} \\ -D^{-1} & AD^{-1} \end{pmatrix}^x_y \right] \begin{pmatrix} A^t & -\delta \\ \delta & A^t \end{pmatrix}^{yz} \partial_{\phi^z}.$$

The hardest term to compute is the second one. First, notice that

$$\frac{d\tilde{\mathcal{F}} \xi_x \tilde{\mathfrak{D}}^{\dagger_H,x} \tilde{\mathcal{F}}^{-1}}{dt} = \xi_x \frac{d}{dt} \begin{pmatrix} AD^{-1} & -D^{-1} \\ D^{-1} & AD^{-1} \end{pmatrix}^x_y \phi^y.$$

And the $\dagger_{\bar{H}}$ operator is given by substituting $\phi^x$ by $-D^{xy}\partial_{\phi^y}$ and transposing the matrix (in this particular case in which $\epsilon$ commutes with the operator). Applying the transformations leads to the desired result.

Once this is computed, simply by using the definitions of (3.21), it follows that

$$\Delta^{xy} = -\left(\begin{array}{cc} A & \delta \\ -\delta & A \end{array}\right)^{xu} D_{uv}^{-1} \left(\begin{array}{cc} A^t & -\delta \\ \delta & A^t \end{array}\right)^{vy}.$$

Then

$$\nabla_t \mathfrak{D}^x = \frac{1}{2}\dot{\Delta}^{xy}\partial_{\phi^y} - \frac{1}{2}\left(\begin{array}{cc} \dot{A} & \dot{\delta} \\ -\dot{\delta} & \dot{A} \end{array}\right)^{xy} \left(\begin{array}{cc} AD^{-1} & -D^{-1} \\ D^{-1} & AD^{-1} \end{array}\right)_y^z \partial_{\phi^z}$$
$$+ \frac{1}{2}\left(\begin{array}{cc} AD^{-1} & D^{-1} \\ -D^{-1} & AD^{-1} \end{array}\right)_y^x \left(\begin{array}{cc} \dot{A}^t & -\dot{\delta} \\ \dot{\delta} & \dot{A}^t \end{array}\right)^{yz} \partial_{\phi^z}. \tag{A.27}$$

For the creation operator, we use the fact $(\nabla_t O)^{\dagger_H} = \nabla_t(O^{\dagger_H})$ and simply substitute $\partial_{\phi^x}$ by $K_{xy}\phi^y$ and transpose the matrices.

Using this expression, we get

$$\tilde{\Gamma} = \frac{1}{2}\phi^x K_{xy}\dot{\Delta}^{yz}\partial_{\phi^z}$$
$$+ \frac{1}{2}\phi^x K_{xy}(\dot{A} - \epsilon\dot{\delta})^{yv}(AD^{-1} + \epsilon D^{-1})_v^u \partial_{\phi^u} - \frac{1}{2}\phi^x[KD^{-1}(A^t - \epsilon\delta)]_{xy}(\dot{A}^t + \epsilon\dot{\delta})^{yv}\partial_{\phi^v}. \tag{A.28}$$

If we particularize to the case in which the diagonal elements of the complex structure are $A = 0$, then $D^{-1} = -\Delta$ and the Fourier transform (5.47) is particularly simple to compute:

$$\tilde{\mathcal{F}}^{\phi\bar{\sigma}} = \exp\left[-\phi^x \epsilon \Delta_{xy}\bar{\sigma}^x\right], \qquad (\tilde{\mathcal{F}}^{\phi\bar{\sigma}})^{-1} = \exp\left[\phi^x \epsilon K_{xy}\bar{\sigma}^x\right].$$

Therefore, we compute

$$\tilde{\mathcal{F}}^{-1}\dot{\tilde{\mathcal{F}}} = \delta_\sigma^\phi \phi^x K_x^{\,y}\dot{\Delta}_y^{\,z}\partial_{\sigma^z}, \tag{A.29}$$

$$(\mathcal{D}\tilde{\mathcal{F}}^{\phi\bar{\gamma}})^{-1}\frac{d(\mathcal{D}\tilde{\mathcal{F}})_{\bar{\gamma}\sigma}}{dt} = \delta_\sigma^\phi \phi^x \delta_{xy}\dot{\delta}^{yz}\partial_{\sigma^z}. \tag{A.30}$$

and denoting $(\dot{\delta}^\circ\delta_\circ)_z^x = \dot{\delta}^{xz}\delta_{zy}$

$$\tilde{\Gamma} = \frac{1}{2}\phi^x K_{xy}\dot{\Delta}^{yz}\partial_{\phi^z} - \frac{1}{2}\phi^x K_{xy}(\dot{\delta}^\circ\delta_\circ)_u^y \Delta^{uz}\partial_{\phi^z} + \frac{1}{2}\phi^x(\dot{\delta}^\circ\delta_\circ)_x^z \partial_{\phi^z}.$$

With this expression, it is easy to check that (A.27) holds.

## A.2.2 Schrödinger connection

We want to explore the connection of (5.70) provided by

$$\nabla_t \mathcal{O} = \frac{1}{2}\tilde{\mathcal{B}}\left[\frac{\partial \mathcal{Q}}{\partial t} + \left(\frac{\partial \mathcal{Q}^{\dagger_s}}{\partial t}\right)^{\dagger_s}\right]\tilde{\mathcal{B}}^{-1}. \tag{A.31}$$

With similar arguments to the previous section, we get

$$\underline{\Gamma}_\sigma^\phi = \frac{1}{2}\left[(\tilde{\mathcal{B}}\dot{\tilde{\mathcal{B}}}^{-1})_\sigma^\phi + (\tilde{\mathcal{B}}\mathcal{K}^s)^{\sigma\varphi}\frac{d(\varDelta^s\tilde{\mathcal{B}}^{-1})_{\varphi\phi}}{dt}\right], \tag{A.32}$$

with $\mathcal{Q} = \tilde{\mathcal{B}}^{-1}\mathcal{O}\tilde{\mathcal{B}}$. Notice that we can write $\tilde{\mathcal{B}}^{-1}$ as an operator: $\tilde{\mathcal{B}}^{-1} = \exp(\sqrt{2}\varphi^x\partial_{\phi^x} - \frac{1}{2}\varDelta^{xy}\partial_{\phi^x}\partial_{\phi^y})$. Using this fact, we compute

$$\tilde{\mathcal{B}}\dot{\tilde{\mathcal{B}}}^{-1} = -\frac{1}{2}\dot{\varDelta}^{xy}\partial_{\phi^x}\partial_{\phi^y}. \tag{A.33}$$

Similarly, using the expressions (5.52) and (5.54), we have

$$(\tilde{\mathcal{B}}\mathcal{K}^s)^{\sigma\varphi}\frac{d(\varDelta^s\tilde{\mathcal{B}}^{-1})_{\varphi\phi}}{dt} =$$

$$-\frac{(\tilde{\mathcal{B}}\mathcal{K}^s)^{\sigma\varphi}}{dt}(\varDelta^s\tilde{\mathcal{B}}^{-1})_{\varphi\phi} =$$

$$-\int_{\mathcal{N}'}D\mu(\varphi^{\mathbf{x}})\left(\sqrt{2}\sigma^x\dot{K}_{xy}\varphi^y - \frac{\sigma^x\dot{K}_{xy}\sigma^y}{2}\right):\exp\left(\sqrt{2}\sigma^x K_{xy}\varphi^y\right):|_{\frac{\Delta}{2}}:\exp\left(\sqrt{2}\varphi^x\delta_{xy}\bar{\phi}^y\right):|_{\frac{\Delta}{2}} =$$

$$-\int_{\mathcal{N}'}D\mu(\varphi^{\mathbf{x}})\left(\sigma^x\dot{K}_{xy}\varDelta^{yz}\partial_{\sigma^z} + \sigma^x\dot{K}_{xy}\sigma^y - \frac{\sigma^x\dot{K}_{xy}\sigma^y}{2}\right)$$

$$:\exp\left(\sqrt{2}\sigma^x K_{xy}\varphi^y\right):|_{\frac{\Delta}{2}}:\exp\left(\sqrt{2}\varphi^x\delta_{xy}\bar{\phi}^y\right):|_{\frac{\Delta}{2}} =$$

$$-\delta_\phi^\sigma\left(\phi^x\dot{K}_{xy}\varDelta^{yz}\partial_{\phi^z} + \frac{\phi^x\dot{K}_{xy}\phi^y}{2}\right).$$

Gathering together all contributions, we have

$$\underline{\Gamma} = \frac{1}{2}\phi^x K_{xy}\dot{\varDelta}^{yz}\partial_{\phi^z} - \frac{1}{4}(\dot{\varDelta}^{xy}\partial_{\phi^x}\partial_{\phi^y} + \dot{K}_{xy}\phi^x\phi^y). \tag{A.34}$$

## A.2.3 Field-Momentum connection

Here we will deal with the connection of (5.73) defined by

$$\nabla_t \mathcal{O} = \frac{1}{2}\tilde{\mathcal{F}}^{-1}\tilde{\tilde{\mathcal{B}}}\left[\frac{\partial \mathcal{O}}{\partial t} + \left(\frac{\partial \mathcal{O}^{\dagger_m}}{\partial t}\right)^{\dagger_m}\right]\tilde{\tilde{\mathcal{B}}}^{-1}\tilde{\mathcal{F}}. \tag{A.35}$$

Similarly to the previous cases, we get

$$\underset{\smile}{\Gamma}{}_\sigma^\phi = \frac{1}{2}\left[\left(\tilde{\mathcal{F}}^{-1}\tilde{\mathcal{B}}\frac{d\tilde{\overline{\mathcal{B}}}^{-1}\tilde{\mathcal{F}}}{dt}\right)_\sigma^\phi + (\tilde{\mathcal{F}}^{-1}\tilde{\mathcal{B}}\mathcal{D}^{m,-1})_{\sigma\varphi}\frac{d(\mathcal{D}^m\tilde{\overline{\mathcal{B}}}^{-1}\tilde{\mathcal{F}})^{\varphi\phi}}{dt}\right] =$$

$$\frac{1}{2}\tilde{\mathcal{F}}^{-1}\left[\left(\tilde{\mathcal{B}}\frac{d\tilde{\overline{\mathcal{B}}}^{-1}}{dt}\right)_\sigma^\phi + (\tilde{\mathcal{B}}\mathcal{D}^{m,-1})_{\sigma\varphi}\frac{d(\mathcal{D}^m\tilde{\overline{\mathcal{B}}}^{-1})^{\varphi\phi}}{dt}\right]\tilde{\mathcal{F}} + \tilde{\mathcal{F}}^{-1}\dot{\tilde{\mathcal{F}}}. \qquad \text{(A.36)}$$

Using the result of the previous section, we see that

$$\underset{\smile}{\Gamma} = \tilde{\mathcal{F}}^{-1}\left[\frac{1}{2}\phi^x D_{xy}^{-1}\dot{D}^{yz}\partial_{\phi^z} + \frac{1}{4}\left(\dot{D}^{xy}\partial_{\phi^x}\partial_{\phi^y} + \dot{D}_{xy}^{-1}\phi^x\phi^y\right)\right]\tilde{\mathcal{F}} + \tilde{\mathcal{F}}^{-1}\dot{\tilde{\mathcal{F}}}. \qquad \text{(A.37)}$$

Notice that we can write

$$\tilde{\mathcal{F}}^{-1}\left[\frac{1}{2}\phi^x D_{xy}^{-1}\dot{D}^{yz}\partial_{\phi^z}\right]\tilde{\mathcal{F}} + \tilde{\mathcal{F}}^{-1}\dot{\tilde{\mathcal{F}}} = \frac{1}{2}\left[\left(\tilde{\mathcal{F}}^{-1}\dot{\tilde{\mathcal{F}}}\right)_\sigma^\phi + (\mathcal{D}\tilde{\mathcal{F}})_{\sigma\bar{\gamma}}^{-1}\frac{d(\mathcal{D}\tilde{\mathcal{F}})^{\bar{\gamma}\phi}}{dt}\right].$$

That coincides with $\tilde{\Gamma}$. The remaining term is the anti-self-adjoint operator

$$\underset{\smile}{\Gamma} =$$

$$\tilde{\Gamma} + \frac{1}{4}\left[(AD^{-1} + \epsilon D^{-1})_u^x \dot{D}^{uv}(AD^{-1} + \epsilon D^{-1})_v^y \partial_{\phi^x}\partial_{\phi^y}\right.$$

$$\left. + (KA - \epsilon K)_x^u (\dot{D}^{-1})_{uv}(KA - \epsilon K)_y^v \phi^x\phi^y\right] =$$

$$\tilde{\Gamma} + \frac{1}{4}\left[(A - \epsilon\delta)^{xu}(\dot{D}^{-1})_{uv}(A^t - \epsilon\delta)^{vy}(K\phi)_x(K\phi)_y - (A^t + \epsilon\delta)^{xu}(\dot{D}^{-1})_{uv}(A + \epsilon\delta)^{vy}\partial_{\phi^x}\partial_{\phi^y}\right].$$

# Appendix B

# Holomorphic Hida Test Functions

Most of the references on white noise [Hid&93, Oba94, Kon&96, Kuo96, Wes03] and Gaussian analysis [Hu16] present the theory using Gaussian measures on some particular real DNF $\mathcal{N}'$ or Banach space $B$. Even though complex chaos is a straightforward generalization in most cases [Hid80], we prepared in this section a quick dictionary of theorems and lemmas that adapt the classical presentation to our needs.

First, let us fix some notation. Consider the system of norms $\|\cdot\|_{p,q}$ introduced above (2.71) and their duals $\|\cdot\|_{-p,-q}$ on the Gel'fand triple $\mathcal{H}_{p,q} \subset \mathcal{H}_{Vol} \subset \mathcal{H}^*_{-p,-q}$. Consider also the system of norms $|\cdot|_{p,q} = \|\cdot\|_{\Gamma\mathcal{H}_{p,q}\otimes\Gamma\mathcal{H}^*_{p,q}}$ of the Fock spaces $\Gamma\mathcal{H}_{p,q} \otimes \Gamma\mathcal{H}^*_{p,q}$. With a slight abuse of notation, we identify this space with $\mathcal{I}^{-1}\left(\Gamma\mathcal{H}_{p,q} \otimes \Gamma\mathcal{H}^*_{p,q}\right)$ and use the same notation for the norm. We can write for a function $\Psi(\phi^{\mathbf{x}}, \bar{\phi}^{\mathbf{x}})$ in this space

$$\left|2^{\frac{qN}{2}}\Psi\right|_{p,0} = \left|\Psi\right|_{p,q},$$

where $N$ is the particle number operator (2.63) of the white noise measure. Recall that the white noise measure is auxiliary, and we can describe an equivalent family of norms using the reference measure in (2.76). We defined our Hida test functions as holomorphic functions over $\mathcal{N}'_{\mathbb{C}}$. We start by stating the characterization theorem in the holomorphic context:

**Theorem B.1.** *(Characterization theorem for holomorphic Hida test functions [Kon&96]) Let $F$ be an entire function (holomorphic for every point in $\mathcal{N}_{\mathbb{C}}$ in the sense of **Definition** 2.8). Then we have $\overline{F} = S_\beta[\Psi]$ where $\Psi \in (\mathcal{N}_{\mathbb{C}})$ if and only if $\forall\, p, q \geq 0$ there is $C_{p,q} > 0$ such that $F$ fulfills the estimate*

$$|F(z\rho_{\mathbf{x}})| \leq C_{p,q} \exp\left(|z|^2\|\rho_{\mathbf{x}}\|^2_{-p,-q}\right), \quad \rho_{\mathbf{x}} \in \mathcal{N}_{\mathbb{C}}, z \in \mathbb{C}.$$

*Proof.* Noticing that $S_\beta[\Psi]$ is antiholomorphic when $\Psi$ is holomorphic, the proof of Theorem 15 in [Kon&96] is directly applicable in this context. $\square$

**Corollary B.1.1.** *(Algebra of Holomorphic Hida Test Functions) $(\mathcal{N}_{\mathbb{C}})$ is an algebra under pointwise multiplication.*

*Proof.* To show this, notice that when $\Psi, \Phi \in (\mathcal{N}_{\mathbb{C}})$ are holomorphic, we have $S_\beta[\Psi\Phi](\bar{\rho}_{\mathbf{x}}) = S_\beta[\Psi](\bar{\rho}_{\mathbf{x}})S_\beta[\Phi](\bar{\rho}_{\mathbf{x}})$, thus

$$|S_\beta[\Psi\Phi](z\bar{\rho}_{\mathbf{x}})| \leq C_{p,q}(\Psi)C_{p,q}(\Phi)\exp\left(|z|^2\|\rho_{\mathbf{x}}\|_{-p,-(q-1)}^2\right).$$

$\square$

For the general space $(\mathcal{N}_{\mathbb{C}}) \otimes (\mathcal{N}_{\mathbb{C}})^*$, this result requires more involved estimates. We provide a sketch of the proof adapting the study found in [Oba]. First, notice that, in a similar way to [Oba94] Lemma 3.5.1:

$$e^{\overline{\rho_z\bar{\phi}^z} + \rho_z\phi^z - \bar{\rho}_u\delta^{uv}\rho_v + \chi_z\overline{\bar{\phi}^z} - \bar{\chi}_u\delta^{uv}\chi_v} = $$
$$e^{\overline{(\rho_z+\chi_z)\phi^z} + (\rho_z+\chi_z)\phi^z - \overline{(\rho_u+\chi_u)}\delta^{uv}(\rho_v+\chi_v)}e^{\overline{\rho_u}\delta^{uv}\chi_v + \overline{\chi_u}\delta^{uv}\rho_v}$$

(B.1)

and (2.31) imply,

$$\mathcal{W}_\delta(\phi^n\bar{\phi}^{\bar{n}})^{\vec{\mathbf{x}}_n\vec{\mathbf{y}}_{\bar{n}}}\mathcal{W}_\delta(\phi^m\bar{\phi}^{\bar{m}})^{\vec{\mathbf{x}}_m\vec{\mathbf{y}}_{\bar{m}}} = $$
$$\sum_{k=0}^{\min(n,\bar{m})}\sum_{\bar{k}=0}^{\min(\bar{n},m)}\binom{n}{k}\binom{\bar{m}}{k}k!\binom{\bar{n}}{\bar{k}}\binom{m}{\bar{k}}\bar{k}!\mathcal{W}_\delta(\phi^{n+m-k-\bar{k}}\bar{\phi}^{\bar{m}+\bar{n}-k-\bar{k}})^{(\vec{\mathbf{x}}_{:}(\vec{\mathbf{y}}}\left[\delta^{\bar{k}}\right]^{\vec{\mathbf{y}}_k)}\left[\delta^k\right]^{\vec{\mathbf{x}}_k)},$$

where the symmetrization is performed over superindices with the same label. We want to adapt the arguments leading to Theorem 3.5.6 in [Oba94] in a similar way as Theorem 64 in [Wes03].

**Theorem B.2.** *(Algebra of Complex Hida Test Functions) From continuity, we get that there is a $K_p \geq 0$ such that $\bar{\rho}_x\delta^{xy}\rho_x = \|\rho_{\mathbf{x}}\|_{0,0}^2 \leq K_p\|\rho_{\mathbf{x}}\|_{p,0}^2$. Choose $\alpha, \beta$ such that, for fixed $q, p \geq 0$, we get $2^{-(\alpha+\beta)/2}K_p^2 - 2^{q-\alpha} - 2^{q-\beta} < 1$, then we estimate*

$$|\Psi\Phi|_{p,q} \leq \frac{1 - 2^{-(\alpha+\beta)/2}K_p^{-2}}{(1 - 2^{-(\alpha+\beta)/2}K_p^2 - 2^{q-\alpha} - 2^{q-\beta})^2}|\Psi|_{p,\alpha}|\Phi|_{p,\beta}.$$

(B.2)

*Hence $(\mathcal{N}_{\mathbb{C}}) \otimes (\mathcal{N}_{\mathbb{C}})^* = \bigcap_{p,q}\mathcal{I}^{-1}\left(\Gamma\mathcal{H}_{p,q} \otimes \Gamma\mathcal{H}_{p,q}^*\right)$ is an algebra under pointwise multiplication.*

**Lemma B.3.** *(Equivalence to the Wiener space) Consider the Gaussian measure defined in (2.10). Because of continuity with respect to the topology of $\mathcal{N}_{\mathbb{C}}$, generated by the family of norms $\{\|\cdot\|_p\}_{p=0}^\infty$, we get that for some $p, q \geq 0$, we must have $|\bar{\rho}_x\Delta^{xy}\rho_y| \leq \|\rho_{\mathbf{x}}\|_{p,q}$. Because of the definition of FN space **Definition** 2.2, the norms are taken such that $\forall q > p$, the inclusion $\mathcal{H}_q \hookrightarrow \mathcal{H}_p$ is Hilbert-Schmidt. As a result, the measure is supported in $\mathcal{H}_q' = \mathcal{H}_{-q} \subset \mathcal{N}_{\mathbb{C}}'$. This is*

$$\mu_c(\mathcal{H}_{-q}) = \mu_c(\mathcal{N}_{\mathbb{C}}') = 1.$$

*Proof.* This is shown in Theorem 3.1 of [Hid80]. □

We will use Theorem B.3 to apply results derived in [Hu16] for Gaussian measures on Banach spaces, that is, using the Gel'fand triple $\mathcal{H}_q \subset \mathcal{H}_\Delta \subset \mathcal{H}_{-q}$. In particular, we get

**Theorem B.4. (Hypercontractivity)** *Let $1 \leq p \leq q < \infty$, then we get $2^{-\alpha\frac{N}{2}}$ is a bounded linear mapping such that*

$$\|2^{-\alpha\frac{N}{2}}\Psi\|_{L^q} \leq \|\Psi\|_{L^p} \tag{B.3}$$

*if and only if $2^{-\alpha}(p-1) \leq (q-1)$.*

*Proof.* Theorem 7.2 of [Hu16]. □

**Corollary B.4.1. (Integrability Properties)** *As a result of hypercontractivity, we get $(\mathcal{N}_{\mathbb{C}}) \subsetneq \bigcap_{p \geq 1, m \geq 0} \mathbb{D}^{p,m}_{\mu_c} = \mathbb{D}_{\mu_c}$.*

*Proof.* Let $p \geq 1$ choose $\lambda \geq 0$ such that $2^{-\lambda}(p-1) \leq 1$, then

$$\begin{aligned}
\|(1+N)^{\frac{m}{2}}\Psi\|_{L^p} &= \|2^{-\lambda\frac{N}{2}}2^{\lambda\frac{N}{2}}(1+N)^{\frac{m}{2}}\Psi\|_{L^p} \\
&\leq \|2^{\lambda\frac{N}{2}}(1+N)^{\frac{m}{2}}\Psi\|_{L^2} \\
&\leq \|2^{\lambda\frac{N}{2}}(1+2^N)^{\frac{m}{2}}\Psi\|_{L^2} \\
&\leq 2^{\frac{m}{2}}\|2^{(\lambda+m)\frac{N}{2}}\Psi\|_{L^2} = 2^{\frac{m}{2}}|\Psi|_{0,\frac{\lambda+m}{2}}.
\end{aligned}$$

Where in the first inequality we used hypercontractivity for $p \geq 2$, and for $1 \leq p \leq 2$, we used Hölder's inequality and the fact that $\|2^{-\lambda\frac{N}{2}}\| \leq 1$ in the operator norm. This proof is valid also using the reference measure $\mu_c$ to define the topology of $(\mathcal{N}_{\mathbb{C}})$. This proves that, defining the auxiliary space $\mathcal{G}_{\mu_c} = \bigcap_q \mathcal{I}^{-1}_{\mu_c}\left(\Gamma\mathcal{H}_{0,q}\right)$,

$$(\mathcal{N}_{\mathbb{C}}) \subsetneq \mathcal{G}_{\mu_c} \subsetneq \bigcap_{p \geq 1, m \geq 0} \mathbb{D}^{p,m}_{\mu_c} = \mathbb{D}_{\mu_c}.$$

Further properties of $\mathcal{G}$ can be found in [PT95]. □

## B.1 Kondratiev Spaces of Stochastic Test Functions

Hida test functions are not the unique notion of *s.q.* test functions. They belong to the wider class of spaces of test functions. Here we present the Kondratiev spaces of test functions $(\mathcal{N})^\rho$ with $\rho \in [0,1]$, useful for a wider class of measures but more limited for many purposes. For instance, *Example* 2.7 cannot be extended to these spaces.

**Definition B.1. (*Entire functions of growth* $k$)** *We denote $\mathcal{E}^k(\mathcal{H}_{-p,-q})$ as the space of all entire functions with domain $\mathcal{H}_{-p,-q}$ and growth $k \in [0,2]$. By entire functions, we refer to holomorphic functions as defined in **Definition** 2.8 at every point of the domain. $\mathcal{E}^k(\mathcal{H}_{-p,-q})$ is a linear space with norm*

$$n_{p,q,k}(\Psi) = \sup_{\phi \in \mathcal{H}_{-p,-q}} |\Psi(\phi)| \exp(-|\phi|^k_{-p,-q}). \tag{B.4}$$

*The space of entire functions of minimal type of growth $k$ over $\mathcal{N}'_{\mathbb{C}}$ is*

$$\mathcal{E}^k_{min}(\mathcal{N}'_{\mathbb{C}}) = pr \lim_{p,q \in \mathbb{N}} \mathcal{E}^k(\mathcal{H}_{-p,-q}). \tag{B.5}$$

We define the chain of Kondratiev spaces denoted $(\mathcal{N})^\rho$ for $0 \leq \rho \leq 1$. The space $(\mathcal{N})^1$ is directly defined as $(\mathcal{N})^1 = \mathcal{E}^1_{min}(\mathcal{N}')$, while the others $(\mathcal{N})^\rho$ are defined by means of an auxiliary Gaussian measure $\mu$. Notice that, in general, using (2.6), $S_\mu[\Psi] : \mathcal{N}_{\mathbb{C}} \to \mathbb{C}$ has a different domain than $\Psi$. To deal with this feature, we introduce the translation instead.

**Definition B.2. (*$C_\mu$ transform*)** *For a Gaussian measure $\mu$ of covariance $\Delta^{xy}$ and mean zero, we define the $C_\mu$ transform as the action of the translation measure parametrized by $\sigma^x$. For a given $\Psi : \mathcal{N}' \to \mathbb{C}$, it takes the form*

$$C_\mu[\Psi](\sigma^x) = \int_{\mathcal{N}'} D\mu(\varphi^x)\Psi(\varphi^x + \bar{\sigma}^x) \tag{B.6}$$

*whenever it converges.*

Both $S_\mu$ and $C_\mu$ coincide in the shared domain using the Gel'fand triple of the Cameron-Martin Hilbert space $\mathcal{H}_\Delta$. The $C_\mu$ transform is suitable for functions $\Psi$ whose domain can be translated by the very large space $\mathcal{N}'_{\mathbb{C}}$ without affecting integrability. This is a necessary condition if we want $\Psi \in L^2(\mathcal{N}'_{\mathbb{C}}, D\nu)$ for a large class of Gaussian measures $\nu$, which is a desirable property for every notion of test function. To finally define the spaces of stochastic test functions $(\mathcal{N})^\rho$, we must choose an auxiliary Gaussian measure $\mu$ to use as a reference. This definition does not really depend on $\mu$ but only on the topology of $\mathcal{N}$ as long as the covariance considered is continuous with respect to it. The common choice in white noise analysis is to choose the so-called white noise measure $\beta$ of (2.69). With this choice, we define the spaces of stochastic Kondratiev test functions as

$$(\mathcal{N})^\rho = C_\beta^{-1}\Big(\mathcal{E}^{\frac{2}{1+\rho}}_{min}(\mathcal{N}'_{\mathbb{C}})\Big). \tag{B.7}$$

Notice that, for the particular case $\rho = 0$, we are describing the hypothesis of Theorem B.1. Then $(\mathcal{N})^0 = (\mathcal{N})$ is the space of Hida test functions. We list here some properties of our interest; for detailed proofs, we refer to [Wes03] and references therein. For $\rho \in [0,1]$:

- $(\mathcal{N})^1 \subset (\mathcal{N})^{\rho_2} \subset (\mathcal{N})^{\rho_1} \subset (\mathcal{N})$ for $0 < \rho_1 < \rho_2 < 1$.

- $(\mathcal{N})^\rho$ are NF spaces.

- $(\mathcal{N})^\rho$ are algebras with the pointwise product.

- Every element $\Psi \in (\mathcal{N})^\rho$ is a strongly continuous function $\Psi : \mathcal{N}'_{\mathbb{C}} \to \mathbb{C}$.

- Every element $\Psi \in (\mathcal{N})^\rho$ is infinitely often Fréchet differentiable.

# Appendix C

# Integration on infinite dimensional spaces: Motivation and technical problems

Our goal in this appendix is to understand what problems arise in functional integration theories. Our way to define measures in infinite dimensions is through projective limits. That is, we will define a measure by projecting the spaces into finite-dimensional measure spaces and study the consistency of the procedure. As we will see, we will start by trying to define this measure over a Hilbert space and show that no such measure exists. After that, we will investigate which TVS is suited for that matter. We will mainly follow [Eld16, Mos19], although there are many sources where the interested reader can find further information. We mention [GJ87, Kuo96, Car&97, Oec12b, GV16, Sam20].

To illustrate the problems that arise on the road to infinite dimensions, let us look at the first definition of integral that we have at hand.

## C.1 From the counting measure over $\mathbb{N}$ to the Lebesgue measure over $\mathbb{R}$

The most natural integral that we can imagine is the integration of functions $(a) : \mathbb{N} \to \mathbb{R}$, which is

$$\int_{\mathbb{N}} a_n d\nu_c(n) := \sum_{n \in \mathbb{N}} a_n \tag{C.1}$$

where $\nu_c$ is the counting measure such that

$$\text{For each } B \subseteq \mathbb{N} \ \ \nu_c(B) = |B| \tag{C.2}$$

where $|B|$ is the cardinality of the set $B$. We would like to generalize this construction to a larger class of functions. The natural choice is to integrate functions

$$f : \mathbb{R} \to \mathbb{R} \tag{C.3}$$

For the sake of simplicity, let us restrict our discussion to the interval $[0,1] \subset \mathbb{R}$. The naive choice of the generalization would be to use again the counting measure

$$\sum_{x \in [0,1]} f(x) \tag{C.4}$$

but this expression cannot be defined in concise mathematical terms. For instance, if we take $f = 1$, the result is the cardinality of the set $[0,1]$. In order to define a well-posed integral, Riemann based its definition of integral on the natural counting measure of $\mathbb{N}$, introducing a lattice $\{x_i^N\}_{i=0}^N \subset [0,1]$ with $x_{i+1} - x_i = \frac{1}{N}$ and then defining

$$\int_0^1 f(x)dx := \lim_{N \to \infty} \frac{1}{N} \sum_{i=0}^N f(x_i^N). \tag{C.5}$$

This definition keeps the cardinality of $\mathbb{R}$ under control by introducing $1/N$ weights in each summand and has proved to be of utmost importance in the history of science. With this definition, the subsets $(a,b)$ of the interval $[0,1]$ are measured using the characteristic function $1_{(a,b)}$, defining the measure $\nu$ as the length

$$\nu((a,b)) = \int_{\mathbb{R}} 1_{(a,b)} = b - a \tag{C.6}$$

This Riemannian construction depends on the notion of lattice $\{x_i^N\}_{i=0}^N$ that heavily exploits the properties of $\mathbb{R}$. Since we are interested in integration in more complicated spaces, we need the Lebesgue notion of measure that can be defined over abstract sets in the following way. Let us focus on the Lebesgue integral over $\mathbb{R}$. In that case, we need three ingredients $(\mathbb{R}, \mathcal{B}(\mathbb{R}), \nu)$ in which

(a) $\mathbb{R}$ is intended as the set of real numbers, forgetting any additional structure other than its topology. In this construction, we forget about the algebraic properties that we used in the Riemannian case for the definition of the lattice.

(b) $\mathcal{B}(\mathbb{R})$ is a $\sigma$-algebra and represents the measurable subsets of $\mathbb{R}$. In general, a $\sigma$-algebra $\Sigma$ is a subset of the power set $\Sigma \subseteq \mathcal{P}(X)$ of another base set $X$ that fulfills

   (i) $X \in \Sigma$

   (ii) if $A \in \Sigma$ then also its complement $X \backslash A \in \Sigma$

   (iii) if $\{A_k\}_{k=0}^\infty \subset \Sigma$ then also its countable union $\cup_{k=0}^\infty A_k \in \Sigma$

   In our example, $\mathcal{B}(\mathbb{R})$ is the Borel $\sigma$-algebra generated by the open subsets of $\mathbb{R}$.

(c) $\nu : \mathcal{B}(\mathbb{R}) \to \mathbb{R}$ is the measure and fulfills

(i) $\nu(E) \geq 0 \ \forall E \in \mathcal{B}(\mathbb{R})$

(ii) $\nu(\emptyset) = 0$

(iii) ($\sigma$-additivity) $\nu\left(\bigcup_{k=1}^{\infty} E_k\right) = \sum_{k=1}^{\infty} \nu(E_k)$ for all countable collections $\{E_k\}_{k=1}^{\infty}$ of pairwise disjoint sets of $\mathcal{B}(\mathbb{R})$.

All the ingredients of Lebesgue measure theory are encoded in that structure. We simply need to substitute in the definitions above $\mathbb{R}$ and $\mathcal{B}(\mathbb{R})$ for any space and $\sigma$-algebra to get the general case.

The Lebesgue integral is built upon characteristic functions $1_B$ with $B \in \mathcal{B}(\mathbb{R})$. By definition, it postulates

$$\nu(B) = \int_{\mathbb{R}} 1_B d\nu \tag{C.7}$$

and extends this construction to any measurable function. An important aspect of this construction, which we will encounter in the infinite-dimensional case, is the following

**Remark C.1.  *Why is this measure not defined over $\mathbb{Q}$?***

*Let $[a, b]_{\mathbb{Q}} \subset \mathbb{Q}$ be the equivalent of an interval over $\mathbb{Q}$ and assume that the measure of this interval is its length*

$$\nu_{\mathbb{Q}}([a, b]_{\mathbb{Q}}) = b - a \tag{C.8}$$

*The fact that $\mathbb{Q}$ is countable implies that its Borel $\sigma$-algebra is the whole power set $\mathcal{P}(\mathbb{Q})$. Also, $\nu_{\mathbb{Q}}(\{a\}) = 0$, then we have*

$$\nu_{\mathbb{Q}}\left(\bigcup_{a \in [0,1]_{\mathbb{Q}}} \{a\}\right) = \nu_{\mathbb{Q}}([0, 1]_{\mathbb{Q}}) = 1 \text{ and } \sum_{a \in [0,1]_{\mathbb{Q}}} \nu_{\mathbb{Q}}(\{a\}) = 0. \tag{C.9}$$

*Since $[0, 1]_{\mathbb{Q}}$ is countable, $\sigma$-additivity (iii) does not hold. As a consequence, to define a proper measure, we should complete the space to $\mathbb{R}$. Furthermore, $\nu(\mathbb{Q}) = 0$ for the Lebesgue measure. As a final comment, we point out that $\sigma$-additivity is a requirement for the application of Fubini's theorem. Therefore, breaking this condition leads to pathological behavior.*

## C.2  From the Lebesgue measure over $\mathbb{R}^n$ to probability measures on infinite dimensional spaces.

Mimicking the discussion of the previous section, our starting point now is the Lebesgue measure over $\mathbb{R}^n$

$$\int_{\mathbb{R}^n} dx^n f(x) \tag{C.10}$$

and our goal is to extend the Lebesgue measure to an infinite-dimensional TVS. As a first attempt, we may use the notion of orthogonality of the separable Hilbert space $\mathcal{H}_\Delta$ to identify it with $\mathbb{R}^\infty$ and investigate its use as a domain of integration. The scalar product of $\mathcal{H}_\Delta$ will be provided by the bilinear $\Delta^{\mathbf{xy}}$ as in the rest of the thesis. In this appendix, to keep the discussion general, we will denote it simply by $\langle, \rangle$. Let

$$g : \mathcal{H}_\Delta \to \mathbb{R}. \tag{C.11}$$

The naive analogue to (C.4) is

$$\int_{\mathcal{H}_\Delta} dx^\infty g(x) := \int_{\mathcal{H}_\Delta} \mathcal{D}x \; g(x). \tag{C.12}$$

Feynman's approach to make sense out of this expression is similar to a projective limit. We consider a sequence of orthogonal projectors $\{P_n\}$ such that $\dim(P_n(\mathcal{H}_\Delta)) < \dim(P_{n+1}(\mathcal{H}_\Delta)) < \infty$ and $\lim_{n \to \infty} P_n(\mathcal{H}_\Delta) = \mathcal{H}_\Delta$. For simplicity, consider $\dim(P_n(\mathcal{H}_\Delta)) = n$, then

$$\int_{\mathcal{H}_\Delta} \mathcal{D}x \; g(x) = \lim_{N \to \infty} \int_{P_N(\mathcal{H}_\Delta)} dx^N g(x)|_{P_N(\mathcal{H}_\Delta)} = \lim_{N \to \infty} \left[ \prod_{k=1}^N \int_\mathbb{R} dx_k \right] g(x_1, \cdots, x_N) \tag{C.13}$$

As we did in the previous example with intervals, we can investigate the measure of an open ball $B(0, 1) = \{h \in \mathcal{H}_\Delta \mid \|h\|_\Delta \leq 1\}$. Then, as shown in [Eld16], we obtain that

$$\int_{B(0,1)} \mathcal{D}x = \infty \tag{C.14}$$

for every translational invariant measure like $\mathcal{D}x$. This means that if we want to compute an integral in this manner, every open subset will have infinite measure. To avoid this problem, as with the Riemann integral, we will try to weight every one-dimensional integral in (C.13).

In more concise terms, let us generate a sequence of projectors $\{Q_i\}_{i=1}^\infty$ defined as above, but in this case, the range of $Q_n(\mathcal{H}_\Delta)$ is spanned by the orthonormal family $e_1, \cdots, e_n \in \mathcal{H}_\Delta$. That is, every projector labeled by $n + 1$ in the sequence just enlarges the basis with a new element $e_{n+1}$ at a time. The $\sigma$-algebra in this case will be induced by $A \in \mathcal{B}(\mathbb{R}^n)$ via the cylinder sets $C_{A,n}$ defined in

$$C_{A,n} = Q_n^{-1}(A) = \left\{ h \in \mathcal{H}_\Delta \; \middle| \; \big(\langle h, e_1 \rangle, \cdots, \langle h, e_n \rangle\big) \in A \subseteq \mathbb{R}^n \right\} \tag{C.15}$$

Those sets generate the whole Borel $\sigma$-algebra $\mathcal{B}(H)$ [Eld16]. In this way, we will focus on them to define our measure[1]. Using (C.13), the characteristic functions $1_{C_{n,A}}$

---

[1] This argument requires a separable Hilbert space to ensure that $\mathcal{B}(\mathcal{H}_\Delta)$ can be generated with the weak topology.

$$\int_{Q_n \mathcal{H}_\Delta} \mathcal{D}x \; 1_{C_{n,A}} = \int_{A \subset \mathbb{R}^n} dx^n = \nu(A) \tag{C.16}$$

$$\int_{Q_m \mathcal{H}_\Delta} \mathcal{D}x \; 1_{C_{n,A}} = \int_{A \subset \mathbb{R}^n} dx^n \int_{\mathbb{R}^{m-n}} dx^{m-n} = \nu(A)\nu(\mathbb{R}^{m-n}) \; \text{ for } m > n \tag{C.17}$$

Thus, again we see that $\int_H \mathcal{D}x \; 1_{C_{n,A}} = \infty$. The source of the infinity in this integral is that $\nu(\mathbb{R}^{m-n}) = \nu(\mathbb{R})^{m-n}$. As a consequence, we should weight this limit with another sequence of measures $\mu_n$ on $\mathbb{R}^n$ such that we have

$$\mu_{n+1}(A \times \mathbb{R}) = \mu_n(A)\mu_1(\mathbb{R}).$$

Thus, in order to have a non-zero finite measure, we must ensure

$$0 < \lim_{N \to \infty} \mu_N(\mathbb{R}^N) = \lim_{N \to \infty} \mu_1(\mathbb{R})^N < \infty \tag{C.18}$$

Therefore, our only choice is $\mu_1(\mathbb{R}) = 1$, i.e., we must choose a probability measure[2].

## C.2.1 Gaussian measures

From now on, we will particularize the discussion to the case in which $\mu_1 = \mu$ denotes a one-dimensional Gaussian measure.

**Definition C.1.** *(**Gaussian measure over** $\mathbb{R}$)*
*A Borel probability measure $\mu$ over $\mathbb{R}$ is Gaussian with variance $\sigma^2$ if*

$$\mu(I) = \int_I \frac{1}{\sqrt{2\pi}\sigma} e^{-x^2/2\sigma^2} dx \tag{C.19}$$

*for $I \in \mathcal{B}(\mathbb{R})$. The case $\sigma^2 = 0$ will also be included, and it represents $d\mu(x) = \delta(x)dx$ in physical notation. An extremely useful tool to take into account all cases is the Fourier transform (characteristic function). It rephrases the information about the variance $\sigma^2$ in the following way:*

$$\hat{\mu}(\lambda) = \int_{\mathbb{R}} e^{i\lambda x} \mu(dx) = e^{-\sigma^2 \lambda^2/2} \tag{C.20}$$

*with $\lambda \in \mathbb{R}$. In this form, the case $\sigma^2 = 0$ is naturally included. In this definition, we had restricted ourselves to measures with mean 0. We can move its value to $y$ by considering:*

$$\mu_y(A) = \mu(A - y), \quad \text{then} \quad \hat{\mu}_y(\lambda) = e^{i\lambda y - \sigma^2 \lambda^2/2} \tag{C.21}$$

---

[2]The argument, in general, is slightly more complicated. Here $Q_j(\mathcal{H}_\Delta)$ is built from a basis in which each element represents independent random variables with respect to the measure $\nu_j$, and the factorization is possible.

At this point, a good candidate for Gaussian measure over $\mathcal{H}_\Delta$, denoted $\mu_H$, is defined over cylinder sets as

$$\int_{C_{A,n}} \mathcal{D}\mu_H(x) = \mu_H\big(\{h \in \mathcal{H}_\Delta \mid (\langle h, e_1 \rangle, \cdots, \langle h, e_n \rangle) \in A \subseteq \mathbb{R}^n\}\big) =$$
$$\lim_{N \to \infty} \int_{Q_N(C_{A,n})} d\mu_N(x) = \mu_n(A) \quad \text{(C.22)}$$

An important feature of this definition is that it does not depend on the sequence of projectors $Q_j$. Therefore, $\mu_H$ is a well-defined function over the cylinder sets; see [Eld16] for a detailed proof. In spite of the apparent success of this procedure, $\mu_H$ does not define a measure for an infinite-dimensional $H$.

**Remark C.2.** *There is no cylinder Gaussian probability measure over a Hilbert space that respects its scalar product*

*To see this, we show below that $\sigma$-additivity (iii) does not hold in this case as in Remark C.1.*

*In the first place, clearly $\mu_H(\mathcal{H}_\Delta) = 1$. We can consider an orthonormal basis $\{e_n\}_{n=1}^\infty$ of $\mathcal{H}_\Delta$ defining the sequence of cylinder sets*

$$A_{n,k} = \{x \in \mathcal{H}_\Delta : |\langle x, e_i \rangle| \le k, i = 1, \ldots, n\} \quad \text{(C.23)}$$

*Also, the open ball $B(0, k)$ of radius $k$ is such that $B(0, k) \subset A_{n,k}$ for any $n > 0$. As a consequence,*

$$\mathcal{H}_\Delta = \cup_{k=1}^\infty A_{n,k} \ \forall \ n > 0.$$

*For each set, we measure $\mu_H(A_{n,k}) = \mu_n([-k, k]^n) = \mu([-k, k])^n$. But, since $\mu$ is a Gaussian probability measure, we have $\mu([-k, k]) < 1 \ \forall k$. As a consequence, we can find $n_k$ for which $\mu([-k, k])^{n_k} < 2^{-k}$, thus*

$$\mu_H\left(\bigcup_{k=1}^\infty A_{n_k,k}\right) = \mu_H(\mathcal{H}_\Delta) = 1, \ \text{and} \ \sum_{k=1}^\infty \mu_H(A_{n_k,k}) < \sum_{k=1}^\infty \frac{1}{2^k} = 1 \quad \text{(C.24)}$$

*Thus, $\sigma$-additivity does not hold, and there is no hope to define a proper Gaussian measure over $\mathcal{H}_\Delta$ using its scalar product.*

The road to define a measure over $\mathcal{H}_\Delta$ relying only on its internal structure leads to an impossibility. However, the solution is again similar to the case of $\mathbb{Q}$. We must endow $\mathcal{H}_\Delta$ with a finer topology and complete it to $\mathcal{H}_\Delta \hookrightarrow W$. In general, there is an infinite family of Banach spaces $W$ that solve this problem. In those cases, the technical part lies in how to recover $\mathcal{H}_\Delta$ from $(W, \mu)$. This construction is usually developed to describe the abstract Wiener space that we will show below.

In order to complete the discussion above, notice that the characteristic function (C.1) allows for an extension to $\mathbb{R}^n$ in a convenient way to study the infinite-dimensional case.

**Definition C.2.** *(Gaussian measure over $\mathbb{R}^n$)*

*Let $\Delta$ be a symmetric and positive semidefinite $n \times n$ matrix. A Borel probability measure $\nu$ over $\mathbb{R}^n$ is Gaussian with $\Delta$ if*

$$\exp\left(-\frac{1}{2}\lambda \cdot \Delta\lambda\right) = \int_{\mathbb{R}^n} d\mu(x) e^{i\langle\lambda, x\rangle} \tag{C.25}$$

*For the translated measure $\nu_y(A) = \nu(A - y)$, the Radon-Nikodym derivative (the Lebesgue equivalent to the Jacobian) is*

$$\frac{d\mu_y}{d\mu}(x) = e^{-\frac{1}{2}\|x\|^2 + \langle x, y\rangle} \tag{C.26}$$

This characterization makes evident the one-to-one correspondence between symmetric positive semidefinite bilinears $\Delta : \mathbb{R}^n \times \mathbb{R}^n \to \mathbb{R}$ and Gaussian measures over $\mathbb{R}^n$. Moreover, the variable that enters in the bilinear is related to $x$ through a dual pairing $\langle \cdot, \cdot \rangle$, and therefore the Gaussian measure is defined over the dual space $(\mathbb{R}^n)^*$ even though it is irrelevant in finite dimensions. In the infinite-dimensional case, we exploit the characteristic function to define:

**Definition C.3.** *(Gaussian measure over TVS)* Let $W$ be a Topological Vector Space (TVS), then a Borel probability measure $\nu$ over $W$ is called Gaussian if $\forall \lambda_{\mathbf{x}} \in W'$

$$\exp\left(-\frac{1}{2}\Delta\left(\lambda_{\mathbf{x}}, \lambda_{\mathbf{x}}\right)\right) = \int_W D\mu(\varphi^{\phi^{\mathbf{x}}}) e^{i\langle\lambda_{\mathbf{x}}, \varphi^{\mathbf{x}}\rangle} \tag{C.27}$$

*Where $\Delta : W' \times W' \to \mathbb{R}$ is a symmetric positive semidefinite bilinear called covariance.*

In this case, not every $\Delta$ produces a Gaussian measure, as we already know from our attempt to construct the Gaussian measure over a Hilbert space $\mathcal{H}_\Delta$. From now on, it will be assumed that $\Delta^{\mathbf{xy}}$ is a non-degenerate bilinear, i.e., $\Delta(\lambda_{\mathbf{x}}, \lambda_{\mathbf{x}}) = \lambda_x \Delta^{xy} \lambda_y > 0$.

**Example C.1.** *(abstract Wiener space and the Cameron Martin Hilbert space)*

*The case where $W$ is a Banach space with Gaussian measure $\mu$ is called abstract Wiener space. With those considerations, three topologies must be considered in $W'$:*

1. *$(W', \|\cdot\|_{W'})$, since it is a Banach space with its own complete topology.*

2. *$(W', \Delta^{\mathbf{xy}})$, since $\Delta^{\mathbf{xy}}$ turns $W'$ into a pre-Hilbert space.*

3.  $(W', \langle \cdot, \cdot \rangle_{L^2(W, D\mu)})$, since it can be shown that $W' \subset L^2(W, D\mu)$ and therefore its scalar product $\langle \cdot, \cdot \rangle_{L^2(W, D\mu)}$ turns $W'$ into a pre-Hilbert space.

It can be shown [Eld16] that the topological completions

$$\overline{W'}^{(W', \Delta^{xy})} \cong \overline{W'}^{L^2(W, D\mu)} := \mathcal{H}_\Delta \tag{C.28}$$

where the equivalence is a metric isomorphism. In this equation, we define the Cameron Martin Hilbert space $\mathcal{H}_\Delta \subsetneq L^2(W, D\mu)$ that corresponds to the linear square integrable functions. Indeed, the scalar product of $\mathcal{H}_\Delta$ is provided by the bilinear $\Delta^{xy}$. We can see it using

$$\langle \zeta_x | \xi_x \rangle_{\mathcal{H}_\Delta} = \int_W D\mu(\varphi^x) \langle \zeta_x, \varphi^x \rangle \langle \xi_x, \varphi^x \rangle = \zeta_x \Delta^{xy} \xi_y, \ \forall \ \zeta_x, \xi_x \in W' \tag{C.29}$$

Also, the inclusion

$$(W', \|\cdot\|_{W'}) \hookrightarrow (\mathcal{H}_\Delta, \|\cdot\|_\Delta) \tag{C.30}$$

is a continuous dense embedding. Moreover, we can consider

$$(\mathcal{H}_\Delta, \|\cdot\|_\Delta) \hookrightarrow (W, \|\cdot\|_W) \tag{C.31}$$

which is a continuous dense embedding too, and we have that $W' \hookrightarrow \mathcal{H}'_\Delta$ is the adjoint of $\mathcal{H}_\Delta \hookrightarrow W$. Identifying $\mathcal{H}_\Delta$ with its dual $\mathcal{H}'_\Delta$, we generate the Gel'fand triple structure that we studied in *section 2.3*

$$W' \hookrightarrow \mathcal{H}_\Delta \hookrightarrow W \tag{C.32}$$

This construction justifies defining the integral over cylinder sets of $\mathcal{H}_\Delta$ or $W'$ and extending the result over $\mathcal{B}(W)$.[3]

Another way of understanding the Cameron-Martin space $\mathcal{H}_\Delta$ is the space in which (C.26) makes sense. For completeness, we state the following theorem:

**Theorem C.1.  (*Cameron Martin theorem for Banach spaces*)**  For $h^x \in \mathcal{H}_\Delta$, $\mu_h$ is absolutely continuous with respect to $\mu$, and

$$\frac{D\mu_h}{D\mu}(\varphi^x) = e^{-\frac{1}{2}\|h^x\|_\Delta^2 + \langle h^x, \varphi^x \rangle_\Delta}. \tag{C.33}$$

For $h^x \in W \backslash \mathcal{H}_\Delta$, $\mu_h$ and $\mu$ are mutually singular.

As a final remark, we notice that the Cameron Martin Hilbert space carries with it all the relevant information about the Gaussian measure. However, it is a set of measure zero $\mu(\mathcal{H}_\Delta) = 0$. This interesting feature turns the space $W$ into an auxiliary but important ingredient of the construction.

---

[3]Separability plays an important role here to identify the $\sigma$-algebras generated by weak and strong topologies of $W$.

### C.2.2 Gaussian measures on DNF spaces

The topology of the Banach space $W$ was not really discussed in the previous section. In this section, we seek the most natural TVS $W$ to complete the picture. We can try to construct a Gaussian measure over a Hilbert space $W = \mathcal{H}_n$, but the Cameron-Martin Hilbert space $\mathcal{H}_\Delta$ must be a dense subspace of $\mathcal{H}_n$, and their norms must satisfy $\|h\|_{\mathcal{H}_n} < C\|h\|_{\mathcal{H}_\Delta}$ in order to make the inclusion continuous. Furthermore, $\mu(\mathcal{H}) = 0$. As a result, the inclusion cannot be open, so their topologies do not coincide. It turns out that the only tool needed to find such a space [Eld16] is a Hilbert-Schmidt operator $A_n : \mathcal{H}_\Delta \to \mathcal{H}_n$ such that $\|h\|_n = \|Ah\|_\Delta$. This is an operator

$$\sum_{i=1}^{\infty} \|A_n e_i\|_\Delta^2 < \infty, \text{ for } \{e_i\} \text{ an orthonormal basis of } \mathcal{H}_\Delta \tag{C.34}$$

and $\mathcal{H}_\Delta \hookrightarrow \mathcal{H}_n$ is a dense embedding where the scalar product of $\mathcal{H}_n$ is given by

$$\langle f, g \rangle_{\mathcal{H}_n} = \sum_{i=1}^{\infty} \sum_{j=1}^{\infty} f_i g_j \langle A_n e_i, A_n e_j \rangle \tag{C.35}$$

while its dual is $\mathcal{H}_{-n}$:

$$\langle f, g \rangle_{\mathcal{H}_{-n}} = \sum_{i=1}^{\infty} \sum_{j=1}^{\infty} f_i g_j \langle A_n^{-1} e_i, A_n^{-1} e_j \rangle \tag{C.36}$$

If we consider the sequence of continuous dense inclusions of Hilbert-Schmidt type that define the Nuclear-Fréchet (NF) space $\mathcal{N}$ in **Definition** 2.2, we have

$$\mathcal{N} = \cap_{i=0}^{\infty} \mathcal{H}_{-i} \hookrightarrow \cdots \hookrightarrow \mathcal{H}_{-1} \hookrightarrow \mathcal{H}_\Delta \hookrightarrow \mathcal{H}_1 \hookrightarrow \cdots \hookrightarrow \mathcal{N}' = \cup_{i=0}^{\infty} \mathcal{H}_i \tag{C.37}$$

This result provides a general framework to construct measures over duals of nuclear spaces, which is the starting point of chapter 2. The next theorem is the generalization of the former discussion.

**Theorem C.2. (Bochner-Minlos theorem)**[*Hid80, GV16*] *Let $\mathcal{N}$ be a nuclear space and $C : N \to \mathbb{C}$ represent the characteristic function of a measure and therefore fulfill*

1. *$C$ is continuous,*

2. *$C$ is positive definite,*

3. *$C(0) = 0$.*

*Then there exists a unique (Radon) probability measure $\mu_C$ on $(\mathcal{N}', \mathcal{B}(\mathcal{N}'), \mu_C)$ such that*

$$\int_{\mathcal{N}'} e^{i\langle \varphi^{\mathbf{x}}, \xi_{\mathbf{x}} \rangle} D\mu_c(\varphi^{\mathbf{x}}) = C(\xi_{\mathbf{x}}). \tag{C.38}$$

Furthermore, in the chain (C.37) of Hilbert-Schmidt inclusions, $\mu_c(\mathcal{H}_n) = 0$.

# Appendix D

# The Wiener-Ito decomposition theorem

## D.1 Introduction

In this appendix, we will follow the presentation of [Hid&93] to provide a detailed analysis of the Wiener-Ito decomposition theorem. This theorem studies the structure of the Hilbert space

$$L^2\left(\mathcal{N}', D\mu\right) \tag{D.1}$$

where $D\mu$ is defined as the measure given through the characteristic functional (2.7) by the Bochner-Minlos theorem. To do so, we must notice that all the information needed to explain this space is encoded into the symmetric and positive definite bilinear $\Delta^{\mathbf{xy}}$, which is a well-defined object:

$$\Delta^{\mathbf{xy}} : \mathcal{N}' \times \mathcal{N}' \to \mathbb{R}. \tag{D.2}$$

Of course, the most general case in which $\mathbb{C}$ is considered is straightforwardly generalized. This bilinear form turns $(\mathcal{N}, \Delta^{\mathbf{xy}})$ into a pre-Hilbert space whose completion will be denoted $\mathcal{H}_\Delta$. This Hilbert space is the Cameron-Martin Hilbert space of the Gaussian measure $\mu$ and encodes all the information necessary to understand (D.1).

The space $\mathcal{H}_\Delta$ can be obtained as follows. Let $\{e_{\mathbf{x}}^n\} \in \mathcal{N}$ be a complete orthogonal family with respect to $\Delta^{\mathbf{xy}}$. Using the projective definition of Appendix C, we have that the integral is projected into a one-dimensional integral with respect to the variable $y = e_x^n \varphi^x$ and, with $\|e_{\mathbf{x}}^n\|^2 = \mathbf{e}_x^n \mathbf{e}_y^n \Delta^{xy}$, it reduces to

$$\int_{\mathcal{N}'} D\mu(\varphi^{\mathbf{x}}) \mathbf{e}_x^n \varphi^x \mathbf{e}_y^m \varphi^y = \delta^{mn} \int_{\mathbb{R}} \frac{1}{\sqrt{2\pi \|e_{\mathbf{x}}^n\|^2}} \exp\left(-\frac{y^2}{2\|e_{\mathbf{x}}^n\|^2}\right) y^2 dy = \mathbf{e}_x^n \mathbf{e}_y^n \Delta^{xy}. \tag{D.3}$$

As a result, we express any element $\xi_{\mathbf{x}} \in \mathcal{H}_\Delta$ with a representative given by $\xi_{\mathbf{x}} = \xi_n \mathbf{e}_{\mathbf{x}}^n$, where summation is intended. In this way, we can integrate a priori meaningless functions such as $\xi_x \varphi^x$ and, in particular,

$$\langle \xi_{\mathbf{x}} | \zeta_{\mathbf{y}} \rangle = \int_{\mathcal{N}'} D\mu(\varphi^{\mathbf{x}}) \xi_x^* \varphi^x \zeta_y \varphi^y = \xi_n^* \zeta_m \int_{\mathcal{N}'} D\mu(\varphi^{\mathbf{x}}) \mathbf{e}_x^n \varphi^x \mathbf{e}_y^m \varphi^y =$$
$$\xi_n^* \zeta_m \mathbf{e}_x^n \mathbf{e}_y^m \Delta^{xy} = \xi_x^* \Delta^{xy} \zeta_y. \tag{D.4}$$

That is, we can obtain the scalar product of $\mathcal{H}_\Delta$ from the integral of the family of quadratic polynomials $\xi_x^* \varphi^x \zeta_y \varphi^y$, which only have meaning under the integral symbol.

Let $\Psi(\varphi^{\mathbf{x}}) \in L^2(\mathcal{N}')$ and consider $S_\mu[\Psi](\xi_{\mathbf{x}})$ given by (2.11). We can understand this transform as

$$S_\mu[\Psi](\xi_{\mathbf{x}}) = \int_{\mathcal{N}'} D\mu(\varphi^{\mathbf{x}})\Psi(\varphi^{\mathbf{x}}) \exp\left(\xi_x \varphi^x - \frac{1}{2}\xi_x \Delta^{xy}\xi_y\right) =$$
$$\int_{\mathcal{N}'} D\mu(\varphi^{\mathbf{x}})\Psi(\varphi^{\mathbf{x}} + \Delta^{\mathbf{x}y}\xi_y). \quad \text{(D.5)}$$

Now, expanding Taylor series around $\xi_{\mathbf{x}} = 0$ the exponential:

$$\exp\left(\xi_y \varphi^y - \frac{1}{2}\xi_y \Delta^{yz}\xi_z\right) = \sum_{k=0}^{\infty} \frac{1}{k!} : \varphi^k :|_\Delta^{x_1, \cdots, x_k} \xi_{x_1} \cdots \xi_{x_k}. \quad \text{(D.6)}$$

In this expression, $: \varphi^k :|_\Delta^{\vec{x}_k}$ is the $k$'th Wick power of (2.39).

This appendix will reinterpret $: \varphi^k :|_\Delta^{\vec{x}_k}$ as the generator of Hermite polynomials in orthogonal directions and extend (D.4) to nonlinear functions.

## D.2 The Bosonic Fock space

In order to prove the theorem, we must introduce Fock spaces. We define the Fock space $\mathcal{F}(\mathcal{H}_\Delta)$ as

$$\mathcal{F}(\mathcal{H}_\Delta) = \sum_{n=0}^{\infty} \mathcal{F}_n(\mathcal{H}_\Delta) = \sum_{n=0}^{\infty} \mathcal{H}_\Delta^{\otimes n} \quad \text{(D.7)}$$

In $\mathcal{F}(\mathcal{H}_\Delta)$, the scalar product $\Delta^{\mathbf{xy}}$ induces a natural scalar product that will be defined by

$$\mathcal{F}(\Delta^{\mathbf{xy}}) = \sum_{n=0}^{\infty} \mathcal{F}_n(\Delta^{\mathbf{xy}}), \qquad \text{for } \psi_{\vec{\mathbf{x}_n}}, \theta_{\vec{\mathbf{x}_n}} \in \mathcal{F}_n(\Delta^{\mathbf{xy}})$$
$$\mathcal{F}_n(\Delta^{\mathbf{xy}})(\psi_{\vec{\mathbf{x}_n}}, \theta_{\vec{\mathbf{x}_n}}) = \psi_{x_1, \cdots, x_n} \Delta^{x_1 y_1} \cdots \Delta^{x_n y_n} \theta_{y_1, \cdots, y_n} \quad \text{(D.8)}$$

An orthonormal basis of $\mathcal{F}(\mathcal{H}_\Delta)$ is easily obtained from an orthonormal basis $\left\{\mathbf{e}_x^i\right\}_{i \in \mathbb{N}_0} \subset \mathcal{N}$ of $\mathcal{H}_\Delta$. Notice that, in this section, we are asking for the condition $\mathbf{e}_x^i \Delta^{xy}\mathbf{e}_y^j = \delta^{ij}$ that we did not consider in the previous section. Using this, we have that

$$\left\{e_{\vec{\mathbf{x}}_n}^{\otimes i} := e_{\mathbf{x}_1}^{i_1} \otimes \cdots \otimes e_{\mathbf{x}_n}^{i_n}, \text{ s.t. } i \in \mathbb{N}_0^n\right\} \text{ is an orthonormal basis of } \mathcal{F}_n(\mathcal{H}_\Delta) \quad \text{(D.9)}$$

Also, the symmetric (Bosonic) Fock space of a Hilbert space $\mathcal{H}_\Delta$ is

$$\Gamma(\mathcal{H}_\Delta) = \sum_{d=0}^{\infty} \Gamma_d(\mathcal{H}_\Delta) = \sum_{d=0}^{\infty} \mathcal{H}_\Delta^{\hat{\otimes} d} \tag{D.10}$$

Here, $\hat{\otimes}$ is the symmetric tensor product, expanded below in (**??**), and the natural Hilbert space structure is induced by $\Delta^{\mathbf{xy}}$ on $\Gamma(\mathcal{H}_\Delta)$. Introducing the space of completely ordered tuples $(\alpha_1, \alpha_2, \cdots, \alpha_d) \in I^d \subset \mathbb{N}_0^d$ such that $\alpha_1 \leq \alpha_2 \leq \cdots \leq \alpha_d$, we define

$$\Gamma(\Delta^{\mathbf{xy}}) = \sum_{d=0}^{\infty} \Gamma_d(\Delta^{\mathbf{xy}}), \text{ for } \psi_{\vec{\mathbf{x_d}}}, \theta_{\vec{\mathbf{x_d}}} \in \Gamma_d(\Delta^{\mathbf{xy}})$$

$$\Gamma_d(\Delta^{\mathbf{xy}})(\psi_{\vec{\mathbf{x_d}}}, \theta_{\vec{\mathbf{x_d}}}) = \psi_{x_1, \cdots, x_d} \left( \sum_{\pi \in S_d} \Delta^{x_1 y_{\pi(1)}} \cdots \Delta^{x_d y_{\pi(d)}} \right) \theta_{y_1, \cdots, y_d}. \tag{D.11}$$

In this expression, $S_d$ denotes the group of permutations of $d$ elements. Thus, this construction is such that $\varphi_{x_1, \cdots, x_d}$ and $\varphi_{(x_1, \cdots, x_d)}$, the symmetrized version given by (2.15), enter equally in the scalar product. A natural orthogonal basis of this space is given by

$$\left\{ e_{\vec{\mathbf{x}}_d}^{\hat{\otimes}\alpha} := e_{\mathbf{x_1}}^{\alpha_1} \hat{\otimes} \cdots \hat{\otimes} e_{\mathbf{x_d}}^{\alpha_d}, \text{ s.t. } \alpha \in I^d \right\} \text{ is an orthogonal basis of } \Gamma_d(\mathcal{H}_\Delta). \tag{D.12}$$

The symmetric tensor product $\hat{\otimes}$ is defined as

$$e_{\mathbf{x_1}}^{\alpha_1} \hat{\otimes} \cdots \hat{\otimes} e_{\mathbf{x_d}}^{\alpha_d} = \frac{1}{d!} \sum_{\pi \in S_d} e_{\mathbf{x_1}}^{\alpha_{\pi(1)}} \otimes \cdots \otimes e_{\mathbf{x_d}}^{\alpha_{\pi(d)}}. \tag{D.13}$$

The basis presented above is orthogonal but not orthonormal for $\Gamma(\mathcal{H}_\Delta)$. This is because the scalar product (D.11) overcounts repeated indices in the labels. To construct an orthonormal basis, we should take into account this degeneracy. Thus, we introduce for each $i \in \mathbb{N}_0$ the function

$$\begin{aligned} n_i: \quad I^d \quad &\longrightarrow \quad \mathbb{N}_0 \\ \alpha \quad &\longmapsto \quad n_i(\alpha) = \# \left\{ \text{entries of } \alpha \text{ equal to } i \right\}. \end{aligned} \tag{D.14}$$

Let $n(\alpha)! = \prod_{i \in \mathbb{N}_0} n_i(\alpha)$. Then an orthonormal basis is provided by

$$\left\{ \frac{1}{\sqrt{n(\alpha)!}} e^{\alpha_1} \hat{\otimes} \cdots \hat{\otimes} e^{\alpha_d}, \text{ s.t. } \alpha \in I^d \right\} \text{ is an orthonormal basis of } \Gamma_d(\mathcal{H}_\Delta) \tag{D.15}$$

Two properties stem from these basis definitions:

1. The Bosonic Fock space can be seen as a linear subspace of the total Fock space $\Gamma(\mathcal{H}_\Delta) \subset \mathcal{F}(\mathcal{H}_\Delta)$.

2. The inclusion $\Gamma(\mathcal{H}_\Delta) \hookrightarrow \mathcal{F}(\mathcal{H}_\Delta)$ is not an isometry. In general, for

$$\Psi \in \Gamma_n(\mathcal{H}_\Delta) = \sum_{\alpha \in I^n} \Psi_\alpha^{(n)} e^{\alpha_1} \hat{\otimes} \cdots \hat{\otimes} e^{\alpha_n} \tag{D.16}$$

it is clear that

$$\sqrt{n!} \|\Psi\|_{\mathcal{F}} = \|\Psi\|_\Gamma. \tag{D.17}$$

## D.3  The Wiener-Ito decomposition

### D.3.1  Hermite Polynomials

Let's focus on the finite-dimensional case. In this context, we can consider a Gaussian measure $dG(x) = dx^d \sqrt{\frac{|K|}{(2\pi)^d}} \exp(-\frac{1}{2} x^i K_{ij} x^j)$. Let's study the decomposition of $\psi \in L^2(\mathbb{R}^d, dG)$ in the basis of Hermite polynomials. On a basis $\{v_m^i$ s.t. $i, m = 1, \cdots, d\}$ in which $v_m^i K_{ij} v_n^j = \delta_{mn}$, we write $x^i = y^m v_m^i$, then

$$dG(y) = \frac{dy^d}{\sqrt{(2\pi)^d}} e^{-\frac{1}{2} y^m \delta_{mn} y^n} \tag{D.18}$$

In those coordinates, any function can be expressed in terms of the basis of Hermite polynomials

$$H^n(t) = \frac{d^n}{dx^n} \exp(tx - \frac{1}{2} x^2) \Big|_{x=0}$$

$$\int_{\mathbb{R}} \frac{dt}{\sqrt{2\pi}} e^{-\frac{1}{2} t^2} H^n(t) H^m(t) = n! \delta^{mn}$$

$$H_0(t) = 1$$

$$H_1(t) = t$$

$$H_n(t) = t H_{n-1}(t) - (n-1) H_{n-2}(t). \tag{D.19}$$

Then we have

$$\psi(y) = \psi_{n_1, \cdots, n_d} H^{n_1}(y_1) \cdots H^{n_d}(y_d). \tag{D.20}$$

This means that the expression

$$\prod_{i=1}^d H^{k_i}(y_i), \text{ s.t. } k_i \in \mathbb{N}_0 \tag{D.21}$$

is an orthogonal basis that spans $L^2(\mathbb{R}^d, dG) = \mathcal{F}_d(L^2(\mathbb{R})) := L^2(\mathbb{R}) \otimes \overbrace{\cdots}^{d \text{ times}} \otimes L^2(\mathbb{R})$.

Now let us assume that $\psi(y) \in \Gamma_d\left(L^2(\mathbb{R})\right) := L^2(\mathbb{R})\hat{\otimes} \overbrace{\cdots}^{d \text{ times}} \hat{\otimes} L^2(\mathbb{R}) \subset L^2(\mathbb{R}^d, dG)$, the symmetrized space, then

$$\psi_{n_1,\cdots,n_d} = \psi_{(n_1,\cdots,n_d)} := \frac{1}{d!}\sum_{\pi \in S_d} \psi_{n_{\pi(1)},\cdots,n_{\pi(d)}}. \tag{D.22}$$

In this case, to avoid degeneracies, the components of $\psi_{n_1,\cdots,n_d}$ are expressed in a unique way if we choose them in the space of completely ordered tuples $(n_1,\cdots,n_d) \in \mathrm{I}^d$.

## D.3.2 The basis of Hermite polynomials in infinite dimensional spaces

We are looking for a basis of polynomials $P : \mathcal{N}' \to \mathbb{R}$ (or $\mathbb{C}$) which are dense in every space $L^2(\mathcal{N}', D\mu)$ as we discussed in [section 2.5](). A polynomial of this type looks like

$$P[\varphi^{\mathbf{x}}] = C^0 + C^1{}_{x_1}\varphi^{x_1} + \cdots + C^n{}_{x_1,\cdots,x_n}\varphi^{x_1}\cdots\varphi^{x_n} \tag{D.23}$$

with $C^i{}_{\vec{\mathbf{x}}_i} \in \Gamma_i(\mathcal{N}) := \mathcal{N}\hat{\otimes} \overbrace{\cdots}^{i \text{ times}} \hat{\otimes}\mathcal{N}$. Using a basis decomposition we can write

$$P[\varphi^{\mathbf{x}}] = \sum_{d=0}^{n} \sum_{\alpha \in I^d} P_\alpha e_{\vec{x}_d}^{\hat{\otimes}\alpha}(\varphi^d)^{\vec{x}_d} \tag{D.24}$$

In this basis, for the dense subset $\varphi^{\mathbf{x}} \in \mathcal{H}_\Delta \simeq \mathcal{H}'_\Delta$, we can approximate $\varphi^{\mathbf{x}} = \varphi^i \mathbf{e}_i^{\mathbf{x}}$ with $\varphi^i \in \mathbb{R}$ and

$$P[\varphi^{\mathbf{x}}] = \sum_{d=0}^{n} \sum_{\alpha \in I^d} P_{\alpha_1,\cdots,\alpha_d}(\varphi^d)^{(\alpha_1,\cdots,\alpha_d)} \tag{D.25}$$

which is a polynomial of $\mathbb{R}^d$ thus can be expressed in terms of Hermite polynomials using the basis ([D.21]()).

$$P[\varphi^{\mathbf{x}}] = \sum_{d=0}^{n} \sum_{\alpha \in I^d} P_{\alpha_1,\cdots,\alpha_d} \sum_{\vec{k}_d \in \mathbb{N}_0^d} B_{k_1,\cdots,k_d} H^{k_1}(\varphi^{\alpha_1})\cdots H^{k_d}(\varphi^{\alpha_d}) =$$

$$\sum_{d=0}^{n} \sum_{\alpha \in I^d} P_{\alpha_1,\cdots,\alpha_d} \sum_{\vec{k}_d \in \mathbb{N}_0^d} B_{k_1,\cdots,k_d} H^{k_1}(\varphi^{\alpha_1})\cdots H^{k_d}(\varphi^{\alpha_d}) \tag{D.26}$$

for some coefficients $B_{\vec{\mathbf{k}}}$. The argument that we followed shows that any polynomial restricted to $\mathcal{H}_\Delta$ can be expressed in the orthogonal basis

$$H^\alpha(\varphi^{\mathbf{x}}) = \prod_{k=0}^{\infty} H^{n_k(\alpha)}[\mathbf{e}_x^k \varphi^x], \text{ for } \alpha \in \mathbf{I}^n. \tag{D.27}$$

Recall $n_k$ from (D.14). These are the products of Hermite polynomials in orthogonal directions such that the total degree is $n$. This construction is, in principle, restricted to $\mathcal{H}_\Delta$. However, it is worth noticing that the Cameron-Martin Hilbert space is not the domain of integration; in fact, it has measure zero as shown in Appendix C. Nonetheless, the projective definition of the Gaussian measure Appendix C indicates that we can extend this basis to an orthogonal basis for the whole Hilbert space. This is, for any $L^2(\mathcal{N}', D\mu)$ we have

$$\Psi(\varphi^{\mathbf{x}}) = \sum_{n=0}^{\infty} \sum_{\alpha \in \mathbf{I}^n(\mathbb{N}_0^d)} \Psi_\alpha^{(n)} H^\alpha(\varphi^{\mathbf{x}}). \tag{D.28}$$

And using the projective definition of Appendix C we have

$$\int_{\mathcal{N}'} D\mu(\varphi^{\mathbf{x}}) H^\alpha(\varphi^{\mathbf{x}}) H^\beta(\varphi^{\mathbf{x}}) = \delta^{\alpha\beta} \prod_{k=0}^{\infty} \int_{\mathbb{R}} dx \frac{e^{-\frac{1}{2}x^2}}{\sqrt{2\pi}} \left( H^{n_k(\alpha)}(x) \right)^2 = n(\alpha)! \tag{D.29}$$

### D.3.3 The Chaos decomposition

If we project the Wick powers of (D.6) over a finite dimensional subspace with $\left\{ \mathbf{e}_{x_1}^{\alpha_1}, \cdots, \mathbf{e}_{x_k}^{\alpha_k} \right\}$, with $\alpha \in \mathbf{I}^k$, then by comparison with (D.19) we have

$$: \varphi^k :|_\Delta^{x_1, \cdots, x_k} \mathbf{e}_{x_1}^{\alpha_1} \cdots \mathbf{e}_{x_k}^{\alpha_k} = H_{\mathbf{e}}^\alpha(\varphi^{\mathbf{x}}). \tag{D.30}$$

By plugging (D.6) and (D.28) into (D.5)

$$S_\mu[\Psi](\xi_{\mathbf{x}}) = \sum_{m,n=0}^{\infty} \sum_{\alpha \in \mathbf{I}^n(\mathbb{N}_0^d)} \Psi_\alpha^{(n)} \int_{\mathcal{N}'} D\mu(\varphi^{\mathbf{x}}) \frac{1}{m!} : \varphi^m :|_\Delta^{x_1, \cdots, x_m} \xi_{x_1} \cdots \xi_{x_m} H_{\mathbf{e}}^\alpha(\varphi^{\mathbf{x}}). \tag{D.31}$$

Together with relations $\xi_x = \xi_i \mathbf{e}_x^i$ and (D.30), it is straightforward to see that

$$: \varphi^m :|_\Delta^{x_1, \cdots, x_m} \xi_{x_1} \cdots \xi_{x_m} =$$
$$\sum_{\beta \in (\mathbb{N}_0)^m} : \varphi^m :|_\Delta^{x_1, \cdots, x_m} \mathbf{e}_{x_1}^{\beta_1} \cdots \mathbf{e}_{x_m}^{\beta_m} \xi_{\beta_1} \cdots \xi_{\beta_m} =$$
$$\sum_{\alpha \in \mathbf{I}^m} \frac{m!}{\prod_{i=1}^{\infty} n_i(\alpha)!} H_{\mathbf{e}}^\alpha(\varphi^{\mathbf{x}}) \xi_{\alpha_1} \cdots \xi_{\alpha_m}. \tag{D.32}$$

Thus, by virtue of (D.29)

$$S_\mu[\Psi](\xi_{\mathbf{x}}) = \sum_{m=0}^{\infty} \sum_{\alpha \in \mathrm{I}^m} \Psi_\alpha^{(m)} \xi_{\alpha_1} \cdots \xi_{\alpha_m} =$$

$$\sum_{m=0}^{\infty} \xi_{x_1} \cdots \xi_{x_m} \Delta^{x_1 y_1} \cdots \Delta^{x_m y_m} \sum_{\alpha \in \mathrm{I}^m} \Psi_\alpha^{(m)} \mathbf{e}_{y_1}^{(\alpha_1} \cdots \mathbf{e}_{y_m}^{\alpha_m)}. \quad \text{(D.33)}$$

This decomposition is such that the coefficients

$$\Psi_{\mathbf{x_m}}^{(m)} = \sum_{\alpha \in \mathrm{I}^m} \Psi_\alpha^{(m)} \mathbf{e}_{\mathbf{x}_1}^{(\alpha_1} \cdots \mathbf{e}_{\mathbf{x}_m}^{\alpha_m)} \in \Gamma_m(H_\Delta). \quad \text{(D.34)}$$

And it is in one-to-one correspondence with (D.28). This is the Segal isomorphism of **Definition** 2.10 that states

$$L^2(\mathcal{N}', D\mu) \cong \Gamma(\mathcal{H}_\Delta). \quad \text{(D.35)}$$

This proves the Wiener-Ito chaos decomposition theorem that states that it can be written as

$$\Psi[\varphi^{\mathbf{x}}] = \sum_{n=0} \psi_{x_1,\cdots,x_n}^{(n)} : \varphi^n :|_\Delta^{x_1,\cdots,x_n}. \quad \text{(D.36)}$$

Finally, the decomposition also provides an isometry with the Bosonic Fock space introduced at the beginning of the appendix such that for another function

$$\Phi[\varphi^{\mathbf{x}}] = \sum_{n=0} \phi_{x_1,\cdots,x_n}^{(n)} : \varphi^n :|_\Delta^{x_1,\cdots,x_n}$$

we obtain that

$$\int_{\mathcal{N}} D\mu(\varphi^{\mathbf{x}}) \overline{\Phi}(\varphi^{\mathbf{x}}) \Psi(\varphi^{\mathbf{x}}) =$$

$$\sum_{n=0}^{\infty} n! \, \overline{\phi^{(n)}}_{\vec{u}_n} \left[\Delta^n\right]^{\vec{u}_n \vec{x}_n} \psi_{\vec{x}_n}^{(n)} = \sum_{n=0}^{\infty} n! \langle \phi_{\vec{\mathbf{x}}_n}^{(n)}, \psi_{\vec{\mathbf{x}}_n}^{(n)} \rangle_{\mathcal{H}_\Delta^n}. \quad \text{(D.37)}$$

From this result and (D.17), then (D.35) is an isometry too.

# List of Figures

# List of Tables

# List of Notational Remarks

# Alphabetical Index